\documentclass[aps,rmp,groupedaddress,11pt]{revtex4-2}
\usepackage{xspace}
\usepackage{enumerate}
\usepackage{csquotes}
\usepackage{graphicx}% Include figure files
\usepackage{subcaption}
\usepackage{tikz}
\usetikzlibrary{shapes}

\usepackage{amsmath, amsfonts, amssymb}
\usepackage{bbold,bm}
\usepackage{siunitx}
\usepackage{pifont} % http://ctan.org/pkg/pifont
\usepackage{dcolumn}% Align table columns on decimal point
\usepackage{array}
\usepackage{multirow}
\usepackage{arydshln}
\usepackage{makecell}
\usepackage{url}
\usepackage[colorlinks=true]{hyperref}

\newcommand{\mb}[1]{\mathbf{#1}}
\newcommand{\lra}[1]{\langle #1 \rangle }
\newcommand{\bds}[1]{\boldsymbol{#1}}
\newcommand{\mc}[1]{\mathcal{#1}}
\newcommand{\dd}[1]{\mathrm{d}#1}
\newcommand{\DD}[1]{\mathrm{D}#1}

\begin{document}

% Use the \preprint command to place your local institutional report
% number in the upper righthand corner of the title page in preprint mode.
% Multiple \preprint commands are allowed.
% Use the 'preprintnumbers' class option to override journal defaults
% to display numbers if necessary
%\preprint{}

\preprint{Proposal submitted to RMP}

%************************************************************************************************************************
\title{The dynamics of discrete particles in turbulent flows: open issues and current challenges in statistical modeling}
%************************************************************************************************************************

\author{Jean-Pierre Minier}
\email{jean-pierre.minier@edf.fr}
\affiliation{EDF R$\&$D, M\'{e}canique des Fluides, Energie et Environnement, 6 quai Watier, 
78400 Chatou, France}

\author{Christophe Henry} 
\email{christophe.henry@inria.fr}
\affiliation{Universit\'e C\^ote d'Azur, Inria, CNRS, Sophia-Antipolis, France}

\date{\today}

\begin{abstract}

This article is an invitation. It is, first, an invitation to consider as a subject worthy of attention the wide range of situations where small discrete elements, either bubbles, droplets or solid particles, are embedded in turbulent flows. Occurring often at a human scale and in our daily environments, these turbulent dispersed two-phase flows display complex behavior due to the interplay of two fundamental interactions, the fluid-particle and particle-particle interactions, compounded by the turbulence of the carrier flow. This is not a domain where the basic laws are unknown but where the huge number of degrees of freedom involved call for reduced, or coarse-grained, statistical descriptions to be developed. Since we are considering transport and collision phenomena or relaxation processes, it would seem that they can be handled by kinetic theory. In the general case of non-fully resolved turbulent flows, we are however dealing with particles influenced by random media with non-zero time and space correlations. The second invitation is therefore to recognize the limitations of kinetic-based descriptions and to address the challenges driving us to extend the classical framework, for fluid-particle as well as particle-particle interactions. Taking the standpoint provided by the modern formulation of stochastic processes and focusing on the description of the particle phase, this review proposes a step-by-step pedagogical presentation of current models while pointing out new directions and remaining uncharted territories. This is done to provide answers to the question `why?' as much as `how?' and to try to kindle interest into these open and fascinating issues.

\end{abstract}

% insert suggested keywords - APS authors don't need to do this
%\keywords{}

\maketitle

\newpage

\tableofcontents

\newpage

%===================================================================================================================
\section{The physics of dispersed turbulent two-phase flows \label{physics of 2phase}}
%===================================================================================================================

What is a dispersed two-phase flow and what are its main characteristics? To provide answers to these questions, it is best to let Nature do the talking through three examples among a wide range of applications. It is actually difficult to do justice to the richness and variety of applications involving dispersed two-phase flows and we have retained situations corresponding mainly to natural phenomena, leaving out a number of interesting industrial processes in which these flows play a key role. 

These examples are helpful to introduce the fundamental interactions (Sec.~\ref{a range of phenomena}), the governing equations (Sec.~\ref{the governing equation}), and the statistical issue which is the main theme of this review in Sec.~\ref{the issue to address in statistical physics}. The probabilistic framework is outlined in Sec.~\ref{statistical modeling} and from then on we follow the modeling road. The issues related to particle transport are analyzed in Sec.~\ref{statistical model transport} and we present current models as well as recent developments for the velocity of the fluid seen in Sec.~\ref{Modeling fluid seen}. Connections with soft matter are discussed in Sec.~\ref{differences soft matter}, before addressing particle collisions in Sec.~\ref{statistical model collision} and proposing some conclusions on the road ahead in Sec.~\ref{conclusion}.

\subsection{A rich tapestry of applications}
%===========================================

\subsubsection{Colloid suspensions and river deltas}\label{colloid suspensions and river deltas}
%---------------------------------------------------

\begin{figure}[h]
\centering
 \begin{subfigure}{0.4\textwidth}
  \centering
  \includegraphics[width=\textwidth]{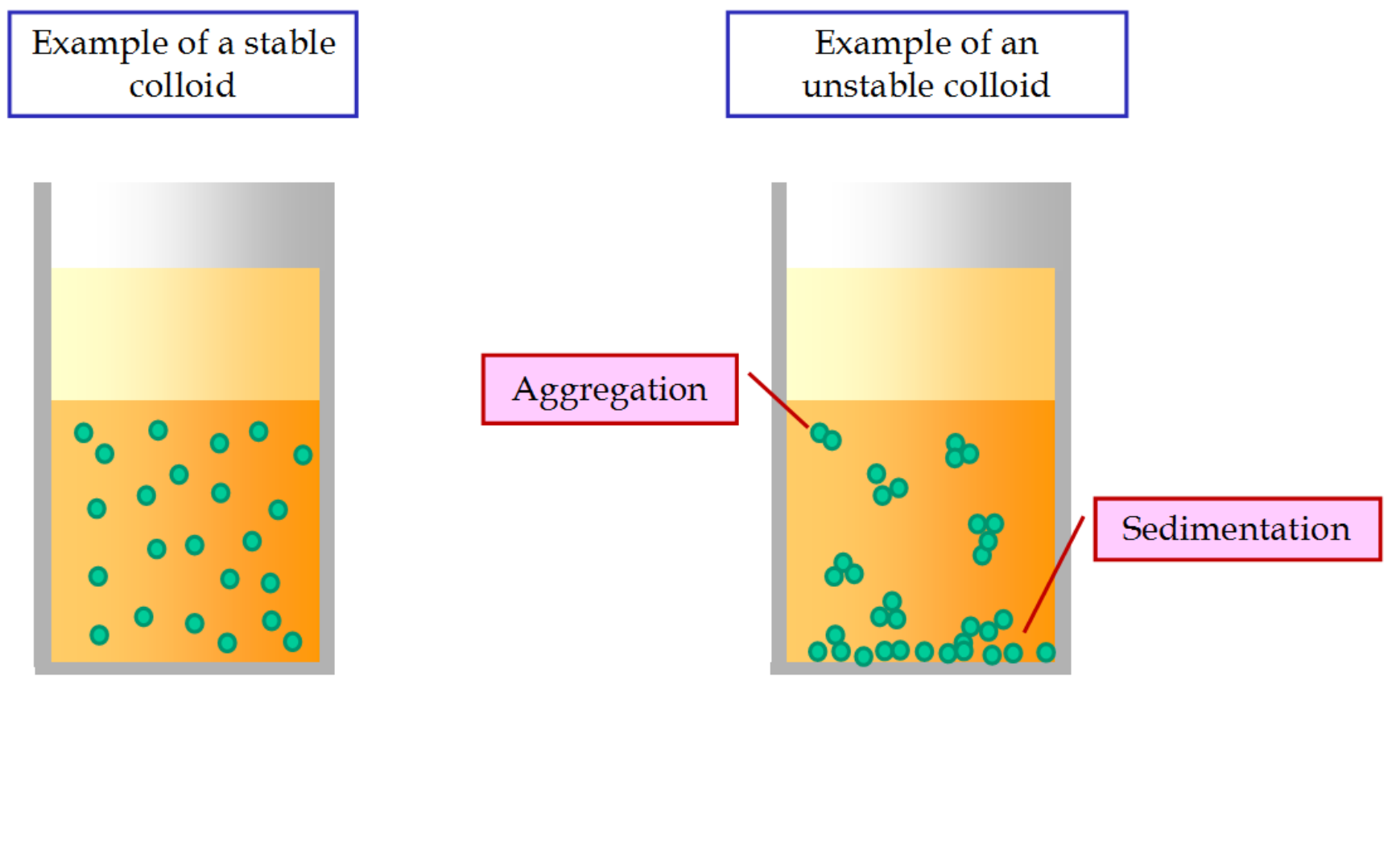}
  \caption{Stability of colloidal suspensions.}
  \label{fig: colloid stability}
 \end{subfigure}
 \hspace{2em}
 \begin{subfigure}{0.54\textwidth}
  \centering
  \includegraphics[width=\textwidth]{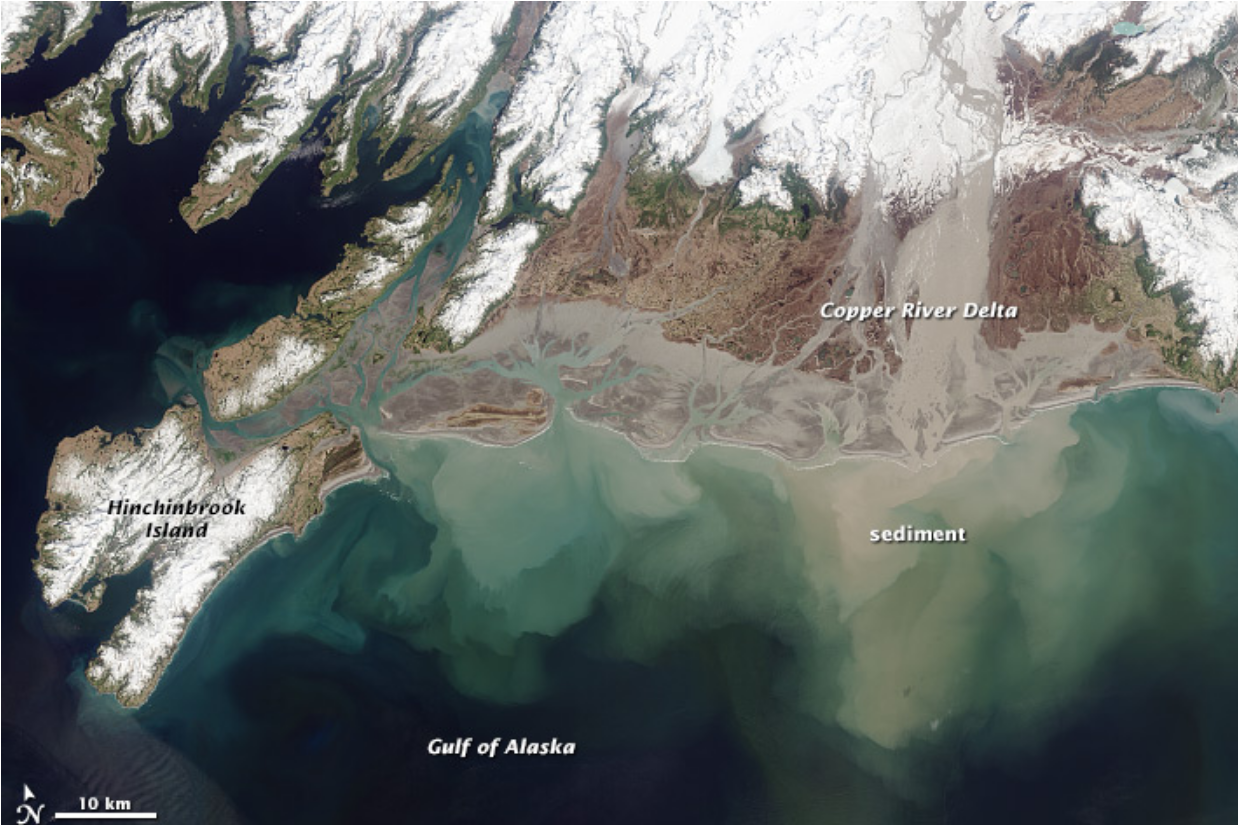}
  \caption{Dynamics of a river delta.}
  \label{fig: river delta}
 \end{subfigure}
\caption{The role of dispersed two-phase flows in the formation of river deltas: (a) The formation of aggregates leads to unstable colloidal suspensions (source: Wikipedia); (b) The resulting deposition forms delta beds, as for the Copper River in the Gulf of Alaska (source: \href{https://earthobservatory.nasa.gov/images/81784/copper-river-delta}{NASA Earth Observatory})).}
\label{fig: delta example}
\end{figure}

The first illustration concerns colloidal suspension stability (detailed accounts can easily be found in textbooks on colloids~\cite{elimelech2013particle,hunter2001foundations,israelachvili2011intermolecular}, among a vast literature on the subject). As sketched in Fig.~\ref{fig: colloid stability}, colloids are small particles sensitive to Brownian effects and which do not sediment (cf. Sec.~\ref{Brownian effects on particles} for a more precise definition). In a liquid solution, these colloids are also subject to forces acting between them. These interactions are well captured by the DLVO theory (see extensive descriptions in~\cite{israelachvili2011intermolecular,jones2002soft} on the formulation of the DLVO theory), which combines attractive forces (typically due to van der Waals forces) when colloids are very close one to another and repulsive ones (typically due to the repulsion between overlapping double layers, which result from an excess of counter-ions in the vicinity of a charged surface). The range of these repulsive forces corresponds to the Debye length, which is a function of the local ion concentration (through the ionic strength of the solvent). When the chemical conditions are such that the Debye length is large, colloids repel one another and the suspension is stable. When the chemical conditions are such that the Debye length is small, colloids can get close enough for the attractive forces to take over. In that case, aggregates begin to form until their size is too large for their weight to be balanced by Brownian effects and the suspension is unstable.

This is the much-simplified but general picture partly explaining the formation of river deltas, as illustrated in Fig.~\ref{fig: river delta} \cite{allen2017sediment}. When colloids are embedded in rivers with a small salt content (thus, a large Debye length), the suspension is stable and colloids are basically carried along by the river flow. When they reach more salty water from the ocean (thus, inducing a small Debye length), colloids start to agglomerate and deposit on the river beds forming river deltas over time. The rate at which this aggregation and sedimentation take place is governed by these chemical conditions but also by hydrodynamical ones, such as currents, recirculations, local turbulence effects or eddies created by the river delta geography. It is also interesting to note that this physical situation implies direct links between microscopic scales (what happens to a colloidal suspension) and macroscopic ones (the formation of river delta and river shores).

\subsubsection{Plume dispersion and bubbly flows}\label{plume dispersion and bubbly flows}
%------------------------------------------------

\begin{figure}[h]
\centering
\begin{subfigure}{0.41\textwidth}
  \includegraphics[width=\textwidth, trim=1cm 0cm 2cm 0cm, clip]{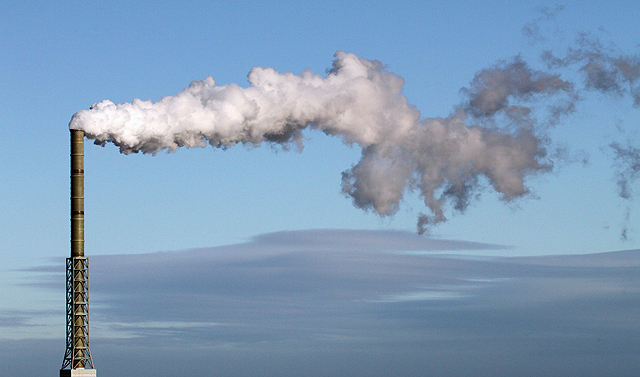}
	\caption{Smoke plume dispersing from an industrial chimney.}
	\label{fig: chimney plume}
\end{subfigure}
\hspace{1em}
\begin{subfigure}{0.335\textwidth}
  \includegraphics[width=\textwidth,trim=0cm 0cm 7cm 0cm, clip]{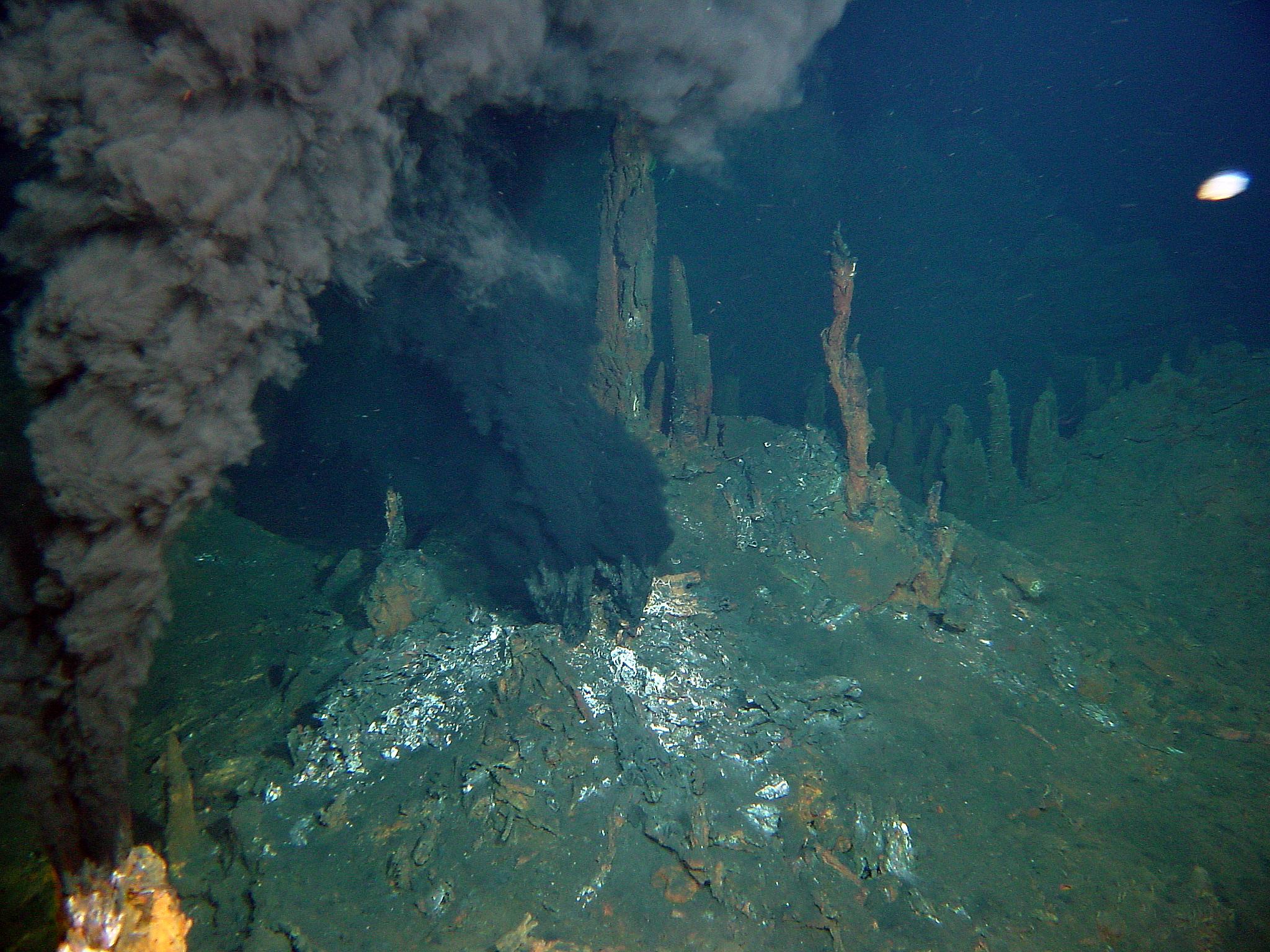}
	\caption{Submarine hydrothermal vent in the Rainbow site.}
	\label{fig: hydrothermal vent}
\end{subfigure}
\hspace{1em}
\begin{subfigure}{0.18\textwidth}
  \includegraphics[width=\textwidth]{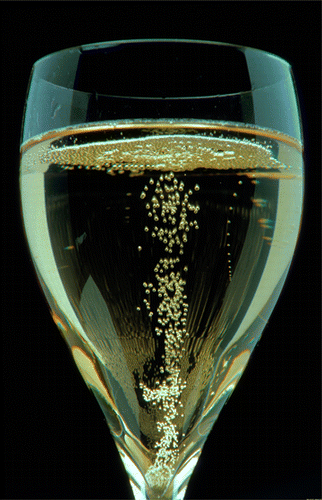}
	\caption{Rising bubbles in a glass.}
	\label{fig: champagne bubble}
\end{subfigure}
\caption{Illustrations showing the dispersion of two-phase flows in marine and atmospheric flows: (a) A plume of smoke from the Dunbar Cement Works chimney dispersing in the air (source: \href{https://www.geograph.org.uk/photo/3220320}{Walter Baxter, geograph.org.uk}); (b) Hydrothermal vent releasing a black smoke due to the precipitation of dissolved minerals in cold seawater (source: \href{https://image.ifremer.fr/data/00568/67987/}{Ifremer}); (c) Bubble of Champagne rising in a glass (Reprinted from \cite{liger2014many} with permission from American Chemical Society).}
\label{fig: plume example}
\end{figure}

For all its familiarity, plume dispersion is one of the best example to exhibit the effects of fluid turbulence. As seen in Fig.~\ref{fig: chimney plume}, a stream of gas material (which is often steam and not necessarily a noxious substance) is released out of a chimney and transported by the wind \cite{said2005experimental}. The plume is rapidly spreading due to the local turbulent fluctuations of the wind flow and the initially-compact form of the released cloud is progressively torn apart by the wind random motions so that it blends in the surrounding flow. Far less familiar, however, is another example of plume dispersion, namely hydrothermal vents through fissures on seabeds as shown in Fig.~\ref{fig: hydrothermal vent}. Discovered in the late 20th century in connection with tectonic plate theory, they are one of the hallmarks of the volcanic activity near mid-ocean ridges and thus related to the formation of oceanic crust (oceanic lithosphere) as well as to under-water life \cite{martin2008hydrothermal}. They involve superheated water, heated by hot rocks in the mantle, rushing through the sea floor. When this hot water mixes with cold sea water, minerals precipitate giving them various colors depending on the type of minerals (the name of black smokers illustrated here is when the plume is black).

At first sight, it is not obvious that we are dealing with a dispersed two-phase flow. However, it is not only possible but appropriate in such situations to describe the flow of released materials as a set of fluid-like individual elements. This corresponds to the important tracer-particle limit obtained by considering particles having vanishing inertia, but still buoyancy forces if there is a temperature difference with the surrounding fluid as in the case of hydrothermal vents. Another similar point source dispersion problem is illustrated in Fig.~\ref{fig: champagne bubble} where a stream of bubbles is rising in a liquid \cite{liger2014many}, clearly showing this time the dispersed two-phase flow nature of the flow. Note that, due to the adsorption of dissolved carbon dioxide, these bubbles grow as they rise from the bottom of the glass, which impacts their dynamics and rising velocities, but remain as individual bubbles up to the free surface.

\subsubsection{The physics of sand dunes}\label{the physics of sand dunes}
%------------------------

\begin{figure}[h]
\centering
\begin{subfigure}{0.47\textwidth}
  \includegraphics[width=\textwidth]{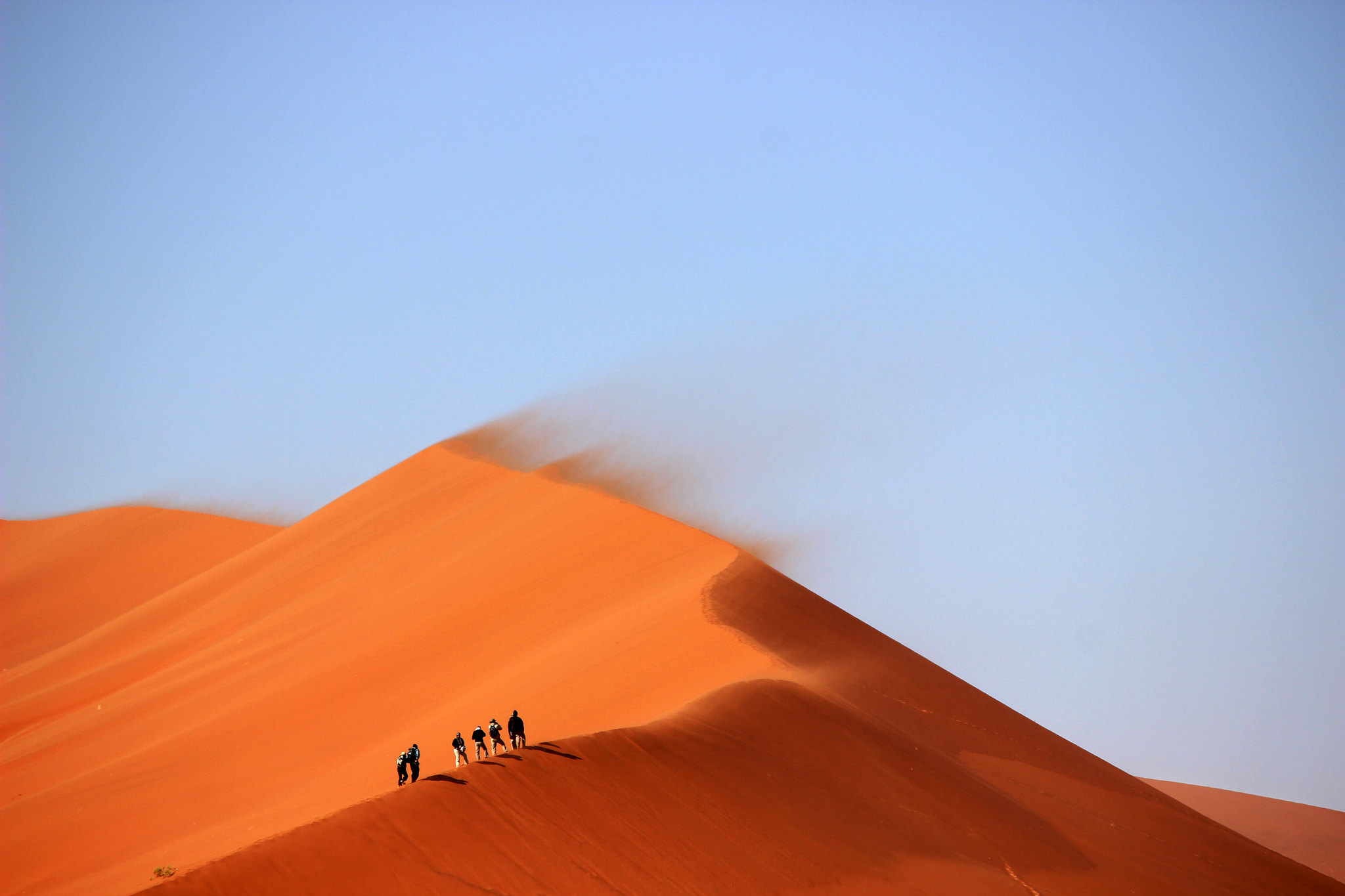}
	\caption{Sand being blown off dunes by the wind.}
	\label{fig: sand resuspension dunes}
\end{subfigure}
\hspace{2em}
\begin{subfigure}{0.47\textwidth}
  \includegraphics[width=\textwidth]{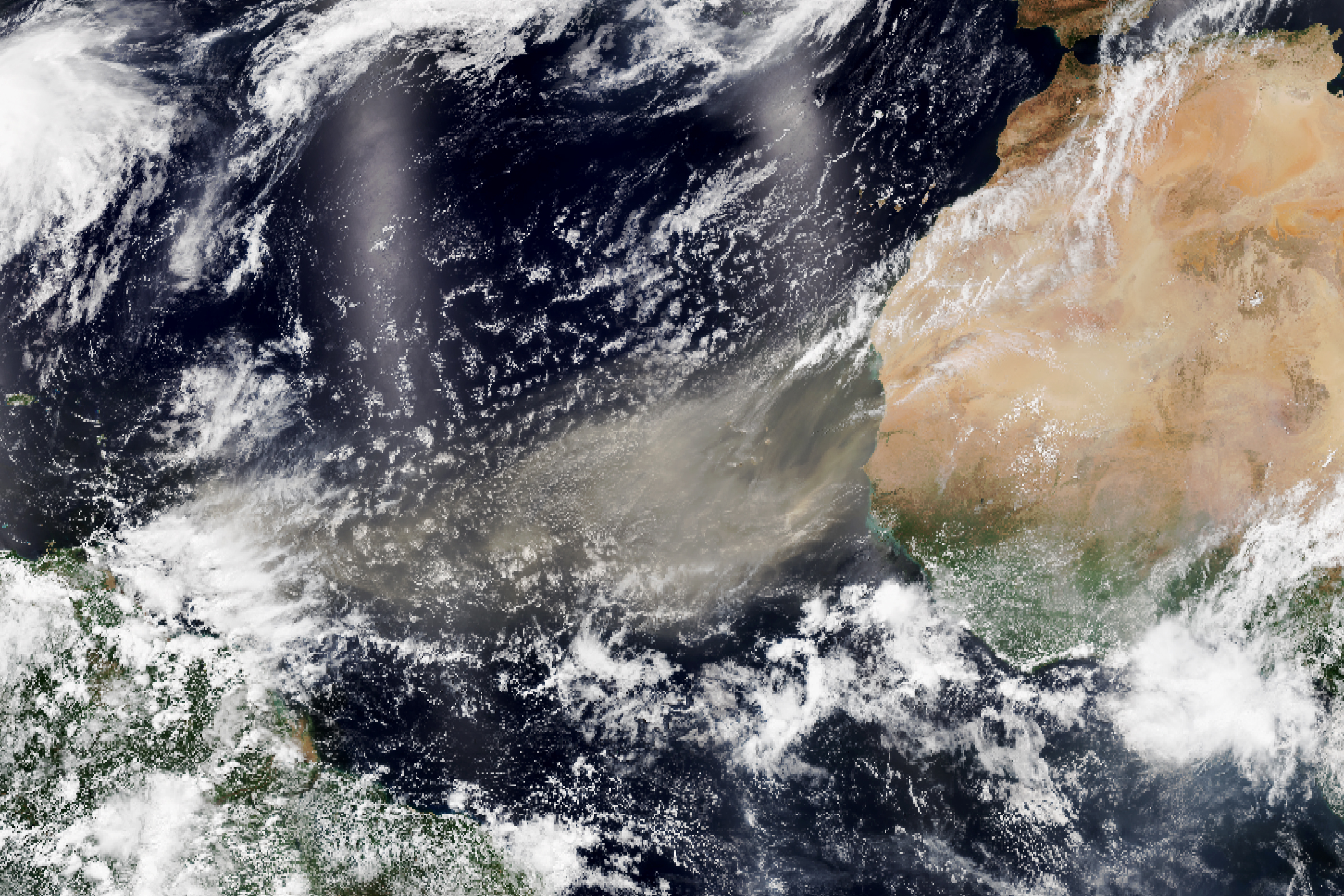}
	\caption{Sahara dust over the Atlantic Ocean.}
	\label{fig: sand over atlantic}
\end{subfigure}
\caption{Illustrations showing how tiny solid particles can be transported across large scales: (a) Picture of sand being blown off dunes (left, source: \href{https://www.flickr.com/photos/image-catalog/18799852813}{Flickr}); (b) Satellite observation of Sahara dust transported in the atmosphere above the Atlantic Ocean (right, source: \href{https://earthobservatory.nasa.gov/images/149918/a-burst-of-saharan-dust}{NASA Earth Observatory}).}
\label{fig: sand example}
\end{figure}

With the publication in 1941 of his book on desert dunes, reprinted in~\cite{bagnold2005physics}, R. A. Bagnold took a much-to-be-admired initiative and can be considered as the founding father of a new topic in physics, namely `the physics of blown sand'. This is a fascinating subject not only for its aesthetic appeal (cf. Fig.~\ref{fig: sand resuspension dunes}) but also for the diversity of the physical processes involved \cite{kok2012physics}. Under the actions of strong-enough winds, sand particles are blown off the edge of sand dunes, as manifested by the layers streaming away in Fig.~\ref{fig: sand resuspension dunes}. This corresponds to the process of particle resuspension by turbulent flows, which is of wide applicability but still contains puzzling issues~\cite{henry2023particle}. Contrary to colloids, sand particles have non-negligible inertia. Therefore, if the wind velocity diminishes or if they were entrained by sudden gusts in an intermittent wind regime, they are transported over some distance but can deposit again. They are then often referred to as saltating particles. When they come in contact with another dune, the kinetic energy imparted by the wind flows means that they hit the surface of previously-deposited layers of sand particles with some force and this can create `splashing effects', inducing a complex interplay between the two mirror processes of resuspension and deposition. This leads also to an ever-changing landscape of sand dunes whose formation is tightly coupled to hydrodynamical conditions. Indeed, when a sand dune is formed or when its slope is modified, there is a back-effect on the strength as well as the orientation of the wind velocity which, in turn, changes the way sand particles are being blown off, and so on.

Sand particles have different morphologies and sizes which can cover a range of values. We are therefore dealing with poly-disperse turbulent two-phase flows and the consequence of the difference in sand particle inertia is that their responses to wind solicitations vary greatly. Above the surface of desert dunes, sand particles interact and collective effects must be considered. On the other hand, once airborne, the tiniest sand particles can be carried over very large distances, in which case the main driving force is the wind currents that disperse them over regions at the scale of a continent or an ocean, as shown in Fig.~\ref{fig: sand example}.

\subsection{An intricate range of phenomena}\label{a range of phenomena}
%===========================================

As it transpires from these examples, dispersed two-phase flows involve two non-miscible thermodynamical phases with one phase, the dispersed one, present as a set of discrete elements (bubbles, droplets, solid particles) embedded in the other phase, the continuous one (also referred to as the carrier phase), which can be a gas or a liquid flow. Dispersed two-phase flows represent an important subclass within the general class of two-phase flows. Note that particle-laden flows are always in that category whereas the topology of the interface between two fluid phases in the dispersed flow regime implies that, for the same volume, the surface of contact is increased compared to the separated one. When bubbles or droplets are present, internal processes (such as recirculation, surface tension) can take place but we consider that, once formed, bubbles and droplets are treated as single entities and we are essentially concerned with their dynamics in turbulent flows. In that sense, apart from their different densities or similar properties, we do not distinguish between solid particles, droplets or bubbles, and refer to them as `discrete particles' or simply `particles'.

\subsubsection{The two fundamental interactions: particle-fluid and particle-particle}
%-------------------------------------------------------------------------------------

These examples are also useful to bring out that there are two fundamental interactions involved: the fluid-particle and the particle-particle interactions. The fluid-particle interaction is at play in particle transport as can be seen from colloids carried by rivers, fluid-like elements dispersing in the atmosphere, etc. In the vast majority of cases where flows are turbulent, we are therefore talking about turbulent convective transport. Yet, even when a fluid is at rest, Brownian motion can induce diffusive transport and we can describe both effects as fluid-induced transport. The particle-particle interaction is at play in the formation (or breakup) of colloid aggregates or when saltating particles hit layers of deposited sand particles. By particle-particle interaction, we are referring to short-range interactions, such as the forces occurring between colloids whose surfaces are separated by distances of the order of the Debye length, and not just to particles bouncing off one another. Nevertheless, compared to hydrodynamical scales, the Debye length is negligible and these particle-particle interactions can be regarded as near-contact ones. 

It follows that we are at the crossroads between several scientific domains since these fundamental interactions involve aspects pertaining to fluid mechanics (in particular, turbulence) but also to contact mechanics and to interface chemistry. A further remark is to note that, for dispersed two-phase flows, we are essentially dealing with a two-step mechanism. The first one is the transport step involved in particle dispersion but also responsible for bringing discrete particles nearby. The second step, which is the collision step (whether it is actual collisions, agglomeration, etc.), can only take place when particles are in the immediate vicinity of one another, as the result of the transport step. This successive mechanism is the reason why issues related to the transport step are given more emphasis in this review and also to pave the way for specific analysis of the open questions related to particle-particle interactions in turbulent flows.

\subsubsection{Addressing complexity}
%------------------------------------

Through these issues, we are touching upon the general question of `how to address complexity?' This is far too vast a subject to be discussed here in details. Nevertheless, this question lies in the background and is connected to some points discussed in later developments. For instance, it will be seen that the selection of the variables retained to characterize particles as mechanical systems is related to our choice to describe as `noise' or `disorder' some external effects. One of the main themes of this review is to emphasize that this is a key point if we are to devise complete and well-based descriptions so that these aspects should be carefully weighted. To put it in more philosophical terms, we should not be too quick in qualifying as noise some effects, since what we may call `chaos' is perhaps an order whose reading we have not yet learned. A second example concerns the modeling approach to follow when trying to model one variable which is displaying `complex behavior'. Should we retain the idea of devising a single model, which is then likely to become complex? Or should we consider that complexity is perhaps best captured by considering random alternation between several models, each of which being simple? This debate reappears in Sec.~\ref{Accounting for structures signature}.

\subsection{The governing equations}\label{the governing equation}
%===================================

A dispersed two-phase flow is a composite system in which the continuous phase is described in terms of fields whereas the dispersed phase is described in terms of particles. In the following, we provide the basic governing equations for the two phases, with more emphasis on the particle one since their statistical treatment is the main theme of the present review, as indicated in Sec.~\ref{the issue to address in statistical physics}.

\subsubsection{The Navier-Stokes equations for fluid flows}\label{The Navier-Stokes equations}
%----------------------------------------------------------

At the continuum level of description, the governing equations of the fluid phase are the continuity, the Navier-Stokes (NS) and the transport equations for a set of $N_s + 1$ scalar fields which gathers the $N_s$ relevant species mass fractions $\bds{\phi}_f(t,\mb{x})=(\phi_{f,l})_{l=1,\ldots, N_s}$ (if chemical reactions are involved) to which the fluid enthalphy (or another thermodynamical potential) is added, along with an equation of state for compressible flows. In the case of constant-property flows, these equations are~\cite{pope2000turbulent}
\begin{subequations}
\label{eq sec2: exact field Navier-Stokes eqs.}
\begin{align}
&\frac{\partial U_{{\rm f},k}}{\partial x_k}=0~, \label{fluid: exact field eqs. rho} \\
&\frac{\partial U_{{\rm f},i}}{\partial t} + U_{{\rm f},k}\, \frac{\partial U_{{\rm f},i}}{\partial x_k}
=-\frac{1}{\rho_{\rm f}}\frac{\partial P_{\rm f}}{\partial x_i} + \nu_{\rm f}\, \frac{\partial^2 U_{{\rm f},i}}{\partial x_k\partial x_k}~, 
\label{fluid: exact field eqs. U} \\
&\frac{\partial \phi_{{\rm f},l}}{\partial t} + U_{{\rm f},k}\, \frac{\partial \phi_{{\rm f},l}}{\partial x_k}=
\Gamma_{\rm f} \, \frac{\partial^2 \phi_{{\rm f},l}}{\partial x_k\partial x_k} + S_{{\rm f},l} \label{fluid: exact field eqs. phi}~,
\end{align}
\end{subequations}
where $\mb{U}_{\rm f}(t,\mb{x})$ and $P_{\rm f}(t,\mb{x})$ are the fluid velocity and pressure fields respectively, $\rho_{\rm f}$ the fluid density, $\nu_{\rm f}$ its kinematic viscosity and $\Gamma_{\rm f}$ the scalar diffusivity. In Eq.~\eqref{fluid: exact field eqs. phi}, the last term on the right-hand side (rhs) is a reactive source term $S_{{\rm f},l}=\hat{S}_{{\rm f},l}(\bds{\phi}_{\rm f}(t,\mb{x}))$ where $\hat{S}_{{\rm f},l}$ are known functions. Eqs.~\eqref{eq sec2: exact field Navier-Stokes eqs.} are the traditional field equations for fluid mechanics. They are generally obtained directly at the hydrodynamical level of description by writing balance equations for fluid mass, momentum and energy densities in which constitutive relations are introduced using the notion of local thermodynamic equilibrium~\cite{bird2002transport} (a much-recommended new presentation is given in~\cite{venerus2018modern}). They can also be derived from the kinetic theory (see~\cite[chapter 7]{ottinger2005beyond} on the subject, among several references), which is more in line with the theme of this review (cf. discussions in Sec.~\ref{differences soft matter} (in particular, in Sec.~\ref{sec: local and non-local closures}) as well as in Sec.~\ref{statistical model collision}). 

In the present work, we are not interested in reactive flows and we do not need to specify the expressions of reactive source terms. The important point is, however, that these terms are obtained from the instantaneous scalar variables $\bds{\phi}_{\rm f}$ through a (usually) non-linear but known relation. This is relevant when modeling reactive flows with similar source terms and the PDF approach to single-phase reactive flows has the key advantage of treating them without approximation (see extensive discussions in~\cite{pope1985pdf,haworth2010progress}). Note that the Navier-Stokes equation in Eq.~\eqref{fluid: exact field eqs. U} is written at locations $\mb{x}$ where no particles are present at time $t$. The equation for scalars, Eq.~\eqref{fluid: exact field eqs. phi}, is given for the sake of completeness and is only referred to (without reactive source terms) when discussing the diffusive limit for passive scalars in Sec.~\ref{sec: soft matter Brownian limit}. From now on, we limit ourselves to dynamical aspects of incompressible constant-property flows represented by Eqs.~\eqref{fluid: exact field eqs. rho}-\eqref{fluid: exact field eqs. U}.

\subsubsection{The equations of motion for particles}\label{hydrodynamical forces}
%----------------------------------------------------

To describe the particle phase, it is natural to adopt a Lagrangian point of view. This approach consists in tracking explicitly each particle by solving the evolution equations for the state vector $\mb{Z}_{\rm p}=(\mb{X}_{\rm p},\mb{U}_{\rm p},\bds{\Omega}_{\rm p})$ where $\mb{X}_{\rm p}$ is the location of the particle center of mass while $\mb{U}_{\rm p}$ and $\bds{\Omega}_{\rm p}$ are the translational and rotational velocities, respectively (more details on how to select the relevant variables entering the state vector are provided in Sec.~\ref{statistical modeling}). Applying the fundamental laws of classical mechanics, a general system of evolution equations for the particle state vector are the following ordinary differential equations (ODE):
\begin{subequations}
\label{eq sec2: particle transport}
\begin{align}
\frac{\dd \mb{X}_{\rm p}}{\dd t}&= \mb{U}_{\rm p}~, \label{particle transport a} \\
m_{\rm p}\frac{\dd \mb{U}_{\rm p}}{\dd t} &= \mb{F}_{\rm f \to p} + \mb{F}_{\rm p \to p} + \mb{F}_{\rm ext}~, \label{particle transport b} \\
I_{\rm p}\frac{\dd \bds{\Omega}_{\rm p}}{\dd t} &= \mb{M}_{\rm f \to p} + \mb{M}_{\rm p \to p}~, \label{particle transport c}
\end{align}
\end{subequations}
where $m_{\rm p}$ the mass of the particle and $I_{\rm p}$ its moment of inertia. In Eqs.~\eqref{particle transport b}-\eqref{particle transport c}, $\mb{F}_{\rm f \to p}$ and $\mb{M}_{\rm f \to p}$ represent the forces and torques exerted by the fluid on each particle, $\mb{F}_{\rm p \to p}$ and $\mb{M}_{\rm p \to p}$ the forces and torques due to particle-particle interactions and $\mb{F}_{\rm ext}$ forces due to external fields (such as gravity).

The derivation of approximate expressions for hydrodynamic forces acting on a particle embedded in a fluid flow has a long history, going back to the 19th century with the Basset-Boussinesq-Oseen (BBO) expression. However, in spite of numerous studies, it cannot be said that this issue has been completely solved, though a state-of-the-art formulation has emerged. Present formulations retain the drag, lift, added-mass and Basset forces \cite{clift1978bubbles, gatignol1983faxen, maxey1983equation, kuerten2016point, brandt2022particle} which are added to the gravity force, but the expressions of these forces vary according to different authors. The lift force is a special case since several expressions are proposed but each one obtained in a very specific or asymptotic situation. The result is a catalog of widely different formulations~\cite{clift1978bubbles, henry2014progress} so that there is, at the moment, no well-established consensus about a unique expression (the `optimal lift force' in McLaughlin's works~\cite{mcLaughlin1991inertial,wang1997role} is perhaps the most general expression available). Approximate expressions for Basset force are also still debated and, consequently, present models for these two forces cannot be regarded as being reliable enough. In the present work, we are more interested in a general and well-accepted form to develop probabilistic approaches and, for these reasons, the lift and Basset forces are left out. When lift forces are not considered, the particle rotation $\bds{\Omega}_p$ does not play an explicit role in the dynamics of spherical particles and can also be left out of the particle state vector which is then reduced to $\mb{Z}_{\rm p}=(\mb{X}_{\rm p},\mb{U}_{\rm p})$. Similarly, we do not consider electro- or thermo-phoresis phenomena which, if present, are accounted for by the addition of corresponding formulations in the particle momentum equation (see discussion in \cite[section 2.3]{minier2015lagrangian}). Nevertheless, it is important to realize that the framework described below is an open one and can easily be extended to account for these effects. Hence, present restrictions should not be seen as limitations but rather as an invitation to widen the range of applications by introducing future developments related to these additional physical phenomena.

For small spherical particles with diameters $d_{\rm p} \sim \eta_{\rm K}$, where $\eta_{\rm K}$ is the Kolmogorov length scale that represents the smallest length scale for fluid motions in a turbulent flow (to be defined in Sec.~\ref{The Kolmogorov theory}), the reference formulation of the particle momentum equation is derived in~\cite{gatignol1983faxen} and writes
\begin{multline}
\label{forces on particle}
m_{\rm p} \frac{\dd \mb{U}_{\rm p}}{\dd t} = \frac{\pi d_{\rm p}^3}{6}\rho_{\rm f} \frac{\DD \mb{U}^{\mc{V}_{\rm p}}_{\rm s}}{\DD t}
+ \frac{\pi d_{\rm p}^3}{6}( \rho_{\rm p} - \rho_{\rm f})\mb{g} \\
+ \frac{1}{2}\frac{\pi d_{\rm p}^2}{4} \rho_{\rm f} C_{\rm D}
\vert \, \mb{U}^{\mc{S}_{\rm p}}_{\rm s} - \mb{U}_{\rm p} \, \vert ( \mb{U}^{\mc{S}_{\rm p}}_{\rm s} - \mb{U}_{\rm p} ) 
+\frac{\pi d_p^3}{6} \rho_{\rm f} C_{\rm A}
     \left( \frac{\DD \mb{U}^{\mc{V}_{\rm p}}_{\rm s}}{\DD t} - \frac{\dd \mb{U}_{\rm p}}{\dd t}\right),
\end{multline}
where $\DD /\DD t$ refers to the derivative along a fluid particle trajectory
\begin{equation}
\frac{\DD}{\DD t}=\frac{\partial}{\partial t} + U_{{\rm f},k}\frac{\partial}{\partial x_k}.
\end{equation}
In Eq.~(\ref{forces on particle}), the first two terms on the rhs are the fluid acceleration and the buoyancy force (involving the difference between the particle density $\rho_{\rm p}$ and the fluid one), the third term is the general form of the drag force written with the drag coefficient $C_{\rm D}$ while the fourth term is the added-mass force written with a coefficient $C_{\rm A}$ (usually taken as $C_{\rm A}=1/2$). The added-mass force depends on the difference between the acceleration of the fluid seen (along its own trajectory) and the particle acceleration while the drag force is expressed in terms of the velocity slip $\mb{U}_{\rm s}^{\mc{S}_{\rm p}}-\mb{U}_{\rm p}$. The last two terms involve $\mb{U}^{\mc{S}_{\rm p}}_{\rm s}$ and $\mb{U}^{\mc{V}_{\rm p}}_{\rm s}$, which are the fluid velocities averaged over the surface $\mc{S}_{\rm p}$ and the volume $\mc{V}_{\rm p}$ of the particle, respectively, i.e.
\begin{align}
\mb{U}^{\mc{S}_{\rm p}}_{\rm s} & = \frac{1}{\mc{S}_{\rm p}} \int_{\mc{S}_{\rm p}} \mb{U}_{\rm f}(t, \mb{r})\, d\mb{r}~, \label{eq sec2: definition U^S}\\
\mb{U}^{\mc{V}_{\rm p}}_{\rm s} & = \frac{1}{\mc{V}_{\rm p}} \int_{\mc{V}_{\rm p}} \mb{U}_{\rm f}(t, \mb{r})\, d\mb{r}~. \label{eq sec2: definition U^V}
\end{align}
These velocities are expressed by a series expansion around the `velocity' at the particle center, which introduces the notion of the velocity of the fluid seen, $\widehat{\mb{U}}_{\rm s}(t)=\mb{U}^{\rm ud}_{\rm f}(t, \mb{X}_{\rm p}(t))$
\begin{align}
\mb{U}^{\mc{S}_{\rm p}}_{\rm s} & \simeq \widehat{\mb{U}}_{\rm s} + \frac{d_{\rm p}^2}{24}\left(\nabla^2 \mb{U}^{\rm ud}_{\rm f}\right)(t, \mb{X}_{\rm p}(t))~, \label{eq: Faxen correction terms S}\\
\mb{U}^{\mc{V}_{\rm p}}_{\rm s} & \simeq \widehat{\mb{U}}_{\rm s} + \frac{d_{\rm p}^2}{40}\left(\nabla^2 \mb{U}^{\rm ud}_{\rm f}\right)(t, \mb{X}_{\rm p}(t))~. \label{eq: Faxen correction terms V}
\end{align}
The notation $\mb{U}^{\rm ud}_{\rm f}(t, \mb{X}_{\rm p}(t))$ designates the `undisturbed velocity field' that would exist if the particle were not present at the location $\mb{x}=\mb{X}_{\rm p}(t)$ at time $t$~\cite{gatignol1983faxen, maxey1983equation} and the last terms added to $\widehat{\mb{U}}_{\rm s}(t)$ on the rhs of Eqs.~\eqref{eq: Faxen correction terms S} and~\eqref{eq: Faxen correction terms V} are the Faxen terms~\cite{gatignol1983faxen, maxey1983equation, clift1978bubbles}.

For small particles (say $d_p \leq \eta_{\rm K}$), the Faxen terms are small enough to be neglected which means that, for the expression of these hydrodynamical forces, particles are considered as mere points. This corresponds to the point-wise approximation for particles and a classical form of the particle momentum equation is
\begin{multline}
\label{forces on particle simple}
m_{\rm p} \frac{\dd \mb{U}_{\rm p}}{\dd t} = \frac{\pi d_{\rm p}^3}{6}\rho_{\rm f} \frac{\DD \mb{U}_{\rm s}}{\DD t}
+ \frac{\pi d_{\rm p}^3}{6} \left( \rho_{\rm p} - \rho_{\rm f} \right)\mb{g} \\
+ \frac{1}{2}\frac{\pi d_{\rm p}^2}{4} \rho_{\rm f} \, C_{\rm D}
\vert \, \mb{U}_{\rm s} - \mb{U}_{\rm p} \, \vert ( \mb{U}_{\rm s} - \mb{U}_{\rm p} )
+\frac{\pi d_{\rm p}^3}{6}\, C_{\rm A}\, \rho_{\rm f} \left( \frac{\DD \mb{U}_{\rm s}}{\DD t} - \frac{\dd \mb{U}_{\rm p}}{\dd t}\right)~.
\end{multline}
The velocity of the fluid seen is now given as the local instantaneous value of the fluid velocity at the same time and at the particle position, i.e. $\mb{U}_{\rm s}(t)=\mb{U}_{\rm f}(t, \mb{X}_{\rm p}(t))$, where $\mb{U}_{\rm f}(t,\mb{x})$ represents the fluid velocity field without having to distinguish undisturbed and disturbed fluid velocity fields anymore.
For particles heavier than the fluid $\rho_p \gg \rho_f$, it can be shown that the drag force is the dominant one~\cite{maxey1983equation,gatignol1983faxen,minier2001pdf} and the particle momentum equation is then further reduced to
\begin{equation}
\label{eq particle momentum ref drag force}
\frac{\dd \mb{U}_{\rm p}}{\dd t} = \frac{\mb{U}_{\rm s}-\mb{U}_{\rm p}}{\tau_{\rm p}}  + \mb{g}
\end{equation}
where the drag force is written so as to bring out the particle relaxation timescale $\tau_{\rm p}$ 
\begin{equation} 
\label{definition taup}
\tau_{\rm p}=\frac{\rho_{\rm p}}{\rho_{\rm f}}\frac{4\,d_{\rm p}}{3\,C_{\rm D} | \mb{U}_{\rm s}-\mb{U}_{\rm p} |}.
\end{equation}
This timescale, which is a measure of particle inertia, is the key notion for particle transport. More precisely, $\tau_{\rm p}$ represents the timescale needed for particle velocities to adjust to the local fluid velocity seen. In the Stokes regime, which is valid when ${\rm Re}_{\rm p} \leq 1$, with ${\rm Re}_{\rm p}=\vert \, \mb{U}_{\rm s}-\mb{U}_{\rm p} \, \vert\, d_{\rm p}/\nu_{\rm f}$ the particle Reynolds number, the drag coefficient is $C_{\rm D}=24/{\rm Re}_{\rm p}$. In that case, we retrieve the classical form $\mb{F}_{\rm f \to p}^{\rm Drag}=3\pi\rho_{\rm f}\, \nu_{\rm f} \, d_{\rm p} \left( \mb{U}_{\rm s}-\mb{U}_{\rm p} \right)$ and the particle relaxation timescale is given by the well-known expression
\begin{equation} 
\label{definition taup_Stokes}
\tau_{\rm p}=\frac{\rho_{\rm p}}{\rho_{\rm f}}\frac{d_{\rm p}^2}{18\nu_{\rm f}}.
\end{equation}
For general values of ${\rm Re}_{\rm p}$, the drag coefficient is obtained through empirical correlations and an often-retained formula is~\cite{clift1978bubbles,Brennen_2005}
\begin{equation}
\label{eq sec2: definition correlation taup} 
C_{\rm D}=
\begin{cases}
\dfrac{24}{{\rm Re}_{\rm p}}\left[\,1 + 0.15\, {\rm Re}_{\rm p}^{0.687}\, \right] & \text{if} \; {\rm Re}_{\rm p} \leq 1000,  \\
0.44 & \text{if} \; {\rm Re}_{\rm p} \geq 1000.
\end{cases}
\end{equation}
This correlation is for isolated particles or when particle concentration is not high enough for collective effects, representing hydrodynamical influences of each particle on its neighbors (the `wake effect'), to be significant. If needed, particle wake effects can be accounted for with modified correlations based, for example, on the local particle volumetric fraction \cite{clift1978bubbles}. 

In most cases involving particles or droplets in a turbulent flow, the preceding expressions provide a satisfactory picture and are sufficient to describe the influence of fluid flows on particle dynamics. For diameters larger than the Kolmogorov length scale, it is seen from the general form of the particle momentum equation, cf. Eq.~\eqref{forces on particle}, and the expressions given in Eqs.~\eqref{eq sec2: definition U^S}-\eqref{eq sec2: definition U^V} that a non-negligible particle size induces a filtering effect and that fluid velocity fluctuations with length scales smaller than $d_{\rm p}$ tend to be smoothed out and act as an underlying noise. Specific developments can then be considered (see for example interesting proposals in~\cite{elperin2002clustering,elperin2007clustering} among other ideas), but since we concentrate essentially on turbulent dispersion and collisional effects, the particle point-wise approximation is retained in the present review from now on (see further comments in Sec.~\ref{the issue to address in statistical physics}).

Even when added-mass forces are neglected, the acceleration of the fluid seen (i.e., $\DD \mb{U}_{\rm s}/\DD t$) is sometimes retained in the particle velocity equation which is then written as
\begin{equation}
\label{eq particle momentum ext drag pressure force zero CA}
\frac{\dd \mb{U}_{\rm p}}{\dd t} = -\frac{1}{\rho_{\rm p}}\nabla P_{\rm f} + \frac{\mb{U}_{\rm s}-\mb{U}_{\rm p}}{\tau_{\rm p}}  + \mb{g}~,
\end{equation}
where we have used the Euler form of the NS equations by neglecting viscosity to relate the fluid pressure gradient and acceleration as
\begin{equation}
\frac{\DD \mb{U}_{\rm s}}{\DD t}= - \frac{1}{\rho_{\rm f}}\nabla P_{\rm f} + \mb{g}~.
\end{equation}
Eq.~\eqref{eq particle momentum ext drag pressure force zero CA}, or variations of it when $C_{\rm A} \neq 0$, is useful for small sediments or bubbles, as discussed below. 

\paragraph{Brownian effects and particle collisions}
%---------------------------------------------------
\label{Brownian effects on particles}

So far, we have only considered forces arising from the continuum description of fluid flows. However, small enough particles are sensitive to Brownian effects due to the molecular nature of the fluid. This vague notion of `small enough' is quantified by introducing the criterion that, in the absence of a fluid flow, the particle settling velocity is counterbalanced by the random velocities imparted by collisions with the molecules of the fluid, so that these particles do not sediment, as the colloids discussed in Sec.~\ref{colloid suspensions and river deltas}. Using a one-dimensional formulation in the direction aligned with gravity and the equipartition theorem of statistical physics, the diameter $d_{\rm p}$ of these particles can be estimated by 
\begin{equation}
\sqrt{\frac{ k_{\rm B}\, \Theta_{\rm f} }{m_{\rm p}}} \sim \tau_{\rm p}\, g \Longrightarrow 
d_{\rm p} \simeq \left( \frac{\rho_{\rm f}^2 \nu_{\rm f}^2}{\rho_{\rm p}^3\, g^2}\, k_{\rm B}\, \Theta_{\rm f} \right)^{1/7}~,
\end{equation}
where $k_{\rm B}$ is the Boltzmann constant and $\Theta_f$ the fluid temperature. This defines more precisely colloidal particles which have therefore a diameter of the order of a few microns ($d_{\rm p} \leq 1-2 \mu m$) while, for $d_{\rm p} \geq 5-10 \mu m$, particles are called inertial and Brownian effects become negligible. In practice, it is best to include Brownian motions in the particle momentum equation whatever the particle diameter so that its effects diminish continuously when considering increasing particle diameters without having to introduce an artificial cut-off between colloidal and inertial particles. Although implementing Brownian effects is straightforward, it is connected to interesting questions and is addressed in more details in Sec.~\ref{sec: soft matter Brownian limit}. We also need to account for particle-particle interactions, typically collisions, in the particle momentum equation. This issue, which is more involved than Brownian effects, is given specific attention in Sec.~\ref{statistical model collision}. Moreover, Brownian motions are typically expressed by Wiener processes while the effects of random collisions are accounted for with Poisson processes. Since these two fundamental stochastic processes are only introduced in Sec.~\ref{statistical modeling}, it is best to address the issues related to Brownian effects and particle collisions in Secs.~\ref{sec: soft matter Brownian limit} and~\ref{statistical model collision}, respectively, after having detailed stochastic models related to turbulent dispersion in Sec.~\ref{Modeling fluid seen}.

\paragraph{Remarks on bubbly flows}

Compared to solid particles and droplets in a gas, the case of small bubbles in a liquid raises specific concerns and a few comments are in order. First, bubbles tend to have larger sizes (due to surface tension and the resulting inner pressure) which implies that the volume fraction occupied by the bubble phase becomes quickly appreciable. In practice, this volumetric back effect from bubbles to the liquid phase is important and the liquid-phase equations need to be modified accordingly (through some interface tracking methods). Second, even if the point-wise approximation is less valid for bubbles, we can still follow their center-of-mass motion. Yet, since bubble density is much smaller than the liquid one ($\rho_{\rm p} \ll \rho_{\rm f}$), we cannot neglect the added-mass force anymore. This leads to the bubble momentum equation
\begin{equation}
\label{eq particle momentum ext drag pressure force}
\frac{\dd \mb{U}_{\rm p}}{\dd t} = -\frac{1}{\rho_{\rm p}}\left( \frac{1 + C_{\rm A}}{1 + C_{\rm A} \rho_{\rm f}/\rho_{\rm p}}\right)\nabla P_{\rm f} + \frac{\mb{U}_{\rm s}-\mb{U}_{\rm p}}{\tau_{\rm p}}  + \mb{g}~.
\end{equation}
Assuming that we are in the Stokes regime, the relaxation timescale in Eq.~\eqref{eq particle momentum ext drag pressure force} is transformed to become
\begin{equation} 
\label{definition taup_AM}
\tau_{\rm p}=\left( \frac{\rho_{\rm p} + C_{\rm A} \rho_{\rm f}}{\rho_{\rm f}} \right)\frac{d_{\rm p}^2}{18\nu_{\rm f}}.
\end{equation}
As a consequence, we must know not only $\mb{U}_{\rm s}$ but also $\DD \mb{U}_{\rm s}/\DD t$. To slightly anticipate on developments to come, it will be seen, however, that models based on a two-step formulation (cf. Sec.~\ref{two-step models for Us}) generate the increments of $\mb{U}_{\rm s}$ (which corresponds to $\dd \mb{U}_{\rm s}/\dd t$, the acceleration along bubble trajectories) as well as the velocity increments along a fluid particle trajectory in the so-called Lagrangian step (which corresponds to $\DD \mb{U}_{\rm s}/\DD t$). This means that new models for $\mb{U}_{\rm s}$ are able to provide the information needed to account for the added-mass force and we can focus on what is at stake from a statistical point of view. 

\subsection{The issue to address in statistical physics} \label{the issue to address in statistical physics}
%=======================================================

With these clarifications, we consider discrete particle dynamics expressed by the system
\begin{subequations}
\label{eq: simplified particle equations}
\begin{align}
\frac{\dd \mb{X}_{\rm p}}{\dd t} & = \mb{U}_{\rm p}~, \label{eq: simplified particle equations a} \\
\frac{\dd \mb{U}_{\rm p}}{\dd t} & = \frac{\mb{U}_{\rm s}-\mb{U}_{\rm p}}{\tau_{\rm p}}  + \mb{g}~, \label{eq: simplified particle equations b}
\end{align}
\end{subequations}
where $\tau_{\rm p}$ is given by Eqs.~\eqref{definition taup} and~\eqref{eq sec2: definition correlation taup}. As explained for bubbles, we have in fact only slightly reduced the applicability of the dynamical formulation. However, by doing so, we have stripped down the problem to its bare statistical essential which is the treatment of the velocity of the fluid seen. Indeed, the advantage of Eq.~\eqref{eq: simplified particle equations b} is to bring out that the real issue is how to model the `driving force', $\mb{U}_{\rm s}(t)$, more than the various forms taken by the particle momentum equation depending on which forces are retained. This can be rephrased using an analogy where the particle momentum equation is regarded as a (non-linear) filter from which we derive statistics of interest which are outputs of the filter. However, the real issue is to model the input (or the driving force) of this filter, which is the velocity of the fluid seen $\mb{U}_{\rm s}(t)$. In the rare situations where the fluid flow is fully resolved and where we have access to the instantaneous velocity field $\mb{U}_{\rm f}(t,\mb{x})$, then $\mb{U}_{\rm s}(t)$ is exactly known and we are dealing with a fully-characterized deterministic time signal as the input of the particle-momentum filter, or a series of such time functions if we are considering a number of discrete particles. In that case, we can proceed with the particle kinetic variables $(\mb{X}_{\rm p},\mb{U}_{\rm p})$ as in classical mechanics. However, for the vast majority of turbulent flows the number of degrees of freedom is so high (cf. estimations in Sec.~\ref{The Kolmogorov theory}) that we have only access to a limited information, usually in the form of the first few moments of the fluid velocity field. The loss of information implied by such reduced, or coarse-grained, descriptions is then reflected by the representation of $\mb{U}_{\rm s}(t)$ as a stochastic process, which means that we enter the world of probabilistic formulations.

%========================================================================================================
\section{Reduced statistical descriptions and the probabilistic framework \label{statistical modeling}}
%========================================================================================================

There are basically two steps in dispersed two-phase flow modeling. The first one deals with the selection of the variables gathered in the particle state vector and with the construction of the stochastic processes used to model particle dynamics. This step encompasses most of the issues related to physics and answers the questions: How do we describe a mechanical system? How do we represent its dynamical evolution? The second step concerns the probabilistic framework guiding us from particle stochastic models to statistics of interest in practical situations. This step is more mathematical in nature  and answers the questions: What are the main stochastic processes? How do we handle them to stay clear of mathematical pitfalls?

To concentrate on the physical issues in the following sections while relying on a safe probabilistic framework, we start with a mere outline of the mathematical background. This is precisely what is not suggested to newcomers in this subject who are encouraged to study in-depth the mathematical aspects of stochastic processes to `learn their trade'. For that purpose, a highly-recommended textbook is~\cite{ottinger2012stochastic} which does a very good job at presenting stochastic processes and, in particular, the interest of formulations in terms of trajectories instead of considering only PDF equations. It is telling that half of this book on polymer models is dedicated to the mathematical properties of stochastic processes.

\subsection{From microscopic to macroscopic descriptions}
%========================================================

\subsubsection{Mass density functions and mean fields}
%-----------------------------------------------------

The PDF approach to single-phase and dispersed two-phase turbulent flows has reached a mature level and detailed accounts can be found in~\cite{pope1985pdf,pope2000turbulent,minier2001pdf,haworth2010progress,minier2015lagrangian,minier2016statistical}. The formalism bears similarities with the classical formulation in terms of distribution functions but there are also differences. First, the normalization is in terms of the mass of the fluid or particle phases instead of the number of particles. Second, this PDF framework grew out of the probabilistic description of turbulent single-phase flows where the `microscopic level of description' (i.e., the Navier-Stokes equations) is made up by fields and, for dispersed two-phase flows, a combination of particles and fields. This is in contrast with classical statistical physics where the microscopic level involves always discrete particles (typically atoms/molecules) whereas the macroscopic one corresponds to fields (the hydrodynamical level). Consequently, a specific distinction is made between Lagrangian and Eulerian PDFs to avoid confusions between descriptions where the position is a variable (Lagrangian) or a parameter (Eulerian).

The methodology can be developed for general $N$-particle PDF descriptions and state vectors (whose selections are addressed later on), but to keep simple notations we consider essentially the one-particle PDF approach. This allows to handle a large number of particles as samples instead of a large number of pairs of particles for two-particle PDF models, etc. Adopting therefore a Lagrangian standpoint and focusing on the particle phase, we introduce the decomposition of the particle state vector $\mb{Z}=(\mb{X}_{\rm p}, \mb{Z}_{\rm c})$ where the particle location is always present and where $\mb{Z}_{\rm c}$ is the complementary part of the state vector. In sample space, the corresponding variables are noted $\mb{z}=(\mb{y}_{\rm p}, \mb{z}_{\rm c})$. Starting from the unconditional Lagrangian PDF $p^{\rm L}(t; \mb{y}_{\rm p}, \mb{z}_{\rm c})$ and using $M_{\rm p}$ for the total mass of the discrete particles in the domain, the Lagrangian and Eulerian mass density functions (MDFs) for the disperse phase are defined by
\begin{subequations}
\label{eq sec4: general definitions of L and E MDFs}
\begin{align}
F_{\rm p}^{\rm L}(t;\mb{y}_{\rm p}, \mb{z}_{\rm c}) &= M_{\rm p}\, p^{\rm L}(t; \mb{y}_{\rm p}, \mb{z}_{\rm c}) ,\\
F_{\rm p}^E(t,\mb{x}; \mb{z}_{\rm c}) &= F_{\rm p}^L(t;\mb{y}_{\rm p}=\mb{x}, \mb{z}_{\rm c})
                             = \int F_{\rm p}^{\rm L}(t;\mb{y}_{\rm p}, \mb{z}_{\rm c})\delta (\mb{y}_{\rm p}-\mb{x})\, \dd \mb{y}_{\rm p}~.
\end{align} 
\end{subequations}
The key relation between Eulerian and Lagrangian descriptions is~\cite{Balescu_1997,minier2001pdf}
\begin{equation}
F_{\rm p}^E(t,\mb{x}; \mb{z}_{\rm c}) = \int p^{\rm L}(t; \mb{x}, \mb{z}_{\rm c} \, \vert \, t_0; \mb{x}_0, \mb{z}_{{\rm c},0} )\, F_{\rm p}^E(t_0,\mb{x}_0; \mb{z}_{{\rm c},0}) \dd \mb{x}_0 \, \dd \mb{z}_{{\rm c},0}~,
\end{equation}
which shows that the Lagrangian transition PDF is the propagator of the information, here the Eulerian MDF from the initial state $(t_0; \mb{x}_0, \mb{z}_{{\rm c},0})$ to the present one $(t; \mb{x}, \mb{z}_{\rm c})$. This central property explains why most of the physics is contained in modeling the Lagrangian transition PDF. Once the Eulerian MDF is defined, particle mean field properties are extracted. For a particle variable written as $H_{\rm p}(t;\mb{Z}_{\rm c})$, its average $\lra{H_{\rm p}}$ (which is a field variable), is defined as
\begin{equation}
\alpha_{\rm p}(t,\mb{x}) \, \rho_{\rm p} \lra{H_{\rm p}}(t,\mb{x})= \int H_{\rm p}(t; \mb{z}_{\rm c}) F_{\rm p}^{\rm E}(t,\mb{x}; \mb{z}_{\rm c})\,\dd \mb{z}_{\rm c}~,
\end{equation}
where $\alpha_{\rm p}(t,\mb{x})$ is the mean particle volumetric fraction. In a discrete sense, with $N_{\rm p}$ stochastic particles, the definitions of Lagrangian and Eulerian MDFs become 
\begin{align}
F_{{\rm p},N_{\rm p}}^{\rm L}(t\,;\,\mb{y}_{\rm p}, \mb{z}_{\rm c}) &=\sum_{i=1}^{N_{\rm p}} m_{\rm p}^{[i]}\delta (\mb{y}_{\rm p}-\mb{x}_{\rm p}^{[i]}) \,\delta (\mb{z}_{\rm c}-\mb{Z}_{\rm c}^{[i]}) \label{eq sec4: discrete LMDF} \\
F_{{\rm p},N_{\rm p}}^{\rm E}(t,\mb{x}\,; \,\mb{z}_{\rm c}) &= F_{{\rm p},N_{\rm p}}^{\rm L}(t\,;\, \mb{y}_{\rm p}=\mb{x}, \mb{z}_{\rm c}) \label{eq sec4: discrete EMDF}
\end{align} 
where $m_{\rm p}^{[i]}$ is the mass of the particle labeled $[i]$. This shows that in a small volume around a point $\mb{x}$ where averages are obtained from Monte Carlo estimations, that is as the ensemble averages over the $N_{\rm p}^{(\mb{x})}$ particles present in that volume, we get the equivalent of Favre, or mass-weighted, averages
\begin{equation} 
\label{eq: mass weighted RANS mean}
  \lra{H_{\rm p}}(t, \mb{x}) \simeq \lra{H_{\rm p}}_{N_{\rm p}} = \frac{\sum_{i=1}^{N_{\rm p}^{(\mb{x})}} m_{\rm p}^{[i]} 
  H_{\rm p}(t; \mb{z}_{\rm c}^{[i]}(t))}{\sum_{i=1}^{N_{\rm p}^{(\mb{x})}} m_{\rm p}^{[i]}}~.
\end{equation}

\subsubsection{A unified approach to statistical operators}
%----------------------------------------------------------

In Eq.~\eqref{eq: mass weighted RANS mean}, we retrieved the traditional ensemble averaging applied in the Reynolds-averaged Navier-Stokes (RANS) equations in turbulence modeling, using $F_{\rm p}^{\rm L}=\lra{F_{{\rm p},N_{\rm p}}^{\rm L}} \simeq F_{{\rm p},N_{\rm p}}^{\rm L}$ for large $N_{\rm p}$. This is an echo of the all-important role of the fine-grained PDF defined as $\delta(\mb{Z}(t) - \mb{z})$, since we have $p(t;\mb{z}) = \lra{\, \delta(\mb{Z}(t) - \mb{z}) \, }$. In recent decades, another commonly encountered approach is the Large-Eddy Simulation (LES) method in which a spatial filtering operation is applied instead of an averaging one. For single-phase turbulent flows where we handle fields, e.g. $Q(t,\mb{x})$, this means that we consider the statistical operator noted $\lra{\cdot}_{\rm ls}$
\begin{equation}
\lra{Q}_{\rm ls} (t,\mb{x}) = \int Q(t,\mb{x}')\, G(\mb{x}'-\mb{x})\, \dd \mb{x}'~,
\end{equation}
where the spatial filter $G$ is taken as a spatially and temporally invariant positive function with a compact support and such that $\int G(\mb{x})\, \dd \mb{x}=1$. In a series of papers, Pope and co-workers~\cite{Gicquel_2002,Sheikhi_2003,Sheikhi_2007} demonstrated that the PDF road can still be followed provided that we consider the filtered density function (FDF) defined as $p_{\rm ls}(t;\mb{z}) = \lra{\, \delta(\mb{Z}(t) - \mb{z}) \, }_{\rm ls}$ (note that this makes sense since the position is always present in the state vector and the decomposition $\mb{Z}=(\mb{X}_{\rm p},\mb{Z}_{\rm c})$ is quite relevant here). This is referred to as the FDF approach. Its original formulation consisted in handling field quantities only but, given the upstream role of the Lagrangian MDF, this suggests to extend the FDF approach to a system made up by either fields or particles by manipulating the discrete Lagrangian MDF. In \cite[section 7]{minier2015lagrangian}, this idea to derive FDF models for LES from a Lagrangian description in terms of particles by starting from the discrete Lagrangian Mass Density Function (LMDF) was developed. 

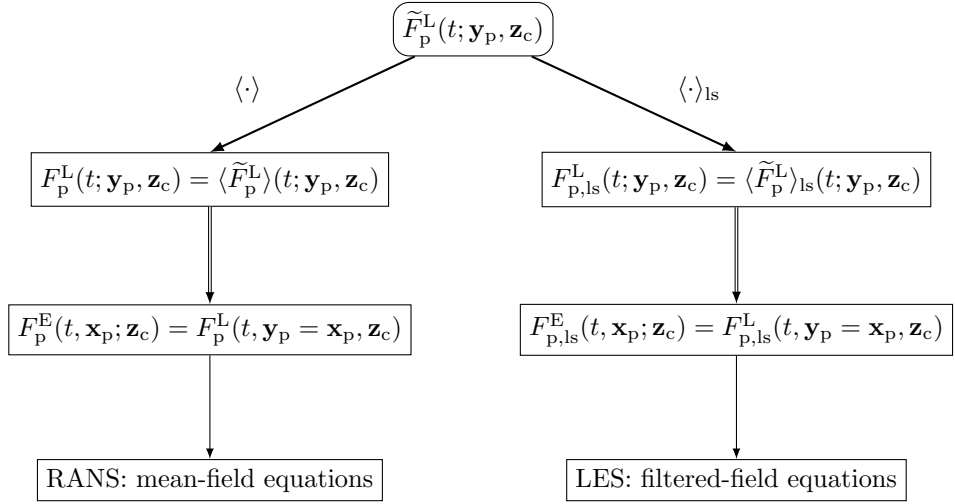
\begin{figure}[htbp]
%\vspace{1em}
\begin{center}
\setlength{\unitlength}{1cm}
\begin{tikzpicture}
\tikzstyle{instruct}=[rectangle,draw]
\tikzstyle{es}=[rectangle,draw,rounded corners=6pt]

\node[es] (discreteLMDF) at (0,0) {$\widetilde{F}_{\rm p}^{\rm L}(t;\mb{y}_{\rm p},\mb{z}_{\rm c})$};
\node[instruct] (LMDF) at (-3.5,-2) {$F_{\rm p}^{\rm L}(t;\mb{y}_{\rm p},\mb{z}_{\rm c})=\lra{ \widetilde{F}_{\rm p}^{\rm L}}(t;\mb{y}_{\rm p},\mb{z}_{\rm c}) $};
\node[instruct] (FLMDF) at (3.5,-2) {$F_{\rm p,ls}^{\rm L}(t;\mb{y}_{\rm p},\mb{z}_{\rm c})=\lra{ \widetilde{F}_{\rm p}^{\rm L} }_{\rm ls}(t;\mb{y}_{\rm p},\mb{z}_{\rm c})$};
\node[instruct] (EMDF) at (-3.5,-4) {$F_{\rm p}^{\rm E}(t,\mb{x}_{\rm p};\mb{z}_{\rm c})=F_{\rm p}^{\rm L}(t,\mb{y}_{\rm p}=\mb{x}_{\rm p},\mb{z}_{\rm c})$};
\node[instruct] (FEMDF) at (3.5,-4) {$F_{\rm p,ls}^{\rm E}(t,\mb{x}_{\rm p};\mb{z}_{\rm c})=F_{\rm p,ls}^{\rm L}(t,\mb{y}_{\rm p}=\mb{x}_{\rm p},\mb{z}_{\rm c})$};
\node[instruct] (RANS) at (-3.5,-6) {RANS: mean-field equations};
\node[instruct] (LES) at (3.5,-6) {LES: filtered-field equations};

\node[] at (-3,-0.8) {$\lra{\cdot}$};
\node[] at (3,-0.8) {$\lra{\cdot}_{\rm ls}$};

\draw[->,>=latex,thick] (discreteLMDF) -- (LMDF.north);
\draw[->,>=latex,thick] (discreteLMDF) -- (FLMDF.north);
\draw[->,>=latex,double] (LMDF) -- (EMDF);
\draw[->,>=latex,double] (FLMDF) -- (FEMDF);
\draw[->,>=latex] (FEMDF) -- (LES);
\draw[->,>=latex] (EMDF) -- (RANS);
\end{tikzpicture}
\end{center}
\caption{Sketch of the central role played by the discrete LMDF $\widetilde{F}_{\rm p}^{\rm L}$ from which the application of different statistical operators yields the Lagrangian and Eulerian MDFs and the resulting macroscopic descriptions.}
\label{fig: general procedure LMDFs}
\end{figure}

We can go one step further and combine the RANS and LES descriptions in a single formulation. Indeed, if we consider the discrete LMDF in Eq.~\eqref{eq sec4: discrete LMDF} with the same decomposition of the state vector, 
\begin{equation}
\widetilde{F}_{\rm p}^{\rm L}(t;\mb{y}_{\rm p},\mb{z}_{\rm c})=\sum_{i=1}^{N_{\rm p}} m_{\rm p}^{[i]}\, \delta(\mb{y}_{\rm p}-\mb{x}^{[i]}_{\rm p}(t))\, \delta(\mb{z}_{\rm c}-\mb{Z}^{[i]}_{\rm c}(t))~,
\end{equation} 
we remark that the LMDF needed for RANS formulations and the Lagrangian Filtered Density Function (LFDF) needed for LES are retrieved by applying the different statistical operators, $\lra{\cdot}$ and $\lra{\cdot}_{ls}$ respectively. This general approach based on the discrete LMDF as the `parent function' is represented in Fig.~\ref{fig: general procedure LMDFs}. 

The great benefit from such a unified formulation is that we do not have to bother anymore about which framework we are evolving in (RANS/LES) or which statistical averages to apply (ensemble or spatial filter). We can concentrate on how the particle state vector $\mb{Z}_{\rm p}$ is represented by a stochastic processes. 

\subsection{Stochastic processes}\label{sec II: Wiener and Poisson}
%==================================================================

\subsubsection{Definition and key characteristics}
%-------------------------------------------------

\paragraph{Definition and key characteristics}
A stochastic process, noted as $\mb{Z}$ or $(\mb{Z}(t), t\geq0)$, is a family of random variables indexed by a  parameter which is usually time (with $t \in [0\,;\,T_{\text{max}}]$, where $T_{\text{max}}$ is an end time with the possible value $T_{\text{max}}=+\infty$). In the following, we consider vector-valued stochastic processes in $\Re^d$ (where $d$ is the dimension) and denote the values in the corresponding sample space by $\mb{z}$. To capture the key characteristics of a stochastic process, it is necessary to resort to the mathematical definition according to which $\mb{Z}$ is a family of measurable functions on an underlying probability space $\Omega$ equipped with a $\sigma$-algebra $\mc{F}$ and a probability measure $\mathbb{P}$. This gives that $\mb{Z}$ as a measurable function of two variables:
\begin{equation}
\begin{gathered}[c]
[0\,;\,T_{\text{max}}] \times (\Omega, \mc{F}, \mathbb{P}) \longrightarrow (\Re^d, \mc{\mb{B}}^d) \\
(t, \omega) \longmapsto \mb{Z}(t, \omega)~,
\end{gathered}
\end{equation}
where $\mc{\mb{B}}^d$ is the Borel $\sigma$-algebra for $\Re^d$ obtained as the $d$-tensorial-product of the Borel $\sigma$-algebra for $\Re$.

In some presentations, this rigorous definition is skipped and stochastic processes are introduced directly in the image space $(\Re^d, \mc{\mb{B}}^d)$. This turns out to be unfortunate. A first reason is that the introduction of $\sigma$-algebras allows to quantify the intuitive notions of the information content and of fine- or coarse-grained descriptions which correspond to different embedded $\sigma$-algebras, or sub-$\sigma$-algebras, embodying the information resolved by one description. A second reason is that the correspondence between two ways to characterize a stochastic process may be overlooked. On the one hand, we can consider the family of vector-valued random variables $\mb{Z}(t)$ at fixed times $t$. This corresponds to the `PDF point of view' where the aim is to derive the PDF equation satisfied by $p(t;\mb{z})$ in sample space. On the other hand, we can consider a family of `elementary events', represented by different values of $\omega$ and handle a (large) number of time functions $t \longmapsto \mb{Z}_{\omega}(t)$. This corresponds to the `trajectory point of view' where the objective is to write the time-evolution equations of these `particles' (samples of the PDF to be used in Monte Carlo estimations). For a stochastic process, there is more information contained in its trajectories than in its PDF. However, since we are interested in approximating statistics from a stochastic process, we can refer to an equivalence between the PDF and trajectory points of view in a weak sense. Said otherwise, considering for instance Langevin type of equations in physical space for a number of trajectories has the same status as the PDF equation in sample space. This is why the trajectory point of view is often adopted in the rest of the article.

\paragraph{The law of a stochastic process}
Knowing a stochastic process is not equivalent to knowing the family of its one-time PDFs $(p(t;\mb{z}), t\geq0)$ since it implies knowing also the joint laws of joint random variables $\mb{Z}(t)$ and $\mb{Z}(t')$ at any times $t$ and $t'$, etc. Using a discrete time setting for the sake of simplicity, this means that knowing the law of a stochastic process is equivalent to the knowledge of the joint PDF $p(t_1,\mb{z}_1;t_2,\mb{z}_2, \ldots; t_N,\mb{z}_N)$ for any set of $N$ times and for any values of the chosen times $(t_1;t_2, \ldots; t_N)$. It is thus clear that the amount of information required is huge and, in particular, much larger than the sole access to the one-time PDF functions $p(t,\mb{z})$. In practice, we need to restrict ourselves to more tractable sub-classes of stochastic processes, such as Markov processes.

\subsubsection{Markovian processes}
%----------------------------------
\label{section: Mathematical background Markov processes}

In the world of ODEs, knowledge of an initial condition (at a time $t_0$) and of the rate of change of a deterministic dynamical system is enough to predict its future. A classical example is Lagrangian/Hamiltonian analytical mechanics. The counterpart of this notion for probabilistic approaches leads to define Markov processes for which knowledge of the present (in a probabilistic sense) and of the law of conditional increments is enough to predict the future (still in a probabilistic sense). To capture this idea, it is sufficient to translate the Markov property in a discrete time setting, which gives
\begin{equation}
\label{Markov property}
p( t_{n+1}, \mb{z}_{n+1} \, \vert \, \left( t_n,\mb{z}_n;t_{n-1},\mb{z}_{n-1};\ldots;t_1,\mb{z}_1 \right))= 
p(t_{n+1}, \mb{z}_{n+1} \, \vert \, t_n,\mb{z}_n )
\end{equation} 
where $t_n$ represents the present time, $t_{n+1}$ the future and $(t_{n-1 }, \ldots, t_1)$ the past, while $\mb{z}_i$ is the value of the process at $t=t_i$ (\textit{i.e.} $\mb{Z}(t_i)=\mb{z}_i$). An important element is that the condition entering the conditional PDF, written as $(t_n,\mb{z}_n)$ in Eq.~\eqref{Markov property}, represents in fact the information known at the present time $t=t_n$.

The fundamental property of Markov processes is that the law of the stochastic process is determined from the knowledge of only two functions: the initial PDF condition $p(t_0,\mb{z}_0)$ and the transition PDF $p(t,\mb{z} \,\vert \, t_0,\mb{z}_0)$ from a state $\mb{z}_0$ at $t=t_0$ to a state $\mb{z}$ at a later time $t$. Indeed, all the $N$-time PDFs are reconstructed from the chain-rule~\cite{ottinger2012stochastic,gardiner2009stochastic,risken1996fokker}
\begin{equation}
\label{eq: chain-rule relation}
p(t_n; \mb{z}_n; t_{n-1}, \mb{z}_{n-1} ; \cdots ; t_1, \mb{z}_1 ; t_0, \mb{z}_0) = p(t_n,\mb{z}_n\,|\,t_{n-1},\mb{z}_{n-1}) \cdots p(t_1,\mb{z}_1\,|\,t_0,\mb{z}_0) p(t_0, \mb{z}_0)
\end{equation}
demonstrating that information on the complete law of the process is derived. In particular, we obtain the Chapman-Kolmogorov equation
\begin{equation}
\label{eq sec4: chapman-kolmogorov relation}
p(t,\mb{z}\,|\,t_0,\mb{z}_0)= \int p(t,\mb{z}\,|\,t_1,\mb{z}_1)\, p(t_1,\mb{z}_1\,|\,t_0,\mb{z}_0)\, d\mb{z}_1~,
\end{equation}
which forms the basis of path-integral formulations~\cite{wiegel1986introduction}. 

In a weak sense, Markov processes are characterized by the infinitesimal operator
\begin{equation}
  \mathcal{L}_t g(\mb{z}) = \lim_{dt \to 0}
  \frac{\langle \left( g(\mb{Z}_{t+dt}) | \mb{Z}_t=\mb{z} \right) \rangle - g(\mb{z})}{dt}~,
\end{equation}
which measures the effect of a conditional increment of the Markov process over a test function $g$. It can then be shown \cite{arnold1974stochastic,gardiner2009stochastic,ottinger2012stochastic} that the transitional PDF $p(t,\mb{z} \,\vert \, t_0,\mb{z}_0)$ is the solution of two equations depending on whether we fix the end condition $(t,\mb{z})$ or the initial one $(t_0,\mb{z}_0)$. In the first case, this corresponds to the Kolmogorov backward equation written as
\begin{equation}
\left\{
\begin{aligned}
& \frac{\partial \, p(t,\mb{z} \,\vert \, t_0,\mb{z}_0)}{\partial t_0} + \mc{L}_{t_0} \left[ \, p(t,\mb{z} \,\vert \, t_0,\mb{z}_0)\, \right] =0~, \label{eq sec4: backward kolmogorov equation}\\
& p(t,\mb{z} \,|\, t_0,\mb{z}_0)=\delta(\mb{z}-\mb{z}_0) \quad t_0 \to t~.
\end{aligned}
\right.
\end{equation}
In the second case, the evolution equation corresponds to the Kolmogorov forward equation
\begin{equation}
\left\{
\begin{aligned}
& \frac{\partial \, p(t,\mb{z} \,\vert \, t_0,\mb{z}_0)}{\partial t} = \mc{L}_t^{\bot} \left[ \, p(t,\mb{z} \,\vert \, t_0,\mb{z}_0)\, \right]~, \label{eq sec4: forward kolmogorov equation}\\
& p(t,\mb{z} \,|\, t_0,\mb{z}_0)=\delta(\mb{z}-\mb{z}_0) \quad t \to t_0~,
\end{aligned}
\right.
\end{equation}
where $\mc{L}_t^{\bot}$ is the adjoint operator of $\mc{L}_t$. From a physical point of view, we often consider the time evolution of a dynamical system from an observed initial state at a given time $t_0$ and we are therefore mostly concerned with Kolmogorov forward equations but the backward one is also of great interest (cf. the celebrated Feynman-Kac formula). When dealing with Markov processes, these equations are central since all the needed information is generated by the transition PDF. Note that the same Kolmogorov forward equation is obtained for the one-time PDF $p(t,\mb{z})$ by integrating over all initial conditions but actual information is in the transition PDF. In practice, the subclass of Markov stochastic processes is the only one for which we can develop a complete description. For non-Markov processes, it is still possible to consider PDF equations for the one-time PDF but it is already clear that the description of the stochastic process is incomplete. It is shown in Sec.~\ref{statistical model transport} that more serious troubles lie ahead for non-Markov processes.

\subsubsection{The two building blocks: the Wiener and Poisson processes}
%------------------------------------------------------------------------
\label{section: the Wiener and Poisson processes}

\paragraph{The Poisson process}

\begin{figure}[htbp]
\centering
\includegraphics[scale = 0.6]{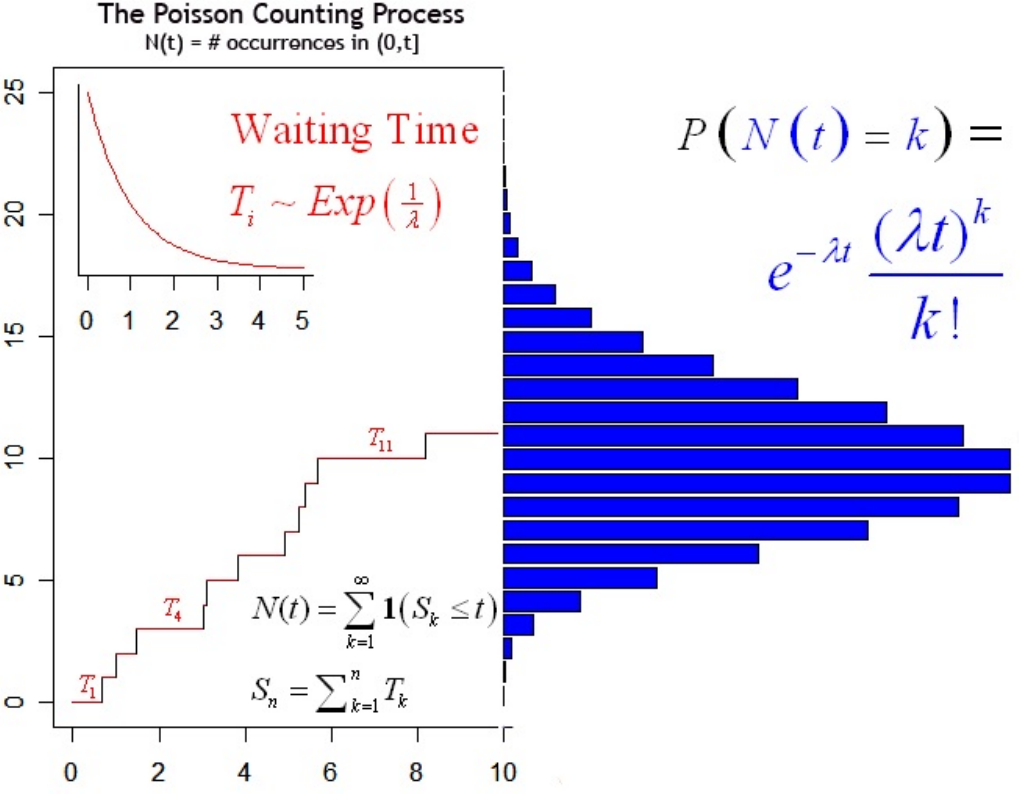}
\caption{Some properties of the Poisson process $N(t)$: example of one trajectory of the process jumping at random times $T_i$ which follows an exponential distribution and the resulting pdf of $N(t)$ at each time $t$
being a Poisson distribution. Reprinted from~\cite{chibbaro2014stochastic} with permission from Springer.}
\label{fig: Poisson_Process}
\end{figure}

For random discrete events, the reference model is the Poisson process where there is a very small probability to have one event but each one implying a change represented by a jump (multiple jumps at one time are not possible). The trajectories of the Poisson process are therefore piecewise constant with jumps (having a step of one unity in the standard Poisson process) occurring at random times~\cite{gardiner2009stochastic,klebaner2012introduction} as illustrated in Fig.~\ref{fig: Poisson_Process}.

A Poisson process is characterized by its intensity $\lambda$ which is the mean value of the number of events per unit time. The number of events in a time interval $[t\, , \, t + \Delta t \,]$, represented by the stationary and independent increments $\Delta N(t)=N(t+\Delta t) - N(t)$, is a Poisson random variable 
\begin{equation}
\label{eq sec4: Poisson distribution for events in intervals}
\mathbb{P}[\, \Delta N(t)=k \,]=\frac{(\lambda \Delta t)^k}{k!}e^{-\lambda \Delta t}
\end{equation} 
from which it follows that the mean and variance of $\Delta N(t)$ are linear with respect to $\Delta t$
\begin{equation}
\lra{ \Delta N(t) }= \lambda \Delta t \qquad 
\lra{ (\Delta N(t)- \lra{ \Delta N(t) })^2}  = \lambda \Delta t.
\end{equation} 
In any finite interval, the times at which the Poisson process jumps are uniformly distributed. In practice, this property explains that deviations of measured particle distributions from the Poisson law are regarded as manifesting an underlying order. There are typically two ways to simulate a Poisson process. The first method is based on waiting times (the time intervals between successive random jumps) and, since values are constant between jumps, the idea is to go directly from one event to the next one. These waiting times are random variables following an exponential distribution with the same intensity $\lambda$:
\begin{equation}
\mathbb{P}[\,  T_i=t \,]=\lambda e^{-\lambda t} \quad \Rightarrow 
\lra{ T_i}= \frac{1}{\lambda}~,
\end{equation} 
and are generated before applying the unit-one jump or more general events when considering generalized Poisson processes (see below in section~\ref{Ito SDE and Fokker-Planck}). This method is often used to simulate collision and agglomeration events in molecular dynamics (cf. Sec.~\ref{sec: sample or physical spaces}) but imposes variable time steps. The second way is to generate the number of events occurring in fixed time intervals (cf. Sec.~\ref{Jump-diffusion processes for collisions}). When the time interval $\Delta t$ is much smaller than the relevant timescale $1/\lambda$ (\textit{i.e.}, $\lambda\, \Delta t \ll 1$), the statistics of the increments, which follow the Poisson distribution given in Eq.~\eqref{eq sec4: Poisson distribution for events in intervals}, have the simplified expression:
\begin{equation}
\mathbb{P}[\, \Delta N(t)=0 \,] \simeq 1 -\lambda \Delta t~, \quad \mathbb{P}[\, \Delta N(t)=1 \,] \simeq \lambda \Delta t~, \quad
\mathbb{P}[\, \Delta N(t)=k \,] \simeq 0 \; (k \geq 2)~. 
\end{equation}

\paragraph{The Wiener process}

\begin{figure}[htbp]
\centering
\includegraphics[width=0.8\textwidth]{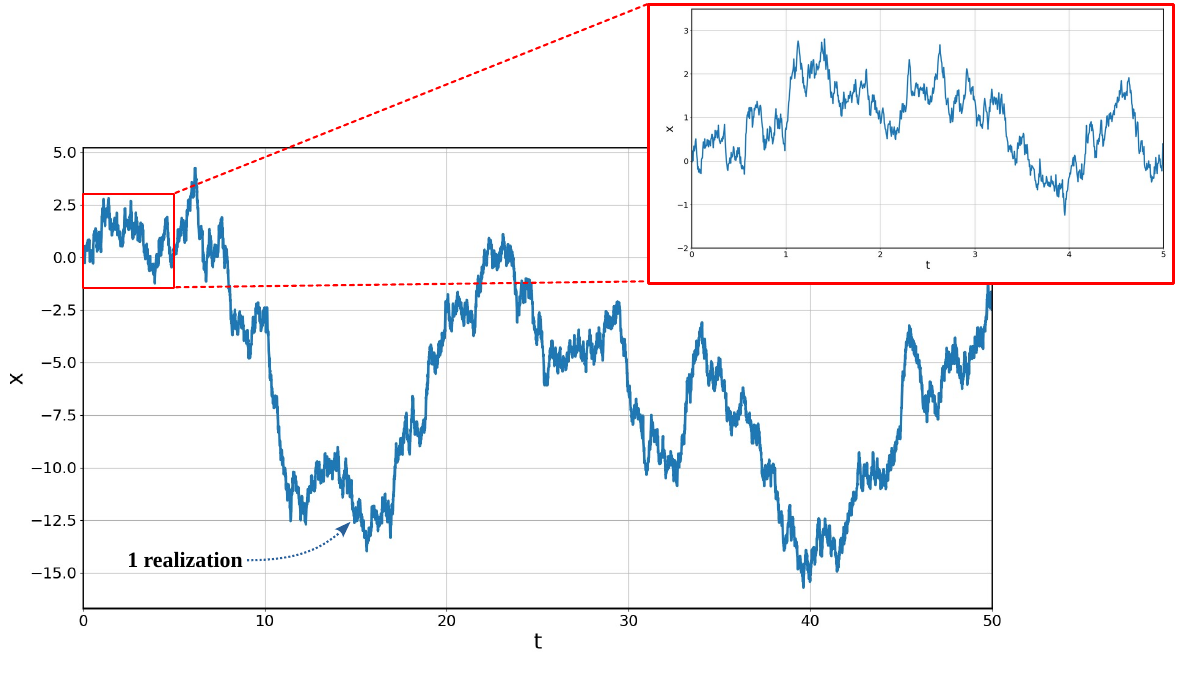}
\caption{One trajectory, or realization, of the Wiener process showing a continuous but ragged profile. The window, with a zoom of the trajectory, reveals the self-similar nature and the fast fluctuations of the trajectories of the Wiener process at all scales.}
\label{fig: Wiener_Process}
\end{figure}

The Wiener process is the canonical model of Brownian motion~\cite{gardiner2009stochastic,ottinger2012stochastic} and can be defined by the following three properties~\cite{karatzas1991brownian}:
\begin{enumerate}[(i)]
\item The process has independent increments, \textit{i.e.} $(W(t_3)-W(t_2))$ and $(W(t_1)-W(t_0))$ are independent when $t_0<t_1<t_2<t_3$;
\item The trajectories of the process are continuous functions (almost everywhere);
\item The increments of the Wiener process $(W(t_2)-W(t_1))$ are centered Gaussian random variables and with a variance equal to $(t_2-t_1)$.
\end{enumerate}
From these characteristics, it results that the Wiener process $W$ is a Gaussian, Markov process, with zero mean and a covariance equal to $\lra{ W(t) W(t') }=\min(t,t')$. Its trajectories are continuous and represent continuous evolutions where there is a near-one chance that modifications happen but with small changes. The most important properties are that the trajectories of the Wiener process are non-differentiable at any point and have even unbounded total variations in any interval. Furthermore, the increments $dW=W(t+dt)-W(t)$ are stationary and independent with successive moments
\begin{equation}
\lra{dW}=0, \;\; \lra{(dW)^2} = dt, \;\; \lra{(dW)^n} = o(dt) \quad \forall n > 2~.
\end{equation}
The linear variation of $\lra{(dW)^2}$ with respect to $dt$ is of paramount importance in the development of stochastic calculus~\cite{arnold1974stochastic,klebaner2012introduction,oksendal2003stochastic}. One trajectory of a Wiener process is displayed in Fig.~\ref{fig: Wiener_Process} illustrating the wild variations at any scale. 

\subsection{Jump-diffusion processes and extended Fokker-Planck equations}\label{Ito SDE and Fokker-Planck}
%=========================================================================

To explain the shift from ODEs to Stochastic Differential Equations (SDEs), we consider a one-dimensional dynamical system $Z(t)$ under the influence of Gaussian white-noise $\zeta(t)$
\begin{equation}
\label{eq: system under white-noise}
\frac{dZ(t)}{dt}= A(t,Z(t)) + B(t,Z(t))\, \zeta(t)~.
\end{equation}
To arrive at a well-posed mathematical definition, the idea is to benefit from the smoothing properties of the integration using the identification of $\int_0^t \zeta(s)\, ds$ with a white-noise process, \textit{i.e.} $\zeta_t\, dt = dW(t)$. Instead of Eq.~\eqref{eq: system under white-noise}, we therefore try to give a meaning to 
\begin{equation}
Z(t) = Z(t_0) + \underbrace{\int_{t_0}^t A(s,Z(s))\, ds}_{\text{regular integral}} + \underbrace{\int_{t_0}^t B(s, Z(s))\, dW(s)}_{\text{stochastic integral}}.
\end{equation}
Due to the infinite total variation in any finite interval of the trajectories of $W$, we cannot obtain the stochastic integral as the limit of classical Riemann-Stieltjes sums. For a non-anticipating process $Z$, the stochastic integral is then defined in the Ito sense as
\begin{equation}
\int_{t0}^t B(s,Z(s))\, dW_s = \text{ms-}\lim_{N \to \infty} \sum_{i=1}^N B(t_i,Z(t_i))\left( W(t_{i+1}) - W(t_i) \right)
\end{equation} 
where $(t_i)_{i=1,N}$ represents a partition of the interval $[0\,;\,t]$ (with $t_1=0$ and $t_{N+1}=t$) and 
where the limit is to be understood in the mean-square sense \cite{arnold1974stochastic,ottinger2012stochastic,oksendal2003stochastic} and not as a convergence trajectory by trajectory. In the physics literature, these equations are written in incremental form as a short-hand notation and are referred to as `Langevin equations'
\begin{equation} 
dZ(t) = A(t,\, Z(t))\, dt + B(t,\, Z(t))\, dW(t)~,
\end{equation}
while their corresponding PDF equation in sample space is the Fokker-Planck (FP) equation
\begin{equation}
\label{eq sec4: one-dimensional FPE simple}
\frac{\partial p(t,z \,\vert \, t_0,z_0)}{\partial t} = -\frac{\partial [\,A(t,z)\, p(t,z \,\vert \, t_0,z_0)\,]}{\partial z} + \frac{1}{2} \frac{\partial^2 [\,B^2(t,z)\, p(t,z \,\vert \, t_0,z_0)\,]}{\partial z^2}~.
\end{equation}
The FP equation is a forward Kolmogorov equation, Eq.~\eqref{eq sec4: forward kolmogorov equation}, with the infinitesimal operator $\mc{L}_t$ being
\begin{equation}
\mc{L}_t[\cdot] = A(t,z)\frac{\partial \,[\,\cdot\,]}{\partial z} + \frac{1}{2} B^2(t,z) \frac{\partial^2 \,[\,\cdot\,]}{\partial z^2}~.
\end{equation}
Since the FP equation appears as a convection-diffusion equation, we speak of stochastic diffusion processes and the functions $A$ and $B$ are referred to as the drift and diffusion coefficients, respectively. Note that, regardless of the sign of $B$, the diffusion coefficient in the PDF equation is $B^2$ and is always positive or null. 

The correspondence between Langevin and Fokker-Planck equations is easily extended to the multi-dimensional case. For instance, the SDEs written for a $d$-dimensional process $\mb{Z}=(Z_1, \ldots, Z_d)$ are
\begin{equation}
\label{eq sec4: general definition Langevin SDEs in d dimensions}
dZ_i(t)= A_i(t,\mb{Z}(t))\, dt + B_{ij}(t,\mb{Z}(t))\, dW_j(t)~,
\end{equation}
with $(W_j)_{j=1,d}$ a set of $d$ independent Wiener processes. In Eq.~\eqref{eq sec4: general definition Langevin SDEs in d dimensions}, the drift $\mb{A}=(A_i)$ is now a vector while the diffusion $\mb{B}=(B_{ij})$ is a matrix. The corresponding Fokker-Planck equation is  
\begin{equation}
\label{eq sec4: general definition PFE in d dimensions}
\frac{\partial p(t,\mb{z} \,\vert \, t_0,\mb{z}_0)}{\partial t} = - \frac{\partial \left[\, A_i(t,\mb{z}) \, p(t,\mb{z} \,\vert \, t_0,\mb{z}_0)\, \right]}{\partial z_i} \\
+ \frac{1}{2}\frac{\partial^2 \left[ \, D_{ij}(t,\mb{z})\, p(t,\mb{z} \,\vert \, t_0,\mb{z}_0)\, \right]}{\partial z_i \partial z_j}~,
\end{equation}
where $\mb{D}=\mb{B}\mb{B}^{\bot}$ (or $D_{ij}=B_{ik}B_{jk}$) is a symmetric definite-positive matrix. As in the one-dimensional case, the positivity of $\mb{D}$ justifies the reference to a diffusive nature induced by the rapidly-varying terms, $B_{ij}(t,\mb{Z}(t))\, dW_j(t)$, in Eq.~\eqref{eq sec4: general definition Langevin SDEs in d dimensions}. The fact that several matrices $\mb{B}$ can correspond to the same diffusion matrix $\mb{D}$ translates the equivalence between the trajectory and PDF points of view in a weak sense. 

\begin{figure}[htbp]
\centering
\includegraphics[width=0.8\textwidth]{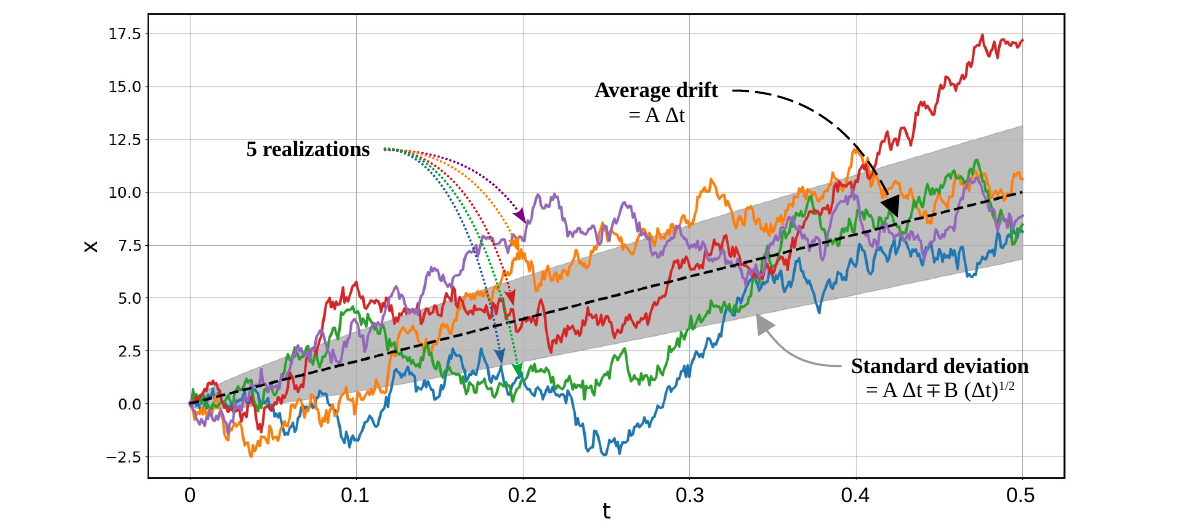}
\caption{Statistics over small time increments on the conditional trajectories of a stochastic diffusion process illustrating the physical meaning of the drift and diffusion coefficients.}
\label{fig: SDE drift-diffusion meaning}
\end{figure}

The physical meaning of $A$ and $B$ in the Langevin and Fokker-Planck equations are revealed by considering the statistics of the conditional increments (using a one-dimensional version)
\begin{subequations}
\label{eq sec4: conditional mean and diffusion meaning}
\begin{align}
\lra{ \Delta Z \,|\, Z(t)=z} &= A(t,z)\, \Delta t~, \\
\lra{ (\Delta Z)^2 \,|\, Z(t)=z} &= B^2(t,z)\, \Delta t~.
\end{align} 
\end{subequations}
As shown in Fig.~\ref{fig: SDE drift-diffusion meaning}, the drift term, $A(t,z)$, governs the mean evolution of the conditional increments while the diffusion coefficient, $B(t,z)$, governs the spread of the conditional increments around its mean value.

There is regularly some confusion about the Gaussian hypothesis in SDEs. As expressed by Eqs.~\eqref{eq sec4: conditional mean and diffusion meaning}, the Gaussian hypothesis applies only to the conditional increments  and, in the general situation when $B(t,z)$ is not constant, the resulting process $Z$ can deviate from Gaussianity. Though this point has been clarified~\cite{minier2001pdf,minier2015lagrangian}, it is worth emphasizing that the observation of non-Gaussian distributions does not invalid models based on stochastic diffusion processes. 

Apart from the case of Brownian motion itself (where $A=0$ and $B=1$), the simplest example of a stochastic diffusion process is the Ornstein-Uhlenbeck (OU) process whose trajectories are
\begin{equation}
\label{eq sec4: definition OU process}
dZ(t) = - \frac{Z(t)}{\tau}\,dt + K\, dW(t)~,
\end{equation}
where the timescale $\tau$ and the diffusion coefficient $K$ are constant. From Ito calculus \cite{gardiner2009stochastic}, it is trivial to show that $K$, $\tau$ and $\lra{Z^2}$ are related through $K^2=2 \lra{Z^2}/\tau$ which corresponds to a fluctuation-dissipation theorem~\cite{ottinger2005beyond}. Another immediate property is that the auto-correlation function of the stationary process $\rho(t'-t)=\lra{Z(t)Z(t')}/\lra{Z^2}$ is $\rho(t'-t)=\exp( -(t'-t)/\tau)$, revealing that the relaxation timescale $\tau$ is also the integral timescale since $\tau=\int_0^{+\infty} \rho(s)\, \dd s$. This is the first hint that timescales play a central role when formulating stochastic models, as exemplified in later sections, in particular Sec.~\ref{Modeling fluid seen}.

\begin{figure}[htbp]
\centering
\includegraphics[width=0.9\textwidth]{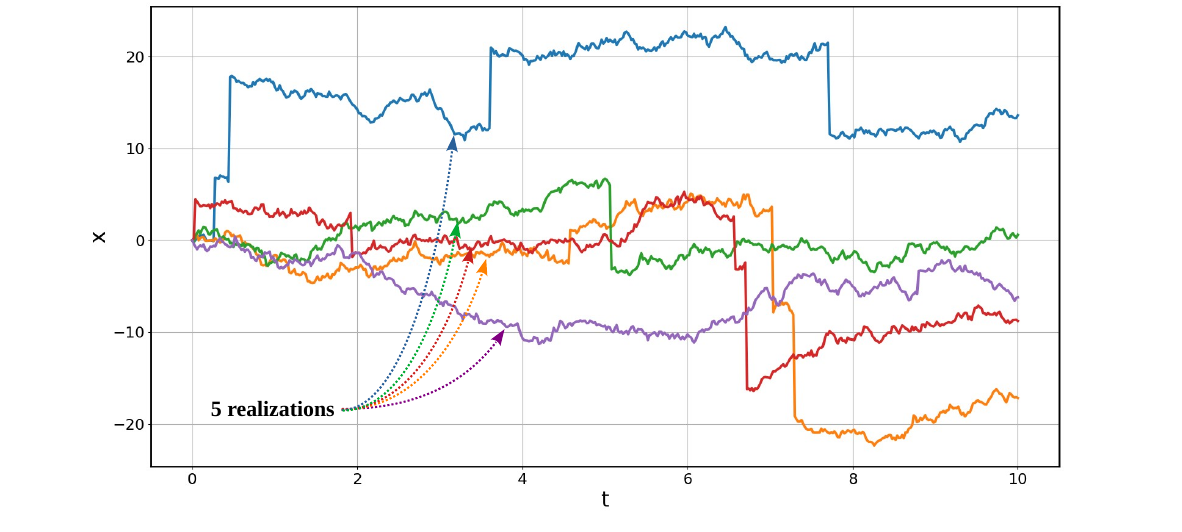}
\caption{Trajectories of a jump-diffusion process with their diffusive behavior in between sudden jumps.}
\label{fig: trajectory jump-diffusion process}
\end{figure}

So far, we have built SDEs based on the Wiener process for diffusion processes with continuous trajectories. Similar steps can be made with the Poisson process to introduce sudden jumps. Still using the trajectory point of view, the SDEs for a jump-diffusion process are written for a one-dimensional process
\begin{equation}
\label{eq sec4: general jump-diffusion trajectory equation}
dZ(t) = A(t,Z(t))\, dt + B(t,Z(t))\, dW(t) + C(t,Z(t))\, dN(t)~,
\end{equation}
where $N$ a Poisson process with intensity $\lambda$ and $C(t,Z(t))$ represents the amplitude of the jumps. The jumps contribute to the statistics of the conditional increments over a time step $\Delta t$ and Eqs.~\eqref{eq sec4: conditional mean and diffusion meaning} become
\begin{subequations}
\label{eq sec4: conditional mean and diffusion with jump meaning}
\begin{align}
& \lra{ \Delta Z \,|\, Z(t)=z} = A(t,z)\, \Delta t + C(t,z)\lambda \, \Delta t~, \\
& \lra{ (\Delta Z)^2 \,|\, Z(t)=z} = B^2(t,z)\, \Delta t + C^2(t,z)\lambda \, \Delta t~.
\end{align}
\end{subequations}
The introduction of jumps is illustrated in Fig.~\ref{fig: trajectory jump-diffusion process}, revealing the discontinuous nature of the trajectories and their diffusive behavior between successive jumps. 

From the properties of the Poisson process, we know that, over an infinitesimal time increment $dt$ and retaining only the first order contributions, the increments of $N(t)$ can take only two possible values
\begin{equation} 
\mathbb{P}[\, dN(t)=k\, ] = (1 - \lambda dt)\,\delta_{k,0} + \lambda \, dt\, \delta_{k,1} 
 \Rightarrow \lra{ (dN(t))^m}= \lambda \,dt \quad \forall m .
 \end{equation}
The Poisson jumps contribute to any order in $dt$ (whereas the increments of the Wiener process contributes only to the second order in $dt$). The SDEs of a jump-diffusion process can be generalized by considering random amplitudes for the jumps (this is referred to as a compound Poisson process), which gives
\begin{equation}
dZ(t) = A(t,Z(t))\, dt + B(t,Z(t))\, dW(t) + C(t,Z(t),Q)\, dN(t)~,
\end{equation}
where $Q$ is an independent random variable. The law of the random jumps becomes then
\begin{equation}
W(y\, |\, t,z)= W(y\, |\, Z(t)=z) = \mathbb{P}[\, C= y \,|\, Z(t)=z \,]\, \lambda~,
\end{equation}
where the probability $\mathbb{P}$ refers to the law of the independent random variable $Q$. In sample space, the equation for the transitional PDF $p(t,z \,\vert \, t_0,z_0)$ is
\begin{multline}
\label{eq sec4: general jump-diffusion PDF equation}
\frac{\partial p(t,z \,\vert \, t_0,z_0)}{\partial t} = 
-\frac{\partial [\, A(t,z)\, p(t,z \,\vert \, t_0,z_0)\,]}{\partial x} +
\frac{1}{2} \frac{\partial^2 [\,B^2(t,z)\, p(t,z \,\vert \, t_0,z_0)\,]}{\partial x^2} \\
+ \int [\, W(z\, |\, t,y)\, p(t,y \,\vert \, t_0,z_0) - W(y\, |\, t,z)\, p(t,z \,\vert \, t_0,z_0)\,]\,dy.
\end{multline}
This form is referred to as the Chapman-Kolmogorov PDF equation in \cite{gardiner2009stochastic}. Jump-diffusion processes resurface in Sec.~\ref{Jump-diffusion processes for collisions} to account for diffusive effects due to random continuous motions of a turbulent flow and sudden velocity jumps due to particle collisions.

\subsection{A double hierarchy: one- or N-particle PDF and particle-attached variables}
%======================================================================================

Having set the probabilistic framework, we return to the first step mentioned at the beginning of this section. Yet, before considering their modeled evolution, we need to define the variables in the state vector. This includes: (a) choosing a one- or a $N$-particle PDF formulation; (b) determining the relevant variables to describe each particle. Point (a) is a very classical issue in statistical physics in relation with the well-known BBGKY hierarchy for any system with (at least) two-particle interactions. At the moment, two-particle PDF models (not to mention $N$-particle ones, with $N \geq 3$) are not developed enough for general inhomogeneous turbulent flows, especially wall-bounded ones, and are therefore far less tractable~\cite{pope1994lagrangian,pope2000turbulent}. For these reasons and though references are made to such approaches (cf. Secs.~\ref{Modeling fluid seen} and~\ref{statistical model collision}), we retain the one-particle PDF level of description and, correspondingly, a one-point Eulerian PDF description of fields. It is then essential to be aware of the inherent loss of information since, for instance, two-point correlations cannot be extracted from one-particle PDF models and particle-particle interactions, such as actions at distance or even collisions, have to be modeled through mean-field formulations. Point (b) turns out to be a challenge to traditional statistical views and requires specific analysis. This is addressed in Sec.~\ref{statistical model transport}.

%==========================================================================================================
\section{Statistical modeling of particle transport in turbulent flows \label{statistical model transport}}
%==========================================================================================================

\subsection{Shortcomings of kinetic-like descriptions}
%=====================================================

The Boltzmann picture is dominating the description of transport phenomena to such an extent that the kinetic framework is rarely questioned. Yet, whether kinetic variables are necessarily the only ones to retain in the particle state vector is not a point over which to pass too quickly. For discrete particles in turbulent flows, there are typically three situations. The first one corresponds to the case of fully-resolved turbulent flows, where the fluid velocity field is known at every point and every time. This means that the velocity of the fluid seen is similar to an external deterministic force field. It is then accounted for without approximation in Boltzmann-like PDF models, i.e. in PDF formulations based on the kinetic state vector $\mb{Z}_{\rm p}^{\rm r}=(\mb{X}_{\rm p}, \mb{U}_{\rm p})$ where we handle a PDF noted $p^{\rm r}(t;\mb{z}_{\rm p}^{\rm r})=p^{\rm r}(t;\mb{y}_{\rm p},\mb{V}_{\rm p})$ in sample space. In these notations, the superscript r indicates a reduced description compared to $\mb{Z}_{\rm p}=(\mb{X}_{\rm p}, \mb{U}_{\rm p},\mb{U}_{\rm s})$ and the PDF $p^{\rm r}$ is the marginal of $p(t;\mb{z}_{\rm p})=p(t;\mb{y}_{\rm p},\mb{V}_{\rm p},\mb{V}_{\rm s})$. The second situation corresponds to high-inertia particles for which the underlying fluid can be regarded as white-noise, leading to a Langevin equation for the discrete particle velocity and a resulting FP equation for $p^{\rm r}$. The third situation corresponds to the usual case of non-fully-resolved turbulent flows, where $\mb{U}_{\rm s}$ is neither deterministic nor white-noise but exhibits memory effects. This is the situation addressed by the kinetic-PDF model. 

By manipulating the fine-grained PDF $\mc{P}(t; \mb{y}_{\rm p},\mb{V}_{\rm p})=\delta( \mb{X}_{\rm p}(t) - \mb{y}_{\rm p})\delta(\mb{U}_{\rm p}(t) - \mb{V}_{\rm p})$ and using standard techniques~\cite{pope2000turbulent} with $ p^{\rm r}(t; \mb{y}_{\rm p},\mb{V}_{\rm p})=\lra{ \mc{P}(t; \mb{y}_{\rm p},\mb{V}_{\rm p})}$, the kinetic PDF equation is derived from the particle equations keeping only the drag force which is the central point. The unclosed PDF equation is
\begin{equation}
\label{open true kinetic-PDF}
 \frac{\partial p^{\rm r}}{\partial t}+ \frac{\partial \left[\, V_{{\rm p},i}\, p^{\rm r}\, \right]}{\partial y_{{\rm p},i}} = 
\left[\,\left\langle \frac{V_{{\rm p},i}}{\tau_{\rm p}}\,|\,(\mb{y}_{\rm p},\mb{V}_{\rm p}) \right\rangle p^{\rm r}\,\right]
- \frac{\partial}{\partial V_{{\rm p},i}}
\left[\,\left\langle \frac{U_{{\rm s},i}}{\tau_{\rm p}}\,|\,(\mb{y}_{\rm p},\mb{V}_{\rm p}) \right\rangle p^{\rm r}\,\right]~.  
\end{equation}
It first appears that the complete expression of the particle relaxation time $\tau_{\rm p}$ cannot be accounted for at this level of description, since it is usually a function of $(\mb{U}_{\rm p},\mb{U}_{\rm s})$. In the kinetic approach, it is assumed that we are only dealing with particles in the Stokes regime so that $\tau_{\rm p} \simeq \tau_{\rm p}^{\rm St}$ becomes independent of $(\mb{U}_{\rm p},\mb{U}_{\rm s})$, which is already a limitation. With this approximation, the unclosed kinetic PDF equation is written as
\begin{equation}
\label{open kinetic-PDF}
 \frac{\partial p^{\rm r}}{\partial t}+ \frac{\partial\left[\, V_{{\rm p},i}\; p^{\rm r}\, \right]}{\partial y_{{\rm p},i}} = 
 \frac{\partial}{\partial V_{{\rm p},i}}\left[\,\frac{V_{{\rm p},i}}{\tau_p^{\rm St}}\, p^{\rm r}\,\right]
- \frac{\partial}{\partial V_{{\rm p},i}}\left[\,\frac{1}{\tau_p^{\rm St}}\lra{U_{{\rm f},i}}\, p^{\rm r}\,\right] 
- \frac{\partial}{\partial V_{{\rm p},i}}\left[\,\frac{1}{\tau_p^{\rm St}}\lra{u^{'}_{{\rm s},i}\,|\,(\mb{y}_{\rm p},\mb{V}_{\rm p})}\, p^{\rm r}\,\right]  
\end{equation}
where the velocity of the fluid seen has been decomposed as $\mb{U}_{\rm s}=\lra{\mb{U}_{\rm f}}(t,\mb{X}_{\rm p}(t)) + \mb{u}_{\rm s}^{'}$. This does not imply, however, that $\lra{\mb{u}_{\rm s}^{'}}=0$ since the set of velocities of the fluid seen represents only a subset of fluid particle velocities at a given location (unless the so-called well-mixed condition is satisfied), and its mean value is called the drift velocity $\mb{U}_{\rm d}$, which is thus equal to $\mb{U}_{\rm d}=\lra{\mb{U}_{\rm s}} - \lra{\mb{U}_{\rm f}}(t,\mb{X}_{\rm p}(t))$. 

Closure of the open flux in sample space is derived from Furutsu-Novikov-Donsker (FND) relation~\cite{furutsu1963statistical,novikov1965functionals,donsker1964function}, which expresses the correlation between a Gaussian field with zero mean and an arbitrary functional of that field. It is applied by assuming that the fluctuating velocity field $\mb{u}_f(t,\mb{x})$ is a random Gaussian field and writes for a functional $F[ t; \mb{u}_f ]$ of $\mb{u}_f(t,\mb{x})$ 
\begin{equation}
\lra{ u_{{\rm f},i}(t,\mb{x})\, F[ t; \mb{u}_{\rm f} ] }=
\int_0^t \int_{\mb{x}'} R_{ik}(t,\mb{x};t',\mb{x}') \left\langle \frac{ \delta F[ t; \mb{u}_{\rm f} ]}{\delta u_{{\rm f},k}(t',\mb{x}')} \right\rangle \dd \mb{x}'\, \dd t'
\end{equation}
where $R_{ik}(t,\mb{x};t',\mb{x}')=\lra{ u_{{\rm f},i}(t,\mb{x})\, u_{{\rm f},k}(t',\mb{x}') }$ is the fluid two-point two-time correlation. Then, applying the FND relation to the fine-grained PDF $\mc{P}$ yields the expression of the flux closure as~\cite{bragg2012drift}
\begin{equation}
\label{kinetic PDF closure a}
\frac{1}{\tau_p^{\rm St}}\lra{U_{{\rm s},i}\,|\,\mb{z}^r_{\rm p}}\, p^r = \frac{1}{\tau_{\rm p}^{\rm St}} \lra{U_{{\rm f},i}}(t,\mb{y}_{\rm p})\, p^r + \frac{1}{\tau_{\rm p}^{\rm St}} U_{{\rm d},i}(t,\mb{y}_{\rm p})\, p^r -\lambda_{ij}\, \frac{\partial \left[\,p^r\,\right]}{\partial y_{{\rm p},j}}- \mu_{ij} \, \frac{\partial \left[\,p^r\,\right]}{\partial V_{{\rm p},j}}
\end{equation}
with $\lambda_{ij}$ and $\mu_{ij}$ given by 
\begin{subequations}
\begin{align}
\lambda_{ij} &= \frac{1}{\tau_{\rm p}^{\rm St}}
\int_0^t \lra{ \Gamma_{jk}(t,t') \, R_{ik}(t,\mb{y}_{\rm p};t',\mb{X}_{\rm p}(t')) }_{(\mb{y}_{\rm p},\mb{V}_{\rm p})}\, dt' \label{kinetic dispersion tensors L} \\
\mu_{ij} &= \frac{1}{\tau_{\rm p}^{\rm St}}
\int_0^t \lra{ \dot{\Gamma}_{jk}(t,t') \, R_{ik}(t,\mb{y}_{\rm p};t',\mb{X}_{\rm p}(t')) }_{(\mb{y}_{\rm p},\mb{V}_{\rm p})} \, dt'
\label{kinetic dispersion tensors M}
\end{align}
\end{subequations}
where the notation $\lra{.}_{(\mb{y}_{\rm p},\mb{V}_{\rm p})}$ indicates the averaged value conditioned on the particle trajectory that `arrives' at $(\mb{y}_{\rm p},\mb{V}_{\rm p})$ at time $t$, which explains that the dispersion tensors are functions of $(\mb{y}_{\rm p},\mb{V}_{\rm p})$ even if the eventual dependence on $\mb{V}_{\rm p}$ is often neglected. In these equations, $\Gamma_{jk}(t,t')$ stands for the response function
\begin{equation}
\label{definition response tensor}
\Gamma_{jk}(t,t')= \frac{ \delta X_{{\rm p},j}(t) }{\delta u_{{\rm f},k}(t',\mb{X}_{\rm p}(t'))}
\end{equation}
that measures the effect of a perturbation of the fluctuating fluid velocity seen at an earlier time $t'$ on the particle position $\mb{X}_{\rm p}(t)$ at time $t$, and $\dot{\Gamma} = \partial \Gamma/\partial t$. With these expressions, the flux can also be written as
\begin{equation}
\label{kinetic PDF closure b}
\frac{1}{\tau_{\rm p}^{\rm St}}\lra{U_{{\rm s},i}\,|\,\mb{z}_{\rm p}}\, p^r = \frac{1}{\tau_{\rm p}^{\rm St}} \lra{U_{{\rm f},i}}(t,\mb{y}_{\rm p})\, p^r + \kappa_i\, p^r -\frac{\partial \left[\,\lambda_{ij}\,p^r\,\right]}{\partial y_{{\rm p},j}}- \frac{\partial \left[\,\mu_{ij}\,p^r\,\right]}{\partial V_{{\rm p},j}}
\end{equation}
with  
\begin{equation}
\kappa_i (t;\mb{z}_{\rm p})= \frac{1}{\tau_{\rm p}^{\rm St}} U_{{\rm d},i}(t,\mb{y}_{\rm p}) + \frac{\partial \lambda_{ij}}{\partial y_{{\rm p},j}} + \frac{\partial \mu_{ij}}{\partial V_{{\rm p},j}}~. 
\label{reduced kinetic dispersion tensors K} 
\end{equation}
Inserting Eq.~\eqref{kinetic PDF closure b} into Eq.~\eqref{open kinetic-PDF}, we obtain the kinetic PDF equation which, in a compact form, is
\begin{equation}
\label{compact kinetic PDF}
 \frac{\partial p^{\rm r}}{\partial t} = - \frac{\partial}{\partial z^{\rm r}_{l}}\left[ A^{\rm KE}_l \, p^{\rm r} \right]
+ \frac{1}{2}\frac{\partial^2}{\partial z^{\rm r}_{l}\partial z^{\rm r}_{m}}\left[B^{\rm KE}_{lm}\,p^{\rm r} \right]~,
\end{equation}
where the components of the drift vector $(A^{\rm KE}_l)_{l=1,6}$ are 
\begin{equation}
A^{\rm KE}_l=
\begin{cases}
V_{p,l} & \; l=1,3,  \\
\dfrac{\lra{U_{{\rm f},l-3}}-V_{{\rm p},l-3}}{\tau_{\rm p}^{\rm St}} + \kappa_{l-3} & \; l=4,6~,
\end{cases}
\end{equation}
and where the symmetrical matrix $B^{\rm KE}_{lm}$ is (using a bloc notation with $i,j=1,3$)
\begin{equation}
\mb{B}^{\rm KE}=
\begin{pmatrix}
0 & \vline & (\lambda_{ij}) \\
--  &  &   ---            \\
(\lambda_{ji}) & \vline & (\mu_{ij})+(\mu_{ji})
\end{pmatrix}~.
\end{equation}
As such, Eq.~\eqref{compact kinetic PDF} looks like a classical convection-diffusion equation with $B^{\rm KE}_{lm}$ in the second-order derivative appearing as a diffusion matrix. However, it is immediate to show that this symmetrical matrix $B^{\rm KE}_{lm}$ is negative definite. Indeed, its determinant is $det(\mb{B}^{\rm KE})=- \left( det(\bds{\lambda}) \right)^2 < 0$ and, consequently, the matrix $\mb{B}^{\rm KE}$ has always at least one negative eigenvalue. This point was first put forward in~\cite{pozorski1999probability} where it was associated with an `anti-diffusive' behavior and the non-Markovian nature of the reduced state vector $\mb{Z}^{\rm r}_{\rm p}$. It was later discussed in a comprehensive analysis of kinetic- and dynamic-PDF models in~\cite{minier2015kinetic} to which readers are referred to for further details. The unavoidable consequence is that this kinetic-PDF equation cannot be solved, unless for very special initial conditions, and, therefore, does not qualify as an acceptable PDF model. 

Is the failure of kinetic descriptions only due to specific closures or to the sole situation of small discrete particles in non-fully-resolved turbulent flows? Or is it related to a poor choice of the variables entering the particle state vector and the loss of Markovianity? In relation to these queries is the remark that, when considering homogeneous situations (for the sake of avoiding more complex notations in general flows, see in-depth discussions of these relations in~\cite{minier2015kinetic}), $\lambda_{ij}$ and $\mu_{ij}$ are simply expressed as the correlations between particle positions and velocities and the variable that is eliminated from the reduced PDF description, that is the velocity of the fluid seen:
\begin{subequations}
\label{homogeneous correlation relations}
\begin{align}
\lambda_{ij} &= \frac{1}{\tau_{\rm p}^{\rm St}}\left\langle u_{{\rm s},i}(t)\, X_{{\rm p},j}(t) \right\rangle~, \\
\mu_{ij}     &= \frac{1}{\tau_{\rm p}^{\rm St}}\left\langle u_{{\rm s},i}(t)\, U_{{\rm p},j}(t) \right\rangle ~.
\end{align}
\end{subequations} 
This raises the question if leaving out $\mb{U}_{\rm s}$ was appropriate. These issues are now investigated.

\subsection{Slow and fast variables and the Markovian approach}\label{statistical model transport: Synergetics}
%========================================================================

\subsubsection{Colored or white noises and well-posed PDF equations}
%-------------------------------------------------------------------

The failure of the kinetic description for discrete particles in non-fully-resolved turbulent flows can be cast in the more general framework of dynamical systems under the influence of `external noises'~\cite{haunggi1994colored,van1989langevin,van1998remarks,van2007stochastic}. To shed light on this issue, we consider a dynamical system characterized by a state vector $\mb{Z}(t)=(Z_1(t),Z_2(t),\ldots,Z_n(t))$ with evolution equations
\begin{equation}
\label{general evolution eq}
\frac{dZ_i}{dt}= A_i(t,\mb{Z}) + B_{ij}(t,\mb{Z})\,\xi_j
\end{equation}
where $\bds{\xi}(t)=(\xi_j(t))_{j=1,n}$ represents an `external noise'. For the sake of simplicity since it does not affect results to come, the dependence of $\mb{A}$ and $\mb{B}$ is simply written as $\mb{A}(t,\mb{Z})$ and $\mb{B}(t,\mb{Z})$ instead of a more general dependence including statistics of the stochastic process (such as its mean, variance, etc.). 

When the external noise $\bds{\xi}$ is a Gaussian process with independent values, we are evolving within the well-established framework of Ito SDEs and Fokker-Planck equations, cf. Sec.~\ref{Ito SDE and Fokker-Planck}. The situation is different for `colored noises', that is when $\bds{\xi}$ has a non-zero correlation timescale. This is found, for example, for stationary Gaussian processes simulated as a set of independent Ornstein-Uhlenbeck (OU) processes
\begin{equation}
\label{OU colored noise}
d\xi_k = -\frac{\xi_k}{\tau}\, dt + \sigma\, dW_k~,
\end{equation}
where $\tau$ is the timescale of $\xi_k$ and $\sigma$ a constant equal to $\sigma^2=2\lra{\xi_k^2}/\tau$ from the classical fluctuation-dissipation theorem~\cite{gardiner2009stochastic}. For the sake of simplicity, $\tau$ and $\sigma$ are retained for each component $\xi_k$ (differences can be accounted for through the choice of $\mb{B}$).  The auto-correlation of this process is an exponential function $\lra{\xi_k(t)\xi_k(s)}=\lra{\xi^2}\exp(- \vert t-s \vert/\tau )$ and is therefore not delta-correlated when $\tau \neq 0$. An important element is that for such external noises, the process $\mb{Z}$ in Eq.~\eqref{general evolution eq} is no longer Markovian~\cite{van1998remarks,van2007stochastic,risken1996fokker} although a classical remark is to note that Markovianity is retrieved by considered the extended process $(\mb{Z},\bds{\xi})$~\cite{van1998remarks,van2007stochastic,risken1996fokker}. Even in a non-Markovian case, we can still consider the equation satisfied by the one-time PDF of the process $p(t;\mb{z})$, from which a set of PDEs are derived for some relevant statistical moments. It is therefore essential that this equation be mathematically well-posed for the resulting macroscopic descriptions to be regarded as acceptable, regardless of the choices of $\mb{A}$ and $\mb{B}$ in Eq.~\eqref{general evolution eq}.

\paragraph{PDF equation and the well-posed criterion}
The PDF equation for Gaussian colored noise is derived by standard techniques~\cite{pope2000turbulent} from the fine-grained PDF $\mc{P}(t;\mb{z})=\delta(\mb{Z}(t) - \mb{z})$ with $p(t;\mb{z})=\lra{ \mc{P}(t;\mb{z}) }$. The exact but unclosed PDF equation for $p(t;\mb{z})$ is
\begin{equation}
\label{open exact PDF}
\frac{\partial p}{\partial t}= - \frac{\partial}{\partial z_i}\left[ A_i(t,\mb{z})\,p\, \right]
- \frac{\partial}{\partial z_i}\left[ B_{ik}(t,\mb{z})\lra{ \xi_k\,\vert\, (t,\mb{z})}\,p\, \right]
\end{equation}
which, using $\lra{X\mc{P}}= \lra{ X | (t,\mb{z})}p(t,\mb{z})$, can be written as
\begin{equation}
\label{open exact PDF bis}
\frac{\partial p}{\partial t}= - \frac{\partial}{\partial z_i}\left[ A_i(t,\mb{z})\,p\, \right]
- \frac{\partial}{\partial z_i}\left[ B_{ik}(t,\mb{z})\lra{ \xi_k\mc{P}} \right]~.
\end{equation}
In Eq.~\eqref{open exact PDF bis}, the open flux $\lra{ \xi_k\mc{P}}$ can be written as $\lra{ \xi_k\mc{F}[t;\bds{\xi}]}$, where $\mc{F}[\cdot]$ stands for a functional dependence since $\mc{P}$ can be seen as a functional of the Gaussian centered process $\bds{\xi}$. As for derivation of the kinetic PDF, we can apply the Furutsu-Novikov-Donsker (FND) relation~\cite{furutsu1963statistical,novikov1965functionals,donsker1964function}
\begin{equation}
\label{FND Lagrange formula}
\lra{ \xi_k(t) F[t;\bds{\xi}]}= \int_{0}^t \lra{ \xi_k(t)\xi_l(t')}\left\langle \frac{\delta F[t;\bds{\xi}]}{\delta \xi_l(t')} \right\rangle dt'.
\end{equation}
Then, using
\begin{subequations}
\begin{align}
\frac{\delta \mc{P}(t;\mb{z})}{\delta \xi_l(t')} &= - \frac{\delta Z_j(t)}{\delta \xi_l(t')}\frac{\partial \mc{P}(t;\mb{z})}{\partial z_j} \\
& = -\frac{\partial }{\partial z_j} \left[ \frac{\delta Z_j(t)}{\delta \xi_l(t')} \mc{P} \right] + \left( \frac{\partial }{\partial z_j}\left[ \frac{\delta Z_j(t)}{\delta \xi_l(t')} \right] \right) \mc{P}
\end{align}
\end{subequations}
and applying the averaging operator, we obtain
\begin{equation}
\label{closure FND}
\lra{ \xi_k\mc{P}} = \alpha_k(t,\mb{z})\, p(t;\mb{z}) - \frac{\partial \left[\, \lambda_{kj}(t,\mb{z})\, p(t;\mb{z})\, \right]}{\partial z_j}
\end{equation}
with
\begin{equation}
\alpha_k(t, \mb{z})= \int_{0}^t \lra{ \xi_k(t)\xi_l(t')}\left\langle \frac{\partial }{\partial z_j} \left[ \frac{\delta Z_j(t)}{\delta \xi_l(t')} \right] \vert (t,\mb{z}) \right\rangle dt'
\end{equation}
and where $\lambda_{kj}$ is given by
\begin{equation}
\label{definition lambda}
\lambda_{kj}(t,\mb{z}) = \int_{0}^t \lra{ \xi_k(t)\xi_l(t')}\left\langle \frac{\delta Z_j(t)}{\delta \xi_l(t')} \vert (t,\mb{z}) \right\rangle dt'~.
\end{equation}
On the other hand, applying directly the FND relation to $Z_j(t)$ in Eq.~\eqref{FND Lagrange formula} leads to
\begin{equation}
\label{property lambda bis}
\lra{ \xi_k Z_j(t)} = \int \lambda_{kj}(t,\mb{z})p(t;\mb{z}) d\mb{z} = \int_{0}^t \lra{ \xi_k(t)\xi_l(t')}\left\langle \frac{\delta Z_j(t)}{\delta \xi_l(t')} \right\rangle dt'~. 
\end{equation}
Combining Eqs.~\eqref{closure FND} and Eq.~\eqref{open exact PDF bis} gives the closed form for the one-time PDF equation 
\begin{equation}
\label{closed PDF}
 \frac{\partial p}{\partial t}= - \frac{\partial}{\partial z_i}\left[\, (A_i+B_{ik}\alpha_k)\,p\, \right]
+ \frac{\partial}{\partial z_i}\left[ B_{ik} \frac{\partial \left( \lambda_{kj}\, p \right)}{\partial z_j} \right]~,
\end{equation}
which can be re-arranged as
\begin{equation}
\label{closed PDF bis}
 \frac{\partial p}{\partial t}= - \frac{\partial}{\partial z_i}\left[ \widetilde{A}_i\,p \right]
+ \frac{1}{2}\frac{\partial^2}{\partial z_i\partial z_j}\left[ \widetilde{D}_{ij}\, p \right]~,
\end{equation}
with $\widetilde{A}_i= A_i + B_{ik}\alpha_k + \lambda_{kj}\partial B_{ik}/\partial z_j$ and $\widetilde{D}_{ij}$ the symmetrical matrix
\begin{equation}
\label{diffusion matrix Dij}
\widetilde{D}_{ij}(t,\mb{z})= B_{ik}(t,\mb{z})\lambda_{kj}(t,\mb{z}) + B_{jk}(t,\mb{z})\lambda_{ki}(t,\mb{z})~.
\end{equation}
The well-posed nature of the PDF equation, Eq.~\eqref{closed PDF bis}, is determined by the matrix $\widetilde{D}_{ij}$ in Eq.~\eqref{diffusion matrix Dij}. Indeed, if $\widetilde{D}_{ij}$ is not positive definite (there is at least one negative eigenvalue), this implies that a marginal of the one-time PDF appears as the `solution' of an `anti-diffusion' PDE which are ill-posed in the sense that they can only be solved for very special initial conditions. Hence, the well-posed criterion is to require that $\widetilde{D}_{ij}(t,\mb{z})$ have only positive (or null) eigenvalues, $\forall \mb{z}$. 

\paragraph{Analysis of the linear case}
In~\cite[section 9.2]{minier2016statistical}, the well-posed property of such PDF equations is investigated, especially for a dynamical system with a general structure containing kinetic descriptions, and it is shown that the positive nature of $D_{ij}$ is only obtained when taking the white-noise limit. In the present context, it is sufficient to consider a simpler situation where the drift vector is linear in $\mb{Z}(t)$ (or linearized around a given point in sample space). Then, Eq.~\eqref{general evolution eq} becomes
\begin{equation}
\label{linear case}
\frac{dZ_i}{dt}=- G_{ik}Z_k + B_{ik}\xi_k
\end{equation}
where $\mb{G}$ is a constant matrix representing return-to-equilibrium effects and where the colored noise $\bds{\xi}$ is a set of independent stationary OU processes as in Eq.~\eqref{OU colored noise}. In the linear case, the `response functions' $\delta Z_i(t)/\delta \xi_l(t')$ in Eq.~\eqref{definition lambda} are independent of the sample space value $\mb{z}$, showing that $\lambda_{ji} = \lra{Z_i\, \xi_j}$. In the stationary state where $\lambda_{ji}$ reach constant values, the correlations are easily derived through 
\begin{equation}
\frac{d\lambda_{ji}}{dt}=0 \Longrightarrow
\left( \delta_{ik} + \tau G_{ik} \right) \lra{Z_k\xi_j}= B_{ij}\tau \lra{\xi^2}~,
\end{equation}
which can be inverted to give $\lambda_{ji}=\tau \lra{\xi^2}\widetilde{G}^{-1}_{ik}B_{kj}$ with $\widetilde{\mb{G}}=\mathbb{1} + \tau\mb{G}$. 

To show that the positive-definite property of $\widetilde{D}_{ij}$ is not automatically satisfied, it is sufficient to consider a specific counter-example. Taking for instance a simple two-dimensional situation where $\mb{Z}=(Z_1,Z_2)$ with an isotropic noise term, \textit{i.e.} $\mb{B}=B\mathbb{1}$, and with a return-to-equilibrium matrix of the form 
\begin{equation}
\mb{G}=
\begin{pmatrix}
1 & -\kappa\\
-\kappa & 1 
\end{pmatrix}~.
\end{equation}
The evolution equations for this system are therefore
\begin{subequations}
\begin{align}
\frac{\dd Z_1}{\dd t} &= - Z_1 + \kappa Z_2 + B \xi_1~, \\
\frac{\dd Z_2}{\dd t} &= - Z_2 + \kappa Z_1 + B \xi_2~,
\end{align}
\end{subequations}
and from Eq.~\eqref{diffusion matrix Dij} we obtain that the determinant of the $(2\times 2)$ matrix $\widetilde{D}_{ij}$ is
\begin{equation}
det(\widetilde{\mb{D}}) = \displaystyle \frac{4 B^2 \left( \tau \lra{\xi^2} \right)^2 }{(1+\tau)^2- \tau^2 \kappa^2}~.
\end{equation}
As soon as $\tau \neq 0$, this determinant is negative for large values of $\kappa$ and, hence, such a system is ill-posed. In other words, as soon as the two components $Z_1$ and $Z_2$ are strongly coupled, the resulting probabilistic description becomes ill-based even for such a trivial system. Note for small values of $\tau$, the inverse matrix can be approximated by $\widetilde{\mb{G}}^{-1}= \mathbb{1} - \tau\mb{G} + \mc{O}(\tau^2)$, which shows that, for the general formulation in Eq.~\eqref{linear case}, the correlations $\lambda_{ji}$ can be written as
\begin{equation}
\lambda_{ji}= \tau \lra{\xi^2}B_{ij} - \tau^2 \lra{\xi^2}G_{ik}B_{kj} + \mc{O}(\tau^2)~.
\end{equation}
Using this approximation in Eq.~\eqref{diffusion matrix Dij}, the symmetrical matrix $\widetilde{D}_{ij}$ is obtained as
\begin{equation}
\label{diffusion matrix Dij approx}
\widetilde{D}_{ij}= 2\tau \lra{\xi^2} (BB^{\bot})_{ij} - \tau^2 \lra{\xi^2} C_{ij} + \mc{O}(\tau^2)~,
\end{equation}
where $\mb{C}= (BB^{\bot})G^{\bot} + G (BB^{\bot})$ is a symmetrical matrix. The first term on the rhs of Eq.~\eqref{diffusion matrix Dij approx} constitutes a positive-definite matrix but the second one is unsure. The important point is that this second term is explicitly dependent upon $\mb{G}$. The only possibility to ensure that the resulting matrix $\widetilde{D}_{ij}$ remains definite positive whatever the choice of $\mb{G}$ is to take the limit $\tau \to 0$. Yet, in order to retain a non-zero $\widetilde{D}_{ij}$ matrix, this limit must be taken as: $\tau \to 0$ with $\lra{\xi^2} \to \infty$, such that $\lim\limits_{\tau\to 0} \left( \tau \lra{\xi^2}\right)= K$ where $K$ is a positive constant. This corresponds to the Markovian approximation which is now introduced.

\subsubsection{The Markovian approach}
%-------------------------------------

It follows from the preceding discussions that there are two main modeling philosophies. The first one consists in keeping the same set of variables, for instance kinetic variables $\mb{Z}_{\rm p}^{\rm r}=(\mb{X}_{\rm p},\mb{U}_{\rm p})$, whatever the context. We have then to handle colored noises when `external forcing' involves non-zero time or space correlations with the risk of ending up with ill-posed PDF formulations, as in the case for discrete particles in non-fully-resolved turbulent flows. The second modeling philosophy consists in adjusting the particle state vector by including additional variables until the eliminated degrees of freedom and/or the `external forcing' can be treated as white-noise terms on the now-extended mechanical system under consideration. The technical derivations of the eigenvalues of the diffusion matrix should not hide the physical issues at stake. In a thermodynamic formulation, the existence of a negative eigenvalue indicates that one is trying to describe a system whose contact with the `external world' cannot be treated as a contact with a heat bath since it contains a (negative) correlation and, thus, an underlying order that needs to be taken into account. On the other hand, positive eigenvalues of the second-order matrix $\widetilde{D}_{ij}$ means that the corresponding effects can be regarded as the sum of uncorrelated `pure noise' perturbations, leading to real diffusive actions on the system. In this second situation, we can regard the extended system as being in contact with the equivalent of `heat bathes' and we are now dealing with Markovian systems. To underline the importance of this notion, we quote~\cite[page 213]{ottinger2005beyond} who gave an excellent presentation of this principle: \enquote{One needs to keep sufficiently many variables and the appropriate non-linearities for achieving a realistic description of a system by \textit{Markovian time-evolution equations}. We here insist on avoiding explicit memory effects and take the standpoint that memory effects always indicate the existence of unrecognized variables relevant to the definition of a proper system for understanding certain phenomena of interest.}

The application of this principle consists in classifying the degrees of freedom of a system as slow and fast variables with respect to an observation time $\Delta t$ which needs to be introduced (in a discrete time version, this observation time interval corresponds to the time step). Variables whose auto-correlation timescale are larger than $\Delta t$ are defined as slow variables while variables with an auto-correlation timescale smaller than $\Delta t$ are defined as fast variables. As such, this is not sufficient since the eliminated fast variables would appear as colored noises in the evolution equations of the retained slow variables. The search is therefore of a scale separation which allows to treat fast variables as white-noise effects while slow variables have not changed appreciably (this is the essence of the `slaving principle'). 

To exemplify this notion, it is instructive to consider a toy model involving a variable $X_{\rm slow}$ whose time-rate-of-change is $X_{\rm fast}$, i.e. $\dd X_{\rm slow}/\dd t=X_{\rm fast}$. In the spirit of the slaving principle, we consider that $X_{\rm fast}$ is a centered process which has reached its equilibrium distribution conditioned on a given value of the slow variable $X_{\rm slow}=x$. Therefore, $X_{\rm fast}$ is a stationary process with variance $\lra{X^2_{\rm fast}}$ and an auto-correlation function $R_{X_{\rm fast}}$, $\lra{X^2_{\rm fast}} R_{X_{\rm fast}}(t'-t) = \lra{ X_{\rm fast}(t) X_{\rm fast}(t')}$, which depends only on the time difference $t'-t$ and whose integral timescale is $T_{X_{\rm fast}}$. Straightforward calculations show that for $X_{\rm slow}(0)=0$
\begin{equation}
\frac{\dd \lra{X^2_{\rm slow}}}{\dd t}= 2 \lra{X^2_{\rm fast}} \int_0^t R_{X_{\rm fast}}(s)\, \dd s~.
\end{equation}
Since by definition $T_{X_{\rm fast}}=\int_0^{+\infty} R_{X_{\rm fast}}(s)\, \dd s$, we get that for `long-enough time lapses'
\begin{equation}
\label{eq: diffusive regime Xslow}
t \gg T_{X_{\rm fast}} \; \Longrightarrow \; \lra{X^2_{\rm slow}}(t) \simeq \left( 2 \lra{X^2_{\rm fast}} T_{X_{\rm fast}} \right) t~,
\end{equation}
which is the linear behavior for the second-order moment of $X_{\rm slow}$ characterizing the diffusive regime. If we wish to obtain the same result at any time $t$ (of the order of the observation time), we need to take the limit of vanishing timescale $T_{X_{\rm fast}}$ and, to retain a non-zero diffusive coefficient for the evolution of $\lra{X^2_{\rm slow}}$ in Eq.~\eqref{eq: diffusive regime Xslow}, we are led to assume that we have
\begin{equation}
T_{X_{\rm fast}} \to 0 \; \text{and} \; \lra{X^2_{\rm fast}} \to +\infty \quad \text{such that} \quad \lra{X^2_{\rm fast}} T_{X_{\rm fast}} \to \mc{D}~,
\end{equation}
where $\mc{D}$ appears as the diffusion coefficient for $X_{\rm slow}$. This corresponds to the white-noise limit for $X_{\rm fast}$ and, in terms of the equation for the trajectories of $X_{\rm slow}$, we have replaced an ODE by a SDE
\begin{equation}
\label{eq: shit ODE to SDE}
\dd X_{\rm slow} = X_{\rm fast}\,\dd t \; \Longrightarrow \; \dd X_{\rm slow} = \sqrt{2 \mc{D}}\, \dd W~,
\end{equation}
where $W$ is a Wiener process. Note that the resulting diffusion coefficient $\mc{D}$ can be written as
\begin{equation}
\label{eq: Green-Kubo formula for D}
\mc{D}= \int_0^{\tau} \lra{ X_{\rm fast}(t) X_{\rm fast}(0)} \dd t~,
\end{equation}
where $\tau$ is an intermediate timescale separating $X_{\rm slow}$ from the rapidly-varying part of its time derivative, which is here $X_{\rm fast}$ (as long as $\tau \gg T_{X_{\rm fast}}$, the upper limit of the integral does not modify significantly the integral compared to $\tau=+\infty$). This is actually a Green-Kubo expression~\cite{ottinger2005beyond}. Following the slaving principle, an heuristic formulation consists in keeping the same results conditioned on a given value $x$ of $X_{\rm slow}$, so that we obtain $\mc{D}(x)$ as in Eq.~\eqref{eq: Green-Kubo formula for D} with the averaging operator being $\lra{ \cdot \, \vert \, X_{\rm slow}=x}$. This turns out to be valid for the diffusion coefficient resulting from the elimination of fast variables but the shift from an ODE to a SDE, as in Eq.~\eqref{eq: shit ODE to SDE}, may involve additional drift terms when $T_{X_{\rm fast}}(x)$ is an explicit function of the slow variable. In the course of the following sections, these questions resurface regularly and they are addressed more rigorously in Sec.~\ref{sec: soft matter Brownian limit}. Going back to the well-posed criterion for PDF equations involving colored noises, it is seen that the conclusion reached to guarantee the positive nature of the diffusion matrix $D_{ij}$ in Eq.~\eqref{diffusion matrix Dij approx} is the same one with $\tau$, $\lra{\xi^2}$ and $K$ being replaced here by $T_{X_{\rm fast}}$, $\lra{X^2_{\rm fast}}$ and $\mc{D}$, respectively. The conclusion is that well-posed formulations are obtained when taking the white-noise limit based on a scale separation to distinguish between slow and (very) fast variables. 

For discrete particles in non-fully-resolved turbulent flows, the Markovian approach leads to include the velocity of the fluid seen in the particle state vector (having similar timescale as $\mb{U}_{\rm p}$ for relatively low-inertia particles, $\mb{U}_{\rm s}$ is clearly a slow variable). However, do we need additional variables with, for instance, the time derivative of $\mb{U}_{\rm s}$? Answers to this question are provided by the Kolmogorov theory.

\subsection{The Kolmogorov theory and Lagrangian models}\label{The Kolmogorov theory}
%=======================================================

\begin{figure}[ht]
\centering
\begin{subfigure}{0.5\textwidth}
  \includegraphics[scale=1.2]{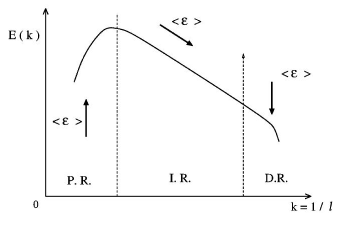}
	\caption{The energy cascade}
	\label{fig: Kolmogorov picture turbulence a}
\end{subfigure}
\hspace{2em}
\begin{subfigure}{0.4\textwidth}
  \includegraphics[scale=0.55]{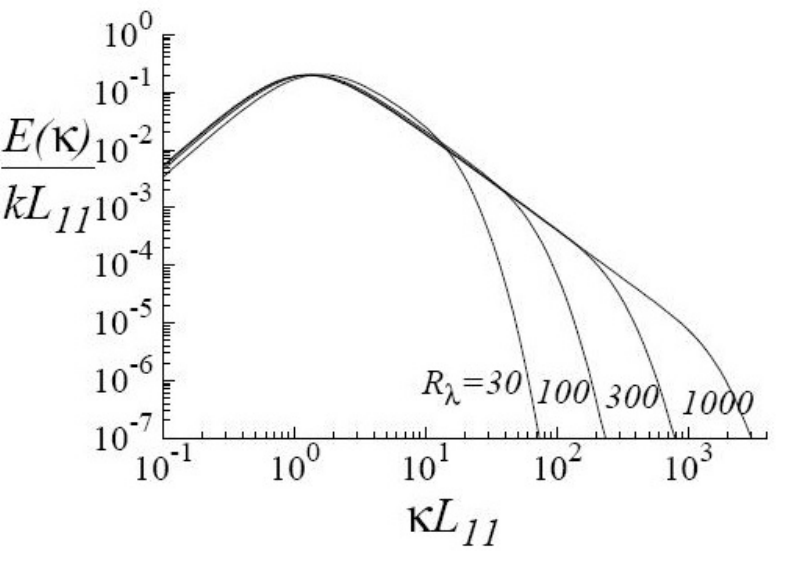}
	\caption{Energy spectra as a function of $Re$}
	\label{fig: Kolmogorov picture turbulence b}
\end{subfigure}
\caption{The Kolmogorov picture of turbulence: (a) Energy is produced at the large scales and is transferred through the inertial range before being dissipated by viscous motions at the same rate $\lra{\epsilon_{\rm f}}$; (b) Evolution of the energy spectra when the Reynolds number based on the Taylor length scale ${\rm Re}_{\lambda} \sim {\rm Re}^{1/2}$ increases (reprinted from~\cite{pope2000turbulent} with permission from Cambridge University Press).}
\label{fig: Kolmogorov picture turbulence}
\end{figure}

In spite of some limitations, the Kolmogorov description of turbulent flows remains the reference theory in turbulence modeling. The first theory was presented in 1941 (the K41 theory) and later refined to account for intermittency in 1962 (the refined or K62 theory) \cite{monin2013statistical, frisch1995turbulence, hunt1991turbulence}. Interestingly, A. N. Kolmogorov, who was one of the most brilliant mathematician of the 20th century, chose a rather qualitative approach based on the image of an energy cascade from which statistical predictions are derived. As in Richardson's first pictorial description in 1922 (\enquote{Big whorls have little whorls, which feed on their velocity, and little whorls have lesser whorls, and so on until viscosity}), the energy cascade corresponds to a description in terms of what is loosely defined as `an eddy' (or a velocity fluctuating `component'), characterized by its `size' $l$, velocity scale $\delta u_{\rm f}(l)$ and timescale $\tau_{\rm f}(l)$: energy is produced at the large scales imposed by the geometry of the flow domain or the boundary conditions and is transferred through the inertial range (eddies for which energy is neither created nor dissipated) until it is dissipated at the smallest scales by viscous motions, see Fig.~\ref{fig: Kolmogorov picture turbulence a}. Given its central role, the Kolmogorov theory has made its way in classical textbooks on turbulence and extensive accounts are available~\cite{monin2013statistical, frisch1995turbulence, pope2000turbulent}. Consequently, we only give a brief outline of its main characteristics with a view towards its application for Lagrangian stochastic models

Since traditional presentations tend to concentrate on spatial correlations, it is worth recalling that the fundamental K41 theory is more general and proceeds from a Lagrangian vision. In that sense, the most comprehensive description remains the one given in~\cite[chapter 8]{monin2013statistical}. It defines the notion of locally isotropic turbulence by considering the fields relative to a chosen fluid particle in a small space-time region around that moving particle. For velocity statistics, this means that we consider the relative field, $\delta \mb{v}_{\rm f}(\tau,\mb{r})= \mb{U}_{\rm f}(t_0+\tau, \mb{x}) - \mb{U}_{\rm f}(t_0,\mb{x}_0)$, in the reference frame moving with the velocity of a chosen fluid particle, $\mb{U}_{\rm f}(t_0,x_0)$, and where the space coordinate $\mb{r}$ is $\mb{r}=\mb{x} - \mb{x}_0 - \mb{U}_{\rm f}(t_0,\mb{x}_0)\, \tau$. Based on this description, the K41 theory states that, for high-Reynolds-number turbulent flows and for small enough $r=\vert \mb{r} \vert$ and $\tau$, turbulence is locally isotropic (in the small space-time region defined by $\mb{r}$ and $\tau$ around the moving fluid particle). Then, the K41 first similarity hypothesis is that statistics within that small space-time region are uniquely determined by $\lra{\epsilon_{\rm f}}$ (which represents also the rate of energy transfer in the inertial range), $\nu_{\rm f}$ (the fluid viscosity) and the space or time coordinates $\mb{r}$ and $\tau$, but not by the specific velocity of the `observation fluid particle ' $\mb{U}_{\rm f}(t_0,\mb{x}_0)$.

The first outcome of the Kolmogorov theory is the expression of the length, velocity and time scales of the smallest scales of turbulence  
\begin{equation}
\label{eq sec6: Kolmogorov time and length scales}
\eta_{\rm K}=\left( \frac{\nu_{\rm f}^3}{\lra{\epsilon_{\rm f}}} \right)^{3/4}, \quad u_{\rm K}= \left( \nu_{\rm f} \lra{\epsilon_{\rm f}} \right)^{1/4}, \quad \tau_{\rm K}= \left( \frac{\nu_{\rm f}}{\lra{\epsilon_{\rm f}}} \right)^{1/2}~.
\end{equation}
If we introduce $L_{\rm f}$ and $u_{\rm f}$ the length and velocity of the large-scale motions and use the estimation $\lra{\epsilon_{\rm f}} \sim u_{\rm f}^3/L_{\rm f}$, we obtain the ratios between the length and time scales of the largest to the smallest scales 
\begin{equation}
\label{eq Kolmogorov time and length ratios}
\frac{L_{\rm f}}{\eta_{\rm K}} \sim {\rm Re}^{3/4}~, \quad \frac{T_{\rm f}}{\tau_{\rm K}} \sim {\rm Re}^{1/2}~,
\end{equation}
with $T_{\rm f}=L_{\rm f}/u_{\rm f}$ and ${\rm Re}=u_{\rm f} L_{\rm f}/\nu_{\rm f}$ the Reynolds number based on the large scales. This provides a way to assess the complexity involved in turbulence. Indeed, a complete spatial resolution of the velocity field (i.e., capturing all the turbulent eddies) in a three-dimensional flow implies that we must handle a number of degrees of freedom that scales as ${\rm Re}^{9/4}$ which have, furthermore, to be tracked in time (just to simulate one large-eddy turnover time $T_{\rm f}$ we must adopt a time resolution of the order of $\tau_{\rm K}$ implying another ${\rm Re}^{1/2}$ factor in the effort required). Given that many flows corresponds to Reynolds numbers of the order of $10^6-10^8$ (for example, in the atmospheric boundary layer), this estimation demonstrates that we are dealing with huge numbers of degrees of freedom so that statistical reduced descriptions are unavoidable. 

The Kolmogorov scales allow to properly define the inertial range as the space and time domain where $\eta_{\rm K} \ll r \ll L_{\rm f}$ and $\tau_{\rm K} \ll \tau \ll T_{\rm f}$. Then, the second Kolmogorov similarity hypothesis assumes that, in the inertial range, statistics do not depend anymore on the fluid viscosity $\nu_{\rm f}$. This yields scaling relations for the eddy velocity and time scales in the inertial range 
\begin{equation}
\label{eq Kolmogorov eddy characteristics}
\delta u_{\rm f}(l) = \left( \lra{\epsilon_{\rm f}} \, l \right)^{1/3} \sim u_{\rm f} \left( l/L_{\rm f} \right)^{1/3}~, \quad \tau_{\rm f}(l) = \left( l^2/\lra{\epsilon_{\rm f}} \right)^{1/3} \sim \frac{L_{\rm f}}{u_{\rm f}} \left( l/L_{\rm f} \right)^{2/3}~.
\end{equation}

\paragraph{Eulerian statistics of velocity differences}

The most usual application is for the Eulerian fluid velocity correlations which are characterized by the tensor 
\begin{subequations}
\label{eq Kolmogorov Eulerian structure functions}
\begin{align}
D_{{\rm f},ij}(t,\mb{x},\mb{r})&=\lra{\, \delta v_{{\rm f},i}(0,\mb{r}) \, \delta v_{{\rm f},j}(0,\mb{r}) \,}~, \\
 & = \lra{ \, \left[ \, U_{{\rm f},i}(t,\mb{x}+\mb{r}) - U_{{\rm f},i}((t,\mb{x}) \, \right]\, \left[\, U_{{\rm f},j}(t,\mb{x}+\mb{r}) - U_{{\rm f},j}((t,\mb{x}) \, \right] \,}.
\end{align}
\end{subequations} 
When isotropy prevails, this velocity-structure tensor does not depend on $\mb{x}$ anymore and is written as \cite{monin2013statistical,pope2000turbulent}
\begin{equation}
\label{eq: isotropic form structure function}
D_{{\rm f},ij}(t,\mb{r})= D_{\rm f, NN}(t,r)\delta_{ij} + \left[ D_{\rm f, LL}(t,r) - D_{\rm f, NN}(t,r) \right] \frac{r_i r_j}{r^2}~,
\end{equation}
where the two scalar functions $D_{\rm f, LL}(t,r)$ and $D_{\rm f, NN}(t,r)$ are the longitudinal and transverse structure functions, respectively. Moreover, the continuity constraint implies that $D_{\rm f, NN}$ is uniquely determined by $D_{\rm f, LL}$ and  $D_{{\rm f},ij}(t,\mb{r})$ is therefore fully characterized by one scalar function $D_{\rm f, LL}(t,r)$. In the inertial range, the second Kolmogorov similarity hypothesis yields then that 
\begin{equation}
D_{\rm f, LL}(t,r)= C_2 \left( \lra{\epsilon_{\rm f}} \, r \right)^{2/3}~,
\end{equation}
where $C_2$ is a constant. This is the same result as the one given in Eqs.~\eqref{eq Kolmogorov eddy characteristics} since $D_{\rm f, LL}$ represents $\left(\delta u_{\rm f}(l)\right)^2$ (using $\lra{\epsilon_{\rm f}} \sim u_{\rm f}^3/L_{\rm f}$). The Fourier transform of $D_{\rm f, LL}(t,r)$ gives the (longitudinal) kinetic energy spectrum with the well-known $-5/3$ variation in the inertia range (see Fig.~\ref{fig: Kolmogorov picture turbulence b}). This point has been the subject of numerous experimental and numerical studies ever since the 1960s and detailed discussions can be found in the references mentioned above. When $\nu_{\rm f} \to 0$ (with fixed $u_{\rm f}$ and $L_{\rm f}$), ${\rm Re} \to +\infty$ but the kinetic energy dissipation rate $\lra{\epsilon_{\rm f}}$ tends towards a finite value which is the only remaining trace of the vanishing viscosity. Smaller and smaller scales are generated and the energy spectrum is stretched to larger and larger wave numbers but with the same slope, as shown in Fig.~\ref{fig: Kolmogorov picture turbulence b}.

\paragraph{Statistics of temporal velocity increments}

The Eulerian velocity difference at a fixed point $\mb{x}_0$ and at two instants $t_0$ and $t_0 + \tau$ can be expressed in terms of
the velocity field $\delta \mb{v}_{\rm f}$ defined in the reference system moving with the velocity $\mb{U}_{\rm f}(t_0,\mb{x}_0)$ as
\begin{equation}
\delta \mb{U}^{(t_0,\mb{x}_0)}_{\rm f} (\tau) \simeq \delta \mb{v}_{\rm f}(\tau, -\mb{U}_{\rm f}(t_0,\mb{x}_0)\tau).
\end{equation}
However, the situation is more complicated than for the Eulerian increments since $\delta \mb{U}^{(t_0,\mb{x}_0)}_{\rm f} (\tau)$ depends explicitly on the reference velocity $\mb{U}_{{\rm f},0}=\mb{U}_{\rm f}(t_0,\mb{x}_0)$. Therefore, we can only conclude from the Kolmogorov theory that there is a conditional probability distribution for $\delta \mb{U}^{(t_0,\mb{x}_0)}_{\rm f} (\tau)$. For a given value of $\mb{U}_{{\rm f},0}$, the tensor $D_{{\rm f},ij}^{(t_0,\mb{x}_0)} (\tau)=\lra{ \delta U^{(t_0,\mb{x}_0)}_{{\rm f},i} (\tau) \, \delta U^{(t_0,\mb{x}_0)}_{{\rm f},j} (\tau)}$ depends on the two functions $D^{(t_0,\mb{x}_0)}_{{\rm f},||}$ and $D^{(t_0,\mb{x}_0)}_{{\rm f},\bot}$ which correspond to the longitudinal and transverse directions, as in Eq.~\eqref{eq: isotropic form structure function}. These two functions depend on $\tau$ as well as on $\mb{r}=\mb{U}_{{\rm f},0} \tau$ and in the inertial range we have
\begin{equation}
D^{(t_0,\mb{x}_0)}_{{\rm f},||} = \lra{\epsilon_{\rm f}}\,\tau \,\alpha_{||}\left( \frac{ \vert \mb{U}_{{\rm f},0} \vert^2}{ \lra{\epsilon_{\rm f}} \tau}\right), \quad D^{(t_0,\mb{x}_0)}_{{\rm f},\bot} = \lra{\epsilon_{\rm f}}\,\tau \,\alpha_{\bot}\left( \frac{ \vert \mb{U}_{{\rm f},0} \vert^2}{ \lra{\epsilon_{\rm f}} \tau}\right),
\end{equation}
where $\alpha_{||}$ and $\alpha_{\bot}$ are universal functions which have to be specified. The dependence on $\mb{U}_{{\rm f},0}$ and the conditional nature of the previous results can be removed by resorting to the Taylor, or frozen turbulence, hypothesis. According to this hypothesis, the turbulent fluctuations are much smaller than the mean velocity, i.e. $\mb{U}_{{\rm f},0} \simeq \lra{ \mb{U}_{{\rm f},0}}$. Over a small time interval $\tau$, the turbulent fluctuations at a fixed point are regarded as being transported without modification at a constant velocity $\lra{ \mb{U}_{{\rm f},0}}$. This frozen-turbulence assumption removes the dependence on $\mb{U}_{{\rm f},0}$ and allows to write the statistics of $\delta \mb{U}^{(t_0,\mb{x}_0)}_{\rm f} (\tau)$ in terms of the velocity-structure functions derived for the Eulerian space increments by replacing $\mb{r}$ by $\lra{ \mb{U}_{{\rm f},0}} \tau$. This leads to
\begin{equation}
D^{(t_0,\mb{x}_0)}_{{\rm f},||} = D_{||} ( \lra{ \mb{U}_{{\rm f},0}} \tau ),
\quad D^{(t_0,\mb{x}_0)}_{{\rm f},\bot} = D_{\bot} (\lra{ \mb{U}_{{\rm f},0}} \tau ),
\end{equation}
and thus in the inertial range to
\begin{equation}
D^{(t_0,\mb{x}_0)}_{{\rm f},||} = C \left( \lra{\epsilon_{\rm f}}\lra{\mb{U}_{{\rm f},0}}\, \tau \right)^{2/3},
\quad D^{(t_0,\mb{x}_0)}_{{\rm f},\bot} = \frac{4}{3}\,C \left( \lra{\epsilon_{\rm f}}\lra{\mb{U}_{{\rm f},0}}\, \tau \right)^{2/3}.
\end{equation}

\paragraph{Lagrangian statistics for particle velocity increments}
The Kolmogorov theory can be directly applied to the velocity increments of a fluid particle $\dd \mb{U}_{\rm f}(\tau)= \mb{U}_{\rm f}(t+\tau)-\mb{U}_{\rm f}(t)$ which, by choosing this particle as the reference one in the Kolmogorov approach, are written as $\dd \mb{U}_{\rm f}(\tau)=\delta \mb{v}_{\rm f}(\tau,0)$. The locally isotropic nature of turbulent flows implies that, for $\tau \ll T_{\rm f}$, the Lagrangian second-order velocity structure function $\lra{\dd U_{{\rm f},i}(\tau)\, \dd U_{{\rm f},j}(\tau)}$ has an isotropic form
\begin{equation}
\lra{\dd U_{{\rm f},i}(\tau)\, \dd U_{{\rm f},j}(\tau)} = D_{\rm f}^{\rm L} (\tau) \, \delta_{ij}~.
\end{equation}
For time differences in the inertial range, i.e. $\tau_{\rm K} \ll \tau \ll T_{\rm f}$, the second Kolmogorov hypothesis gives that 
\begin{equation}
\label{eq sec6: Lagrangian structure function C0}
D_{\rm f}^{\rm L} (\tau) = C_0 \, \lra{\epsilon_{\rm f}} \, \tau~,
\end{equation}
with $C_0$ a constant. This is a significant result since the linear variation in $\tau$ of the second-order moment of velocity increments is a signature of white-noise effects and of a diffusive behavior of fluid particle velocities. 

While there is no well-marked separation in terms of length scales, it is important to note that there is one in terms of timescales. More precisely, it is seen that $\delta u_{\rm f}(l)$ diminishes as $l$ becomes smaller but without any sharp decrease between two comparable scales (the decrease is continuous) whereas there is a clear-cut distinction between the correlation timescales of a fluid particle velocity $\mb{U}_{\rm f}$ and its acceleration $\mb{A}_{\rm f}$. Fluid particle velocities are governed by the large-scale motions of a turbulent flow and scale as $(\mb{U}_{\rm f})^2 \sim u_{\rm f}^2$ with a timescale $T_{\rm L} \sim T_{\rm f}=u_{\rm f}^2/\lra{\epsilon_{\rm f}}$, while fluid particle accelerations are governed by the small-scale motions and scale as $(\mb{A}_{\rm f})^2 \sim \lra{\epsilon_{\rm f}}/\tau_{\rm K}$ with a timescale $\tau_{\rm A}$ which is of the order of the Kolmogorov timescale $\tau_{\rm A} \simeq \tau_{\rm K}$. This situation plays a key role in the analysis of stochastic models. Indeed, we can use the result obtained with the toy model presented above, with $X_{\rm slow}$ being the fluid particle velocity $U_{\rm f}$ and $X_{\rm fast}$ its acceleration $A_{\rm f}$ (in a one-dimensional notation). In high-Reynolds-number flows where $\tau_{\rm A} \ll T_{\rm L}$ and with $\lra{\epsilon_{\rm f}}$ remaining finite as indicated above, we retrieve the white-noise limit since we have
\begin{equation}
\tau_{\rm A} \to 0 \; \text{and} \; \lra{A_{\rm f}^2} \to +\infty~, \quad \text{such that} \quad \tau_{\rm A} \lra{A_{\rm f}^2} \to \lra{\epsilon_{\rm f}}~.
\end{equation}
This is in line with the prediction already given in Eq.~\eqref{eq sec6: Lagrangian structure function C0} and this indicates that, in the inertial range, we expect fluid particle velocity increments to be described by a stochastic model containing a white-noise term such as $\sqrt{C_0 \, \lra{\epsilon_{\rm f}}}\, \dd \mb{W}$, as will be seen in more detail in Sec.~\ref{Modeling fluid seen}. 

\paragraph{Intermittency and refined hypothesis}

So far, this outline of the Kolmogorov theory has essentially followed the K41 picture in that the rate of kinetic transfer and dissipation $\lra{\epsilon_{\rm f}}$ is taken as a parameter rather than as a random variable. This is related to the intermittency of the flow produced by the inherent transfer processes of turbulence, referred to as `inner intermittency' (to be distinguished from external intermittency due, for example, to large-scale mixing between laminar and turbulent flows) and we have assumed that $\epsilon_{\rm f} \simeq \lra{\epsilon_{\rm f}}$. While the above picture can be kept substituting $\lra{\epsilon_{\rm f}}$ with $\epsilon_{\rm f}$ , especially for second-order moments where the impact of intermittency is small, a refined description based on a log-normal distribution for the dissipation of kinetic energy was developed in the K62 theory. For this question, the basic presentation of the refined similarity hypothesis is~\cite[section 25.2]{monin2013statistical} and this question has been investigated in several studies, in particular in~\cite{frisch1995turbulence}. Without addressing in detail this issue, related comments are proposed in connection with coherent structures in Sec.~\ref{Accounting for structures signature}.

\paragraph{Particle state vectors and Lagrangian models}

From these results, it appears that the Kolmogorov theory provides clear indications as to which state vectors should be selected to follow the Markovian approach. For tracer particles, the relevant state vector is $\mb{Z}_{\rm f}=(\mb{X}_{\rm f},\mb{U}_{\rm f})$ while for discrete particles the velocity of the fluid seen should be included, leading to $\mb{Z}_{\rm p}=(\mb{X}_{\rm p},\mb{U}_{\rm p},\mb{U}_{\rm s})$, since we can expect the `acceleration' of the fluid seen (or its time-rate-of-change) to be also a fast variable. In terms of the trajectories of the stochastic process $\mb{Z}_{\rm p}$, the general structure of one-particle PDF models for discrete particles in turbulent flows is
\begin{subequations}
\label{eq: general structures Zp}
\begin{align}
\dd \mb{X}_{\rm p} & = \mb{U}_{\rm p}\, \dd t~, \label{eq: general structures Zp a} \\
\dd \mb{U}_{\rm p} & = \frac{ \mb{U}_{\rm s} - \mb{U}_{\rm p}}{\tau_{\rm p}}\, \dd t + \mb{F}_{\rm f \to p}\, \dd t~, \label{eq: general structures Zp b} \\
\dd \mb{U}_{\rm s} & = (\text{stochastic model})~, \label{eq: general structures Zp c}
\end{align}
\end{subequations}
where $\mb{F}_{\rm f \to p}$ in Eq.~\eqref{eq: general structures Zp b} stands for additional forces due to the fluid acting on particles (i.e. other than drag). The task ahead is to work out appropriate stochastic models for $\mb{U}_{\rm s}$ in Eq.~\eqref{eq: general structures Zp c}.

%================================================
\section{Modeling the velocity of the fluid seen \label{Modeling fluid seen}}
%================================================

\subsection{The physics of particle dispersion} \label{Physics Particle Dispersion}
%==============================================

\subsubsection{The crossing-trajectory effect} \label{the crossing-trajectory effect}
%---------------------------------------------

In a nutshell, particle dispersion in turbulent flows is due to the fluctuating fluid velocities encountered by discrete particles and is, thus, governed by the statistical properties of the velocity of the fluid seen $\mb{U}_{\rm s}(t)$. Our purpose, in this section, is to bring out the main physical phenomena at play and the relevant variables that need to be accounted for when modeling $\mb{U}_{\rm s}(t)$. 

\begin{figure}[ht]
\includegraphics[width=0.9\textwidth]{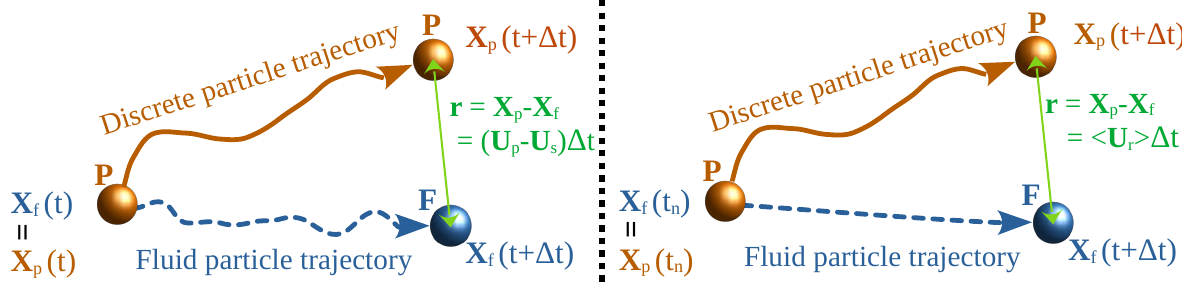}%
\caption{Illustration of the CTE and of its modeled representation. The CTE occurs as soon as there is a relative velocity between discrete particles and the fluid, due to particle inertia and mean velocity drifts (left). In the present formulations, it is accounted for only due to mean relative velocities (right).}
\label{fig: instantaneous separation Particle-Fluid}
\end{figure}

The problem can be described as follows (see also Fig.~\ref{fig: instantaneous separation Particle-Fluid}). At time $t$, we consider a discrete particle (P) located at a position $\mb{X}_{\rm p}(t)$, with a velocity $\mb{U}_{\rm p}(t)$ and a velocity of the fluid seen $\mb{U}_{\rm s}(t)$. Then, after a small time interval $\Delta t$, the discrete particle has a probability to move to a downstream location $\mb{X}_{\rm p}(t+\Delta t)$, while a fluid particle (F) starting from $\mb{X}_{\rm p}(t)$ with the velocity $\mb{U}_{\rm s}(t)$ at time $t$ has a probability to move to another position $\mb{X}_{\rm f}(t+\Delta t)$. The modeling issue is to estimate $\mb{U}_{\rm s}(t+\Delta t)$, which is the velocity of another fluid particle than (F), say (F'), at time $t+\Delta t$. In most cases, due to particle inertia and/or external forces inducing mean velocity slips $\lra{\mb{U}_{\rm r}}$ (with $\mb{U}_{\rm r}=\mb{U}_{\rm s} - \mb{U}_{\rm p}$ the relative velocity), the trajectories of the discrete particle (P) and of the fluid particle (F) located at the same point at time $t$ separate during the time interval $\Delta t$. This is referred to as the crossing-trajectory effect (CTE). As such, the CTE is related to particle inertia and the various effects mentioned above but its precise definition in the present modeling context will be narrowed down below. When we re-express the velocity of the fluid seen in terms of the instantaneous fluid velocity field $\mb{U}_{\rm f}(t,\mb{x})$, it appears that the modeling issue is to simulate $\mb{U}_{\rm s}(t+\Delta t)=\mb{U}_{\rm f}(t+\Delta t, \mb{X}_{\rm p}(t+\Delta t))$ knowing $\mb{U}_{\rm s}(t)=\mb{U}_{\rm f}(t, \mb{X}_{\rm p}(t))$. This clearly involves a two-time two-point conditional fluid velocity correlation, which reveals that it can only be treated without approximation at the level of two-particle PDF models, i.e. when following pairs of fluid particles or when generating spatial information about the fluid velocity field in the vicinity of every discrete particle being tracked. At the level of one-particle PDF models, this constitutes an open issue and additional assumptions have to be made.

\vspace*{1em}

To pave the way for the more precise developments to come, we first outline the expected statistical characteristics of the velocity of the fluid seen. As pictured in Fig.~\ref{fig: instantaneous separation Particle-Fluid}, two main physical processes are involved, namely transport and relaxation processes. However, once $\mb{U}_{\rm s}(t)$ is properly defined (as the fluid velocity sampled along discrete particle trajectory), transport processes are implicitly accounted for and we can concentrate on describing how $\mb{U}_{\rm s}(t)$ relaxes towards mean fluid velocities (assuming, therefore, that it does). This means that we essentially consider the relaxation timescale(s) of $\mb{U}_{\rm s}(t)$, which is taken as the same as its integral timescale noted $T_{\rm L}^{*}$. In non-homogeneous situations, relaxation is expressed by a matrix in the return-to-equilibrium term but, in the present context, it is sufficient to handle an isotropic matrix involving only one coefficient whose inverse is the correlation timescale $T_{\rm L}^{*}$. We need therefore to assess how $T_{\rm L}^{*}$ varies as a function of $T_{\rm L}$ and $T_{\rm E}$, the Lagrangian and Eulerian timescales, respectively, and particle-related statistics, such as  $\lra{\mb{U}_{\rm r}}$,  $\lra{\mb{U}^2_{\rm p}}$,  $\lra{\mb{U}_{\rm p}\, \mb{U}_{\rm s}}$, etc. To that effect, it proves useful to follow the analysis carried out in~\cite{minier2001pdf,minier2004pdf} and to evaluate separately the effects of particle inertia on $T_{\rm L}^{*}$: (a) in the absence of a mean velocity slip; and (b) when a mean drift is present. 

\begin{figure}[ht]
\includegraphics[width=0.9\textwidth]{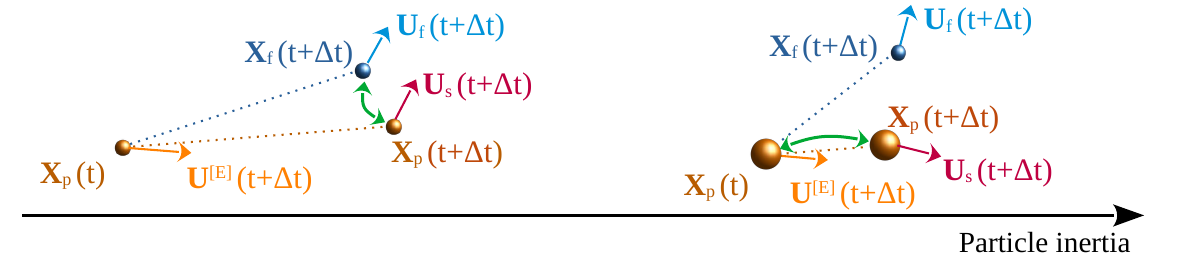}%
\caption{Estimations of the integral (or correlation) timescale of the velocity of the fluid seen with respect to the Lagrangian and Eulerian timescales in the absence of a mean velocity drift between discrete particles and the fluid. Low-inertia particles tend to follow fluid particle trajectories so that the statistics of $\mb{U}_{\rm s}(t)$ are close to those of a fluid particle velocity $\mb{U}_{\rm f}(t)$, which means that $T_{\rm L}^{*} \simeq T_{\rm L}$ (left side). High-inertia particles do not move very much over a time interval $\Delta t$ so that the statistics of $\mb{U}_{\rm s}(t)$ are close to those of the fluid velocity $\mb{U}^{[\rm E]}(t)$ at a fixed location, which means that $T_{\rm L}^{*} \simeq T_{\rm E}$ (right side).}
\label{fig: particle inertia effect}
\end{figure}

(a) Even in the absence of mean velocity differences, there is a CTE due to particle inertia, measured by the Stokes number $St=\tau_{\rm p}/T_{\rm L}$ which is the ratio of the particle relaxation timescale and the Lagrangian timescale of fluid velocities. As depicted in Fig.~\ref{fig: particle inertia effect}, low-inertia discrete particles tend to follow the fluid, which means that we have $T_{\rm L}^{*} \simeq T_{\rm L}$ when $St \ll 1$. Since $\lra{\mb{u}_{\rm r}^2} \simeq St/(1+St)$, there is always a separation when particle inertia is not negligible and this has led to the introduction of an additional step to try to correlate the velocities of the two fluid particles located at $\mb{X}_{\rm f}(t+\Delta t)$ and at $\mb{X}_{\rm p}(t+\Delta t)$ at time $t+\Delta t$ (this is referred to as the Eulerian step) once the velocity of the first fluid particle, i.e. $\mb{U}_{\rm f}(t+\Delta t)$, is generated from $\mb{U}_{\rm s}(t)$ (this is referred to as the Lagrangian step). This is, however, misleading since we should not build on one sample (we cannot actually say that `the fluid particle goes there' or `has this velocity' but simply that it has a probability to do so) and, more importantly, these two steps are not independent. This was nevertheless treated as such in earlier works which led to spurious de-correlation effects in the resulting timescale $T_{\rm L}^{*}$. This is revealed by considering the other limit case of high-inertia particles. In that case, we expect discrete particles not to travel a great distance compared to nearby fluid particles, since $\lra{\mb{u}^2_{\rm p}} \simeq \lra{\mb{u}^2_{\rm f}}/(1 + St) \ll \lra{\mb{u}^2_{\rm f}}$ and, thus, to be essentially correlated with the fluid velocity at time $t+\Delta t$ at the initial discrete particle location $\mb{X}_{\rm p}(t)$ which is noted $\mb{U}^{[\rm E]}(t+\Delta t)$ in Fig.~\ref{fig: particle inertia effect}. In short, in that case, the velocity of the fluid seen is similar to Eulerian fluid velocities at a given position and we expect that $T_{\rm L}^{*} \simeq T_{\rm E}$ when $St \gg 1$. In between these two limits, we can derive from this description that $T_{\rm L}^{*}$ remains of the order of $T_{\rm L}$ and $T_{\rm E}$, and is a function of the Stokes number without being able to work out the exact dependency in the one-particle PDF framework. Note that this does not imply that $T_{\rm L}^{*}$ be a monotonous function of $St$ between the two limit values of $T_{\rm L}$ and $T_{\rm E}$ (discrete particles can be trapped in specific coherent structures whose timescales are different, cf. Sec.~\ref{sec: particle preferential concentration effect}). In most cases, and although detailed information is not always available, it is estimated that $T_{\rm L}$ and $T_{\rm E}$ are comparable. As a first approximation and leaving out the possible effects of coherent structures (which, if need be, can be introduced explicitly as discussed in Sec.~\ref{Accounting for structures signature}), we retain the simplified notion that, in the absence of mean velocity slips, we have $T_{\rm L}^{*} \simeq T_{\rm L}$.

(b) In the case when a mean velocity slip $\lra{\mb{U}_{\rm r}}$ is present at the discrete particle location at time $t$, the potential trajectories of the discrete particle and the fluid ones starting from $\mb{X}_{\rm p}(t)$ at time $t$ with the velocity $\mb{U}_{\rm s}(t)$ always separate, cf. the rhs in Fig.~\ref{fig: instantaneous separation Particle-Fluid}. Moreover, since these fluid particles are not affected by the mean velocity drifts, a systematic de-correlation is induced between the velocity of the fluid seen at the time $t+\Delta t$, $\mb{U}_{\rm s}(t+\Delta t)$, and the `equivalent fluid particle' whose velocity timescale is between $T_{\rm L}$ and $T_{\rm E}$ (as a result of the first step) and which is represented by (F) at time $t+\Delta t$ on the rhs in Fig.~\ref{fig: instantaneous separation Particle-Fluid}. The essential point is that, since it is due to mean velocity differences, this effect is independent of the first one (the `inertia effect') and can be introduced successively in a step-by-step model construction.

Based on this approach in terms of, first, particle-inertia effects in the absence of mean velocity drifts and, second, mean-velocity-slip effects inducing a de-correlation between the velocities of discrete particles and nearby fluid ones, the emerging picture is one of a two-step process where each step can be addressed independently. We deduce from this analysis that the Lagrangian timescale of the velocity of the fluid seen can be written as $T_{\rm L}^{*}=T_{\rm L}^{*}(\lra{\mb{U}_{\rm r}}=0) \times F (\lra{\mb{U}_{\rm r}};T_{\rm L};T_{\rm E},k_{\rm f};\cdots)$ where $F$ is a correcting factor accounting for the effects of mean drifts and is a function of particle and fluid statistics (thus, with $F(\lra{\mb{U}_{\rm r}}=0;T_{\rm L};T_{\rm E};k_{\rm f},\cdots)=1$). As indicated above, a further simplification consists in neglecting the influence of particle inertia on $T_{\rm L}^{*}(\lra{\mb{U}_{\rm r}}=0)$ and in writing $T_{\rm L}^{*}(\lra{\mb{U}_{\rm r}}=0)\simeq T_{\rm L}$. This means that in the absence of mean velocity slips, the statistics of the velocity of the fluid seen are the same as those of a fluid particle and that the CTE is considered from now on as being due only to the existence of mean velocity drifts, cf. Fig.~\ref{fig: instantaneous separation Particle-Fluid}. Since $F=T_{\rm L}^{*}/T_{\rm L}$ is a non-dimensional function, it is best expressed as a function of non-dimensional variables, i.e. $F=F( \lra{\mb{U}_{\rm r}}/u_{\rm f}; C_{\rm T}; \cdots)$,  with $u_{\rm f}=(2/3 k_{\rm f})^{1/2}$ the characteristic fluid turbulent velocity and where $C_{\rm T}$ is the ratio of the Lagrangian to Eulerian timescales, $C_{\rm T}=T_{\rm L}/T_{\rm E}$. Given that the difference between $T_{\rm L}$ and $T_{\rm E}$ is neglected in $T_{\rm L}^{*}(\lra{\mb{U}_{\rm r}}=0)$, it would be consistent to have $C_{\rm T}=1$. Yet, to capture the correct behavior in the limit of high mean velocity drifts when the frozen-turbulence hypothesis is applied with (small) differences between $T_{\rm L}$ and $T_{\rm E}$, we retain the parameter $C_{\rm T}$, with $C_{\rm T} \sim \mc{O}(1)$.

\subsubsection{Stochastic diffusion models for the velocity of the fluid seen}
%-----------------------------------------------------------------------------

The description of the crossing-trajectory effect in terms of the mean velocity slip between particles and the fluid allows us to recast the issue of modeling the velocity of the fluid seen into the framework of the Kolmogorov theory. Indeed, the increments of $\mb{U}_{\rm s}$ over a small time interval $\dd t$ can be written as
\begin{equation}
\label{eq: dUs in Kolomogorov theory}
\dd \mb{U}_s = \mb{\delta v}_{\rm f} (\dd t, \lra{ \mb{U}_{\rm r} } \, \dd t ),
\end{equation}
where $\delta \mb{v}_{\rm f}(t,\mb{x})$ represents the fluid velocity field relative to the motion of a fluid particle that was located at the particle location at time $t$ (cf. Sec.~\ref{The Kolmogorov theory}), as sketched on the rhs of Fig.~\ref{fig: instantaneous separation Particle-Fluid}.

According to the Kolmogorov theory, the statistics of $\dd \mb{U}_{\rm s}$ do not depend on the instantaneous velocity of the fluid particle selected to express $\dd \mb{U}_{\rm s}$ through the observation velocity field $\delta \mb{v}_{\rm f}(t,\mb{x})$ in Eq.~\eqref{eq: dUs in Kolomogorov theory} but only on some statistics and, in high-Reynolds-number flows and for a time increment $\dd t$ in the inertial range, it follows from the locally isotropic hypothesis that the second-order structure functions of the velocity of the fluid seen, $D_{{\rm s}, ij}(\dd t)=\lra{\dd U_{{\rm s},i} \dd U_{{\rm s},j}}$, is determined by the two scalars functions $D_{{\rm s}, ||}$ and $D_{{\rm s}, \bot}$ through
\begin{equation}
D_{{\rm s},ij} = D_{{\rm s}, \bot}\delta_{ij} + \left[ D_{{\rm s}, ||} - D_{{\rm s}, \bot} \right] \widehat{r}_i \, \widehat{r}_j~,
\end{equation}
with the separation unit vector $\widehat{\mb{r}}$ being in the direction of the mean relative velocity, i.e. $\widehat{\mb{r}}=\lra{\mb{U}_{\rm r}}/ \vert \lra{\mb{U}_{\rm r}} \vert $. The functions $D_{{\rm s}, ||}$ (resp. $D_{{\rm s}, \bot}$) represents the correlations for the velocity components aligned with the separation vector $\mb{r}$ (resp. transverse to $\mb{r}$), similarly to the situation with Eulerian velocity differences~\cite{monin2013statistical}. For a time increment in the inertial range, the Kolmogorov theory implies that $D_{{\rm s}, ij}$ depends only on the mean dissipation rate $\lra{\epsilon_{\rm f}}$, $\dd t$ and $\lra{\mb{U}_{\rm r}}$. Then, in the same manner as what was done in Sec.~\ref{The Kolmogorov theory}, we obtain
\begin{equation}
\label{eq: Us velocity structure functions}
D_{{\rm s}, ||}(dt) = \lra{\epsilon_{\rm f}}\, \dd t\, \alpha_{||}\left( \, \frac{ \vert \lra{\mb{U}_{\rm r}} \vert^2}{\lra{\epsilon_{\rm f}} \dd t }\, \right), \qquad D_{{\rm s}, \bot}(\dd t) = \lra{\epsilon_{\rm f}}\, \dd t\, \alpha_{\bot}\left( \, \frac{ \vert \lra{\mb{U}_{\rm r}} \vert^2}{\lra{\epsilon_{\rm f}} \dd t }\, \right),
\end{equation}
where $\alpha_{||}(x)$ and $\alpha_{\bot}(x)$ are regarded as two universal functions of a single parameter $x=\vert \lra{\mb{U}_{\rm r}} \vert^2/\left(\lra{\epsilon_{\rm f}}\dd t\right)$, whose form can be obtained in two limit cases. First, when the mean relative velocity is small, the statistics of the velocity of the fluid seen should be close to the fluid ones which means that we have
\begin{equation}
\frac{\vert \lra{\mb{U}_{\rm r}} \vert^2}{\lra{\epsilon_{\rm f}} \dd t } \ll 1 \quad \Longrightarrow \quad \alpha_{||} \simeq \alpha_{\bot} \simeq C_0 . 
\end{equation}
Second, when the relative mean velocity is large, we can use the frozen turbulence hypothesis to obtain, similarly to the case of temporal velocity increments (cf. Sec.~\ref{The Kolmogorov theory}), that
\begin{equation}
\frac{\vert \lra{\mb{U}_{\rm r}} \vert^2}{\lra{\epsilon_{\rm f}} \dd t } \gg 1 \quad \Longrightarrow \quad D_{{\rm s}, ||}(dt) \simeq C (\lra{\epsilon_{\rm f}}\, \lra{\mb{U}_{\rm r}} \, \dd t)^{2/3}, \quad D_{{\rm s}, \bot}(dt) \simeq \frac{4}{3} \,C (\lra{\epsilon_{\rm f}}\, \lra{\mb{U}_{\rm r}} \, dt)^{2/3},
\end{equation}
showing that, in this limit, the two functions $\alpha_{||}(x)$ and $\alpha_{\bot}(x)$ vary as $x^{1/3}$. Apart from the limit case of fluid particles, the variation of these functions and their dependencies on the  time increment implies that $D_{{\rm s}, ||}(\dd t)$ and $D_{{\rm s}, \bot}(\dd t)$ are not linear in $\dd t$ and, therefore, do not exhibit the typical signature of stochastic diffusion processes. An approximation can, however, be worked out by freezing the values of the functions $\alpha_{||}$ and $\alpha_{\bot}$ at a certain value of the time interval, say $\Delta t_{\rm r}$, which gives
\begin{equation}
D_{{\rm s}, ||}(\dd t) \simeq \lra{\epsilon_{\rm f}}\, \dd t\, \alpha_{||}\left( \, \frac{ \vert \lra{\mb{U}_{\rm r}} \vert^2}{\lra{\epsilon_{\rm f}} \Delta t_{\rm r} }\, \right), \qquad D_{{\rm s},\bot}(\dd t) \simeq \lra{\epsilon_{\rm f}}\, \dd t\, \alpha_{\bot}\left( \, \frac{ \vert \lra{\mb{U}_{\rm r}} \vert^2}{\lra{\epsilon_{\rm f}} \Delta t_{\rm r} }\, \right)~.
\end{equation}
A physically-sound choice for $\Delta t_{\rm r}$ is the Lagrangian timescale which is the timescale over which fluid velocities remain correlated. Using the simple estimation $\Delta t_{\rm r} \simeq T_{\rm L} \simeq k_{\rm f}/\lra{\epsilon_{\rm f}}$, we get
\begin{equation}  
\label{structure modified}
D_{{\rm s}, ||}(\dd t) \simeq \lra{\epsilon_{\rm f}}\, \dd t\, \alpha_{||}\left( \, \frac{ \vert \lra{\mb{U}_{\rm r}} \vert^2}{k_{\rm f}}\, \right),
\qquad D_{{\rm s},\bot}(\dd t) \simeq \lra{\epsilon_{\rm f}}\, \dd t\, \alpha_{\bot}\left( \, \frac{ \vert \lra{\mb{U}_{\rm r}} \vert^2}{k_{\rm f}}\, \right).
\end{equation}

As for the case of fluid particles, the linear-in-time variation of the second-order moments of the increments of the velocity of the fluid seen in Eqs.~\eqref{structure modified} suggests to model $\mb{U}_{\rm s}$ by a stochastic diffusion process. It is also seen that this modeling step has less support than in the fluid case since additional assumptions have to be made, especially concerning how spatial correlations are accounted for. In spite of these limitations, it appears nevertheless that there is sufficient support to model $\mb{U}_{\rm s}$ by a stochastic diffusion process. With this choice, the Langevin model for the velocity of the fluid seen can be written as 
\begin{equation}
\label{eq: general Langevin model for Us}
\dd U_{{\rm s},i}= A_{{\rm s},i}(t,\mb{Z}_{\rm p},\lra{\mc{G}(\mb{Z}_{\rm p})},\bds{\Phi}_{\rm f}(\mb{X}_{\rm p}))\, \dd t + B_{{\rm s},ij}(t,\mb{Z}_{\rm p},\lra{\mc{G}(\mb{Z}_{\rm p})},\bds{\Phi}_{\rm f}(\mb{X}_{\rm p}))\, \dd W_j,
\end{equation}
where the drift vector $\mb{A}_{\rm s}$ and the diffusion matrix $\mb{B}_{\rm s}$ have to be modeled. In Eq.~\eqref{eq: general Langevin model for Us}, a general notation is used to indicate that the drift and diffusion coefficients depend not only on the particle state vector $\mb{Z}_{\rm p}$ but also on the value of mean fluid fields at the particle positions noted $\bds{\Phi}_{\rm f}(\mb{X}_{\rm p})$ as well as on statistics derived from the particle set, written $\lra{\mc{G}(\mb{Z}_{\rm p})}$ (e.g., the mean particle velocity field). Yet, for the sake of simplicity, these functional dependencies are omitted from now on. The selection of a Langevin model for $\mb{U}_{\rm s}$ means that the particle state vector $\mb{Z}_{\rm p}$ is modeled as a stochastic diffusion process with the form
\begin{subequations} \label{eq:sde}
\begin{align}
\dd X_{{\rm p},i} & = U_{{\rm p},i}\, \dd t, \\
\dd U_{{\rm p},i} & = A_{{\rm p},i}\, \dd t, \\
\dd U_{s,i} & = A_{{\rm s},i}\, \dd t + B_{{\rm s},ij}\, \dd W_j,
\end{align}
\end{subequations}
where the particle acceleration is often limited to $\mb{A}_{\rm p}=(\mb{U}_s - \mb{U}_p)/\tau_p + \mb{g}$ (i.e. considering only contributions from hydrodynamic drag and gravity). This formulation is equivalent to a Fokker-Planck equation for the corresponding PDF $p(t;\mb{y}_{\rm p},\mb{V}_{\rm p},\mb{V}_{\rm s})$ in sample space
\begin{equation} \label{fokker-planck:fluid seen}
\frac{\partial p}{\partial t} + \frac{\partial }{\partial y_{{\rm p},i}}\left[ V_{{\rm p},i}\, p \, \right] = - \frac{\partial }{\partial V_{{\rm p},i}}\left[ A_{{\rm p},i}\, p \, \right] - \frac{\partial }{\partial V_{{\rm s},i}}\left[ A_{{\rm s},i}\, p \, \right] + \frac{1}{2}\frac{\partial ^2}{\partial V_{{\rm s},i}\partial V_{{\rm s},j}}\left[ (B_{\rm s} B_{\rm s}^T)_{ij}\, p \, \right]. 
\end{equation}

It can be noted that the approximations leading to Eqs.~\eqref{structure modified} are not only useful to point to stochastic diffusion processes in order to model $\mb{U}_{\rm s}$ but also to bring out the non-dimensional mean velocity slip, $\vert \lra{\mb{U}_{\rm r}} \vert^2/k_{\rm f}$, which enters typical closures of the Lagrangian timescale of the velocity of the fluid seen, such as the Csanady's formulas to be introduced in Sec.~\ref{Langevin models for Us} and in new proposals presented in Sec.~\ref{sec: new macroscopic approaches}.

\subsubsection{Remarks on the particle preferential concentration effect}\label{sec: particle preferential concentration effect}
%------------------------------------------------------------------------

In a particle-based approach, the incompressibility of the flow has a double manifestation for fluid particles or fluid-like elements. It requires that the mean or filtered velocity field derived from particle instantaneous velocities be of zero divergence but, also, that the distribution of particle positions be uniform. Since each fluid particle represents the same fixed amount of mass, a deviation of their position PDF from uniformity signifies that there is an accumulation or a depletion of mass in some areas, at variance with the mass continuity equation. This is not necessarily the case for discrete particles (either solid particles or bubbles) which can concentrate in certain zones of a turbulent flow, often in connection with the existence of specific flow patterns. This phenomenon is referred to as the particle preferential concentration \cite{eaton1994preferential,balachandar2010turbulent}. 

This effect has been recognized for some time with, for example, the first observations of particle concentration at the outskirts of the large-scale vortices forming downstream of the separating plate in a turbulent mixing layer (see, for instance, an historical account in~\cite{monchaux2012analyzing}). It has then received increasing attention and the literature is vast on the subject particularly since the advent of high-resolution DNS which give access to statistics that are difficult to measure, such as the dissipation rate near a wall. In practice, particle preferential concentration is essentially addressed either in wall boundary layers or in homogeneous isotropic situations. The former is in relation with particle transport in the near-wall region and the all-important practical concern of particle deposition while the latter considers flows in which it was believed that preferential concentration was not likely to occur. 

\vspace*{1em}

\begin{figure}[ht]
\includegraphics[scale=0.9]{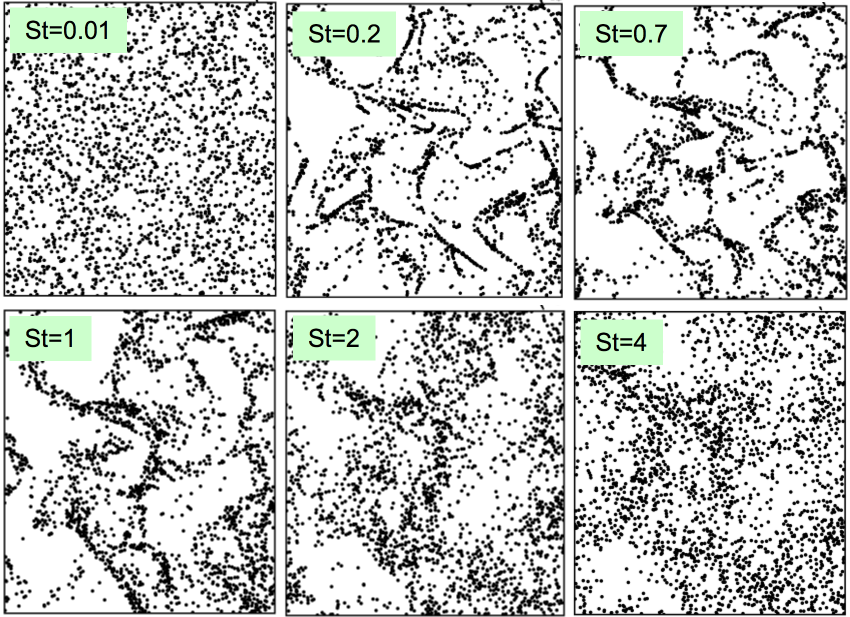}%
\caption{Snapshots of instantaneous particle positions in HIT obtained from tracking discrete particles in well-resolved DNS for different Stokes numbers based on the Kolmogorov timscale: (a) $St_{\eta_{\rm K}}=0.01$; (b) $St_{\eta_{\rm K}}=0.2$; (c) $St_{\eta_{\rm K}}=0.7$; (d) $St_{\eta_{\rm K}}=1$; (e) $St_{\eta_{\rm K}}=2$ and $St_{\eta_{\rm K}}=4$. Reprinted with permission from \cite{monchaux2012analyzing}. Copyright Elsevier, 2012.}
\label{fig: preferential concentration DNS}
\end{figure}

Homogeneous isotropic turbulence (HIT) is an interesting situation to consider since all fluid statistics are uniform and it could be expected that an initially uniform distribution of discrete particles would remain so. As revealed by instantaneous snapshots, cf. Fig.~\ref{fig: preferential concentration DNS}, concentration buildups can however be observed in some zones while other areas are nearly devoid of particles. This effect depends on particle inertia which is measured here by the Stokes number $St_{\eta_{\rm K}}=\tau_{\rm p}/\tau_{\rm K}$ based on the Kolmogorov timescale $\tau_{\rm K}$. As mentioned above, particles with negligible inertia (fluid-like elements) remain uniformly distributed and, at the other extreme of the range of Stokes numbers, heavy particles show the same tendency since they are insensitive to fluid structures and, in that sense, filter out any flow patterns that can be formed. Preferential concentration is most marked for particles with Stokes numbers $St_{\eta_{\rm K}}= \mc{O}(1)$, which can be attributed to a `resonance effect'. Indeed, insightful analyses have revealed that homogeneous fluid flows in the statistical sense can nevertheless contain characteristic flow features, exhibiting an organized pattern locally in space and time and which, for that reason, are called `coherent structures'. Typical examples of such fluid flow topology are vortex or saddle points. Many efforts have been devoted to characterize the geometry and dynamics of these coherent structures. When discrete elements have the `right inertia', they can correlate with some specific structures or be expelled from others leading to the observed concentration buildup or depletion. For instance, particles heavier than the fluid are preferentially found in low vorticity and high strain-rate regions while bubbles are captured in low-pressure zones \cite{eaton1994preferential}. A noteworthy consequence is that the statistics of the `turbulence seen by discrete particles' can be different from the fluid ones (this bias effect is significant for bubbles but is considered less noticeable for small solid particles).

Particle preferential concentration is also interesting in that it challenges traditional statistical views. Once a statistical averaging operator is introduced, it is tempting to regard fluctuations as a manifestation of `disorder' (noise, heat, etc.) and averages as an expression of `order' (forces, work, etc.). This turns out to be misleading in some situations such as the one discussed here, which is helpful to remind us that disorder can turn out to be an order whose understanding was eluding us. From a modeling standpoint, the notion of coherent structures is attractive to physicists since they are related to small scales which, according to the Kolmogorov theory, are the only features susceptible to be universal. This is, however, a simplification and these structures cannot be classified as belonging to the small-scale world only. Furthermore, developments are presently hindered by the case-by-case approach mentioned above and a quantitative theory that goes beyond mere observations and that is not limited to specific turbulent flows has yet to emerge. Finally, it is worth recalling that dispersion effects remain governed by the large energy-containing scales which are usually flow dependent and that predicting the mean energy transfer rate or capturing the anisotropy of the Reynolds stresses remain major issues~\cite{pope2000turbulent}. In that sense, it is perhaps best to regard models developed in terms of coherent structures as complements to traditional statistical formulations. Further comments to that effect are proposed in Sec.~\ref{consistency issues} and, more specifically, in Sec.~\ref{Accounting for structures signature}, after having presented state-of-the-art stochastic models based on Langevin equations in Secs.~\ref{Fluid GLM models} and~\ref{Langevin models for Us}.

\subsection{Stochastic models for fluid particles} \label{Fluid GLM models}
%=================================================

Since two-phase PDF models are based on the ones developed for single-phase turbulence, it is useful to introduce first the basic features of Langevin models used for single-phase turbulent flows. This has been achieved essentially by Pope and co-workers through a series of papers, and detailed presentations can be found in several publications~\cite{pope1985pdf,pope1994lagrangian,pope2000turbulent,haworth2010progress}. A reference stochastic model is the GLM (Generalized Langevin Model) which represents fluid particle velocities by a stochastic diffusion process with the form~\cite{haworth1986generalized,pope1994relationship,pope2000turbulent}
\begin{subequations}
\label{model PDF fluid particle GLM}
\begin{align}
dX_{{\rm f},i} &= U_{{\rm f},i}\, dt~, \label{model PDF fluid particle GLM a}\\
dU_{{\rm f},i} &= -\frac{1}{\rho_{\rm f}}\frac{\partial \lra{ P_{\rm f} }}{\partial x_i}\, dt +
G_{ij} \left( U_{{\rm f},j}- \lra{U_{{\rm f},j}} \right)\, dt + \sqrt{C_0\lra{\epsilon_{\rm f}}}\, dW_i~. \label{model PDF fluid particle GLM b}
\end{align}
\end{subequations}
where the matrix $G_{ij}$ depends on the particle location and on statistics of the fluid flow but not on the particle velocity, \textit{i.e.} $G_{ij}=G_{ij}(t,\mb{X}_{\rm f}(t),\mc{F}[(\mb{X}_{\rm f},\mb{U}_{\rm f})])$ where the notation $\mc{F}[(\mb{X}_{\rm f},\mb{U}_{\rm f})]$ refers to fluid mean quantities given or calculated from the set of particles and interpolated at the particle position. Therefore, in homogeneous flows, $G_{ij}$ depends only on time and Eq.~\eqref{model PDF fluid particle GLM b} is a linear model for fluid particle velocities, while the complete particle system in Eqs.~\eqref{model PDF fluid particle GLM} is non-linear in general non-homogeneous situations~\cite{pope1994lagrangian,minier2001pdf}. For the sake of simplicity, these dependencies are considered as implicit and are not kept from now onwards. Similarly, the mean terms entering these equations are to be understood as the values at the particle location, e.g. $\lra{\mb{U}_{\rm f}}=\lra{\mb{U}_{\rm f}}(t, \mb{X}_{\rm f}(t))$.

As indicated by its name, the simplest model is the SLM (Simplified Langevin Model) where
\begin{equation}
G_{ij} = - \left( \frac{1}{2} + \frac{3}{4}C_0 \right)\frac{\lra{\epsilon_{\rm f}}}{k_{\rm f}}\, \delta_{ij}~,
\end{equation}
with $C_0$ the Kolmogorov constant (cf. Sec.~\ref{The Kolmogorov theory}) as in the diffusion coefficient in Eq.~\eqref{model PDF fluid particle GLM b}. 

A classical decomposition of the matrix $G_{ij}$ is \cite{pope1994lagrangian,pope1994relationship}
\begin{equation}
\label{GLM decomposition drift matrix G}
G_{ij} = - \left( \frac{1}{2} + \frac{3}{4}C_0 \right)\frac{\lra{\epsilon_{\rm f}}}{k_{\rm f}} \delta_{ij} + G_{ij}^{\rm a}~,
\end{equation}
where the matrix $G_{ij}^{\rm a}$ represents anisotropic effects and, to be consistent with the kinetic energy budget, is subject to the condition $\text{Tr}(\mb{G}^{\rm a} \cdot \mb{R}_{\rm f})=0$, with $\mb{R}_{\rm f}$ the Reynolds-stress tensor (i.e., $R_{{\rm f},ij}=\lra{u_{{\rm f},i} u_{{\rm f},j}}$ with $u_{{\rm f},i}=U_{{\rm f},i} - \lra{U_{{\rm f},i}}$ the fluctuating velocity component) if the diffusion coefficient is written as in Eq.~\eqref{model PDF fluid particle GLM b}. There are, in fact, several ways to express the GLM which are considered as equivalent (they belong to the same class of models) if they yield the same Reynolds-stress equations. For example, relaxing the constraint $\text{Tr}(\mb{G}^{\rm a} \cdot \mb{R}_{\rm f})=0$, it is possible to write a stochastic diffusion process for the particle velocity $\mb{U}_{\rm f}$ as
\begin{align}
\label{model PDF fluid particle GLM bis} 
dU_{{\rm f},i} = -\frac{1}{\rho_{\rm f}}\frac{\partial \lra{ P_{\rm f} }}{\partial x_i}\, dt
&- \left( \frac{1}{2} + \frac{3}{4}C_0 \right)\frac{\lra{\epsilon_{\rm f}}}{k_{\rm f}} \left( U_{{\rm f},i}- \lra{U_{{\rm f},i}} \right)\, dt + \sqrt{C_0\lra{\epsilon_{\rm f}}}\, dW^{(1)}_i \nonumber \\
& + G^a_{ij} \left( U_{{\rm f},j}- \lra{U_{{\rm f},j}} \right)\, dt + \sqrt{-2/3 \, \text{Tr}(\mb{G}^{\rm a}\,\mb{R}_{\rm f})}\, dW^{(2)}_i~,
\end{align}
where $\mb{W}^{(1)}$ and $\mb{W}^{(2)}$ are two independent Wiener processes. In a weak formulation, these two increments of Wiener processes can be added to give
\begin{equation}
\label{model PDF fluid particle GLM ter} 
dU_{{\rm f},i} = -\frac{1}{\rho_{\rm f}}\frac{\partial \lra{ P_{\rm f} }}{\partial x_i}\, dt
+ G_{ij} \left( U_{{\rm f},j}- \lra{U_{{\rm f},j}} \right)\, dt + \sqrt{\lra{\epsilon_{\rm f}}\left(C_0 -2/3 \frac{\text{Tr}(\mb{G}^{\rm a}\,\mb{R}_{\rm f})}{\lra{\epsilon_{\rm f}}}\right)}\, dW_i~.
\end{equation}
For example, if we consider that $G_{ij}^{\rm a}$ reproduces effects due to the fluid mean shear and is
\begin{equation}
\label{Ga LRR-IP}
G_{ij}^{\rm a}= C_2 \, \frac{\partial \lra{U_{{\rm f},i}}}{\partial x_j}~,
\end{equation}
where $C_2$ is a constant, we have then
\begin{equation}
dU_{{\rm f},i} = -\frac{1}{\rho_{\rm f}}\frac{\partial \lra{ P_{\rm f} }}{\partial x_i}\, dt
+ G_{ij} \left( U_{{\rm f},j}- \lra{U_{{\rm f},j}} \right)\, dt + \sqrt{\lra{\epsilon_{\rm f}}\left(C_0 + 2/3 \frac{\mc{P}_{\rm f}}{\lra{\epsilon_{\rm f}}}\right)}\, dW_i~,
\end{equation}
where $\mc{P}_{\rm f}$ is the turbulent kinetic energy production term, i.e., $\mc{P}_{\rm f}=1/2 \mc{P}_{{\rm f},kk}$ with $\mc{P}_{{\rm f},ij}$ the production tensor whose components are source/sink terms in the transport equation for the Reynolds stress components $R_{{\rm f},ij}$ and have the following expression  
\begin{equation} 
\label{eq: kinetic energy production term}
\mc{P}_{{\rm f},ij} = - \lra{u_{{\rm f},j} u_{{\rm f},k}} \frac{\partial \lra{U_{{\rm f},i}}}{\partial x_k} - \lra{u_{{\rm f},i} u_{{\rm f},k}} \frac{\partial \lra{U_{{\rm f},j}}}{\partial x_k}~, \; \mc{P}_{\rm f}=1/2 \mc{P}_{{\rm f},kk}~.
\end{equation}
With the form of $G^{\rm a}_{ij}$ in Eq.~\eqref{Ga LRR-IP}, the resulting matrix $G_{ij}$ given in Eq.~\eqref{GLM decomposition drift matrix G} retrieves the LRR-IP (Launder, Reece and Rodi model based on the isotropization of production (IP) model) second-order model~\cite{pope2000turbulent}. A word of caveat: while the diffusion coefficient in the first formulation, cf. Eq.~\eqref{model PDF fluid particle GLM b}, ensures its realizability (we are taking the square root of a positive quantity), the second formulation in Eq.~\eqref{model PDF fluid particle GLM ter} implies that the expression within the square root remains positive (which is generally the case). Therefore, it should be considered that diffusion coefficients written as $\sqrt{D}$ must, in fact, be read as $\sqrt{\min(0,D)}$ for the various functions $D$ that are considered in all the models presented in this section. 

\subsection{Langevin models for the velocity of the fluid seen}\label{Langevin models for Us}
%==============================================================

\subsubsection{State-of-the-art two-phase SLM} \label{two-phase SLM}
%---------------------------------------------

The current Langevin model for the velocity of the fluid seen is an extension of the SLM for fluid particle velocity and has the form~\cite{minier2001pdf,minier2004pdf,minier2014guidelines}
\begin{equation}
\label{complete Langevin stochastic model}
dU_{{\rm s},i} = -\frac{1}{\rho_{\rm f}}\frac{\partial \lra{P_{\rm f}}}{\partial x_i}\, dt
 + \left( \lra{U_{{\rm p},j}} - \lra{U_{{\rm f},j}} \right) \frac{\partial \lra{U_{{\rm f},i}}}{\partial x_j} dt 
 + G^{*}_{ij}\left( U_{{\rm s},j} - \lra{U_{{\rm f},j}} \right) dt + B_{{\rm s},ij}\, dW_j~.
\end{equation}
The matrix $G^{*}_{ij}$ is built from $G_{ij}$ used in the fluid-SLM and is given by
\begin{equation}
\label{eq matrix G* drift Us}
G^{*}_{ij}= -\left( \frac{1}{2} + \frac{3}{4} C_0 \right) \frac{\lra{\epsilon_{\rm f}}}{k_{\rm f}}\, H_{ij},
\end{equation}
where the matrix $H_{ij}$ accounts for the crossing-trajectory effect and is expressed by 
\begin{equation}
\label{simple expression Hij}
H_{ij} = b_{\bot}\delta_{ij} + \left[ \, b_{\vert \vert} - b_{\bot} \right]\, \widehat{r}_i \, \widehat{r}_j,
\end{equation}
with $(\widehat{r}_i)_{i=1,3}$ the components of the unit vector $\widehat{\mb{r}}$ aligned with the mean relative velocity, $\widehat{\mb{r}}=\lra{\mb{U}_{\rm r}}/ \vert \lra{\mb{U}_{\rm r}} \vert$. The coefficients $b_{\vert \vert}$ and $b_{\bot}$ represent the Csanady factors which stand for the ratio between the timescale of fluid particle velocities $T_{\rm L}$ and the timescale of the fluid velocities seen by discrete particles $T_{{\rm L},||}^{*}$ and $T_{{\rm L},\bot}^{*}$, in the direction parallel to the mean relative velocity or transverse to it, respectively. Using the Csanady formulas for these timescales (see \cite{minier2001pdf})
\begin{equation}
\label{Csanady timescales}
T_{{\rm L},||}^{*}= \frac{T_{\rm L}}{\sqrt{ 1 + C_{\rm T}^2 \displaystyle \frac{| \lra{\mb{U}_{\rm r}}|^2}{2k_{\rm f}/3}}} ~,
\qquad
T_{{\rm L},\bot}^{*}=\frac{T_{\rm L}}{\sqrt{ 1 + 4\, C_{\rm T}^2 \displaystyle \frac{| \lra{\mb{U}_{\rm r}}|^2}{2k_{\rm f}/3}}}~, 
\end{equation}
(remember that $C_{\rm T}=T_{\rm L}/T_{\rm E}$), the Csanady factors are obtained as $b_{\vert \vert}=T_{\rm L}/T_{{\rm L},||}^{*}$ and $b_{\bot}=T_{\rm L}/T_{{\rm L},\bot}^{*}$
\begin{subequations}
\label{definition coefficient b_l et b_t}
\begin{align}
b_{\vert \vert} & = \sqrt{ 1 + C_{\rm T}^2 \, \displaystyle \frac{| \lra{\mb{U}_{\rm r}}|^2}{2k_{\rm f}/3}}~,\\
b_{\bot}        & = \sqrt{ 1 + 4 \, C_{\rm T}^2 \, \displaystyle \frac{| \lra{\mb{U}_{\rm r}}|^2}{2k_{\rm f}/3}}~.
\end{align}
\end{subequations} 
Using the classical expression for $T_{\rm L}$ from the fluid SLM, whereby
\begin{equation}
\label{definition TL fluide}
T_L=\left( \frac{1}{1/2 + 3/4\,C_0}\right) \, \frac{k_{\rm f}}{\lra{\epsilon_{\rm f}}}
\end{equation} 
the matrix $G^{*}_{ij}$ can be re-expressed as
\begin{equation}
\label{eq matrix G* drift Us 2}
G^{*}_{ik}= - \frac{1}{T_{{\rm L},\bot}^{*}}\, \delta_{ik} - \left[ \frac{1}{T_{{\rm L},||}^{*}} - \frac{1}{T_{{\rm L},\bot}^{*}} \right]\, \widehat{r}_i\, \widehat{r}_k~.
\end{equation}
In Eq.~\eqref{complete Langevin stochastic model}, the diffusion matrix $B_{{\rm s},ij}$ is the square root of the matrix $L_{ij}$ (i.e. $\mb{B}_{\rm s}^{}\, \mb{B}_{\rm s}^{T}=\mb{L}$) given by
\begin{equation}
\label{formulation of Lij}
L_{ij} = L_{\bot}\delta_{ij} + \left[ \, L_{\vert \vert} - L_{\bot} \right]\, \widehat{r}_i\, \widehat{r}_j,
\end{equation}
where the coefficients $L_{\vert \vert}$ and $L_{\bot}$ are
\begin{subequations}
\label{expression Lij}
\begin{align}
L_{\vert \vert} &= \lra{\epsilon_{\rm f}} \left( C_0 b_{\vert \vert} \widetilde{k_{\rm f}}/k_{\rm f} + 
\frac{2}{3}( b_{\vert \vert}\widetilde{k_{\rm f}}/k_{\rm f} -1) \right),\\
L_{\bot} &= \lra{\epsilon_{\rm f}} \left( C_0 b_{\bot} \widetilde{k_{\rm f}}/k_{\rm f} + 
\frac{2}{3}( b_{\bot}\widetilde{k_{\rm f}}/k_{\rm f} -1) \right).
\end{align}
\end{subequations}
In these expressions, a new kinetic energy $\widetilde{k_{\rm f}}$ is introduced and is defined as
\begin{equation}
\label{eq sec7: dynamic PDF definition tilde k}
\widetilde{k_{\rm f}}=\frac{3}{2}\frac{\text{Tr}(\mb{H}\cdot \mb{R}_{\rm f})}{\text{Tr}(\mb{H})},
\end{equation}
where $\text{Tr}(\mb{H})=H_{ii}$ denotes the trace of the matrix $\mb{H}$. As indicated at the end of Sec.~\ref{Fluid GLM models}, the eigenvalues of $\mb{L}$ must remain positive so that we are actually handling $\min(0 ; L_{\vert \vert})$ and $\min(0 ; L_{\bot})$ with $L_{\vert \vert}$ and $L_{\bot}$ given in Eqs.~\eqref{expression Lij}. Note that the matrix equation $\mb{B}_{\rm s}=\mb{L}^{1/2}$ does not yield a unique solution but, since we are only interested in weak solutions, different solutions $\mb{B}_{\rm s}$ give the same statistics and, in a weak sense, are equivalent. It is also worth noting that the decomposition of $\mb{L}$ in Eq.~\eqref{formulation of Lij} follows the same one as $\mb{G}^{*}$ in Eqs.~\eqref{eq matrix G* drift Us} and~\eqref{simple expression Hij} since we can rewrite Eqs.~\eqref{formulation of Lij}-~\eqref{expression Lij} as
\begin{equation}
\label{eq: Lij as a function of Hij}
L_{ij}= \left( 1 + \frac{3}{2} C_0 \right) \frac{\lra{\epsilon_{\rm f}}}{k_{\rm f}} \frac{\text{Tr}(\mb{H}\cdot \mb{R}_{\rm f})}{\text{Tr}(\mb{H})} H_{ij} - \frac{2}{3} \lra{\epsilon_{\rm f}} \, \delta_{ij}~.
\end{equation}
Therefore, $\mb{G}^{*}$ and $\mb{L}$ are directly formulated in terms of the matrix $\mb{H}$ which appears as the key operator to go from the fluid-SLM to the two-phase-SLM resulting expressions. 

In~\cite{minier2014guidelines}, an analysis was performed to assess various modeling formulations with respect to a set of criteria. It appeared that only the two-phase SLM outlined above met all the requirements set forth, which is not a satisfactory situation. However, do we need more elaborate models than the SLM? Answers to that question depend on the statistical context in which these models are applied, that is whether we consider LES or RANS approaches.

\subsubsection{Applications and extensions in LES}
%-------------------------------------------------

For sufficiently well-resolved LES, not only the large energy-containing scales but also a substantial fraction of the energy spectrum are explicitly resolved, leaving small scales as the unresolved ones. The distinction between the large (resolved) and small (unresolved) scales is not always clear-cut since there is no physically well-justified scale separation (see Sec.~\ref{The Kolmogorov theory}) but we can consider that the unresolved ones are well inside the inertial range where the Kolmogorov theory prevails, at least as the reference framework. This does not mean that we assume these scales to be isotropic but, according to the Kolmogorov theory, we can expect that the driving mechanism towards isotropy is the dominant one. This explains that the SLM is often considered as an adequate model in LES, as reflected by a series of developments for dynamical and reactive flow applications~\cite{Gicquel_2002,Sheikhi_2003,Sheikhi_2007,Sheikhi_2009}.

Yet, the same reasoning suggests to consider also viscous terms since the unresolved velocity field can be a low-enough Reynolds-number flow to be impacted by molecular viscosity. The introduction of molecular transport coefficients in PDF/FDF formulations raises specific issues which have, however, not received much attention. To the authors' knowledge, only two proposals have been put forward and it is interesting to discuss the ideas behind these developments as well as to point to related challenges. The first proposition~\cite{dreeben1997probability,dreeben1998probability} consists in introducing a random-walk model added to convective displacements which, with Ito stochastic calculus and further modeling choices, yields the following model for fluid-particle positions and velocities
\begin{subequations}
\label{eq: viscous terms in PDF model DP}
\begin{align}
\dd X_{{\rm f},i} & = U_{{\rm f},i}\, \dd t + \sqrt{2\, \nu_{\rm f}}\, \dd W'_i~, \label{eq: viscous terms in PDF model DP Xf} \\
\dd U_{{\rm f},i} & = - \frac{1}{\rho_{\rm f}}\frac{\partial \lra{P_{\rm f}}}{\partial x_i}\, \dd t + \sqrt{2\, \nu_{\rm f}}\, \frac{\partial \lra{U_{{\rm f},i}}}{\partial x_k}\, \dd W'_k + G_{ik} \left( U_{{\rm f},k} - \lra{U_{{\rm f},k}} \right)\, \dd t + \sqrt{C_0\, \lra{\epsilon_{\rm f}}}\, \dd W_i~, \label{eq: viscous terms in PDF model DP Uf}
\end{align}
\end{subequations}
where $\mb{W}$ and $\mb{W}'$ are two independent vector Wiener processes. An important feature of this model is that the white-noise term, $\dd\mb{W}'$, used in Eq.~\eqref{eq: viscous terms in PDF model DP Xf} intervenes also in the particle velocity equation, cf. the second term on the rhs of Eq.~\eqref{eq: viscous terms in PDF model DP Uf}, leading to cross-correlations that need to be properly ascertained by carefully writing the corresponding Fokker-Planck equation~\cite{dreeben1997probability,dreeben1998probability,pope2000turbulent}.
The second proposition~\cite{waclawczyk2004probability} retains the random-walk formulation for particle positions but considers that the only remaining viscous effect to account for is the dissipation of the kinetic energy of the mean flow field. This leads to the model
\begin{subequations}
\label{eq: viscous terms in PDF model 1}
\begin{align}
\dd X_{{\rm f},i} & = U_{{\rm f},i}\, \dd t + \sqrt{2\, \nu_{\rm f}}\, \dd W'_i~, \label{eq: viscous terms in PDF model 1 Xf} \\
\dd U_{{\rm f},i} & = - \frac{1}{\rho_{\rm f}}\frac{\partial \lra{P_{\rm f}}}{\partial x_i}\, \dd t +  \left[ A_{{\rm vis}, ik} + G_{ik} \right] \left( U_{{\rm f},k} - \lra{U_{{\rm f},k}} \right)\, \dd t + \sqrt{C_0\, \lra{\epsilon_{\rm f}}}\, \dd W_i \label{eq: viscous terms in PDF model 1 Uf}
\end{align}
\end{subequations}
where the matrix $\mb{A}_{\rm vis}$ is such that we have 
\begin{equation}
A_{{\rm vis}, ik} R_{{\rm f},kj}= - \nu_{\rm f} \left( \frac{\partial \lra{U_{{\rm f},i}}}{\partial x_k} \frac{\partial \lra{U_{{\rm f},j}}}{\partial x_k} \right)~,
\end{equation}
so that this term is written as a classical dissipation term. Both models rely therefore on extending particle convective transport by a viscosity-governed random walk using the exact correspondence between white-noise shifts, $\sqrt{2\, \nu_{\rm f}}\,\dd \mb{W}$, and the viscous diffusion term, $\nu_{\rm f}\, \partial^2 p/(\partial x_k\partial x_k)$, in the Fokker-Planck equation. It is worth noting that this in line with similar ideas used in Vortex Methods~\cite{chorin2013vorticity}. From a physical standpoint, it may be considered that we are not dealing anymore with fluid particles, that is as elements of continuum mechanics, but with particles like macro-molecules and that the random-walk model is the remaining non-vanishing trace of molecular collisions. 

To retrieve the correct description of velocity first and second-order moments, other formulations are also possible. For example, one can propose the following model 
\begin{subequations}
\label{eq: viscous terms in PDF model 2}
\begin{align}
\dd X_{{\rm f},i} & = U_{{\rm f},i}\, \dd t~, \label{eq: viscous terms in PDF model 2 Xf} \\
\dd U_{{\rm f},i} & = - \frac{1}{\rho_{\rm f}}\frac{\partial \lra{P_{\rm f}}}{\partial x_i}\, \dd t +  F_{{\rm vis},i}\, \dd t + G_{ik} \left( U_{{\rm f},k} - \lra{U_{{\rm f},k}} \right)\, \dd t + \sqrt{C_0\, \lra{\epsilon_{\rm f}}}\, \dd W_i \label{eq: viscous terms in PDF model 2 Uf}
\end{align}
\end{subequations}
where viscous effects are now treated through the new acceleration term $\mb{F}_{\rm vis}$. To provide a closed expression of $\mb{F}_{\rm vis}$, a first possibility is to retain a deterministic expression as a function of $\mb{Z}_{\rm f}=(\mb{X}_{\rm f},\mb{U}_{\rm f})$, such as
\begin{equation}
\label{eq: viscous model first Fv}
F_{{\rm vis},i}(\mb{U}_{\rm f})= \nu_{\rm f} \, \frac{\partial^2 \lra{U_{{\rm f},i}}}{\partial x_k \partial x_k} + \widetilde{A}_{{\rm vis},ik} \left( U_{{\rm f},k} - \lra{U_{{\rm f},k}} \right)~,
\end{equation}
where the matrix $\widetilde{A}_{\rm vis}$ satisfies the following equation
\begin{equation}
\widetilde{A}_{{\rm vis}, ik} R_{{\rm f},kj}= \frac{\nu_{\rm f}}{2} \, \frac{\partial^2 R_{{\rm f},ij}}{\partial x_k \partial x_k}~.
\end{equation}
It is also possible to propose a time-evolution model for $\mb{F}_{\rm vis}$ having the form
\begin{equation}
\dd F_{{\rm vis},i}= - \frac{ F_{{\rm vis},i} - \lra{ F_{{\rm vis},i}\, \vert \, \mb{U}_{\rm f}=\mb{V}_{\rm f}}}{\tau_{\rm vis}}\, \dd t + K\, \dd W_{{\rm vis},i}~,
\end{equation}
with $K$ is a diffusion coefficient and $\tau_{\rm vis}$ a timescale that we can both tune, $\mb{W}_{\rm vis}$ a vector of independent Wiener processes, and where the conditional mean value noted $\lra{ F_{{\rm vis},i}\, \vert \, \mb{U}_{\rm f}=\mb{V}_{\rm f}}$ stands for the rhs of Eq.~\eqref{eq: viscous model first Fv}. When the timescale $\tau_{\rm vis}$ is very small compared to the Lagrangian one for particle velocity, $\mb{F}_{\rm vis}$ can be regarded as a fast-variable and (at the moment, heuristic) manipulations lead to approximating the effect of $\mb{F}_{\rm vis}$ over small time increments as
\begin{equation}
\mb{F}_{\rm vis}\, \dd t \simeq \lra{ \mb{F}_{\rm vis}\, \vert \, \mb{U}_{\rm f}=\mb{V}_{\rm f}}\, \dd t + \left( K\, \tau_{\rm vis} \right) \dd \mb{W}_{\rm vis}~.
\end{equation}
This can be seen as an extension of the first expression given in Eq.~\eqref{eq: viscous model first Fv} in which $\mb{F}_{\rm vis}$ appears now as a conditional Gaussian random variable. 

Although it may be surprising to model viscous transport by a local term, we are here sticking to the notion of macroscopic fluid particles (rather than `macro-molecules') for which viscous length scales are considered small for moderate- or high-Reynolds-number turbulent flows. It is also important to note that formulations such as the one in Eqs.~\eqref{eq: viscous terms in PDF model 2} are more attractive than the one in Eqs.~\eqref{eq: viscous terms in PDF model DP} or Eqs.~\eqref{eq: viscous terms in PDF model 1} since the unchanged position evolution equation, cf. Eq.~\eqref{eq: viscous terms in PDF model 2 Xf}, makes it easier to carry out such developments to the case of discrete particles. In fact, discrete particles are also affected by molecular random motions but through Brownian effects rather than through the same viscosity-governed random walk. Clearly, this is an area where further analysis is needed. In the present context, this formulation is nevertheless already useful in that it introduces discussions in terms of viscous/Brownian effects and local/non-local closures in relation with fast-variable elimination. These points are addressed in more details in Sec.~\ref{sec: soft matter Brownian limit}. 

\subsubsection{Proposals for general two-phase GLM} \label{two-phase GLM}
%--------------------------------------------------

In RANS approaches, viscous terms can be disregarded for high Reynolds-number turbulent flows (with the exception of the remaining finite value of the mean turbulent kinetic energy dissipation rate $\lra{\epsilon_{\rm f}}$). On the other hand, retaining only the SLM to account for turbulent fluctuations is clearly limited and, at least, direct responses to mean fluid velocity gradients should be included (see discussions in terms of slow and rapid pressure in~\cite{pope2000turbulent}). This was recognized very early in turbulence modeling and is the reason behind the formulation of a wide range of models, as manifested in our context by the development of GLMs. Though the LRR-IP model is often considered as the basic model~\cite{pope2000turbulent}, no particular proposition can be singled out and we need to consider various formulations. This has direct bearing on two-phase flow modeling, indicating that what is needed is a general methodology to go from single-phase GLMs to two-phase flow ones without relying on the specific characteristics of a given GLM (as was done for the SLM). Such a methodology was proposed recently~\cite{minier2021methodology} and can be presented as follows:
\begin{enumerate}[(i)]
\item A general operator, represented by a matrix $\mb{H}$, is introduced and transforms fluid particle response functions to those of the fluid seen. Conditioned on a given location $\mb{x}$, the response function is defined as the derivative of mean conditional increment of the velocity over a small time interval $\Delta t$ with respect to the velocity at time $t$. For a fluid particle located at $\mb{x}$ at time $t$, this is translated by
\begin{equation}
\frac{1}{\Delta t} \, \frac{\delta }{\delta \mb{U}_{\rm f}} \left\langle \Delta \mb{U}_{\rm f}[ \mb{U}_{\rm f}] \vert \, \mb{X}_{\rm f}=\mb{x} \right\rangle~,
\end{equation}
and it is therefore assumed that we have 
\begin{equation}
\frac{1}{\Delta t} \, \frac{\delta }{\delta \mb{U}_{\rm s}} \left\langle \Delta \mb{U}_{\rm s}[ \mb{U}_{\rm s}] \vert \, \mb{X}_{\rm p}=\mb{x} \right\rangle = \mb{H} \left( \frac{1}{\Delta t} \, \frac{\delta }{\delta \mb{U}_{\rm f}} \left\langle \Delta \mb{U}_{\rm f}[ \mb{U}_{\rm f}] \vert \, \mb{X}_{\rm f}=\mb{x} \right\rangle \right)~.
\end{equation}
\item At the moment, the specific form given in Eq.~\eqref{simple expression Hij} is retained but this is not regarded anymore as the very definition of $\mb{H}$ but as one proposition among a class of possible expressions;
\item The two-phase GLM must be an extension of the fluid one built with minimum additions and so that it reverts to the form given above in the tracer-particle limit;
\end{enumerate}
Based on these guidelines, the general form of the proposed two-phase GLM is:
\begin{equation}
\label{new GLM complete}
dU_{{\rm s},i} = -\frac{1}{\rho_{\rm f}}\frac{\partial \lra{P_{\rm f}}}{\partial x_i}\, dt
 + \left( \lra{U_{{\rm p},j}} - \lra{U_{{\rm f},j}} \right) \frac{\partial \lra{U_{{\rm f},i}}}{\partial x_j} dt 
 + G^{*}_{ij}\left( U_{{\rm s},j} - \lra{U_{{\rm f},j}} \right) dt + B_{{\rm s},ij}\, dW_j~.
\end{equation}
With the fluid GLM given in Eq.~\eqref{model PDF fluid particle GLM b}, the local value of the fluid response function is the matrix $\mb{G}$, while Eq.~\eqref{new GLM complete} indicates that the response function of the velocity of the fluid seen is $\mb{G}^{*}$. From the first principle stated above, it follows therefore that the matrix $G^{*}_{ij}$ is built from $G_{ij}$ as
\begin{equation}
\label{eq matrix G* GLM}
G^{*}_{ij}= (\mb{H}\,\mb{G})_{ij}= H_{ik}\, G_{kj}~.
\end{equation}
In that sense, we are evolving in the frame of linear response theories. Note that the operator transformation is applied only on the `relaxation timescale' characterizing the return-to-equilibrium term (the matrix $\mb{G}$) and not on the mean part of the drift coefficients (the first two terms on the rhs of Eq.~\eqref{new GLM complete}) which is transformed as in the two-phase SLM presented in Sec.~\ref{two-phase SLM}. In Eq.~\eqref{new GLM complete}, the diffusion matrix $B_{{\rm s},ij}$ is still obtained as the square root of the matrix $L_{ij}$ (i.e. $\mb{B}_s^{}\, \mb{B}_s^{T}=\mb{L}$) expressed by Eq.~\eqref{formulation of Lij} but where the coefficients $L_{\vert \vert}$ and $L_{\bot}$ are now given by
\begin{subequations}
\label{new expression Lij}
\begin{align}
L_{\vert \vert} &= \lra{\epsilon_{\rm f}} \left[ - \, \frac{4}{3} \, \frac{\text{Tr}( \mb{H} \widetilde{\mb{G}} \mb{R}_{\rm f})}{\text{Tr}(\mb{H}\mb{R}_{\rm f})} \, \frac{\widetilde{k_{\rm f}}}{k_{\rm f}} \, b_{\vert \vert} - \frac{2}{3} \, \right],\\
L_{\bot} &= \lra{\epsilon_{\rm f}} \left[ - \, \frac{4}{3} \, \frac{\text{Tr}( \mb{H} \widetilde{\mb{G}}\mb{R}_{\rm f})}{\text{Tr}(\mb{H}\mb{R}_{\rm f})} \, \frac{\widetilde{k_{\rm f}}}{k_{\rm f}} \, b_{\bot} - \frac{2}{3} \, \right],
\end{align}
\end{subequations}
in which $\widetilde{\mb{G}}$ is the normalized matrix $\mb{G}$ defined by
\begin{equation}
\label{definition Gtilde}
\widetilde{G}_{ij}= \frac{k_{\rm f}}{\lra{\epsilon_{\rm f}}}\, G_{ij}~.
\end{equation}
As in Eq.~\eqref{eq: Lij as a function of Hij}, these expressions for the components of the matrix $\mb{L}$ rely on the same decomposition as $\mb{G}^{*}$ and involve the same transformation operator $\mb{H}$ as in Eq.~\eqref{eq matrix G* GLM} since we have 
\begin{equation}
L_{ij} = - 2 \frac{\lra{\epsilon_{\rm f}}}{k_{\rm f}} \, \frac{\text{Tr}( \mb{H} \widetilde{\mb{G}} \mb{R}_{\rm f})}{\text{Tr}(\mb{H})} H_{ij} - \frac{2}{3} \lra{\epsilon_{\rm f}}\, \delta_{ij}~.
\end{equation}

To carry out a few simple checks on the new formulation, it is useful to use the same decomposition of $\mb{G}$ given in Eq.~\eqref{GLM decomposition drift matrix G} from which we have 
\begin{equation}
\widetilde{G}_{ij}= -\left( \frac{1}{2} + \frac{3}{4} C_0 \right) \, \delta_{ij} + \frac{k_{\rm f}}{\lra{\epsilon_{\rm f}}}\, G^{\rm a}_{ij} = -\left( \frac{1}{2} + \frac{3}{4} C_0 \right) \, \delta_{ij} + \widetilde{G^{\rm a}_{ij}}~.
\end{equation}
Introducing this decomposition into Eqs.~\eqref{new expression Lij}, we get
\begin{subequations}
\label{new expression Lij bis}
\begin{align}
L_{\vert \vert} &= \lra{\epsilon_{\rm f}} \left[ \left( C_0 + \frac{2}{3} \right)\, \frac{\widetilde{k_{\rm f}}}{k_{\rm f}} \, b_{\vert \vert} - \frac{2}{3} \right] - \lra{\epsilon_{\rm f}} \left[ \, \frac{4}{3} \, \frac{\text{Tr}( \mb{H} \widetilde{\mb{G^{\rm a}}} \mb{R}_{\rm f})}{\text{Tr}(\mb{H}\mb{R}_f)} \, \frac{\widetilde{k_{\rm f}}}{k_{\rm f}} \, b_{\vert \vert} \, \right],\\
L_{\bot} &= \lra{\epsilon_{\rm f}} \left[ \left( C_0 + \frac{2}{3} \right)\, \frac{\widetilde{k_{\rm f}}}{k_{\rm f}} \, b_{\bot} - \frac{2}{3} \right] - \lra{\epsilon_{\rm f}} \left[ \, \frac{4}{3} \, \frac{\text{Tr}( \mb{H} \widetilde{\mb{G^{\rm a}}} \mb{R}_{\rm f})}{\text{Tr}(\mb{H}\mb{R}_{\rm f})} \, \frac{\widetilde{k_{\rm f}}}{k_{\rm f}} \, b_{\bot} \, \right].
\end{align}
\end{subequations}
As mentioned, the same clipping condition by zero applies to $L_{\vert \vert}$ and $L_{\bot}$ in Eqs.~\eqref{new expression Lij} or~\eqref{new expression Lij bis}.

The formulation of the diffusion coefficients in Eqs.~\eqref{new expression Lij bis} is convenient to study the fluid limit case, that is when discrete particles become fluid ones in the absence of any remaining drift velocity ($\mb{U}_{\rm r}=0$). In this tracer-particle limit, $H_{ij}=\delta_{ij}$ while $b_{\vert \vert}=b_{\bot}=\widetilde{k_{\rm f}}/k_{\rm f}=1$. From Eqs.~\eqref{new expression Lij bis}, it is then seen that $\mb{L}$ is isotropic so that we retrieve also an isotropic formulation for the diffusion coefficients $B_{ij}$, since
\begin{equation}
L_{ij}= \left( C_0 \lra{\epsilon_{\rm f}} - \frac{2}{3} \text{Tr}( \mb{G}^{\rm a}\mb{R}_{\rm f} ) \right)\, \delta_{ij}
\end{equation}
which, with $\mb{G}^{*}=\mb{G}$, is indeed the general GLM formulation for single-phase turbulence, cf. Eq.~\eqref{model PDF fluid particle GLM ter}.

As a second verification, we can check that the two-phase GLM reverts to the two-phase SLM when $\mb{G}^a=0$. Indeed, Eq.~\eqref{eq matrix G* GLM} is then the same as Eq.~\eqref{eq matrix G* drift Us} and since the first terms on the rhs of Eqs.~\eqref{new expression Lij bis} are identical to the ones on the rhs of Eqs.~\eqref{expression Lij}, this proves that, when $\mb{G}^a=0$, we retrieve the current form of the two-phase SLM. To emphasize that the two-phase GLM in Eqs.~\eqref{new GLM complete}-\eqref{definition Gtilde} is actually an extension of the current two-phase SLM, we can express $\mb{G}^{*}$ as the sum of two components. From Eq.~\eqref{eq matrix G* GLM}, we have
\begin{equation}
G^{*}_{ij}= G^{\rm slm,*}_{ij} + G^{{\rm a},*}_{ij}
\end{equation}
with $G^{\rm slm,*}_{ij}$ as in Eq.~\eqref{eq matrix G* drift Us} and $G^{{\rm a},*}_{ij}=H_{ik}G^{\rm a}_{kj}$. The decomposition of the diffusion coefficients $L_{ij}$ given in Eqs.~\eqref{new expression Lij bis} shows that we can rewrite Eq.~\eqref{new GLM complete} as the sum of two Langevin models
\begin{align}
\label{new GLM complete bis}
dU_{{\rm s},i} = -\frac{1}{\rho_{\rm f}}\frac{\partial \lra{P_{\rm f}}}{\partial x_i}\, dt
 + \left( \lra{U_{{\rm p},j}} - \lra{U_{{\rm f},j}} \right) \frac{\partial \lra{U_{{\rm f},i}}}{\partial x_j} dt &+ G^{\rm slm,*}_{ij}\left( U_{{\rm s},j} - \lra{U_{{\rm f},j}} \right) dt + B^{\rm slm}_{{\rm s},ij}\, dW^{(\rm slm)}_j \nonumber \\
 & + G^{{\rm a},*}_{ij}\left( U_{{\rm s},j} - \lra{U_{{\rm f},j}} \right) dt + B^{\rm a}_{{\rm s},ij}\, dW^{(\rm a)}_j~,
\end{align}
with $\mb{W}^{\rm slm}$ and $\mb{W}^{\rm a}$ two independent Wiener vector processes. In Eq.~\eqref{new GLM complete bis}, the first line corresponds to the current formulation with $B^{\rm slm}_{s,ij}$ obtained as the square root of the matrix $L_{ij}^{\rm slm}$ given by Eqs.~\eqref{expression Lij}, while the second line accounts for new effects of the two-phase GLM with $G^{{\rm a},*}_{ij}$ defined as above and $B^{{\rm a}}_{{\rm s},ij}$ obtained as the square root of the matrix $L_{ij}^{{\rm a}}$ defined from Eq.~\eqref{formulation of Lij} based on the coefficients $L^{\rm a}_{\vert \vert}$ and $L^{\rm a}_{\bot}$ given by
\begin{subequations}
\label{new expression L^aij}
\begin{align}
L^{\rm a}_{\vert \vert} &= - \lra{\epsilon_{\rm f}} \left[ \, \frac{4}{3} \, \frac{\text{Tr}( \mb{H} \widetilde{\mb{G^{\rm a}}} \mb{R}_{\rm f})}{\text{Tr}(\mb{H}\mb{R}_{\rm f})} \, \frac{\widetilde{k_{\rm f}}}{k_{\rm f}} \, b_{\vert \vert} \, \right],\\
L^{\rm a}_{\bot} &= - \lra{\epsilon_{\rm f}} \left[ \, \frac{4}{3} \, \frac{\text{Tr}( \mb{H} \widetilde{\mb{G^{\rm a}}} \mb{R}_{\rm f})}{\text{Tr}(\mb{H}\mb{R}_{\rm f})} \, \frac{\widetilde{k_{\rm f}}}{k_{\rm f}} \, b_{\bot} \, \right].
\end{align}
\end{subequations}
It is then seen that, since $\text{Tr}(\mb{L}^{\rm a})=-2\text{Tr}( \mb{H} \mb{G^{\rm a}} \mb{R}_{\rm f})=-2\text{Tr}( \mb{G^{{\rm a},*}} \mb{R}_{\rm f})$, the deviatoric Langevin model from the SLM one, represented by the second line in Eq.~\eqref{new GLM complete bis}, does not yield any contribution to the kinetic energy budget. This is the counterpart of what happens in the fluid case and is in line with the criteria set forth in \cite{minier2014guidelines}.

A few remarks can be made. First, this new methodology rests upon the proposition of a mapping operator $\mb{H}$. The form given in Eq.~\eqref{simple expression Hij} has support in locally isotropic turbulence but, since we are mapping turbulent fluctuations rather than small-scale components, this is not necessarily the only possibility. New ideas, based on underlying first principles, are called for to obtain extended formulations. Second, if direct mapping through the operator $\mb{H}$ turns out to be difficult to express, it is also possible to build two-step models with an explicit step dedicated to the CTE. Third, a general observation is that, in all the previously-discussed expressions, the timescales of the velocity of the fluid seen rely only on the Csanady's formulas. Given the central role played by these timescales, it is surprising that so little attention has been devoted to clarifying their physical justification. These open issues are addressed in Sec.~\ref{sec: new macroscopic approaches}. 

\subsubsection{Consistency issues and limitations}\label{consistency issues}
%-------------------------------------------------

From these developments, it transpires that present stochastic models for two-phase flows are extensions of the ones for single-phase flows which are retrieved when particle inertia vanishes. In that sense, single-phase PDF models are contained in two-phase PDF models. This is true for the dynamical variables, such as the velocity of fluid particles and the resulting mean or filtered moments of the velocity field as well as the mean or filtered pressure gradient (derived from them through the mass continuity equation), but not for the kinetic energy dissipation rate which has no equivalent in the set of variables retained in discrete-particle state vectors. It is interesting to note that the difference between the RANS and LES frameworks hinges precisely on how the dissipation rate is modeled. 

If the dissipation rate is obtained as the solution of a physically-based transport equation, we are evolving in the RANS framework and the ensemble averages obtained from the particle set correspond to mean field quantities. To obtain a self-contained PDF description (also referred to as stand-alone methods, for obvious reasons), an additional particle-attached variable $\epsilon_{\rm f}$ is added to the fluid-particle state vector. In practice, models are devised in terms of the turbulent frequency, defined as $\omega_{\rm f}=\epsilon_{\rm f}/k_{\rm f}$, but this does not change the information content ($k_{\rm f}$ being a mean quantity, $\epsilon_{\rm f}$ and $\omega_{\rm f}$ are basically the same stochastic processes, although the physical signification of $\omega_{\rm f}$ is perhaps clearer). To express the evolution equation of the particle instantaneous frequency, it is natural to rely on the refined K62 theory and model $\omega_{\rm f}$ so that it follows a log-normal distribution in homogeneous situations while being applicable to general non-stationary and non-homogeneous flows. This leads to the following expression for $\omega_{\rm f}$~\cite{pope1991application,pope2000turbulent}
\begin{equation}
\label{eq: long-normal model for omega}
\dd \omega_{\rm f} = - \omega_{\rm f} \, \lra{\omega_{\rm f}} \left( S_{\omega_{\rm f}} + C_{\chi} \left[ \ln \left( \frac{\omega_{\rm f}}{\lra{\omega_{\rm f}}} \right) - \left\langle \frac{\omega_{\rm f}}{\lra{\omega_{\rm f}}} \ln \left( \frac{\omega_{\rm f}}{\lra{\omega_{\rm f}}} \right) \right\rangle \right] \right)\, \dd t + \omega_{\rm f} \sqrt{ 2 C_{\chi} \lra{\omega_{\rm f}}\, \sigma^2 }\, \dd W~,
\end{equation}
where $\sigma^2$ and $C_{\chi}$ are constants of the model ($\sigma^2$ is the variance of $\chi=\ln(\omega_{\rm f}/\lra{\omega_{\rm f}})$ whose correlation timescale is written $T_{\chi}^{-1}=C_{\chi} \lra{\omega_{\rm f}}$). The mean value of the frequency is the solution of the transport equation 
\begin{equation}
\frac{\partial \lra{\omega_{\rm f}}}{\partial t} + \lra{U_{{\rm f},k}}\frac{\partial \lra{\omega_{\rm f}}}{\partial x_k} = - S_{\omega_{\rm f}}\, \lra{\omega_{\rm f}}^2~, 
\end{equation}
where the driving term $S_{\omega_{\rm f}}$ in this equation as well as in Eq.~\eqref{eq: long-normal model for omega} is specified as 
\begin{equation}
S_{\omega_{\rm f}} = - C_{{\omega_{\rm f}},1} \frac{\mc{P}}{\lra{\epsilon_{\rm f}}} + C_{{\omega_{\rm f}},2}~,
\end{equation}
with $\mc{P}$ the kinetic energy production term, cf. Eq.~\eqref{eq: kinetic energy production term} , and $C_{{\omega_{\rm f}},1}$ and $C_{{\omega_{\rm f}},2}$ constants of the model. It is worth noting that even if $\epsilon_{\rm f}$ is included as an instantaneous variable and described with the refined K62 theory, thereby accounting for internal intermittency, the fact that only $\lra{\epsilon_{\rm f}}$ (through $\lra{\epsilon_{\rm f}}=k_{\rm f}\, \lra{\omega_{\rm f}}$) enters the GLM equations, as in Eq.~\eqref{model PDF fluid particle GLM}, implies that fluid-particle velocities are modeled with the K41 theory. As such, the role of the log-normal model for the dissipation rate or the turbulent frequency in Eq.~\eqref{eq: long-normal model for omega} is merely to provide the mean value $\lra{\epsilon_{\rm f}}$ used in the various Langevin formulations in Secs.~\ref{Fluid GLM models} and~\ref{Langevin models for Us}. The evolution toward a velocity-frequency PDF description fully consistent with the refined K62 theory seems straightforward and consists in replacing $\lra{\epsilon_{\rm f}}$ with $\epsilon_{\rm f}$ in the SDEs, for example by substituting the diffusion coefficient $\sqrt{C_0\, \lra{\epsilon_{\rm f}}}\, \dd W$ with $\sqrt{C_0\, \epsilon_{\rm f}}\, \dd W$ and, similarly, in the drift term. This leads to the Refined Langevin Model, as described in~\cite{pope1991application}. On the one hand, such steps emphasize the great malleability of the PDF modeling framework. On the other hand, to maintain the basic property that the velocity PDF should be Gaussian in homogeneous situations, additional terms must be added to the drift term, increasing the complexity of the model formulation and making it harder to handle~\cite{pope1991application,pope2000turbulent}. We are thus faced with a first hint that addressing situations of increasing complexity with a single formulation could lead to more intricate model expressions. The philosophy behind the different approaches to complexity is investigated in more details in Sec.~\ref{Accounting for structures signature}. 

If the dissipation rate is obtained from a parameter-based local equation, we are evolving in the LES framework and the ensemble averages obtained from the particle set correspond to filtered field quantities. While the stochastic models for particle velocity have the same expression, the turbulent kinetic energy becomes the residual kinetic energy $k_{\rm f}^r$ and, more importantly, the dissipation rate noted $\widehat{\epsilon_{\rm f}}$ is now estimated locally using the cut-off parameter $\Delta$ separating the resolved from the unresolved scales. A model applied in several studies is~\cite{Gicquel_2002,Sheikhi_2003,Sheikhi_2007}
\begin{equation}
\widehat{\epsilon_{\rm f}} = C_{\epsilon_{\rm f}} \left( k_{\rm f}^r \right)^{3/2}/\Delta~,
\end{equation}
where $C_{\epsilon_{\rm f}}$ is a constant. In most practical applications, the parameter $\Delta$ is the grid size used in the computations, which means that there is a rather indistinct, if not blurred, separation between the continuous and numerical formulations (a characteristic feature of LES and a subject of discussions on the foundations of the approach~\cite{pope2000turbulent}). More recent proposals~\cite{Sheikhi_2009,johnson2018predicting} introduce the notion of a cascade time delay between production and dissipation of energy, which leads to (re-)introduce an instantaneous particle-attached value of the dissipation rate, say $\epsilon_{\rm f}^r$, which is the solution of a time-evolution equation such as
\begin{equation}
\dd \epsilon_{\rm f}^r = - \left( \frac{ \epsilon_{\rm f}^r - \widehat{\epsilon_{\rm f}}}{\tau_{\epsilon_{\rm f}^r}} \right)\dd t~,
\end{equation}
where different proposals can be found for the timescale $\tau_{\epsilon_{\rm f}^r}$ governing this delay~\cite{Sheikhi_2009,johnson2018predicting}.

These considerations bring us back to the issue mentioned at the beginning, namely the consistency between the description of the fluid phase on the one hand and the description obtained from the dispersed-phase model in the tracer-particle limit on the other hand. When concentrating on PDF models for discrete particles, the mean or filtered fields which characterize the fluid phase are considered as available (a complete PDF description of the two phases based on two-particle PDF models, i.e. models treating pair of particles made up by one fluid particle and one discrete particle, is not within the scope of the present review but can be found in~\cite{minier2001pdf,peirano2002probabilistic}). Yet, one unfortunate error is to assume that the descriptions retained for each phase are unrelated. This is not so. Indeed, once a choice is made for the dissipation rate, the corresponding RANS or LES frameworks are selected for both phases. Furthermore, the limit of a two-phase GLM when particle inertia goes to zero is a GLM model for the fluid phase. It follows that, in order to be consistent in the tracer-particle limit, both should be identical. For example, if we select a two-phase SLM coupled to a description of the fluid phase, that description of the fluid-phase must be based on the Rotta second-order model with the same constants~\cite{chibbaro2011note,minier2015lagrangian}. In that sense, it is of importance to note that a stochastic model based on a particle state vector which includes the particle velocity corresponds necessarily to a second-order turbulence model~\cite{pope1994relationship} and not to one based on an `eddy'- or `turbulent-' viscosity. It is inconsistent to simulate the fluid phase with such a turbulence model and then track discrete particles with a stochastic model for the velocity of the fluid seen. Unfortunately, this is a mistake impacting several simulations of discrete particles in LES since the majority of LES formulations rely on the Smagorinsky model based on the notion of an effective, or sub-grid scale (SGS), viscosity to model the effects of the residual fluid motions. Note that coupling an `ideal LES' (obtained by filtering a DNS) with a stochastic model designed to mimic SGS fluctuations is still inconsistent since, if the ideal LES is free of modeling error for single-phase predictions, the complete model formulation is not since there is a discrepancy between the `ideal LES' description and the one corresponding to the tracer-particle limit of the stochastic model used for the velocity of the fluid seen. When using a two-phase GLM for discrete particles in LES, then the components of the SGS tensor (the equivalent of the Reynolds-stress tensor in the context of filtered fields) must be obtained as the solution of their own transport equations based on the corresponding single-phase GLM~\cite{Gicquel_2002}). On the other hand, when an eddy-viscosity formulation is retained for the fluid phase, it can be wondered which stochastic formulations for discrete particles are consistent. Such questions resurface in Sec.~\ref{sec: local and non-local closures}.

Even when consistency is ensured in the tracer particle limit, it is also useful to be aware of inherent limitations of present Langevin-based stochastic models as revealed, for instance, in studies of particle concentration effects~\cite{Pozorski_2009,marchioli2017large}. Actually, Langevin-type of models are devised so as to reproduce essentially one-particle statistics, such as the correct level of SGS subgrid kinetic energy, velocity auto-correlations, dispersion coefficients, etc., but not necessarily instantaneous distributions due to captures by some underlying fluid structures which are not present in the mean or filtered fields. For all their merits, it is also fair to recall that basic stochastic models, such as the SLM, live up to their names and remain, in fact, crude pictures. Indeed, looking back at the general structure of a Langevin model, as in Eq.~\eqref{model PDF fluid particle GLM b}, it is seen that it involves three terms: a deterministic one, a purely ordered term (since all fluctuations are driven back to the local mean value in the same manner) and a purely disordered term manifested by the White-noise increments. Thus, it may not be an altogether surprise if such a formulation remains limited. In that respect, new ideas are called for and suggestions are made in Sec.~\ref{Accounting for structures signature}.

\subsection{New macroscopic approaches and microscopic models}\label{sec: new macroscopic approaches}
%=============================================================

\subsubsection{Extended Kolmogorov hypotheses for Lagrangian timescales}\label{EKH for Lagrangian timescales}
%-----------------------------------------------------------------------

\begin{figure}[ht]
\includegraphics[scale=0.8]{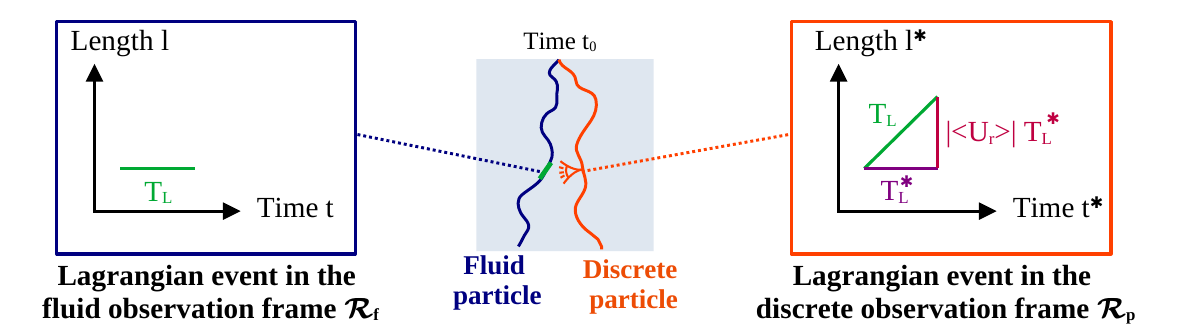}%
\caption{Graphical representation of a fluid Lagrangian event $T_{\rm L}$ as seen from fluid particles, where it appears as $(T_{\rm L} \, , \,0)_{\mc{R}_{\rm f}}$ in the observation frame $\mc{R}_{\rm f}$ (left side) and as seen from discrete particles, where it appears as a time-length event $(T_{\rm L}^{*}\, , \, \vert \lra{U_{\rm r}} \vert \, T_{\rm L}^{*})_{\mc{R}_{\rm p}}$ in the observation frame $\mc{R}_{\rm p}$ (right side).\label{Sketch of EKH for timescales}}
\end{figure}

As recalled in Sec.~\ref{The Kolmogorov theory}, the classical Kolmogorov similarity hypotheses are developed from a Lagrangian standpoint by considering space-time fluid correlations in a small domain described as locally isotropic around fluid particles (which form an equivalence class). In the following, we propose to extend this hypothesis to discrete particles having a mean velocity drift $\lra{\mb{U}_{\rm r}}$ with respect to the fluid. This means that we are observing similar space and time fluid correlations but from the standpoint of discrete particles (which form also an equivalence class, though different from the one made up by fluid particles). Said otherwise, we have two different points of view to describe the same fluid statistical events, as pictured in Fig.~\ref{Sketch of EKH for timescales}. 

To formulate the relations between these two different reference systems, it is useful to introduce the notion of (fluid) statistical events: we consider fluid velocity statistical events $(\tau,l)$ having a typical timescale $\tau$ and a typical lengthscale $l$. In this time-length statistical space, the two equivalent standpoints (i.e., astride a fluid particle or a discrete one) imply that the corresponding coordinates are different but related. To illustrate that point, let us consider a purely Lagrangian event, as sketched in Fig.~\ref{Sketch of EKH for timescales}, and limit ourselves to a one-dimensional spatial setting for the sake of simplicity. If we follow fluid particles, we have a purely time event and since we are interested in the characteristic timescale of fluid velocities, this statistical event can be written as $(T_{\rm L} \, , \,0)_{\mc{R}_{\rm f}}$ in this observation frame. However, if we observe the same statistical event from the standpoint of discrete particles, we have a time-length statistical event which we can write as $(T_{\rm L}^{*}\, , \, \vert \lra{U_{\rm r}} \vert \, T_{\rm L}^{*})_{\mc{R}_{\rm p}}$ where $\mc{R}_{\rm p}$ is used to denote the discrete particle observation frame. Obviously, the time-component $T_{\rm L}^{*}$ is the characteristic timescale of the velocity of the fluid seen and we have the equivalence
\begin{equation}
(T_{\rm L} \, , \,0)_{\mc{R}_{\rm f}} \equiv (T_{\rm L}^{*}\, , \, \vert \lra{U_{\rm r}} \vert \, T_{\rm L}^{*})_{\mc{R}_{\rm p}}~.
\end{equation}
Loosely speaking, the timescale of the velocity of the fluid seen appears as a projection of the fluid Lagrangian timescale onto discrete particle trajectories and we need to retain both the time and length components to properly describe the same statistical event (namely, the characteristic timescale of fluid particle velocities).

To work out explicit closure relations, the next step consists in introducing a measure of such statistical events and requiring that this measure be an invariant when switching from one observation standpoint to another. If we retain the classical measure using the time and length coordinates, we can write that the invariant measure $\Delta s$ of an event $(\tau \, , \, l)_{\mc{R}}$ in any observation frame $\mc{R}$ is
\begin{equation}
\label{eq: definition measure time-length 1}
(\Delta s)^2 = \left( \frac{\tau}{T_{\rm L}} \right)^2 + \left( \frac{l}{L_{\rm E}} \right)^2~,
\end{equation}
where time increments are normalized by the Lagrangian timescale $T_{\rm L}$ whereas length increments are normalized by the Eulerian lengthscale $L_{\rm E}$. In the example illustrated in Fig.~\ref{Sketch of EKH for timescales}, the statistical even $(T_{\rm L} \, , \,0)_{\mc{R}_{\rm f}}$ is then of unit measure. The same remains valid when observed from discrete particles and this implies that the coordinates $(T_{\rm L}^{*}\, , \, \vert \lra{U_{\rm r}} \vert \, T_{\rm L}^{*})_{\mc{R}_{\rm p}}$ are such that we have
\begin{equation}
\label{eq: measure TL* Csanady 1}
\left( \frac{T_{\rm L}^{*}}{T_{\rm L}} \right)^2 + \left( \frac{\vert \lra{U_{\rm r}} \vert \, T_{\rm L}^{*}}{L_{\rm E}} \right)^2 = 1~,
\end{equation}
from which we get that 
\begin{equation}
T_{\rm L}^{*} = \frac{T_{\rm L}}{\sqrt{ 1 + \left(  \dfrac{\vert \lra{U_{\rm r}} \vert T_{\rm L}}{L_{\rm E}} \right)^2 }}~.
\end{equation}
By using $L_{\rm E}= u \, T_{\rm E}$ (with $u=2/3k$, in the locally isotropic formulation), we retrieve therefore the Csanady expression for the Lagrangian timescale of the velocity of the fluid seen (cf. Eqs.~\eqref{Csanady timescales} in Sec.~\ref{two-phase SLM})
\begin{equation}
T_{\rm L}^{*} = \frac{T_{\rm L}}{\sqrt{ 1 + C_{\rm T}^2 \dfrac{\vert \lra{U_{\rm r}} \vert^2 }{u^2} }}~.
\end{equation}

An interesting outcome of such reasoning is to bring out that the typical dependence of the Csanady expression, which scales as $(1 + x^2)^{-1/2}$, with $x=\vert \lra{U_{\rm r}} \vert/u$ the normalized mean velocity drift, results from the choice of the measure used to evaluate $\Delta s$ for each statistical event. Other measures would result in different relations for $T_{\rm L}^{*}$. For example, if we consider
\begin{equation}
\label{eq: definition measure time-length 2}
\Delta s = \frac{\tau}{T_{\rm L}} + \frac{l}{L_{\rm E}}
\end{equation}
where all quantities are positive, we would obtain for the timescale of the velocity of the fluid seen 
\begin{equation}
\label{New Csanady direct sum}
T_{\rm L}^{*} = \frac{T_{\rm L}}{1 + C_{\rm T} \dfrac{\vert \lra{U_{\rm r}} \vert }{u} }~.
\end{equation}
By re-expressing Eq.~\eqref{New Csanady direct sum}, it is seen that we are directly adding the inverse of the timescales
\begin{equation}
\frac{1}{T_{\rm L}^{*}} = \frac{1}{T_{\rm L}} + \frac{ (\vert \lra{U_{\rm r}} \vert/ u) }{T_{\rm E}} = \frac{1}{T_{\rm L}} + \frac{ \vert \lra{U_{\rm r}}\vert }{L_{\rm E}}~.
\end{equation}
Such expressions are more in line with stochastic models based on Langevin formulations since the return-to-equilibrium term is written with a friction or resistance coefficient which is the inverse of a timescale. Then, when devising a two-step Langevin model by adding two successive independent steps (a Lagrangian one followed by an Eulerian one), we are adding resistances as in Ohm's law. In comparison, it is seen that the Csanady expressions consist in adding the square of the same resistance coefficients since we have from Eq.~\eqref{eq: measure TL* Csanady 1}
\begin{equation}
\left( \frac{1}{T_{\rm L}^{*}} \right)^2 = \left( \frac{1}{T_{\rm L}} \right)^2 + \left( \frac{ \vert \lra{U_{\rm r}} \vert }{L_{\rm E}} \right)^2~.
\end{equation}

The formulation of the present extended Kolmogorov hypothesis can be further developed by writing transformation rules for the time-length coordinates of the same fluid statistical event observed from two different standpoints, written as $\mc{R}'$ and $\mc{R}$, corresponding to two different mean slip velocities $\lra{U_{\rm r}'}$ and $\lra{U_{\rm r}}$ with respect to the fluid (the relative velocity between $\mc{R}'$ and $\mc{R}$ is therefore $\Delta \lra{U_{\rm r}}=\lra{U_{\rm r}}-\lra{U_{\rm r}'}$). By assuming a linear mapping between coordinates in $\mc{R}'$ and $\mc{R}$, the relations between $(t',x')$ in $\mc{R}'$ and $(t,x)$ in $\mc{R}$ are obtained from the requirement that the measure in Eq.~\eqref{eq: definition measure time-length 1} remains invariant. This yields 
\begin{subequations}
\begin{align}
x & = \gamma \left[\, x' + \left( \Delta \lra{U_{\rm r}} \right) t' \, \right]~, \\
t & = \gamma \left[ \, t' - C_{\rm T}^2 \left( \Delta \lra{U_{\rm r}} \right) \frac{x'}{u^2} \, \right]~,
\end{align}
\end{subequations} 
where $\gamma$ is the Csanady factor
\begin{equation}
\gamma = \frac{1}{\sqrt{ 1 + C_{\rm T}^2 \, \dfrac{\vert \Delta \lra{U_{\rm r}} \vert^2}{u^2} }}~.
\end{equation}
Then, if we revisit the example introduced above, the first standpoint following fluid particles corresponds to $\lra{U'_{\rm r}}=0$ (i.e., $\mc{R}'=\mc{R}_{\rm f}$) and the transformation of a time event $(\Delta t \, , \, 0)$ is now given by the equivalence 
\begin{equation}
(\Delta t \, , \, 0)_{\mc{R}_{\rm f}} \Longleftrightarrow (\gamma \Delta t \, , \, \gamma \lra{U_{\rm r}} \Delta t)_{\mc{R}_{\rm p}}~,
\end{equation}
from which we retrieve $T_{\rm L}^{*}=\gamma T_{\rm L}$ by taking $\Delta t=T_{\rm L}$ in the above relation. Interestingly, in the limit of high velocity slips, $\vert \lra{U_{\rm r}} \vert /u \gg 1$, a time event $(\Delta t' \, , \, 0)$ in $\mc{R}_{\rm f}$ is transformed into a spatial one in $\mc{R}_{\rm p}$, since
\begin{subequations}
\label{eq: EKH frozen turbulence}
\begin{align}
\Delta t & \simeq 0~, \label{eq: EKH frozen turbulence a}\\
\Delta x & \simeq \frac{1}{C_{\rm T}} \, u \, \Delta t'= \frac{T_{\rm E}}{T_{\rm L}} \, u \, \Delta t'~, \label{eq: EKH frozen turbulence b}
\end{align}
\end{subequations} 
which is the translation of the frozen-turbulence hypothesis~\cite{pope2000turbulent}. Indeed, by taking $\Delta t'=T_{\rm L}$, it is seen that, when $\vert \lra{U_{\rm r}} \vert /u \gg 1$, the statistical event $(T_{\rm L} \, , \, 0)$ in $\mc{R}_{\rm f}$ is transformed into $(0 \, , \, L_{\rm E})$ in $\mc{R}_{\rm p}$.

As indicated above, the previous relations have been worked out in a one-dimensional spatial setting and we need to consider the more realistic three-dimensional spatial version. In the Kolmogorov picture, turbulence is described as isotropic in small time and space domains around fluid particles. This means that we can still consider an isotropic Lagrangian timescale tensor and retain $T_{\rm L}$ as a scalar timescale that provides the necessary time information. However, this is not the case for fluid velocity increments over a separation increments $\bm{r}$ since the lengthscale for longitudinal velocity components is twice the one for transverse components. In other words, we must now consider that $L_{\rm E}$, as well as $T_{\rm E}$ which is deduced from it, become second-order tensors which are, however, isotropic functions of the separation vector $\bm{r}$. In the general three-dimensional spatial setting, the space effects are due to the mean velocity drift between discrete particles and the fluid and the relevant separation vector is $\mb{r}= \lra{\mb{U}_{\rm r}} \Delta t$ over a time interval $\Delta t$. This means that the Eulerian time tensor can be expressed as
\begin{equation}
\label{eq: Eulerian time tensor 3D}
T_{\rm E, ij}= T_{\rm E, \bot}\delta_{ij} + \left( T_{\rm E, ||} - T_{\rm E, \bot} \right) \frac{r_i\, r_j}{\vert \mb{r} \vert}~,
\end{equation}
where $T_{\rm E, ||} = 2 T_{\rm E, \bot}$. It is then appropriate to express directly the timescale of the velocity of the fluid seen in the local reference system aligned with the mean velocity drift and to follow the previous developments. This leads therefore to retrieve the Csanady expressions, cf. Eqs.~\eqref{Csanady timescales}. 

A few remarks are in order. First, the previous relations must not be regarded as a new kinematics but merely as a tentative formulation in terms of time and length scales to provide new insights into the closure issue of the timescale of the velocity of the fluid seen. Should they prove useful, the ideas introduced in the present section are meant to reveal what the functional form entering Csanady expressions represents and how we can improve them or propose alternative closures (e.g., as in Eq.~\eqref{New Csanady direct sum}). Second, these considerations, which follow a relativity-like principle, still constitute a top-down approach where the timescales of the velocity of the fluid seen are obtained from guiding rules and are input into Langevin models. Another, more microscopic, approach consists in formulating stochastic models from which the timescales $T_{\rm L}^{*}$ would be an outcome instead of being an input. In that respect, some new modeling roads are now suggested. 

\subsubsection{Capturing space and time correlations through time changes} \label{two-step models for Us}
%-------------------------------------------------------------------------

\begin{figure}[ht]
\includegraphics[scale=0.7]{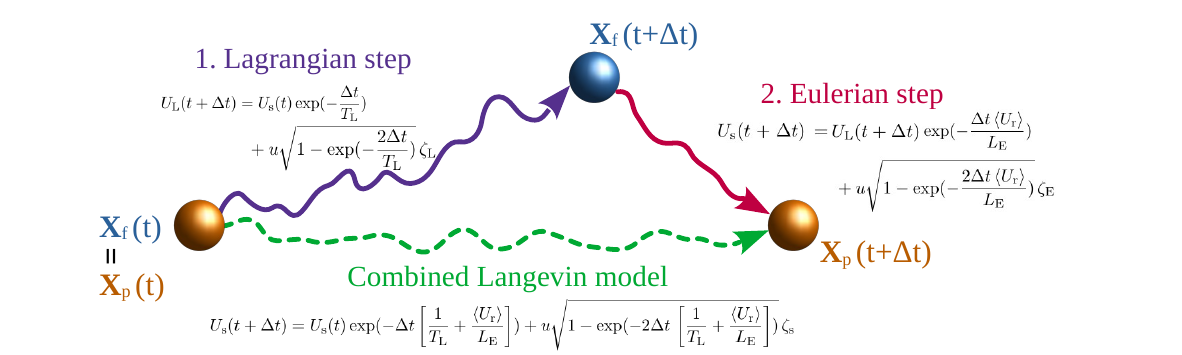}%
\caption{Graphical representation of a two-step formulation of the stochastic model for the velocity of the fluid seen with a Lagrangian step followed by an Eulerian one governed by the mean velocity drift.}
\label{fig: Two-step-model_1}
\end{figure}

Early stochastic models for turbulent dispersion were often developed by considering a Lagrangian step followed by an Eulerian one. Although these models suffered from shortcomings, in particular due to the fact that they used instantaneous relative velocities and were not properly written as stochastic diffusion processes in continuous time, it is interesting to revisit such two-step formulations, however with the Eulerian step governed by the mean velocity drift as indicated in Sec.~\ref{Physics Particle Dispersion}. The basic situation is represented in Fig.~\ref{fig: Two-step-model_1} and the purpose of this section is to suggest new modeling possibilities. 

\paragraph{Basic ideas and current limitations of two-step stochastic models}
For this discussion, we limit ourselves to a one-dimensional setting where $T_{\rm L}$ and $T_{\rm E}$ are constant scalars (this amounts to considering stationary isotropic turbulence). In this case, we retain two simple Langevin models, for the Lagrangian step
\begin{equation}
\label{eq: SLM in time with t}
\dd U_{\rm L}= - \frac{U_{\rm L}}{T_{\rm L}}\, \dd t + \sqrt{\frac{2\, u^2}{T_{\rm L}}}\, \dd W_{\rm L}~,
\end{equation}
as well as for the Eulerian one
\begin{equation}
\label{eq: SLM in space with r}
\dd U_{\rm E}= - \frac{U_{\rm E}}{L_{\rm E}}\, \dd r + \sqrt{\frac{2\, u^2}{L_{\rm E}}}\, \dd W'_{\rm E}~,
\end{equation}
where $\dd W_{\rm L}$ and $\dd W'_{\rm E}$ are two independent Wiener processes (note that here $\lra{(\dd W'_{\rm E})^2}=\dd r$). Since the Eulerian step is due to the mean velocity drift, we have $dr=\lra{U_{\rm r}}\, dt$ and we can reformulate Eq.~\eqref{eq: SLM in space with r} as 
\begin{equation}
\dd U_{\rm E}= - \frac{U_{\rm E} \, \lra{U_{\rm r}}}{L_{\rm E}}\, \dd t + \sqrt{\frac{2\, u^2 \lra{U_{\rm r}}}{L_{\rm E}}} \dd W_{\rm E}~,
\end{equation}
with $W_{\rm E}$ a Wiener process in time (i.e., $\lra{(\dd W_{\rm E})^2}=\dd t$). For OU processes, exact integration of the trajectories is straightforward and we can proceed with the values of velocities at successive time intervals $\Delta t$, which corresponds to handling Markov chains. Starting from $U_{\rm s}(t)$ at time $t$, we have
\begin{subequations}
\label{eq: MC two-step model first}
\begin{align}
U_{\rm L}(t+\Delta t) &= U_{\rm s}(t)\exp( -\frac{\Delta t}{T_{\rm L}} ) + u\sqrt{ 1 - \exp( -\frac{2\Delta t}{T_{\rm L}} )} \, \zeta_{\rm L}~, \label{eq: MC two-step model first L} \\
U_{\rm E}(t+\Delta t) &= U_{\rm E}(t)\exp( -\frac{\Delta t\, \lra{U_{\rm r}}}{L_{\rm E}} ) + u\sqrt{ 1 - \exp( -\frac{2\Delta t\, \lra{U_{\rm r}}}{L_{\rm E}} )} \, \zeta_{\rm E}~, \label{eq: MC two-step model first E}
\end{align}
\end{subequations}
where $\zeta_{\rm L}$ and $\zeta_{\rm E}$ are sampled in two independent centered normalized Gaussian random variables, i.e., both $\zeta_{\rm L}$ and $\zeta_{\rm E}$ are $\mc{N}(0,1)$. By construction of this two-step model, we have $U_{\rm s}(t+\Delta t)=U_{\rm E}(t+\Delta t)$ and $U_{\rm E}(t)=U_{\rm L}(t+\Delta t)$ (due to its transformation from $r$, time is actually fictitious in the Eulerian step). The Lagrangian and Eulerian steps can then be combined to yield
\begin{multline}
U_{\rm s}(t+\Delta t)= U_{\rm s}(t)\exp( -\Delta t \left[ \frac{1}{T_{\rm L}} + \frac{\lra{U_{\rm r}}}{L_{\rm E}} \right] ) \\ +  u\sqrt{ 1 - \exp( -\frac{2\Delta t\, \lra{U_{\rm r}}}{L_{\rm E}} )} \, \zeta_{\rm E} + u \exp( -\frac{\Delta t\, \lra{U_{\rm r}}}{L_{\rm E}} ) \sqrt{ 1 - \exp( -\frac{2\Delta t}{T_{\rm L}} )} \, \zeta_{\rm L}~.
\end{multline}
The sum of the two independent Gaussian number can be lumped into a third one and straightforward calculation gives 
\begin{equation}
\label{eq: first two-step model combined}
U_{\rm s}(t+\Delta t)= U_{\rm s}(t)\exp( -\Delta t \left[ \frac{1}{T_{\rm L}} + \frac{\lra{U_{\rm r}}}{L_{\rm E}} \right] ) +  u\sqrt{ 1 - \exp( -2\Delta t\,\left[ \frac{1}{T_{\rm L}} + \frac{\lra{U_{\rm r}}}{L_{\rm E}} \right] ) }\, \zeta_{\rm s}~,
\end{equation}
where $\zeta_{\rm s}$ is another $\mc{N}(0,1)$ random variable. In continuous time, this corresponds to the combined simple Langevin model for the velocity of the fluid seen $U_{\rm s}$ 
\begin{equation}
\dd U_{\rm s} = - \frac{U_{\rm s}}{T_{\rm L}^{*}}\, \dd t + \sqrt{\frac{2\, u^2}{T_{\rm L}^{*}}}\, \dd W~,
\end{equation}
with $T_{\rm L}^{*}$ given as in Eq.~\eqref{New Csanady direct sum}. From this simple two-step construction, we retrieve as an outcome the expression of the Lagrangian timescale of the fluid seen where the inverse of the timescales are added. 

Even in this simplified case of turbulent flows, the weak point of the previous two-step model is clearly the formulation of the Eulerian step. Regardless of whether the resulting expression of the timescale of the fluid seen, cf. Eq.~\eqref{New Csanady direct sum}, is satisfactory or not, there are difficulties with respect to the predicted correlation in space. Indeed, from the assumption of a simple Langevin, as in Eq.~\eqref{eq: SLM in space with r}, we have
\begin{equation}
\lra{ U_{\rm E}(r_0+r)\, U_{\rm E}(r_0)}= u^2\, \exp( - \frac{r}{L_{\rm E}} )
\end{equation}
which yields for the second-order moment of velocity differences over $r$, $\delta U_{\rm E}(r)=U_{\rm E}(r_0+r)-U_{\rm E}(r_0)$, 
\begin{equation}
\lra{\left( \delta U_{\rm E}(r) \right)^2}= 2\, \biggl[ u^2 - \lra{ U_{\rm E}(r_0+r)\, U_{\rm E}(r_0)} \biggr]
= 2\, u^2 \left[ 1 - \exp( - \frac{r}{L_{\rm E}} ) \right]~.
\end{equation}
In the inertial range, where $r \ll L_{\rm E}$, this implies $\lra{\left( \delta U_{\rm E}(r) \right)^2} \simeq 2u^2\, r/L_{\rm E}$, whereas the proper Kolmogorov scaling should be $\lra{\left( \delta U_{\rm E}(r) \right)^2} \simeq \left( \lra{\epsilon}\, r \right)^{2/3}$. To address this issue, it is instructive to consider first a simple model for two-particle relative motion. 

\paragraph{A reminder on relative dispersion}
\begin{figure}[ht]
\includegraphics[scale=0.6]{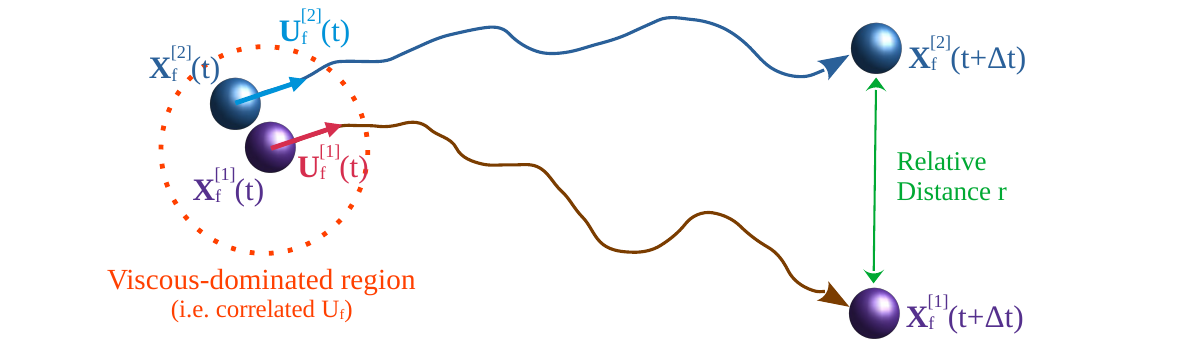}%
\caption{Representation of the two-fluid-particle relative dispersion situation. At an initial time $t$, two fluid particles are in a zone of strong interaction so that their velocities are identical. Over a time interval $\Delta t$, these two fluid particles leave this zone and evolve by independent simple Langevin models.}
\label{fig: two-fluid-particle-dispersion}
\end{figure}
This situation is displayed in Fig.~\ref{fig: two-fluid-particle-dispersion}: at a time $t_0=0$, taken as an initial time, two fluid particles are located in a small region indicated as a zone of strong interaction (whose size is typically of the order of the Kolmogorov scale $\eta$). This means that viscous forces take over so that the two fluid particles have the same initial velocity, $U_{\rm f}^{[1]}(0)= U_{\rm f}^{[2]}(0)=U_0$. In terms of particle velocities, this amounts to a `perfectly inelastic collision' (particles do not stick but their velocities relax immediately to the same value). Outside of this small viscous-dominated region, it is assumed that the two particle velocities are driven by two simple Langevin models
\begin{equation}
\label{eq: two-particle SLMs}
\dd U_{\rm f}^{[i]}= - \frac{U_{\rm f}^{[i]}}{T_{\rm L}}\, \dd t + \sqrt{C_0\, \lra{\epsilon_{\rm f}}}\, \dd W^{[i]}~,
\end{equation}
where the two Wiener processes, $W^{[i]}$, with $i=1,2$, are independent. The relative velocity, $\delta U^{[1-2]}_{\rm f}=U_{\rm f}^{[2]}- U_{\rm f}^{[1]}$, follows therefore a similar Langevin model, which is
\begin{equation}
\dd (\delta U^{[1-2]}_{\rm f}) = -\frac{ \delta U^{[1-2]}_{\rm f} }{ T_{\rm L}}\, \dd t + \sqrt{2\, C_0\, \lra{\epsilon_{\rm f}}}\, \dd W~,
\end{equation}
where $W$ is another Wiener process. Given that the initial condition is $\delta U^{[1-2]}_{\rm f}(0)=0$, the exact solution is
\begin{equation}
\delta U^{[1-2]}_{\rm f} (t) = \sqrt{2\, C_0\, \lra{\epsilon_{\rm f}}} \exp(- \frac{t}{T_{\rm L}}) \int_0^t \exp( \frac{s}{T_{\rm L}}) dW(s)~,
\end{equation}
from which we obtain 
\begin{equation}
\lra{ (\delta U^{[1-2]}_{\rm f})^2 (t) } = C_0\, \lra{\epsilon_{\rm f}} T_{\rm L} \left[ 1 - \exp(-  \frac{2 t}{T_{\rm L}} ) \right] = 2 u^2 \left[ 1 - \exp(-  \frac{2 t}{T_{\rm L}} ) \right]~.
\end{equation}
The relative distance $r^{[1-2]}$ between the two fluid particles is obtained by integrating $\delta U^{[1-2]}_{\rm f} (t)$, which gives
\begin{equation}
r^{[1-2]}(t)= \int_0^t \delta U^{[1-2]}_{\rm f} (s)\, ds =\sqrt{2\, C_0\, \lra{\epsilon_{\rm f}}} \left( T_{\rm L} W_t - T_{\rm L} \exp( - \frac{t}{T_{\rm L}} ) \int_0^t \exp(\frac{s}{T_{\rm L}}) dW_s \right)~,
\end{equation}
from which a tedious but straightforward calculation shows that
\begin{equation}
\label{eq: variance of relative positions}
\lra{ (r^{[1-2]})^2(t) }= 2\, C_0\, \lra{\epsilon_{\rm f}} T_{\rm L}^2 \left\{ t - \frac{T_{\rm L}}{2}\left[ 1 - \exp(-\frac{t}{T_{\rm L}}) \right]\, \left[ 3 - \exp(-\frac{t}{T_{\rm L}}) \right] \right\}~.
\end{equation}
When the two fluid particles are outside the zone of strong interaction but still within the initial range, i.e., $t \ll T_{\rm L}$, the interesting result is that 
\begin{equation}
\label{eq: scaling relative dispersion}
\lra{ (\delta U^{[1-2]}_{\rm f})^2 (t) } \simeq 2\, C_0\, \lra{\epsilon_{\rm f}}\, t \quad \text{and} \quad \lra{ (r^{[1-2]})^2(t) } \simeq \frac{2}{3}\, C_0 \lra{\epsilon_{\rm f}}\, t^3 \Longrightarrow \lra{ (\delta U^{[1-2]}_{\rm f})^2 } \sim \left( \lra{\epsilon_{\rm f}} \, r^{[1-2]} \right)^{2/3}~,
\end{equation}
which shows that we retrieve the correct scaling for Eulerian statistics within the inertial range. 

In physical terms, these results indicate that a diffusive model, in which fluid particles interact strongly when located nearby (within distances of the order of $\eta$) and evolve with a simple dynamical model involving a mean-field term (i.e., the return-to-equilibrium term of the Langevin models in Eqs.~\eqref{eq: two-particle SLMs}) and independent random forcing (i.e., the diffusive term driven by independent Wiener processes), is consistent with the expected Kolmogorov scaling both in time and space. 

\paragraph{A new two-step stochastic model}
Coming back to the difficulty encountered with the correlated velocities generated by the Eulerian step with Eq.~\eqref{eq: SLM in space with r}, it is seen that we are in fact trying to hold together two different regimes: on the one hand, a convective regime (where $r=\lra{U_{\rm r}}\, t$) and, on the other hand, a diffusive one (where $r =\left( \lra{\epsilon_{\rm f}}\, t^3 \right)^{1/2}$). Since the Eulerian step is governed by the mean drift (see Sec.~\ref{Physics Particle Dispersion}), the linear relation between $r$ and $t$ is relevant but the use of a Langevin model is more appropriate to reproduce velocity correlations in the diffusive regime. This suggests to consider a Langevin model formulated in terms of a rescaled time so that separation distances in the two regimes are statistically identical. More precisely, starting from a given initial time (to be discussed below), we introduce a rescaled time $\widehat{t}$ so that
\begin{equation}
\label{eq: rescaled time relation}
r = \lra{U_{\rm r}}\, t \simeq \left[ \lra{\epsilon_{\rm f}}\, \widehat{t}^{\phantom{,}3} \right]^{1/2} \Longrightarrow 
\widehat{t}= C\left( \frac{\lra{U_{\rm r}}^2}{\lra{\epsilon_{\rm f}}} \right)^{1/3} t^{2/3}
\end{equation}
where $C$ is a constant (for example, we get from Eq.~\eqref{eq: scaling relative dispersion} that $C=\left( 3/(2C_0) \right)^{1/3}$). 

\begin{figure}[ht]
\includegraphics[scale=0.6]{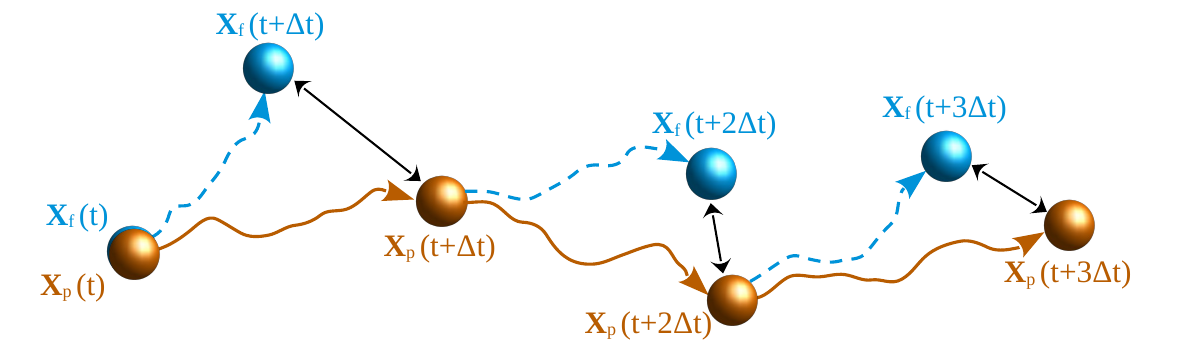}%
\caption{Sketch of the two-step model in discrete time. As indicated by the black arrows, the correct spatial correlation is obtained at each time step between the velocity of the fluid seen $\mb{U}_{\rm s}(t+k\Delta t)$ at the particle locations $\mb{X}_{\rm p}(t+k\Delta t)$ and the velocity of its companion fluid-particle $\mb{U}_{\rm f}(t+k\Delta t)$ at the positions $\mb{X}_{\rm f}(t+k\Delta t)$ which are drifting away from the discrete particle trajectory due to the mean relative velocity slip. At each time step, the companion fluid-particle is re-initialized, as indicated by the dashed blue lines.}
\label{fig: two-step-model_2}
\end{figure}

In the discrete-time and Markov-chain setting, a modified two-step model can then be proposed. It consists in the same Lagrangian step, cf. Eq.~\eqref{eq: MC two-step model first L}, and in an Eulerian step expressed by
\begin{equation}
U_{\rm E}(t+\Delta t) = U_{\rm E}(t)\exp\left( -\frac{\Delta \widehat{t}}{T_{\rm E}} \right) + u\sqrt{ 1 - \exp\left( -\frac{2\Delta \widehat{t}}{T_{\rm E}} \right)} \, \zeta_{\rm E}~,
\end{equation}
where $\Delta \widehat{t}= C \left( \lra{U_{\rm r}}^{2/3}/\lra{\epsilon_{\rm f}}^{1/3} \right) (\Delta t)^{2/3}$. As in Eq.~\eqref{eq: first two-step model combined}, the two steps can be fused into one to yield
\begin{equation}
U_{\rm s}(t+\Delta t)= U_{\rm s}(t)\exp\left[ - \left(\frac{\Delta t}{T_{\rm L}} + \frac{\Delta \widehat{t}}{T_{\rm E}} \right) \right]
+  u\sqrt{ 1 - \exp\left[ -2\,\left( \frac{\Delta t}{T_{\rm L}} + \frac{\Delta \widehat{t}}{T_{\rm E}} \right) \right] }\, \zeta_{\rm s}~,
\end{equation}
whose behavior is more in line with the expected scaling laws in time and in space. Indeed, by replacing directly in the classical expression of the auto-correlation of first-order Markov chains, we get
\begin{equation}
\label{eq: rescaled time MC for correl}
\lra{U_{\rm s}(t)\, U_{\rm s}(t+\Delta t)}= u^2 \exp\left( - \frac{\Delta t}{T_{\rm L}} - \frac{ A \left(\Delta t \right)^{2/3}}{T_{\rm E}} \right)~,
\end{equation}
where we have used $A=C \left( \lra{U_{\rm r}}^{2/3}/\lra{\epsilon_{\rm f}}^{1/3} \right)$. In the inertial range where ($\Delta t \ll T_{\rm L}, T_{\rm E}$), this yields for the second-order moment of the velocity increments of $\delta U_{\rm s}(\Delta t)= U_{\rm s}(t+\Delta t) - U_{\rm s}(t)$
\begin{equation}
\lra{ \left( \delta U_{\rm s}(\Delta t) \right)^2 } \simeq 2 u^2 \left( \frac{\Delta t}{T_{\rm L}} + \frac{ A \left(\Delta t \right)^{2/3}}{T_{\rm E}} \right) \simeq C_0 \lra{\epsilon} \Delta t + C_{\rm E}\, C \biggl( \lra{\epsilon} \lra{U_{\rm r}} \, \Delta t \biggr)^{2/3}~,
\end{equation}
where $C_{\rm E}$ is the constant entering the closure of $T_{\rm E}$ (i.e., $T_{\rm E}=2 u^2/(C_{\rm E}\lra{\epsilon_{\rm f}})$, while it is recalled that $T_{\rm L}=2u^2/(C_0 \lra{\epsilon_{\rm f}})$). It is seen that we retrieve the linear relation in $\Delta t$ for the Lagrangian correlations and the $2/3$ law for the Eulerian ones in terms of the separating distance $(\Delta r)_{\rm drift}=\lra{U_{\rm r}} \, \Delta t$ induced by the mean velocity drift over the time increment $\Delta t$. At this stage, it is tempting to rewrite Eq.~\eqref{eq: rescaled time MC for correl}, as 
\begin{equation}
\label{eq: rescaled time MC for correl bis}
\lra{U_{\rm s}(t)\, U_{\rm s}(t+\Delta t)}= u^2 \exp( -\frac{\Delta t}{\widetilde{T}_{\rm L}^{*}} ) \quad \text{with} \quad
\widetilde{T}_{\rm L}^{*}= \frac{T_{\rm L}}{1 + C_{\rm T} \left( \dfrac{\lra{U_{\rm r}}^2}{\lra{\epsilon_{\rm f}} \Delta t}\right)^{1/3} }~.
\end{equation}
However, though we are treating $\Delta t$ as a parameter in the time-discrete setting, it does not mean that $\widetilde{T}_{\rm L}^{*}$ is the expression of a proper timescale since it is an explicit function of the time increment. Actually, a specific feature of this model is that, in contrast with previous formulations, the timescale of the velocity of the fluid seen is not an input but an output (see below). Another noteworthy point is that we have used each discrete time, $t_0 + k\Delta t$ (with $k \in \mathbb{N}$), as the initial time for the companion fluid-particle used in the reformulation of the Eulerian step, which means that the time step $\Delta t$ is the elapsed times from the moment of strong interaction used in the two-particle relative dispersion problem, (i.e. $t \equiv \Delta t$ as displayed in Fig.~\ref{fig: two-step-model_2}). This explains that we have used the same relation between $\Delta \widehat{t}$ and $\Delta t$ as $\widehat{t}$ and $t$ in Eq.~\eqref{eq: rescaled time relation}.

It appears therefore that the previous considerations have been developed somewhat loosely and need to be put on a more rigorous basis by formulating the corresponding stochastic differential equations in continuous time. Using the time rescaling given in Eq.~\eqref{eq: rescaled time relation} for the Eulerian step, the proposed equation for $U_{\rm s}$ is obtained by applying the time change in Eq.~\eqref{eq: SLM in space with r} which, combined by Eq.~\eqref{eq: SLM in time with t}, leads to
\begin{equation}
\label{eq: rescaled time Langevin for Us}
\dd U_{\rm s}= - \frac{U_{\rm s}}{T_{\rm L}}\, \dd t - \frac{U_{\rm s}}{T_{\rm E}} \frac{2}{3} A (t-t_0)^{-1/3} \, \dd t 
+ \sqrt{ \lra{\epsilon} \left( C_0 + \frac{2 C_{\rm E}}{3}\, A \, (t-t_0)^{-1/3} \right) }\, \dd W~.
\end{equation}
However, the signification of the time(s) written as $t_0$ in this equation requires careful consideration. 

\begin{figure}[ht]
\includegraphics[scale=0.6]{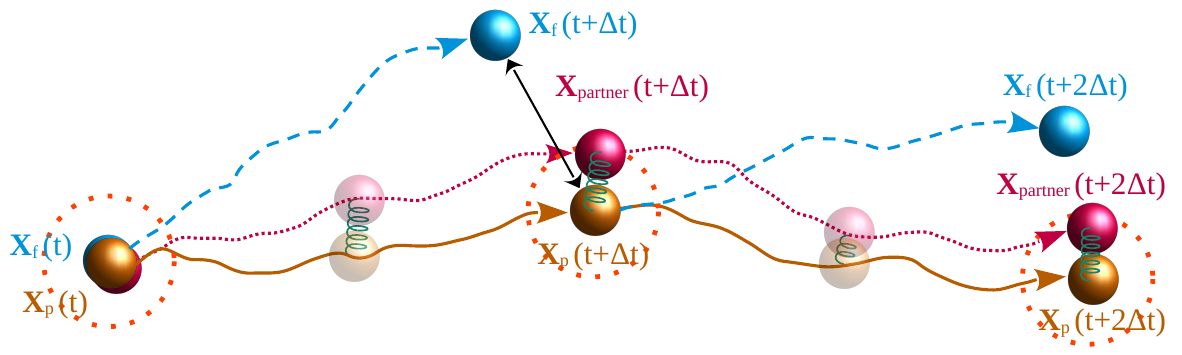}%
\caption{Sketch of the two-step model in continuous time. A fluid-particle partner is associated to each discrete particle trajectory, cf. Eq.~\eqref{eq: spring term for FP partner}, and is used to generate new companion fluid particles for the two-step model. The correct spatial correlations are still between the velocity of the fluid seen and the velocity of the companion fluid particle, as shown by the black arrows. This companion fluid partner is unchanged as long as the distance between the discrete particle and its partner remains larger than the size of the zone of strong interactions. However, whenever the particle partner enters a zone of strong interaction, a new companion fluid-particle is generated, as indicated by the dashed blue lines.}
\label{fig: two-step-model_3}
\end{figure}

From the two-particle relative dispersion study, it follows that the Eulerian step is devised to retrieve the correct correlation between the velocity of the fluid seen (the fluid velocity at the discrete particle location) and the velocity of a companion fluid particle whose location is moving away from the discrete particle according to the mean velocity drift. In Eq.~\eqref{eq: rescaled time Langevin for Us}, the time $t_0$ refers to the choice of the `initial time' at which this companion fluid particle is released (this is the time at which these two fluid elements leave the `strong interaction zone', see Fig.~\ref{fig: two-fluid-particle-dispersion}) and its value reflects a modeling choice for the size of the domain in which we wish to capture, at each time, the velocity correlations between these two fluid elements. Since the relative distance between the trajectory of a discrete particle and its companion fluid particle increases due to the mean velocity drift, it can grow unbounded and after a long-enough time lapse (i.e., $t-t_0 \gg L/\lra{U_{\rm r}}$), these two fluid elements are so far apart that their velocities become independent. This is reflected in Eq.~\eqref{eq: rescaled time Langevin for Us} by the fact that, if $t_0$ is a constant, then the drift and diffusion coefficients of the Eulerian step go to zero as $t-t_0$ increases. It is then more appropriate to consider that $t_0$ becomes a random variable whose value is somehow updated and attached to each discrete particle. For instance, in the discrete time setting presented above, we have implicitly assumed that this companion fluid particle is `re-initialized' at each time step, while $t-t_0=\Delta t$. In the continuous time setting, this can be monitored by considering a fluid particle partner attached to the trajectory of each discrete particle. The evolution equation for the relative distance between the discrete particle and this fluid partner, say $\delta r_{\rm p}^{\rm f}$, is defined as
\begin{equation}
\label{eq: spring term for FP partner}
\frac{\dd \left( \delta r_{\rm p}^{\rm f}(t) \right)}{\dd t}= \lra{U_{\rm r}} - k_{\rm spring} \, \delta r_{\rm p}^{\rm f} (t)~,
\end{equation}
where $k_{\rm spring}$ is the stiffness of an associated spring that links the trajectory of the discrete particle to the one of its fluid particle partner. Note that when $k_{\rm spring}=0$ this fluid particle partner is identical to the companion fluid particle described above. On the other hand, when $k_{\rm spring}$ is very large, we are in the situation where the fluid particle partner is quickly pulled back towards the discrete particle, which corresponds to the case considered in the discrete time setting (re-initialization of the fluid particle partner at each time step). For intermediate value of $k_{\rm spring}$, the evolution of $\delta r_{\rm p}^{\rm f}(t)$ governs how $t_0$ is updated, since we have $t_0=\tau^{\eta}_{\delta r_{\rm p}^{\rm f}}$ where $\tau^{\eta}_{\delta r_{\rm p}^{\rm f}}$ are the instants when the discrete particle and its fluid particle partner enter a zone of strong interaction, i.e., when $\vert \delta r_{\rm p}^{\rm f}(\tau^{\eta}_{\delta r_{\rm p}^{\rm f}}) \vert \leq \eta$. This two-step stochastic model is sketched in Fig.~\ref{fig: two-step-model_3}. 

With the complete formulation of the new two-step stochastic model, we can express the timescale of the velocity of the fluid seen that results from it. First, the evolution equation for the auto-correlation of the velocity of the fluid seen, $R_{\rm s}(t)=\lra{U_{\rm s}(\tau+t)\, U_{\rm s}(\tau)}$, is derived from Eq.~\eqref{eq: rescaled time Langevin for Us} and is
\begin{equation}
\frac{\dd R_{\rm s}(t)}{\dd t} = - \frac{R_{\rm s}(t)}{T_{\rm L}} - \frac{R_{\rm s}(t)}{T_{\rm E}} \frac{2}{3} A t^{-1/3}~,
\end{equation}
where we have taken $t_0=0$. By direct integration, we get 
\begin{equation}
\log \left( \frac{R_{\rm s}}{R_{\rm s}(0)} \right)= - \frac{t}{T_{\rm L}} - \frac{A\, t^{2/3}}{T_{\rm E}}~.
\end{equation}
Using $R_{\rm s}(0)=u^2$, we obtain therefore the same expression as the one used directly with the Markov chains with $\Delta t$ as the elapsed time, cf. Eq.~\eqref{eq: rescaled time MC for correl}. However, to work out the timescale of the velocity of the fluid seen $T_{\rm L}^{*}$, we need to go back to its proper definition as the integral of the auto-correlation function for the velocity of the fluid seen, $\rho_{R_{\rm s}}(t)=R_{\rm s}(t)/u^2$. This gives
\begin{equation}
T_{\rm L}^{*} = \int_0^{+\infty} \rho_{R_{\rm s}}(t') dt' = \int_0^{+\infty} \exp\left( - \frac{t'}{T_{\rm L}} - \frac{A\, t'^{2/3}}{T_{\rm E}} \right) dt'~,
\end{equation}
which, by making the change of variable $s=t'/T_{\rm L}$, can be written as
\begin{equation}
\label{eq: two-step model predicted TL*}
T_{\rm L}^{*} = T_{\rm L} \int_0^{+\infty} \exp\left( - s - \frac{C_{\rm T} \, A}{T_{\rm L}^{1/3}} \, s^{2/3} \right) ds~.
\end{equation}
To use the same notation as in Sec.~\ref{the crossing-trajectory effect}, this means that the equivalent of the Csanady factor, $T_{\rm L}^{*}/T_{\rm L}$ is
\begin{equation}
F( \lra{U_{\rm r}}, T_{\rm L}, T_{\rm E},\lra{\epsilon_{\rm f}})= \int_0^{+\infty} \exp\left( - s - \frac{C_{\rm T} \, A}{T_{\rm L}^{1/3}} \, s^{2/3} \right) ds~.
\end{equation}
From the above expressions of the coefficient $A$ and of $T_{\rm L}$, it is seen that $A/T_{\rm L}^{1/3}=(C_0/2)^{1/3} \left( \lra{U_{\rm r}}/u \right)^{2/3}$ and the above integral is actually only a function of the ratio $\lra{U_{\rm r}}/u$
\begin{equation}
\label{eq: Csanady factor two-step model}
F( \lra{U_{\rm r}}/u)= \int_0^{+\infty} \exp\left( - s - C_{\rm T} (C_0/2)^{1/3} \left(\frac{\lra{U_{\rm r}}}{u}\right)^{2/3} \, s^{2/3} \right) ds~,
\end{equation}
which is in line with the functional forms used in the various Csanady-like formulas presented previously. Some additional properties of the timescale of the velocity of the fluid seen resulting from this two-step model, cf. Eq.~\eqref{eq: two-step model predicted TL*}, can be given. First, since the integrand function is smaller than $\exp(-s)$, whose integral over $\left[0,+\infty \right]$ is equal to $1$, then we get the following bounding interval $0\leq F( \lra{U_{\rm r}}/u) \leq 1$. When the mean velocity drift $\lra{U_{\rm r}}$ is negligible (i.e., $\lra{U_{\rm r}}=0$ in Eq.~\eqref{eq: two-step model predicted TL*}), we retrieve the fluid-like behavior and $T_{\rm L}^{*}(\lra{U_{\rm r}}=0)=T_{\rm L}$, or $F(\lra{U_{\rm r}}=0)=1$. Furthermore, $F$ is obviously a decreasing function of $\lra{U_{\rm r}}/u$ and its behavior in the limit of a very large velocity drift can be assessed. In that limit, we can indeed neglect the first term in the exponential function in Eq.~\eqref{eq: two-step model predicted TL*}, which corresponds to the Lagrangian step. This gives
\begin{equation}
T_{\rm L}^{*}( \lra{U_{\rm r}}/u \gg 1) \simeq T_{\rm L} \int_0^{+\infty} \exp\left( - \frac{C_{\rm T} \, A}{T_{\rm L}^{1/3}} \, s^{2/3} \right) ds~.
\end{equation}
Writing $A=\widetilde{C} \left( \lra{U_{\rm r}}^{2/3}/\lra{\epsilon_{\rm f}}^{1/3} \right)$, where $\widetilde{C}$ is a constant which is now to be determined, we obtain 
\begin{equation}
T_{\rm L}^{*}( \lra{U_{\rm r}}/u \gg 1) \simeq \frac{u \, T_{\rm E}}{\lra{U_{\rm r}}} \frac{1}{\widetilde{C}^{3/2}} \left( \frac{2}{C_{\rm E}} \right)^{1/2} \int_0^{+\infty} \exp\left( - v^{2/3} \right) dv~.
\end{equation}
Since $\int_0^{+\infty} \exp\left( - v^{2/3} \right) dv=3\sqrt{\pi}/4$, we can therefore choose the constant $\widetilde{C}$ so that
\begin{equation}
\frac{1}{\widetilde{C}^{3/2}} \left( \frac{2}{C_{\rm E}} \right)^{1/2} \frac{3\, \sqrt{\pi}}{4} =1~,
\end{equation}
which means that we retrieve the expected scaling of $T_{\rm L}^{*}$ in the limit of very large velocity slips, i.e., 
\begin{equation}
T_{\rm L}^{*}( \lra{U_{\rm r}}/u \gg 1) \simeq \frac{u \, T_{\rm E}}{\lra{U_{\rm r}}}=\frac{L_{\rm E}}{\lra{U_{\rm r}}}~.
\end{equation}
This corresponds to the limit situation of frozen turbulence, already stated with the formalism developed from the Extended Kolmogorov Hypothesis and given in Eq.~\eqref{eq: EKH frozen turbulence b}, since $\lra{U_{\rm r}} \, T_{\rm L}^{*}( \lra{U_{\rm r}}/u \gg 1)=L_{\rm E}$. 

If the limit $T_{\rm L}^{*}(\lra{U_{\rm r}}=0)$ is correct, the expression of $T_{\rm L}^{*}$ from the two-step model needs to be refined for very small values of $\lra{U_{\rm r}}$ (note that, as such, Eq.~\eqref{eq: Csanady factor two-step model} predicts that the derivative of $F$ with respect to $\lra{U_{\rm r}}$ is infinite at $\lra{U_{\rm r}}=0$). Indeed, in the derivations leading to Eq.~\eqref{eq: two-step model predicted TL*}, we not only retained a single zone of strong interaction but we also assumed that the Eulerian step was applied as soon as $t > 0$. This amounts to neglecting the time spent in the zone of strong interaction. Yet, by assuming that the size of the zone of strong interaction is of the order of the Kolmogorov scale $\eta$, there is always a minimum time, which we can estimate as $t_{\rm min} \simeq \eta/\lra{U_{\rm r}}$, to consider before switching on the additional terms coming from the time-rescaling model (when $t \leq t_{\rm min}$, only the Lagrangian step can be retained). In the previous derivations, we implicitly assumed very high Reynolds-number turbulent flows, where $\eta$ becomes vanishingly small, and non-zero values of $\lra{U_{\rm r}}$, so that $t_{\rm min} \simeq 0$. It is, however, much sounder to regard $\eta$ as a small but non-zero value, which means that $t_{\rm min}$ increases as $\lra{U_{\rm r}} \to 0$. When $t_{\rm min}$ becomes, for instance, of the order of the Lagrangian timescale $T_{\rm L}$, then the Eulerian part of the two-step model would only be applied when velocities are already nearly uncorrelated and would only slightly modify the value of $T_{\rm L}^{*}$ in comparison to $T_{\rm L}$. Said otherwise, the two-step model is applied and is effective for the time range which is typically between $t_{\rm min}$ and $T_{\rm L}$ (or another similar timescale of large-scale motions) and the above derivations are valid provided that we have $U_{\rm r} \geq \eta/T_{\rm L}$, while we can expect that $T_{\rm L}^{*} \simeq T_{\rm L}$ when $\lra{U_{\rm r}} \leq \eta/T_{\rm L}$. It results that $F$, or $T_{\rm L}^{*}/T_{\rm L}$, should have a zero-slope behavior in the immediate vicinity of $\lra{U_{\rm r}} \simeq 0$.

Another remark is in order. Even though it is difficult to work out an analytical formula for the resulting timescale $T_{\rm L}^{*}$, it must be noted that the above expressions were obtained in the limit where $k_{\rm spring}$ is small so that $t_0$ was taken as a constant. When this is not the case, it follows from the description of the model that Eq.~\eqref{eq: rescaled time Langevin for Us} involves successive terms such as $t-\tau_{\eta}(\delta r_{\rm p}^{\rm f})$ where $\tau_{\eta}(\delta r_{\rm p}^{\rm f})$ are random numbers (similar to threshold or first-passage times), which makes the above derivations intractable. Only numerical estimations of $T_{\rm L}^{*}$ seem accessible, which is why the model was also described in the discrete-time setting.

\paragraph{Extension to the general 3D non-homogeneous case}
Should they prove worth pursuing, these developments need to be generalized to three-dimensional situations. Actually, the same remarks made at the end of Sec.~\ref{EKH for Lagrangian timescales} for the propositions from the Extended Kolmogorov hypothesis apply here. Given that the Eulerian time scales are different depending on whether we consider the direction aligned with the mean relative velocity or a transverse one, cf. Eq.~\eqref{eq: Eulerian time tensor 3D}, care should be taken to distinguish between directions when using the two-step model, as in Eq.~\eqref{eq: rescaled time Langevin for Us}. The safest way is probably to make a (local) change in the frame of reference so as to have directions aligned or transverse with respect to the (local) mean relative velocity. Note that this implies a local change in the frame of reference that is, however, not dependent on each particle (it depends only on mean properties) and is the same for all the particles contained in a small volume around a point of interest. There is, therefore, no need to account for non-inertial effects.

\subsubsection{Accounting for the statistical signature of coherent structures}\label{Accounting for structures signature}
%------------------------------------------------------------------------------

When investigating turbulence characteristics, two main standpoints are typically followed. The first one, referred to as the statistical approach, is represented by the Kolmogorov theories (K41 and K62, see Sec.~\ref{The Kolmogorov theory}) as well as similar analyses centered around finding proper scaling for quantities of interest. With the development of numerical tools (DNS) regarded as numerical experiments and used in complement of actual experiments, another viewpoint has emerged in which instantaneous flow patterns are identified and analyses performed in terms of the dynamics of these structures. This corresponds to the geometrical approach, centered around isolating a set of structures whose characteristics could explain key turbulence properties (e.g., the so-called hairpin vortices in near-wall boundary layers on which there is now a vast literature). Surprisingly, a middle-of-the-road formulation has rarely been considered. In this third approach, one starts with a given picture of turbulence including models for a selected number of fully-characterized coherent structures and the aim is to develop a statistical model where these structures are explicitly treated. In that sense, the objective is neither to average out their presence (as in the statistical approach) nor to calculate or predict their formation or dynamical role (as in the geometrical approach) but to account for them in a statistical description. In a nutshell, we wish to mimic given coherent structures using stochastic models.

To illustrate these ideas, we develop the following two-fold description in which turbulent flows are regarded as composed of two regions having different statistical behavior:
\begin{enumerate}
\item a random background flow, occupying most of the flow domain, well-described by the K41 theory (including, perhaps, some features of the K62 refined theory), and where the PDF models presented in the previous sections provide a satisfactory description of one-particle and one-point statistics; 
\item a collection of coherent structures (exhibiting stable, ordered and long-life patterns) taken to be intense vortex filaments, whose statistical description needs to be detailed and which can account for intermittency and deviations of high-order moments from Kolmogorov predictions. 
\end{enumerate}

Since we are drawing a roughly-cut sketch to exemplify the new approach, we limit ourselves to simple scaling arguments to support the selection of vortex filaments and express their characteristic features. 

\begin{figure}[ht]
 \centering
 \includegraphics[width=0.75\textwidth]{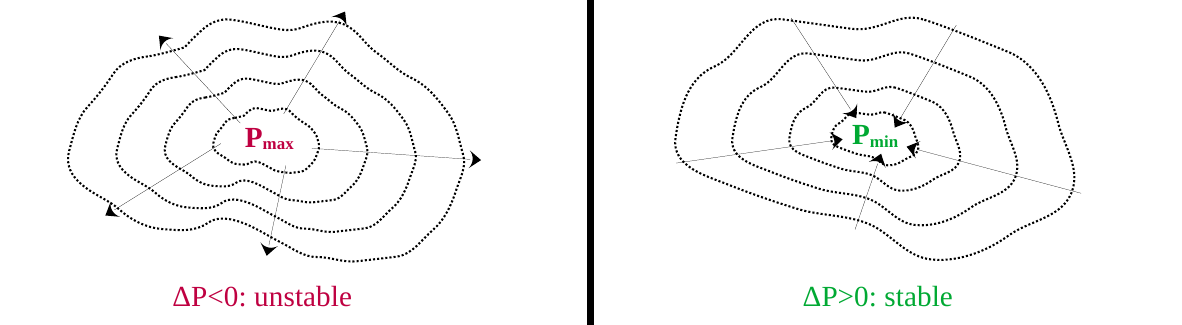}
	\caption{Representation of an unstable flow structure with positive pressure (left side) at the center and where the pressure-gradient force is oriented outward and a stable one with negative pressure (right side) at the center and where the pressure-gradient force is oriented inward.}
 \label{fig: structure pressure}
\end{figure}

Retaining vortex tubes is related to simple equilibrium reasoning on the pressure-gradient force and, thus, on the minimum of potentials (recalling that pressure is basically a potential), cf. Fig.~\ref{fig: structure pressure}.

Therefore, the positive pressure side (i.e., $P \geq \lra{P}$) is not stable and follows approximately a Gaussian distribution. On the other hand, the negative pressure side (i.e., $P \leq \lra{P}$) corresponds to more stable flow patterns and this part of the pressure distribution deviates from Gaussianity~\cite{vincent1991spatial}. In short, stable structures are associated with high vorticity, often under the form of vortex tubes. In the present description, we do not consider the classical vortex structures, such as vortex sheets, whose study received a great deal of attention in past decades~\cite{saffman1995vortex} but some intense events inherent to the turbulence cascade although they do not explain the mean turbulent kinetic energy dissipation rates (see detailed analyses in~\cite{jimenez1993structure,jimenez1998characteristics} about this difference). The most intense vorticity events are obtained by considering velocity differences of the order of the large-scale fluctuating velocity $u$ over the smallest size which is of the order of the Kolmogorov length scale $\eta_{\rm K}$. This yields the typical vorticity of vortex tubes as $\omega_{\rm vt} \sim u/\eta_{\rm K}$. This estimate is also obtained by considering that the kinetic energy of large scales $u^2$ is converted entirely into small-scale rotational energy:
\begin{equation}
r^2 \omega^2 \simeq u^2~, \; \text{which for} \; r \simeq \eta_{\rm K} \; \Longrightarrow \omega_{\rm vt} \sim \frac{u}{\eta_{\rm K}}~.
\end{equation}
From the equation satisfied by the enstrophy $\omega^2$ (with $\omega^2=\bds{\omega}\cdot \bds{\omega}$)
\begin{equation}
\frac{D \omega^2}{Dt} = 2 \, \omega_i \, S_{ij}\, \omega_j + \nu_{\rm f}\,\omega_i \, \Delta \omega_i = 2 \, \omega_i \, S_{ij}\, \omega_j - \nu_{\rm f} \left( \frac{\partial \omega_i}{\partial x_j}\right) \left( \frac{\partial \omega_i}{\partial x_j} \right) + \nu_{\rm f} \Delta (\omega^2)~,
\end{equation}
where $S_{ij}=1/2\left( \partial U_{{\rm f},i}/\partial x_j + \partial U_{{\rm f},j}/\partial x_i\right)$ is the flow strain rate, we see that the stretching term $\omega_i \, S_{ij}\, \omega_j$ must be positive to balance sink terms due to molecular viscosity and maintain the structure for a longer-than-average life time. If we regard these vortex tubes as relatively stable structures maintained out of equilibrium due to the energy flux from the background flow, we are referring to `dissipative structures' that are by-products of the turbulence cascade~\cite{jimenez1998characteristics} and we can apply the theorem of minimum entropy production rate~\cite{kjelstrup2008non,kjelstrup2010non} for such structures (with the `entropy' here embodied by the enstrophy $\omega^2$). This means that, for $\omega^2$ remaining approximately constant, the entropy production or stretching term, $\omega_i \, S_{ij}\, \omega_j$, is minimum inside the vortex filaments. It follows that the vorticity $\bds{\omega}$ is essentially aligned, or correlated, with the intermediate eigenvalue of the strain rate tensor $S_{ij}$ and that, in these structures, this eigenvalue is positive. This is in line with results from DNS studies~\cite{jimenez1993structure,jimenez1998characteristics}.

\begin{figure}[ht]
 \centering
 \includegraphics[width=0.9\textwidth]{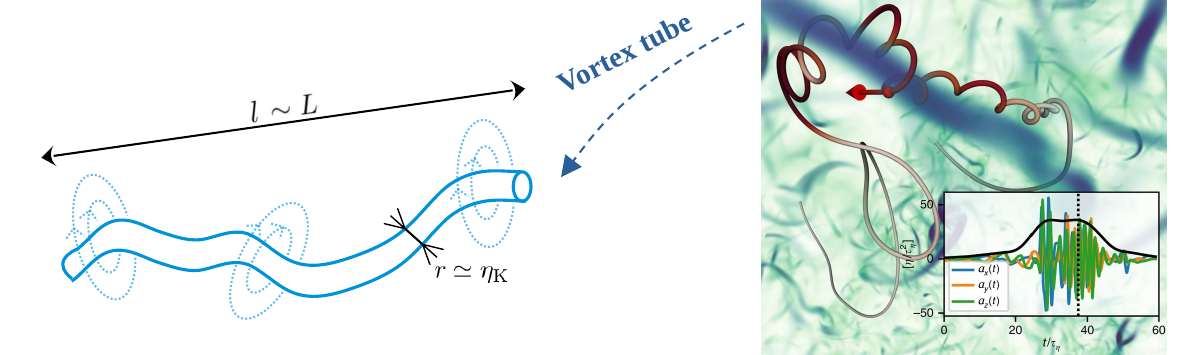}
 \caption{Representation of a vortex filament with its characteristic spatial scales (left) and an illustration from DNS simulations showing the interaction between tracer particles and vortex filaments (trajectories are colored according to the particle acceleration, shown in the inset on the right). Reprinted with permission from \cite{bentkamp2019persistent}. Copyright Springer Nature, 2019.}
 \label{fig: structure vortex}
\end{figure}

When velocities are rescaled to the velocity $u$ of the large scales, the typical length of dissipative structures is the Taylor lengthscale $\lambda$. Pursuing the idea of transforming kinetic energy coherently into rotational energy, we can estimate the length of a vortex tube by imagining that an initial sphere of radius $\lambda$ is squeezed into a tube until the cross-section radius of that tube is of the order of $\eta_{\rm K}$. From the incompressibility of the flow, which entails that volume is conserved, we have that the typical length of a vortex tube is $l \sim \lambda^3/\eta_{\rm K}^2 \sim L$ and is therefore of the order of the length of the large-scale velocity fluctuations, as found in fundamental analyses~\cite{jimenez1998characteristics}. The emerging simplified picture is sketched in Fig.~\ref{fig: structure vortex} with a region of high vorticity (but small dissipation) inside the tubes and a region of high dissipation (but reduced vorticity) around these tubes. The life-time, or typical time scale, of these structures is harder to estimate and, to the authors' knowledge, has not been considered in the fundamental DNS studies dedicated to the role of vortex tubes, reflecting that probing turbulence from a Lagrangian viewpoint has only emerged in recent years. For the present concern, we refer to it as $\tau_{\rm vt}$, whose value needs to be derived from first principles or extracted from DNS investigations (a first, and rather rough, guess could be to assume that $\tau_{\rm vt} \sim L/u$, which is the eddy-turn-over time of large scales).

To assess the macroscopic role played by the vortex tubes, the key point is to estimate the number of these tubes in a box of volume $L^3$ or, conversely, the probability to encounter such a structure. If we consider a locally homogeneous flow domain, this probability is the same as the volume fraction, $\alpha_{\rm vt}$, occupied by the vortex tubes. In the spirit of Kolmogorov-like theories, we imagine then a process of successive breakdowns whereby a large-scale particle with size $L$ and velocity $u$ yields a number $n_{\rm vt}$ of small-scale particles having a rotation velocity $\omega_{\rm vt}$ and, at least, one typical length of the order of $\eta_{\rm K}$ (as for the cross-section radius of the intense vortex filaments). From the conservation of angular momentum $r^2 \omega$, we get that 
\begin{equation}
L^2 \left( \frac{u}{L} \right) = u \, L \simeq n_{\rm vt} \left( \eta_{\rm K}^2 \, \omega_{\rm vt}\right) = n_{\rm vt} \left( \eta_{\rm K} \, u \right) \; \Longrightarrow n_{\rm vt} \sim \left( \frac{L}{\eta_{\rm K}} \right) \;\; \text{and} \;\; \alpha_{\rm vt}=n_{\rm vt} \left( \frac{\eta_{\rm K}^2 L}{L^3} \right) \sim \frac{\eta_{\rm K}}{L}~.
\end{equation}
We can then evaluate the contribution of the vortex tubes to the budget of the first moments of the dissipation rate (or the enstrophy). Since the dissipation rate associated to a vortex tube can be written as $\epsilon_{\rm vt}= \nu_{\rm f} (u^2/\eta_{\rm K}^2)=\lra{\epsilon_{\rm f}}(\lambda^2/\eta_{\rm K}^2)$, we obtain that the average contribution of the dissipation rate due to vortex tubes to the budget of the mean dissipation rate is negligible since
\begin{equation}
\lra{\epsilon_{\rm vt}} = \lra{\epsilon_{\rm f}} \, \frac{\lambda^2}{\eta_{\rm K}^2} \, \frac{\eta_{\rm K}}{L} \; \Longrightarrow \; \lra{\epsilon_{\rm vt}} \sim \lra{\epsilon_{\rm f}} \left( \frac{L}{\eta_{\rm K}} \right)^{-1/3}~.
\end{equation} 
However, for the second-order moment of the dissipation rate, we get that
\begin{equation}
\lra{\epsilon_{\rm vt}^2} = \lra{\epsilon_{\rm f}}^2 \, \frac{\lambda^4}{\eta_{\rm K}^4} \, \frac{\eta_{\rm K}}{L} \; \Longrightarrow \; \lra{\epsilon_{\rm vt}^2} \sim \lra{\epsilon_{\rm f}}^2 \left( \frac{L}{\eta_{\rm K}} \right)^{1/3}~,
\end{equation}
which shows that the vortex tubes provide the main contribution to the variance of the dissipation rate (similar results hold for the enstrophy). Note that we retrieve a scaling relation in line with Kolmogorov K62 hypothesis with an exponent $\mu=1/3$. To modify the crude estimation of the number of vortex tubes and its volumetric fraction to account for a Kolmogorov exponent $\mu$ different from the value above (usually $\mu$ is estimated at a slightly lower value than $1/3$, often $\mu \simeq 0.2-0.25$), it can be proposed that $\alpha_{\rm vt}$ scales as $(\eta_{\rm K}/L)^{4/3 - \mu}$, from which the same estimations yield that 
\begin{equation}
\lra{\epsilon_{\rm f}^2} \sim \lra{\epsilon_{\rm f}}^2 \left( L/\eta_{\rm K} \right)^{\mu}~.
\end{equation}
Additional results of interest concern Lagrangian statistics, e.g. the acceleration of a fluid particle $a=\dd u/\dd t$. As already indicated, the K41 scaling for the variance of a fluid particle acceleration is $a_{\rm K}^2 \sim \left( \lra{\epsilon_{\rm f}}^3/\nu_{\rm f} \right)^{1/2}$. When a fluid particle is caught by a vortex tube, its acceleration is $a_{\rm vt} \sim u^2/\eta_{\rm K}$, from which we get that $a_{\rm vt}^2 \sim a_{\rm K}^2 \left( \lambda\, L \right)/\eta_{\rm K}^2$. The second-order moment of a fluid particle acceleration deviates therefore from the K41 prediction $\lra{a^2} \sim a_{\rm K}^2$, since we obtain for the vortex tube contribution that
\begin{equation}
\lra{ a_{\rm vt}^2 } \sim a_{\rm K}^2 \, \frac{\lambda}{\eta_{\rm K}} \sim a_{\rm K}^2 \, Re_{\lambda}^{1/2}~,
\end{equation}
a relation found in~\cite{yeung1989lagrangian} but not explained by classical scaling~\cite{yeung2002lagrangian}. Note that if we retain the more general estimate $\alpha_{\rm vt} \simeq (\eta_{\rm K}/L)^{4/3 - \mu}$, the predicted scaling becomes $\lra{ a_{\rm vt}^2 } \sim a_{\rm K}^2 \, Re_{\lambda}^{3\mu/2}$, with a smaller exponent, in line with more recent probes~\cite{lanotte2013new}). Drawing on these results, it is interesting to note that the effects of coherent structures are different on Eulerian and Lagrangian statistics. At a fixed location, these structures are swept by, making little marks on Eulerian statistics, whereas fluid particles can be trapped in them for longer times (of the order of $\tau_{\rm vt}$) leaving therefore potentially more significant traces on Lagrangian statistics (see an interesting account in~\cite{lanotte2013new}).

In spite of its crudely-cut features (for instance, we only considered one coherent structure and only one type of vortex tubes instead of a distribution of sizes, intensity, length, etc.), this two-fold model is useful to reveal that, not only classical signatures of intermittency are retrieved, but also that complex scaling of fluid particle statistics can be captured. This is relevant for stochastic modeling since descriptions such as the one presented above are easily implemented in the PDF framework. In practice, it means that we consider two stochastic models for the velocity of the fluid seen by discrete particles, one for the background `structure-less flow' based typically on a Langevin model as introduced in preceding sections (we call this model, `model BG') and one model for the flow in the near vicinity of a vortex tube typically by generating a Burgers-like flow once the vortex tube orientation and intensity are generated (we call this model, `model VT'). To that effect, we introduce a parent process, $S(t)$, attached to each particle, which governs the random switches from one flow region to the other: say $S(t)=1$ when the particle is considered in the background flow and where the model BG is applied for $\mb{U}_{\rm s}$, and $S(t)=2$ when the particle is captured by one vortex tube and where the model VT is applied for $\mb{U}_{\rm s}$. This parent process can be modeled by a generalized Poisson process jumping at random times between $S(t)=1$ and $S(t)=2$ at the end of duration times (i.e., the times spent in each flow type) which, according to the properties of Poisson processes, are random variables following exponential laws determined by two timescales: $\tau_{\rm BG}$ the mean duration within the background flow and $\tau_{\rm vt}$ the mean duration within a vortex tube. Note that $\tau_{\rm vt}$ was already introduced above, while $\tau_{\rm BG}$ is derived from the volumetric fraction of the vortex tubes $\alpha_{\rm vt}$ as for inter-collision times in the kinetic theory (this is, thus, a foretaste of developments to come in Sec.~\ref{statistical model collision}). These conditions define the $2\times 2$ transition matrix $\mb{M}(\Delta t)$ which governs the Markov chain $S(k \Delta t)$ at discrete times $t=k\Delta t$ ($k \in \mathbb{N}$) with $M_{ij}(\Delta t)$ the probability to have $S(t+\Delta t)=j$ at time $t+\Delta t$ conditioned on $S(t)=i$ at time $t$:
\begin{equation}
\mb{M}(\Delta t) = 
\begin{pmatrix}
e^{-\Delta t/\tau_{\rm BG}} & 1 - e^{-\Delta t/\tau_{\rm BG}} \\
1 - e^{-\Delta t/\tau_{\rm vt}} & e^{-\Delta t/\tau_{\rm vt}}
\end{pmatrix}~.
\end{equation}

\begin{figure}[ht]
 \centering
 \includegraphics[width=\textwidth]{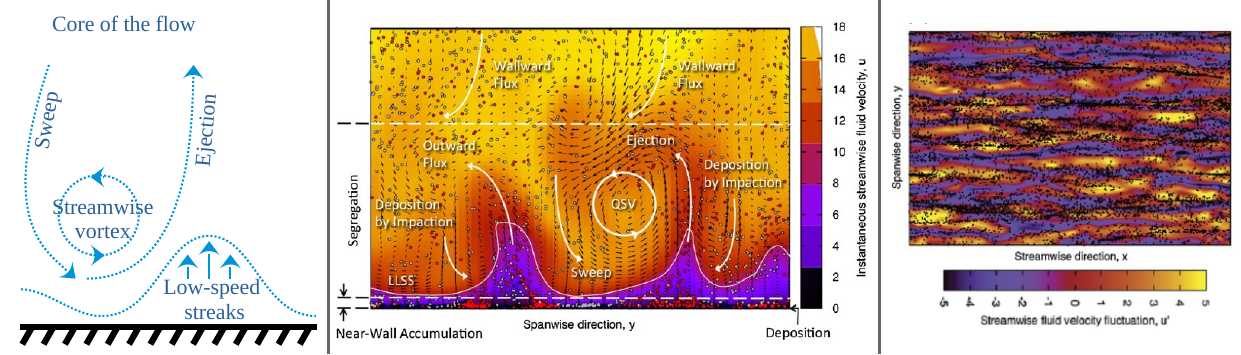}
 \caption{Representation of near-wall turbulent structures (left) together with side view (middle) and top view (right) of a DNS simulation showing the vorticity regions as well as the particle dynamics in the near-wall region. Reprinted with permission from \cite{marchioli2017large}. Copyright Springer, 2017.}
 \label{fig: structure near-wall}
\end{figure}

To the best of the authors' knowledge, the development of such structure-based stochastic models is still at an early stage (though there are similarities with ideas put forward in~\cite{kerstein1999one}), apart from one model of particle transport in wall-boundary layers for deposition problems~\cite{guingo2008stochastic}. In this model, a random combination of three typical structures of the near-wall region, namely sweeps, ejections and diffusion, was used to capture the particle flux towards the wall where a sticking condition was enforced for particles touching the surface. The model represents a simplified picture based on insights provided by DNS results about particle transport in wall-boundary layers, cf. Fig.~\ref{fig: structure near-wall}. 

\begin{figure}[ht]
 \centering
 \includegraphics[width=\textwidth]{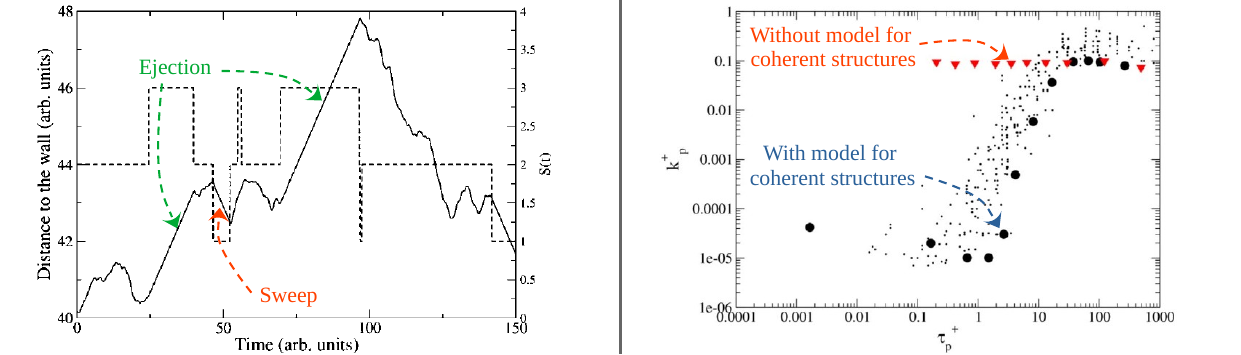}
	\caption{Results obtained with a stochastic model for near-wall turbulent structures showing the effect of the random switches between coherent structures on particle trajectories (left) and on the deposition rate (right). Reprinted with permission from \cite{guingo2008stochastic}. Copyright Elsevier, 2008.}
 \label{fig: structure near-wall 2}
\end{figure}

The construction of this model is in line with the two-fold description introduced at the beginning of this section which is applied using a zonal decomposition: the background `structure-less' flow is identified with the bulk of the flow while the three selected structures are explicitly simulated only in a thin zone in the immediate vicinity of a wall surface (say between $0\leq z^{+} \leq 100$, with $z^{+}=z u_{{\rm f},*}/\nu_{\rm f}$ the wall-distance $z$ normalized by the so-called wall units with $u_{{\rm f},*}$ the fluid friction velocity, see accounts in~\cite{henry2012towards,minier2015lagrangian}). The random interplay of structures is clearly visible in the trajectories of discrete particles and satisfactory statistics for the particle deposition rate are retrieved over the whole range of particle inertia, as shown in Fig.~\ref{fig: structure near-wall 2}.

In more general terms, it is worth underlying that these structure-based models represent a new way to address complexity: instead of developing a single model which is likely to become complex itself, the leading idea is to capture complexity as the result of the random changes between the sub-models, each of which remaining simple, which are used to describe the components of an heterogeneous system. In that sense, this introduces us to similar concerns and approaches in so-called complex fluids, which are now addressed.

%=========================================================================================
\section{Similarities and differences with complex fluids \label{differences soft matter}}
%=========================================================================================

Though they are also referred to as soft matter, it is more appropriate for our discussion to designate by `complex fluids' fluids whose microstructure involves specific components (the solute) having sizes and relaxation timescales much larger than those of the atoms/molecules of the `simple fluid' (the solvent) in which they are embedded. In simple fluids, the microstructure remains regular and locally well-described by classical equilibrium statistical mechanics regardless of whether the material is flowing or at rest. At the hydrodynamical level of description, this is reflected by the linear force-flux relations entering the balance equations, such as the Newton, Fick and Fourier laws, where the only trace of molecular details is in the transport coefficients, i.e., the fluid viscosity, diffusivity and conductivity. In complex fluids, the slow and non-equilibrium response of the solute is manifested by more involved rheological laws. Typical examples include polymeric fluids, colloidal suspensions, liquid crystals, etc. (granular or glassy materials can also be considered). To overcome the difficulty of formulating the rheological properties of such complex fluids directly at the hydrodynamical or macroscopic level of description, additional structural variables are selected and a simplified (i.e., a coarse-grained) dynamical model is proposed to capture the solute response to external forces. This corresponds, for instance, to the Rouse model for polymers \cite{rouse1953theory}, Brownian dynamics for colloids \cite{chen2004brownian} or the Maier-Saupe theory for liquid crystals \cite{stephen1974physics}.

\begin{figure}[ht]
 \centering
 \includegraphics[width=0.9\textwidth]{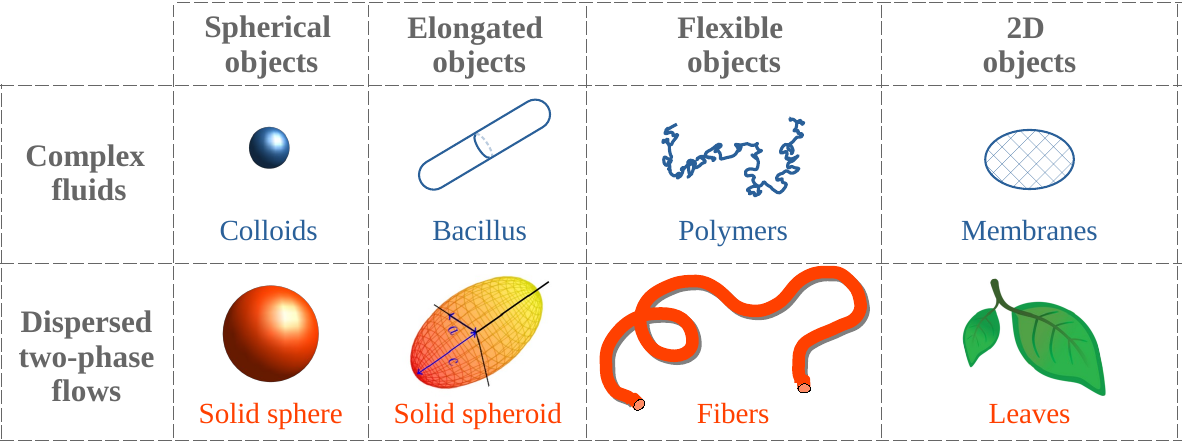}
 \caption{Correspondence between objects in complex fluids and in dispersed two-phase flows.}
 \label{fig: correspondence objects}
\end{figure}
From this brief description, it follows that complex fluids and dispersed turbulent two-phase flows have marked similarities. First, we are considering similar physical entities, with the ones handled in complex fluids appearing as smaller-size versions of the ones in dispersed two-phase flows. This correspondence is displayed in Fig.~\ref{fig: correspondence objects}, which exhibits the obvious couples: polymers and flexible fibers; colloids and discrete particles; rod-like molecules and rod-like particles. Second, there are also similarities between the main physical processes: fluctuations and dispersive effects on the one hand, and, particle-particle interactions and tendency to self-assemble on the other hand. For complex fluids, fluctuations are related to thermal noise while `turbulence noise' is present in one-particle PDF descriptions of dispersed two-phase flows (note that repulsive particle-particle interactions have also a dispersive effect). The tendency to self-assemble is manifested by phase transitions in soft matter systems, e.g., solid/liquid/gas transitions in fluids, isotropic/nematic liquid crystals, paramagnetic/ferromagnetic states (if we regard them as a complex fluid), while attractive forces govern the formation of aggregates in dispersed two-phase flows.

There are, nevertheless, differences in the overall objective being pursued. Indeed, even if a specific model is applied to describe the solute dynamics, the purpose in soft matter is to close the equations governing the evolution of a complex fluid treated as a single entity. This is typically the case of polymeric fluids where the simulation of polymer motion is used to obtain the additional stress tensor, through the Kramers expression, which is added to the (Newtonian) stress tensor of the solvent. In other words, the aim is to model the complete system, solvent+solute, regarded as one equivalent (but now complex) fluid. This is not what is being done in dispersed two-phase flow modeling where the continuous (the solvent) and the dispersed (the solute) phases are treated separately as two physical systems in interaction through the exchange of mass, momentum and energy. Said otherwise, complex fluids are handled as an equivalent `homogeneous two-component system' in which particle inertia is not considered significant. In effect, most, if not all, formulations of complex fluids neglect the solute inertia and rely on the Brownian limit. This raises, however, interesting issues in the proper formulation of such fast-variable techniques as well as in the resulting expression of random terms and corresponding Fokker-Planck type of closures.

In this section, we investigate the connections between complex fluids and dispersed turbulent two-phase flows by analyzing how fluctuations and random terms are accounted for. Key challenges are first brought out in Sec.~\ref{sec: soft matter noise}. We then concentrate on discrete particles and colloids and address the core issue of fast-variable elimination in space-dependent Langevin equations in Sec.~\ref{sec: soft matter Brownian limit}, before proposing some conclusions.

\subsection{Neither complete order nor disorder: thermal versus turbulence noise} \label{sec: soft matter noise}
%===============================================================================

To bring out the differences related to how random terms are introduced in complex fluids and in turbulent dispersed two-phase flows, it is interesting to consider polymers and flexible fibers. We start with the chosen mechanical description before discussing the stochastic model applied in each case. 

\begin{figure}[ht]
 \centering
 \includegraphics[width=\textwidth]{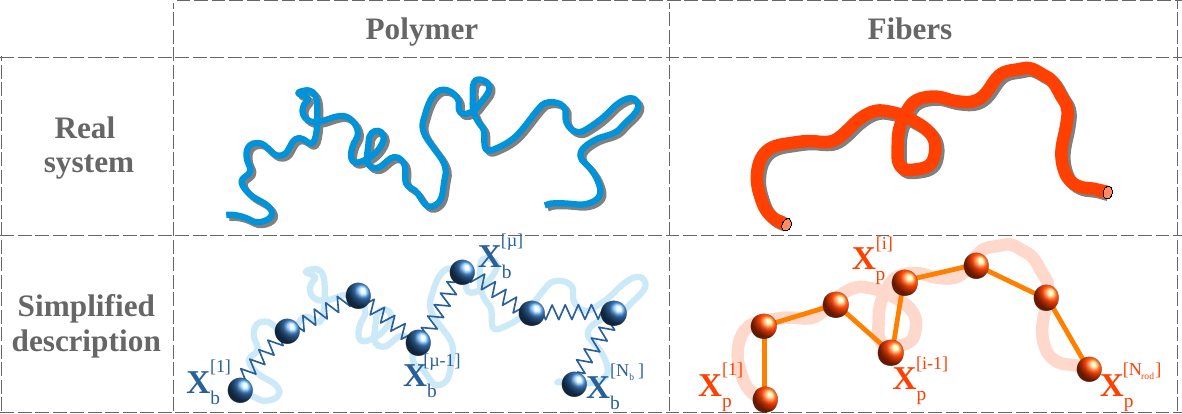}
 \caption{Comparison between equivalent models for flexible objects: polymer are represented by a chain of $N_{\rm b}$ beads connected by Hookean springs (in complex fluids) while fibers can be simplified to $N_{\rm rod}$ connected rigid rods (in dispersed two-phase flows).}
 \label{fig: flexible model}
\end{figure}

For polymers, it is sufficient to retain the classical Rouse model in which a polymer is described by a chain of $N_{\rm b}$ beads connected by $N_{\rm b}-1$ Hookean springs (see Fig.~\ref{fig: flexible model}) \cite{rouse1953theory}. This is a phenomenological model where the beads represent large segments of monomers so that a continuum description of the solvent is justified while the springs represent `entropic forces' used to account for the elimination of many degrees of freedom from the actual polymer.

Even though a continuous representation of a flexible fiber using curvilinear coordinates (as in Slender Body Theory or in Cosserat equations \cite{lindner2015elastic}) is possible, a typical approach consists in representing a flexible fiber as a chain of $N_{\rm rod}$ connected rigid rods (see Fig.~\ref{fig: flexible model}). This allows to capture fiber bending and twisting motions while preserving their connectivity through constraints applied at the hinges between two adjacent rigid rods. In a way, this description amounts to a discretized version of the continuous one, each rod being a stiffened segment of the fiber. Note that a description in terms of beads and connecting FENE springs is also possible, with the rigid-segment description obtained as the limit of infinite spring stiffness \cite{somasi2002brownian, henry2018tumbling}. To capture the evolution of such a rod chain, we can choose to follow the center of mass of each segment, in which case each elementary mechanical object is a non-spherical rigid particle. Similarly to the polymer bead-spring description, another choice consists in following the equivalent of beads which are here the hinges around which two adjacent rods can rotate. For the sake of having two similar descriptions and without any loss of generality for the present analysis, we retain the latter mechanical representation (see Fig.~\ref{fig: flexible model}).

In polymer theory, bead inertia is usually neglected and the Rouse model is formulated in the frame of Brownian dynamics. This means that the particle state vector attached to each polymer $\mb{Z}_{\rm b}$ is made up by the bead positions, i.e., $\mb{Z}_{\rm b}=(\mb{X}_{\rm b}^{[\mu]})_{\mu=1,\ldots, N_{\rm b}}$. In the absence of external forces and leaving out the interactions between polymers (dilute polymer systems), the evolution equations for $\mb{Z}_{\rm b}$ are
\begin{equation}
\label{eq: model polymer Rouse}
\dd\mb{X}_{\rm b}^{[\mu]} = \mb{U}_{\rm f}(t,\mb{X}_{\rm b}^{[\mu]}(t))\, \dd t + \frac{\tau_{\rm p}}{m_{\rm p}}\mb{F}_{\rm spring}^{\rm R, [\mu]}\, \dd t + \sqrt{\frac{2 k_{\rm B} \, \Theta_{\rm f}\, \tau_{\rm p}}{m_{\rm p}}}\, \dd\mb{W}^{[\mu]}~, \; \mu=1,\ldots, N_{\rm b}~,
\end{equation}
with $m_{\rm p}$ the bead mass and where the friction coefficient is written as $m_{\rm p}/\tau_{\rm p}$. In this equation, $\mb{F}_{\rm spring}^{\rm R}$ represent the forces acting on a bead due to the connecting springs, as expressed in the Rouse model. We introduce first a formulation in which springs can have different stiffness (i.e., $\mb{k}_{\rm spring}=(k_{\rm spring}^{[\mu]})$ for $\mu=1,\ldots, N_{\rm b}-1$), which for the general spring forces noted $\mb{F}_{\rm spring}(\mb{k}_{\rm spring})$ gives the following expression for the force $\mb{F}_{\rm spring}^{[\mu]}$ acting on bead $[\mu]$
\begin{equation}
\label{eq: definition spring Rouse}
\mb{F}_{\rm spring}^{[\mu]}(\mb{k}_{\rm spring})=
\begin{cases}
\phantom{-} k_{\rm spring}^{[1]} \left( \mb{X}_{\rm b}^{[2]} - \mb{X}_{\rm b}^{[1]} \right) & \text{if} \; \mu=1,  \\
\phantom{-} k_{\rm spring}^{[\mu]} \left( \mb{X}_{\rm b}^{[\mu+1]} - \mb{X}_{\rm b}^{[\mu]} \right) - k_{\rm spring}^{[\mu-1]} \left( \mb{X}_{\rm b}^{[\mu]} - \mb{X}_{\rm b}^{[\mu-1]} \right) & \text{if} \; 1 < \mu < N_{\rm b},  \\
- k_{\rm spring}^{[N_{\rm b} -1]} \left( \mb{X}_{\rm b}^{[N_{\rm b}]} - \mb{X}_{\rm b}^{[N_{\rm b}-1]} \right) & \text{if} \; \mu=N_{\rm b}.
\end{cases}
\end{equation}
In the Rouse model, all springs have the same stiffness $k_{\rm spring}^{\rm R}$ so that the vector $\mb{k}_{\rm spring}^{\rm R}$ is a vector of identical values (i.e., $k_{\rm spring}^{\rm R, [\mu]}=k_{\rm spring}^{\rm R}, \forall \mu$), which means that in Eq.~\eqref{eq: model polymer Rouse} we have $\mb{F}_{\rm spring}^{\rm R}=\mb{F}_{\rm spring}(k_{\rm spring}^{\rm R})$.

For flexible fibers, rod inertia is not necessarily neglected leading to a particle state vector with bead positions and velocities, i.e., $\mb{Z}_{\rm p}=(\mb{X}_{\rm p}^{[\rm i]},\mb{U}_{\rm p}^{[\rm i]})_{{\rm i}=1, N_{\rm rod}}$. A simple dynamical model has then the form
\begin{subequations}
\label{eq: flexible fiber model}
\begin{align}
\dd\mb{X}_{\rm p}^{[\rm i]} &= \mb{U}_{\rm p}^{[\rm i]}~, \label{eq: flexible fiber model a} \\
\dd\mb{U}_{\rm p}^{[\rm i]} &= - \frac{ \mb{U}_{\rm p}^{[\rm i]} - \mb{U}_{\rm f}(t,\mb{X}_{\rm p}^{[\rm i]}(t)) }{\tau_{\rm p}}\, \dd t + \frac{1}{m_{\rm p}} \mb{F}_{\rm spring}^{[\rm i]}(\bds{\lambda})\, \dd t + \sqrt{ \frac{2 k_{\rm B} \, \Theta_{\rm f}}{m_{\rm p}\, \tau_{\rm p}}}\, \dd\mb{W}^{[\rm i]}~, \label{eq: flexible fiber model b}
\end{align}
\end{subequations}
where the vector $\bds{\lambda}/m_{\rm p}$ can be thought of as a set of spring stiffness. When the flexible fiber is made up by connected rigid rods, $\bds{\lambda}$ represents the Lagrange multipliers used to enforce the non-extensibility of each rod, obtained from the constraint that $\vert \mb{X}_{\rm p}^{[\rm i+1]} - \mb{X}_{\rm p}^{[\rm i]} \vert$ remains constant (for $i=1,\ldots, N_{\rm rod}-1$). Note that Brownian effects are retained in Eqs.~\eqref{eq: flexible fiber model} but are expressed by a white-noise term appearing in the velocity equation, cf. Eq.~\eqref{eq: flexible fiber model b}, rather than directly in the position equation, cf.~\eqref{eq: flexible fiber model}, as in Brownian dynamics (see further discussion in Sec.~\ref{sec: soft matter Brownian limit}).

Whatever the selected way to account for thermal noise, it is seen from Eq.~\eqref{eq: model polymer Rouse} and Eq.~\eqref{eq: flexible fiber model b} that its introduction is straightforward and consists in an isotropic diffusion term in front of a Wiener process. A crucial point is that, at the bead or particle level of description, thermal noise is independent between distinct locations. This means that the formulation given, for instance, in Eq.~\eqref{eq: model polymer Rouse} remains exact regardless of the fact that we track one or a set of $N$ particles.  

Far different is the situation with respect to turbulent flows in the case of non-fully resolved velocity fields. Note that this situation does not arise for polymers transported by laminar flows (most of the time, even simple laminar shear flows), where the velocities at the bead locations, $(\mb{U}_{\rm f}(t,\mb{X}_{\rm b}^{[\mu]}(t)))_{\mu=1,N_{\rm b}}$ in Eq.~\eqref{eq: model polymer Rouse}, are known. The same remains valid for the velocity equation of each segment or bead of the flexible fiber, cf. Eq.~\eqref{eq: flexible fiber model b}, but only provided that we are dealing with fully-resolved turbulent flows by which we have access to the values of the instantaneous velocities at the bead locations, $(\mb{U}_{\rm f}(t,\mb{X}_{\rm p}^{[\rm i]}(t)))_{i=1,N_{\rm rod}}$. In the case of high Reynolds-number turbulent flows or when we have only access to a reduced statistical description of the velocity field, we are then faced with the difficulty of having to reconstruct the set of the $N_{\rm rod}$ fluid velocities seen, $(\mb{U}^{[\rm i]}_{\rm s}(t)=\mb{U}_{\rm f}(t,\mb{X}_{\rm p}^{[\rm i]}(t)))_{i=1,N_{\rm rod}}$. In that case, the Langevin model studied in preceding sections cannot be applied since it corresponds to a one-particle PDF model and, therefore, to a one-point type of closure. As already emphasized, there is no spatial information in present one-particle model for the velocity of the fluid seen. It is thus clear that a set of Langevin models for $(\mb{U}^{[\rm i]}_{\rm s}(t))_{i=1,N_{\rm rod}}$, with a drift vector $\mb{A}^{[\rm i]}_{\rm s}$ and diffusion matrix $\mb{B}^{[\rm i]}_{\rm s}$ (which need not be precised for the present discussion), such as
\begin{equation}
\dd\mb{U}^{[\rm i]}_{\rm s}(t) = \mb{A}^{[\rm i]}_{\rm s}\, \dd t + \mb{B}^{[\rm i]}_{\rm s}\, \dd\mb{W}^{[\rm i]}~,
\end{equation}
but with independent Wiener process, $(\mb{W}^{[\rm i]}(t))_{i=1,N_{\rm rod}}$, between each rod segment or bead $[\rm i]$ would grossly misrepresent the spatial correlations which are one of the hallmarks of turbulent flows. In other words, we are dealing with `turbulence noise' which is neither order nor complete disorder. For general non-homogeneous turbulent flows where velocities can deviate from Gaussianity, generating a set of $N$ fluid velocities seen, $(\mb{U}^{[\rm i]}_{\rm s}(t))_{i=1,N}$, that respect known correlations in time (this is already achieved) but also in space (this is an open issue) appears as a quite challenging task. 

Comparing polymers in laminar flows and flexible fibers in non-fully resolved turbulent flows is relevant to illustrate that modeling `turbulence noise' is more complicated than adding a $k_{\rm B}\,\Theta_{\rm f}$ diffusion term to a deterministic mechanical model. In the present review where we concentrate on small discrete particles whose counterpart among soft matter systems is colloidal suspensions, the above issue is less sensitive. We can then ask: is the description of colloidal suspensions, in particular Brownian motion and the diffusive limit (Brownian dynamics), contained in dispersed two-phase flow models? To provide answers to this question, we consider first how Brownian motion is accounted for and, then, discuss recent results on the overdamped limit of Langevin equations. These results are of interest for non-equilibrium dynamics of soft matter.

\subsection{Fast-variable elimination and the Brownian limit}\label{sec: soft matter Brownian limit}
%============================================================

\subsubsection{Accounting for Brownian motion}\label{sec: accounting for Brownian motion}
%---------------------------------------------

The introduction of Brownian effects in particle stochastic models is fairly well-established and goes back to the contributions of \cite{einstein1905motion} and \cite{langevin1908theory}. We benefit now from the rigorous framework of Ito calculus for stochastic differential equations (see Sec.~\ref{Ito SDE and Fokker-Planck}) and it is therefore interesting to revisit its main points since they are useful to introduce fast-variable elimination. To account for Brownian motion, it is best to consider first Brownian particles in a fluid at rest. Following Langevin's original point of view \cite{langevin1908theory}, the classical formulation consists in writing the particle velocity equation as the sum of a linear return-to-equilibrium term due to fluid friction and a random one written in terms of the increments of a Wiener process. In a one-dimensional notation, this gives
\begin{subequations}
\label{eq: Brownian motion simple X-U}
\begin{align}
\dd X_{\rm p} & = U_{\rm p}\, \dd t~, \label{eq: Brownian motion simple X-U a}\\
\dd U_{\rm p} & = -\frac{U_{\rm p}}{\tau_{\rm p}}\, \dd t + K\, \dd W~, \label{eq: Brownian motion simple X-U b}
\end{align}
\end{subequations}
where $K$ is a diffusion coefficient to be determined. From classical statistical mechanics and the equipartition of energy, we know that $1/2\, m_{\rm p} \lra{(U_{\rm p})^2}$ reaches a constant value in the long-time limit which is equal to $1/2\, k_{\rm B}\,\Theta_{\rm f}$. Then, an immediate application of Ito calculus yields that 
\begin{equation}
\label{eq: F6D theorem basic Brownian}
\frac{\dd \lra{(U_{\rm p})^2}}{\dd t}=0  \quad \Longrightarrow \quad K^2 = \frac{ 2 \lra{(U_{\rm p})^2}}{\tau_{\rm p}} = \frac{2 k_{\rm B} \, \Theta_{\rm f}}{m_{\rm p}\, \tau_{\rm p}}~,
\end{equation}
which is the form already given in Eq.~\eqref{eq: flexible fiber model b}. Note that this a straightforward expression of what is referred to as a fluctuation-dissipation theorem, as already noted in Sec.~\ref{Ito SDE and Fokker-Planck}. 

When $\tau_{\rm p}$ and $K$ are constants, $U_{\rm p}$ is a simple OU process and the exact behavior of the position of Brownian particles can be worked out before taking the limit $\tau_{\rm p} \to 0$ (for example by substituting $r$ with $X_{\rm p}$, $T_{\rm L}$ with $\tau_{\rm p}$, and $\sqrt{C_0\, \lra{\epsilon_{\rm f}}}$ with $K$ in Eq.~\eqref{eq: variance of relative positions}). There is, however, a short-cut method which is often used and which consists in neglecting the particle acceleration in the limit of small particle inertia (manifested here by small values of the relaxation timescale $\tau_{\rm p}$). For example, if we consider a more general evolution equation with the fluid velocity seen in the drag force when the fluid is not at rest and a force $F_{\rm p}$ due to external actions or to other particles,
\begin{equation}
\label{eq: Brownian particle momentum equation}
\dd U_{\rm p} = \frac{U_{\rm s} - U_{\rm p}}{\tau_{\rm p}}\, \dd t + F_{\rm p}\, \dd t + K\, \dd W~,
\end{equation}
we get from enforcing $\dd U_{\rm p} \simeq 0$ that the particle position equation $\dd X_{\rm p} = U_{\rm p}\, \dd t$ becomes
\begin{equation}
\label{eq: diffusive limit short-cut}
\dd X_{\rm p}= U_{\rm s}\, \dd t + \tau_{\rm p} F_{\rm p}\, \dd t + \left( \tau_{\rm p} K \right) \dd W~.
\end{equation}
This corresponds to the purely diffusive behavior of Brownian particles, from which we retrieve the classical Einstein relation for the position diffusion coefficient since
\begin{equation}
\mc{D}= \frac{1}{2} \left( \tau_{\rm p} K \right)^2 = \frac{k_{\rm B} \, \Theta_{\rm f}}{3\pi d_{\rm p} \rho_{\rm f} \nu_{\rm f}}~,
\end{equation}
and the classical formulation for the Brownian particle position equation
\begin{equation}
\label{eq: diffusive limit naive derivation}
\dd X_{\rm p}= U_{\rm s}\, \dd t + \tau_{\rm p} F_{\rm p}\, \dd t + \sqrt{2 \mc{D}}\, \dd W~.
\end{equation}

\begin{figure}[ht]
 \centering
 \includegraphics[width=0.9\textwidth]{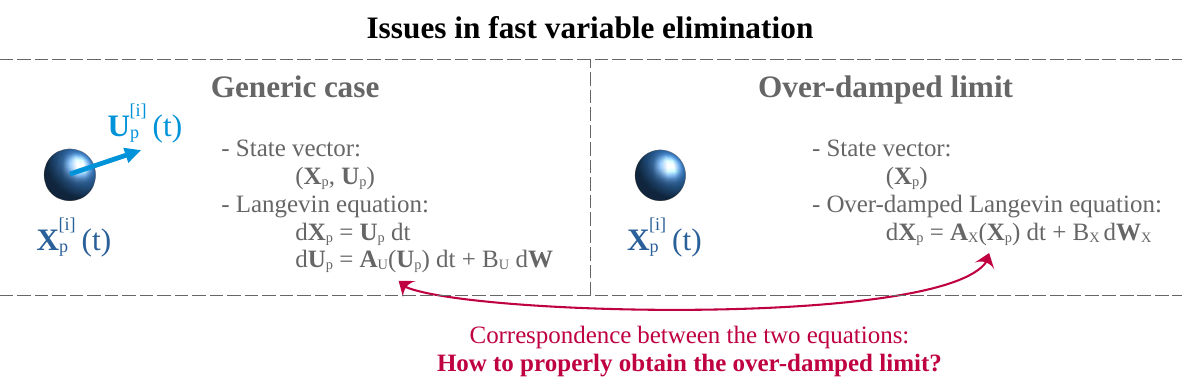}
 \caption{Representation of the key issue encountered when formulating the over-damped Langevin.}
 \label{fig: Overdamped Langevin}
\end{figure}

This is the basis of the slaving principle (introduced in Sec.~\ref{statistical model transport: Synergetics}), obtained by this heuristic formulation of fast-variable elimination techniques (see also Fig.~\ref{fig: Overdamped Langevin}). When the degrees of freedom are separated into slow and fast variables (here, the particle position and velocity, respectively), and even when the drift and diffusion coefficients are functions of the slow variables, the same reasoning is assumed to remain valid leading to the same diffusive limit for the slow variable. In the present case, this still gives Eq.~\eqref{eq: diffusive limit short-cut} with $\tau_{\rm p}=\tau_{\rm p}(t, X_{\rm p})$ and $K=K(t, X_{\rm p})$. Yet, though intuitive, this short-cut method of the slaving principle lacks rigor and raises two concerns.

The first one is the very loose manner with which the limit $\tau_{\rm p} \to 0$ is taken and the proper sense with which terms such as $(\tau_{\rm p} K)$ and $\tau_{\rm p} F_{\rm p}$ are to be regarded in that limit. To get around this difficulty, we introduce a small parameter $\chi$ and consider the following stochastic model
\begin{subequations}
\begin{align}
\dd X_{\rm p}^{(\chi)} & = U_{\rm p}^{(\chi)} \, \dd t~, \\
\dd U_{\rm p}^{(\chi)} & = \frac{U_{\rm s}- U_{\rm p}^{(\chi)}}{\chi \, \tau_{\rm p}}\, \dd t + \frac{1}{\chi} F_{\rm p}\, \dd t + \frac{K}{\chi}\, \dd W~,
\end{align}
\end{subequations}
where the timescale varies as $\chi \, \tau_{\rm p}$ while the force and the diffusion coefficient scale as $F_{\rm p}/\chi$ and $K/\chi$, respectively. This allows to formalize the previous manipulations and give sense to the limit $\tau_{\rm p} \to 0$ and $K \to +\infty$ with $\tau_{\rm p}\, K$ remaining constant. Other parametrizations have been proposed but have been shown to be equivalent or even of a less general application for time-dependent force, i.e. when $F_{\rm p}=F_{\rm p}(t,X_{\rm p})$ (see a comprehensive discussion of this point in~\cite[section 9.3]{minier2016statistical}). The short-cut method of fast-variable elimination takes now a more satisfactory expression since we have 
\begin{equation}
U_{\rm p}^{(\chi)} \dd t \simeq U_{\rm s}\, \dd t + \tau_{\rm p} F_{\rm p}\, \dd t + \left( K \tau_{\rm p} \right) \dd W~,
\end{equation}
leading, in a more rigorous manner, to the diffusive limit of $\dd X_{\rm p}^{(\chi)}$ when $\chi \to 0$, which is Eq.~\eqref{eq: diffusive limit short-cut}. Note that the particle mean kinetic energy $\lra{ (U_{\rm p}^{(\chi)})^2}$ scales now as $1/2 K^2/(\chi \, \tau_{\rm p})$, cf. Eq.~\eqref{eq: F6D theorem basic Brownian}, and we retrieve that it is the diffusive coefficient, $\lra{ (U_{\rm p}^{(\chi)})^2}\times (\chi \, \tau_{\rm p})$ that tends to a finite limit, which is, of course, the Einstein diffusion coefficient.

The second concern is about the validity of the assumption that the diffusive limit can be derived simply by neglecting the particle acceleration, i.e. by enforcing directly $\dd U_{\rm p}^{(\chi)}=0$. In particular, does this short-cut method remain valid when the coefficients are not constant but are functions of the particle position and, therefore, of the slow variable itself? To answer this question, we consider now the diffusive limit of `turbulence noise' for fluid particles (the tracer-particle limit without Brownian motion) where the timescale and the diffusion coefficient in general Langevin models are indeed space dependent.

\subsubsection{Brownian dynamics in turbulent flows}\label{sec: Brownian dynamics and turbulence}
%---------------------------------------------------

In~\cite[section 9.3]{minier2016statistical}, it is shown that the most general way to derive the limit of over-damped Langevin equations is to consider for $\mb{Z}=(\mb{X},\mb{U})$ the system
\begin{subequations}
\label{eq: general overdamped Langevin}
\begin{align}
\dd X_i & = U_i \, \dd t~, \\
\dd U_i & = G_{ij}(t,\mb{X}(t))\left[ U_j - \Phi_j(t,\mb{X}(t)) \right]\, \dd t + \sigma_{ij}(t,\mb{X}(t))\, \dd W_j~,
\end{align}
\end{subequations}
and to study the limit when $\chi \to 0$ of a parametrized version where a small parameter $\chi$ is introduced 
\begin{subequations}
\label{eq: general overdamped Langevin chi}
\begin{align}
\dd X^{(\chi)}_i & = U^{(\chi)}_i \, \dd t~, \label{eq: general overdamped Langevin chi X} \\
\dd U^{(\chi)}_i & = \frac{1}{\chi} G_{ij}(t,\mb{X}^{(\chi)}(t))\left[ U^{(\chi)}_j - \Phi_j(t,\mb{X}^{(\chi)}(t)) \right]\, \dd t + \frac{1}{\chi}\sigma_{ij}(t,\mb{X}^{(\chi)}(t))\, \dd W_j~. \label{eq: general overdamped Langevin chi U}
\end{align}
\end{subequations}
As indicated for Brownian particles, the introduction of the parameter $\chi$ formalizes the limit we wish to take (in a scalar version, we consider $G^{-1} \to 0$ and $\sigma \to +\infty$ so that $G^{-1} \sigma$ remains constant). This is done by multiplying directly the drift and diffusion coefficients appearing in Eq.~\eqref{eq: general overdamped Langevin} by $1/\chi$ and the same notations are kept for these coefficients in Eq.~\eqref{eq: general overdamped Langevin chi U}. In a physics-oriented analysis, it is best to identify first a small parameter and describe then how the drift and diffusion coefficients scale with it, as shown below for the case of fluid particles in turbulent flows. However, we are first concerned with the correct mathematical limit of Eqs.~\eqref{eq: general overdamped Langevin chi}, with $\mb{G}$ and $\bds{\sigma}$ now regular functions. In Eqs.~\eqref{eq: general overdamped Langevin chi}, where the specific dependencies of the drift and diffusion coefficients were kept for clarity but are left out from now on for the sake of simplicity, the matrix $\mb{G}$ is assumed to be invertible and $\bds{\Phi}(t,\mb{x})$ is a known field that remains regular.

It is then demonstrated that, in the limit $\chi \to 0$, the over-damped Langevin limit model is a stochastic diffusion process for particle positions which writes
\begin{equation}
\label{eq: diffusive limit first expression}
\dd X_i = \Phi_i\, \dd t - \frac{\partial G^{-1}_{ij}}{\partial x_k} A_{kj}\dd t + \left( G^{-1}\sigma \right)_{ij} \dd W_j~,
\end{equation}
where the matrix $\mb{A}$ stands for $A_{ij}=\lim_{\chi \to 0} \left( \chi \lra{U^{(\chi)}_i U^{(\chi)}_j} \right)$ and is the solution of the matrix equation
\begin{equation}
\mb{G}\mb{A} + \mb{A}\mb{G}^{\bot} = - \bds{\sigma}\bds{\sigma}^{\bot}~.
\end{equation}
To apply these results to the case of fluid particles in turbulent flows, for which we use the notations $\mb{Z}_{\rm f}=(\mb{X}_{\rm f},\mb{U}_{\rm f})$, the relevant choice for the field $\Phi(t,\mb{x})$ is to take
\begin{equation}
\phi_i = \lra{ U_{{\rm f},i} } + G_{ij}^{-1} \left( \frac{1}{\rho_{\rm f}} \frac{\partial \lra{P_{\rm f}}}{\partial x_j} \right)~,
\end{equation}
which means that both $\lra{\mb{U}_{\rm f}}$ and $\lra{ P_{\rm f}}$ must remain regular whereas the finite nature of the matrix $\mb{A}$ shows that the kinetic energy grows unbounded as $1/\chi$ (this is in line with the remark above concerning Brownian motion). This corresponds to a situation where the mean velocity field remains regular but where the second-order moments of the fluctuating velocity components grow as $1/\chi$, which implies that $\mb{A}$ is also given by $A_{ij}=\lim_{\chi \to 0} \left( \chi \lra{u^{(\chi)}_i u^{(\chi)}_j} \right)$. This explains the introduction of the mean-pressure gradient in $\bds{\phi}$ which, according to Eq.~\eqref{eq: general overdamped Langevin chi U}, scales as $1/\chi$ and compensates the divergence of the Reynolds stresses in the mean Navier-Stokes equations to ensure that the mean velocity field remains regular. This important point is detailed in~\cite[section 9.3]{minier2016statistical} and is also revisited below. With these conditions, the proper over-damped Langevin limit model for fluid particles is
\begin{align}
\dd X_{{\rm f},i} & = \lra{ U_{{\rm f},i} }\, \dd t + G_{ij}^{-1} \left( \frac{1}{\rho_{\rm f}} \frac{\partial \lra{ P_{\rm f}}}{\partial x_j} \right)\, \dd t - \frac{\partial G^{-1}_{ij}}{\partial x_k} A_{kj}\dd t + \left( G^{-1}\sigma \right)_{ij} \dd W_j~, \label{eq: overdamped Langevin limit turbulence a} \\
                  & = \lra{ U_{{\rm f},i} }\, \dd t + G_{ij}^{-1}\left\{ \frac{1}{\rho_{\rm f}} \frac{\partial \lra{ P_{\rm f}}}{\partial x_j} + \frac{\partial A_{jk}}{\partial x_k} \right\}\dd t - \frac{\partial}{\partial x_k}\left[ G^{-1}_{ij} A_{jk} \right] \dd t  + \left( G^{-1}\sigma \right)_{ij} \dd W_j~, \label{eq: overdamped Langevin limit turbulence b} \\
									& = \lra{ U_{{\rm f},i} }\, \dd t - \frac{\partial}{\partial x_k}\left[ G^{-1}_{ij} A_{jk} \right] \dd t  + \left( G^{-1}\sigma \right)_{ij} \dd W_j~, \label{eq: overdamped Langevin limit turbulence c}
\end{align}
since the term between the brackets on the rhs of Eq.~\eqref{eq: overdamped Langevin limit turbulence b} vanishes due to the enforced condition of the mean pressure gradient. In the corresponding Fokker-Planck equation, the diffusion matrix, $\mb{D}$, is then 
\begin{equation}
\label{eq: definition general diffusivity matrix D}
\mb{D}=\frac{1}{2} \left( \mb{G}^{-1} \bds{\sigma} \right) \left( \mb{G}^{-1} \bds{\sigma} \right)^{\bot}~.
\end{equation}
With $\mb{B}$ a matrix such that $\mb{B}\mb{B}^{\bot}=2 \mb{D}$, we obtain after some tedious but straightforward manipulations the final resulting form of the diffusion equation for fluid particle positions in the over-damped limit as
\begin{equation}
\label{eq: correct limit diffusion with tensor D}
\dd X_{{\rm f},i} = \lra{ U_{{\rm f},i} }\, \dd t + \frac{\partial D_{ik}}{\partial x_k}\, \dd t + B_{ik}\, \dd W_k. 
\end{equation}
In the case where $\mb{G}$ and $\bds{\sigma}$ are isotropic tensors, $G_{ij}= - 1/T_{\rm L}\delta_{ij}$ and $\sigma_{ij}=\sigma \delta_{ij}$, we have the simple form
\begin{equation}
\label{eq: diffusive limit simple form isotropic}
\dd X_{{\rm f},i} = \lra{ U_{{\rm f},i} }\, \dd t + \frac{\partial}{\partial x_i} \left[ \frac{1}{2}\left( T_{\rm L}\, \sigma \right)^2 \right]\, \dd t + \left( T_{\rm L}\, \sigma \right)\, \dd W_i. 
\end{equation}

These results call for some comments which, for the sake of simplicity, are based on the case of a scalar diffusivity to avoid more cumbersome tensorial notations. 

\begin{enumerate}[(1)]
\item The derivation of the diffusive limit shows that the elimination of the fast-variable $\mb{U}_{\rm f}$ leads to both a diffusive and a drift term in the particle position equation
\begin{equation}
\label{eq: complete diffusive limit}
\dd X_{{\rm f}, i} = \lra{U_{{\rm f}, i}}\, \dd t + \frac{\partial \Gamma_{\rm ft}}{\partial x_i} + \sqrt{2 \Gamma_{\rm ft}}\, \dd W_i~.
\end{equation}
where we have used $\Gamma_{\rm ft}=\left( T_{\rm L}\, \sigma \right)^2/2$. This is to be compared to the naive formulation which gives
\begin{equation}
\label{eq: naive diffusive limit}
\dd X^{(\rm naive)}_{{\rm f}, i} = \lra{U_{{\rm f}, i}}\, \dd t + \sqrt{2\Gamma_{\rm ft}}\, \dd W_i~,
\end{equation}
where the drift term $\partial \Gamma_{\rm ft}/\partial x_i$ is missing. 
\item When we consider a conserved scalar attached to each particle, i.e. $\dd \phi_{\rm f} =0$, then $\lra{\phi_{\rm f}}$ represents the mean particle concentration. In sample space, the Fokker-Planck equation for $p(t;\mb{y},\psi_{\rm f})$ is
\begin{equation}
\label{eq: diffusive limit F-P equation}
\frac{\partial p}{\partial t}= - \frac{ \partial \left[\, \lra{U_{{\rm f}, i}}\, p \, \right]}{\partial y_i} - \frac{\partial}{\partial y_i}\left[ \left( \frac{\partial \Gamma_{\rm ft}}{\partial x_i} \right)\, p \, \right] + \frac{\partial^2 \left[\, \Gamma_{\rm ft}\, p \, \right]}{\partial y_i \partial y_i}~,
\end{equation}
where the diffusion matrix involves only diagonal (even isotropic) terms since we are dealing with a scalar diffusivity (the general case with a full second-order diffusivity tensor is addressed below). By integration over the sample space variable $\psi_{\rm f}$ corresponding to $\phi_{\rm f}$, we see that the mean concentration $\lra{c_{\rm f}}(t,\mb{x})$ satisfies the same equation in physical space
\begin{equation}
\frac{\partial \lra{c_{\rm f}}}{\partial t}= - \frac{ \partial \left[\, \lra{U_{{\rm f}, i}}\, \lra{c_{\rm f}} \, \right]}{\partial x_i} - \frac{\partial}{\partial x_i}\left[ \left( \frac{\partial \, \Gamma_{\rm ft}}{\partial x_i} \right)\, \lra{c_{\rm f}} \, \right] + \frac{\partial^2 \left[\, \Gamma_{\rm ft}\, \lra{c_{\rm f}} \, \right]}{\partial x_i \partial x_i}~.
\end{equation}
Both forms show that an uniform particle concentration field is conserved, as it should be for incompressible flows (since $\partial \lra{U_{{\rm f}, i}}/\partial x_i=0$). In contrast, the naive diffusive limit, expressed by Eq.~\eqref{eq: naive diffusive limit}, would produce a spurious drift term due to a non-compensated variable diffusion coefficient $\Gamma_{\rm ft}(\mb{x})$. Since each particle represents the same amount of mass, we would then induce artificial accumulations/depletions of mass at variance with the basic property of incompressible flows.
\item By considering the position pdf $p(t;\mb{y})$, obtained as the marginal of the position-scalar one $p(t;\mb{y},\psi_{\rm f})$, and rearranging the last two terms on the rhs of Eq.~\eqref{eq: diffusive limit F-P equation}, we obtain
\begin{equation}
\label{eq: diffusive limit F-P equation 2}
\frac{\partial p}{\partial t}= - \frac{ \partial \left[\, \lra{U_{{\rm f}, i}}\, p \, \right]}{\partial y_i} + \frac{\partial}{\partial y_i}\left[ \Gamma_{\rm ft} \frac{\partial p}{\partial y_i} \right] ~,
\end{equation}
and by further combinations 
\begin{equation}
\label{eq: diffusive limit F-P equation 3}
\frac{\partial p}{\partial t}= - \frac{\partial}{\partial y_i}\left[ \, \left( \lra{U_{{\rm f}, i}} - \Gamma_{\rm ft} \frac{\partial \ln(p)}{\partial y_i} \right)\, p \,\right] ~.
\end{equation}
As such, the random term due to the elimination of the fast-variable $\mb{U}_{\rm f}$ appears as an equivalent drift term in the equation for the slow-variable $\mb{X}_{\rm f}$. If we remember that such drift terms are obtained as $(T_{\rm L}\, F_{\rm rand})$ for the case of an applied force $F_{\rm rand}$ in the particle velocity equation, cf. Eqs.~\eqref{eq: Brownian particle momentum equation} and~\eqref{eq: diffusive limit short-cut} substituting $T_{\rm L}$ with $\tau_{\rm p}$ and $F_{\rm rand}$ with $F_{\rm p}$, this formulation presents random effects as being due to an equivalent `entropic force' which, per unit mass, is
\begin{equation}
\label{eq: analogy entropic forces}
F_{\rm rand} = \frac{\Gamma_{\rm ft}}{T_{\rm L}}\, \frac{\partial \ln(p)}{\partial x_i} \quad \text{or} \quad F_{\rm rand} = \frac{\Gamma_{\rm ft}}{T_{\rm L}}\, \frac{\partial \ln(\lra{c_{\rm f}})}{\partial x_i}~,
\end{equation}
where the second expression makes use of the proportionality between $p(t,\mb{x})$ and $\lra{c_{\rm f}}(t,\mb{x})$. Nevertheless, this correspondence should be taken with some care as the two formulations do not have the same mathematical support. Indeed, the first formulation, in Eqs.~\eqref{eq: complete diffusive limit} or~\eqref{eq: diffusive limit F-P equation}, corresponds to a well-established SDE in which the drift and diffusion coefficients are functions of the stochastic process $\mb{X}_{\rm f}$ or, perhaps, of statistics derived from the PDF in the extended sense referred to as McKean SDEs (see an interesting discussion, in physical terms, of the differences in~\cite[section 3.3.4]{Ottinger_1996} and the related important issue of propagation of chaos in processes with mean-field interactions). However, the second formulation, given by $F_{\rm rand}$ on the left side of Eq.~\eqref{eq: analogy entropic forces}, is written in terms of the PDF itself and does not fall into the category of mathematically well-defined drift or diffusion terms in SDEs.
\item To underline the previous point, we can revert to Brownian particles and substitute $\tau_{\rm p}$ with $T_{\rm L}$, knowing that $\left(k_{\rm B} \, \Theta_{\rm f}\right)/m_{\rm p}$ is the particle velocity variance. We retrieve the Einstein formula for the diffusivity since $\Gamma_{\rm B}=\lra{U_{\rm p}^2}\, \tau_{\rm p}=\left(k_{\rm B} \, \Theta_{\rm f}\, \tau_{\rm p}\right)/m_{\rm p}$ (with $m_{\rm p}/\tau_{\rm p}$ the friction coefficient). This leads to the position-pdf equation for Brownian particles
\begin{equation}
\label{eq: Brownian F-P equation 3}
\frac{\partial p}{\partial t}= - \frac{\partial}{\partial y_i}\left[ \, \left( \lra{U_{{\rm f}, i}} - \frac{k_{\rm B} \, \Theta_{\rm f}\, \tau_{\rm p}}{m_{\rm p}} \, \frac{\partial \ln(p)}{\partial y_i} \right)\, p \,\right] ~,
\end{equation}
involving what is referred to in polymer physics as a `smoothed Brownian force', $\mb{F}_{\rm B}=k_{\rm B} \, \Theta_{\rm f}\, \bds{\nabla}\ln(p)$ (see discussions in~\cite{Ottinger_1996,doi1988theory}). The direct introduction of such `smoothed Brownian forces' appears, however, difficult to grasp at first sight and it is believed that the present approach offers more justification. Indeed, compared to the `raw Brownian term' which is the diffusive coefficient involving the Wiener process $\sqrt{2 \Gamma_{\rm ft}}\, \dd W$, the smoothed version arises from the existence of a corresponding non-zero drift term which is $\partial \Gamma_{\rm ft}/\partial x_i$ (or $\partial \Gamma_{\rm B}/\partial x_i$): it is a combination of `complete disorder' (diffusion) and `order' (drift) that leads to these smoothed forces. 
\item It is worth noting that the diffusive limit of overdamped Langevin models has been obtained here by following the trajectory point of view and using Ito stochastic calculus. In other words, we have not relied on the projection-operator formalism, which is often believed to be needed (see the discussion in~\cite[chapter 8]{gardiner2009stochastic}) but which, although relevant, appears as not mandatory.

Yet, these results are in line with the way fluctuations are proposed to be added in the GENERIC framework (see~\cite[section 1.2.5]{ottinger2005beyond}) and with subsequent derivations based on the (rather formal) projection-operator techniques (see~\cite[section 6.3.3]{ottinger2005beyond}). The present approach works the other way around and obtains the Fokker-Planck equation for the slow variables after having derived their SDEs. In physical terms, present ideas are perhaps best cast with the general terminology of force-flux relations, as in thermodynamics. To illustrate that point, let us consider again the simple toy model used in Sec.~\ref{statistical model transport: Synergetics} with a slow variable $X_{\rm slow}$ whose time-rate-of-change is a fast-variable $X_{\rm fast}$, i.e. $\dd X_{\rm slow}/\dd t=X_{\rm fast}$. The exact but unclosed equation for the marginal pdf $p(t;y)$ is
\begin{equation}
\frac{\partial\, p(t;y)}{\partial t}= - \frac{\partial}{\partial y}\left[ \, \lra{ \, X_{\rm fast} \, \vert_{X_{\rm slow}=y} } \, p(t;y) \, \right]~.
\end{equation}
When $X_{\rm fast}$ becomes a fast variable in the sense studied above, that is with $\lra{X_{\rm fast}^2} \, T_{X_{\rm fast}}$ tending towards a finite coefficient $\Gamma_{X_{\rm fast}} (x)$ which can depend on the local value of the slow variable $X_{\rm slow}$, we have  
\begin{equation}
\lra{ \, X_{\rm fast} \, \vert_{X_{\rm slow}=y} } \, p(t;y) = - \Gamma_{\rm X_{\rm fast}} (y) \frac{\partial \, p(t;y)}{\partial y} \quad \Longrightarrow \quad \frac{\partial \, p(t;y)}{\partial t}=  \frac{\partial}{\partial y}\left[ \Gamma_{\rm X_{\rm fast}} (y) \frac{\partial \, p(t;y)}{\partial y} \right]~,
\end{equation}
with the same form of the Fokker-Planck equation. Therefore, the `flux' $\lra{ \, X_{\rm fast} \, \vert_{X=y} }$ is 
\begin{equation}
\lra{ \, X_{\rm fast} \, \vert_{X_{\rm slow}=y} } = - \Gamma_{\rm X_{\rm fast}} (y) \frac{\partial \ln \left[ \, p(t;y)\, \right]}{\partial y}~,
\end{equation}
and is proportional to the gradient of the `entropic force' , since the statistical entropy is $S \simeq \ln(p)$. 
\end{enumerate}

\subsubsection{Non-local and local closures}\label{sec: local and non-local closures}
%-------------------------------------------

There is an interesting connection between the introduction of white-noise effects in the evolution equations of the selected variables and local/non-local closures in the transport equations satisfied by moments of these variables. To exemplify that connection, we consider first the example of a conserved scalar $\phi_{\rm f}$ and two sets of fluid-particle-attached variables. When the particle velocity is retained in the particle state vector, which becomes $\mb{Z}_{\rm f}=(\mb{X}_{\rm f},\mb{U}_{\rm f},\phi_{\rm f})$, and is modeled, for instance, by a Langevin model, we have 
\begin{subequations}
\label{eq: scalar-GLM non-local}
\begin{align}
\dd X_{{\rm f},i} & = U_{{\rm f},i}\, \dd t~, \label{eq: scalar-GLM non-local Xf} \\
\dd U_{{\rm f},i} & = - \frac{1}{\rho_{\rm f}}\frac{\partial \lra{P_{\rm f}}}{\partial x_i}\, \dd t + G_{ij}\left( U_{{\rm f},j} - \lra{U_{{\rm f},j}} \right)\, \dd t + \sqrt{C_0\, \lra{\epsilon_{\rm f}}}\, \dd W_i~, \label{eq: scalar-GLM non-local Uf} \\
\dd \phi_{\rm f} &= 0~. \label{eq: scalar-GLM non-local phi}
\end{align}
\end{subequations}
In that case, the fluid-particle acceleration is regarded as a fast variable and is eliminated, leaving the modeled Langevin terms in Eq.~\eqref{eq: scalar-GLM non-local Uf}. We are then handling a velocity-scalar PDF, $p(t;\mb{x},\mb{V}_{\rm f},\psi_{\rm f})$, which follows a Fokker-Planck equation. From that equation, it is easy to derive the equation satisfied by the mean scalar field $\lra{\phi_{\rm f}}(t,\mb{x})$, which is
\begin{equation}
\frac{\partial \lra{\phi_{\rm f}}}{\partial t} + \lra{U_{{\rm f},i}}\frac{\partial \lra{\phi_{\rm f}}}{\partial x_i} = 
- \frac{\partial }{\partial x_i}\left[ \lra{u_{{\rm f},i} \phi'_{\rm f}} \right]~,
\end{equation}
where $\phi'_{\rm f}=\phi_{\rm f} - \lra{\phi_{\rm f}}$ is the scalar fluctuation and $u_{{\rm f},i}$ is the velocity fluctuation. In this equation, $\lra{u_{{\rm f},i} \phi'_{\rm f}}$ stands for the scalar flux and is obtained as the solution of the equation
\begin{equation}
\label{eq: scalar flux transport equation}
\frac{\partial \lra{u_{{\rm f},i} \phi'_{\rm f}}}{\partial t} + \frac{ \partial \lra{u_{{\rm f},k} u_{{\rm f},i} \phi'_{\rm f}}}{\partial x_k} = - \lra{ u_{{\rm f},k} \phi'_{\rm f}}\frac{\partial \lra{U_{{\rm f},i}}}{\partial x_k} - \lra{ u_{{\rm f},i} u_{{\rm f},k}}\frac{\partial \lra{\phi_{\rm f}}}{\partial x_k} + G_{ij} \lra{u_{{\rm f},j} \phi'_{\rm f}}~,
\end{equation}
which is also extracted from the Fokker-Planck equation. The important point is that the scalar flux $\lra{u_{{\rm f},i} \phi'_{\rm f}}$ is a non-local quantity in physical space since it results from a transport equation. 

\begin{figure}[ht]
 \centering
 \includegraphics[width=0.9\textwidth]{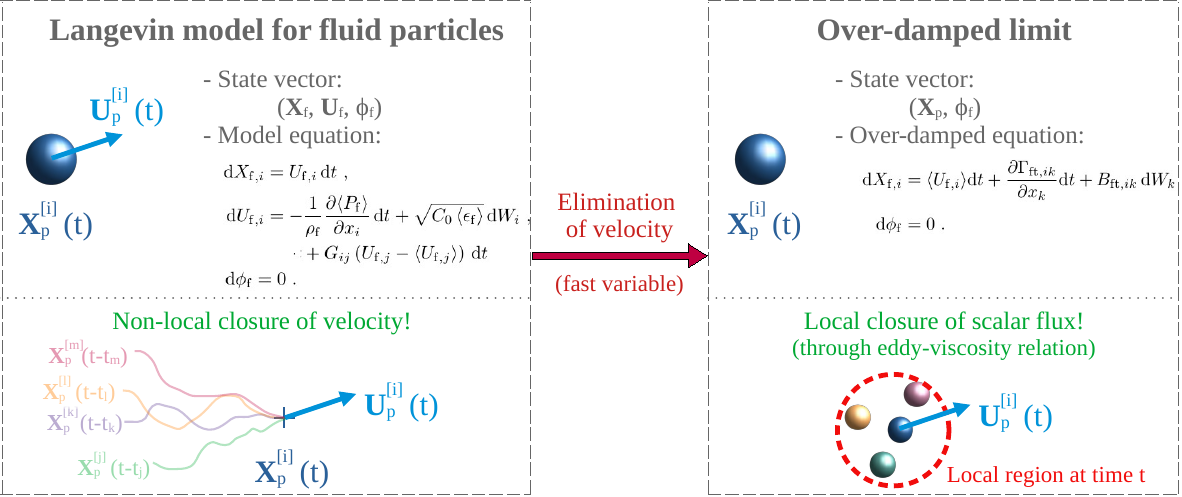}
 \caption{Representation of non-local versus local closures from a Langevin model for fluid particles.}
 \label{fig: Overdamped Langevin local}
\end{figure}

When the particle velocity is regarded as a fast-variable and eliminated with the techniques outlined above, the particle state vector is reduced to $\mb{Z}_{\rm f}=(\mb{X}_{\rm f},\phi)$ and the evolution equations have the form
\begin{subequations}
\label{eq: scalar diffusive complete}
\begin{align}
\dd X_{{\rm f},i} & = \lra{U_{{\rm f},i}} \dd t + \frac{\partial \Gamma_{{\rm ft}, ik}}{\partial x_k}\dd t + B_{{\rm ft},ik}\, \dd W_k~, \label{eq: scalar diffusive complete Xf} \\
\dd \phi_{\rm f}  & = 0~.
\end{align}
\end{subequations}
In Eq.~\eqref{eq: scalar diffusive complete Xf}, the diffusive matrix $\mb{B}_{\rm ft}$ corresponds to $(2 \bds{\Gamma}_{\rm ft})^{1/2}$, which means that $\mb{B}_{\rm ft}\mb{B}^{\bot}_{\rm ft}=2\bds{\Gamma}_{\rm ft}$, while the diffusivity tensor is $\bds{\Gamma}_{\rm ft}=1/2 \mb{G}^{-1}\bds{\sigma} \bds{\sigma}^{\bot} (\mb{G}^{\bot})^{-1}$ with $\bds{\sigma}=\sqrt{C_0\lra{\epsilon_{\rm f}}}\, \bds{\delta}$ (note that this is the same expression as the general diffusivity tensor given above, cf. Eq.~\eqref{eq: definition general diffusivity matrix D}, but we use $\bds{\Gamma}_{\rm ft}$ for this specific choice of $\bds{\sigma}$). The Fokker-Planck equation for the PDF $p(t;y_{\rm f},\psi_{\rm f})$ is then
\begin{equation}
\label{eq: scalar diffusive complete F-P}
\frac{\partial p}{\partial t}= - \frac{ \partial \left[\, \lra{U_{{\rm f}, i}}\, p \, \right]}{\partial y_i} + \frac{\partial}{\partial y_i}\left[ \Gamma_{{\rm ft},ij} \frac{\partial p}{\partial y_j} \right] ~.
\end{equation}
Note that Eqs.~\eqref{eq: complete diffusive limit} and~\eqref{eq: diffusive limit F-P equation} or Eq.~\eqref{eq: diffusive limit F-P equation 2} correspond to the case where $\mb{G}$ and $\bds{\sigma}$ are isotropic tensors. From Eq.~\eqref{eq: scalar diffusive complete F-P}, it follows that the mean scalar field $\lra{\phi_{\rm f}}(t,\mb{x})$ evolves now as
\begin{equation}
\frac{\partial \lra{\phi_{\rm f}}}{\partial t} + \lra{U_{{\rm f},i}}\frac{\partial \lra{\phi_{\rm f}}}{\partial x_i} = 
\frac{\partial}{\partial x_i}\left[ \Gamma_{{\rm ft},ij} \left( \frac{\partial \lra{\phi_{\rm f}}}{\partial x_j} \right) \right]~. 
\end{equation}
The important point now is that the scalar flux $\lra{u_{{\rm f},i} \phi'_{\rm f}}$ is no longer the solution of a transport equation, as in Eq.~\eqref{eq: scalar flux transport equation}, but is expressed locally in physical space through an eddy-diffusivity relation
\begin{equation}
\lra{u_{{\rm f},i} \phi'_{\rm f}} = - \Gamma_{{\rm ft},ij} \left( \frac{\partial \lra{\phi_{\rm f}}}{\partial x_j} \right)~. 
\end{equation}
We have moved from a non-local closure when the particle velocity is explicitly modeled to a local closure when it is eliminated (as displayed in Fig.~\ref{fig: Overdamped Langevin local}). As such, this seems to reflect essentially the selection of the variables retained in the particle state vector. However, since this selection follows the idea of decomposing between slow and fast variables and that the trace of fast variables is manifested by white-noise terms (thus by increments of Wiener processes in the SDEs), the connection is actually with white-noise effects.

A similar analysis can be carried out for the mean velocity field but requires careful bookkeeping of the unbounded terms that cancel each other out. To that effect, it proves useful to follow up the ideas introduced in~\cite[section 9.3.3]{minier2016statistical} and consider for $\mb{Z}_{\rm f}=(\mb{X}_{\rm f},\mb{U}_{\rm f})$ the stochastic system written as
\begin{subequations}
\label{eq: fast variable mean velocity}
\begin{align}
\dd X^{(\chi)}_{{\rm f},i} & = U^{(\chi)}_{{\rm f},i}\, \dd t~,   \label{eq: fast variable mean velocity Xf}\\
\dd U^{(\chi)}_{{\rm f},i} & = - \frac{1}{\rho_{\rm f}}\frac{\partial \lra{P^{(\chi)}_{\rm f}}}{\partial x_i}\,\dd t - \frac{1}{2} \frac{\lra{\epsilon^{(\chi)}_{\rm f}}}{k^{(\chi)}_{\rm f}} \left[ U^{(\chi)}_{{\rm f},i} - \lra{U^{(\chi)}_{{\rm f},i}} \right]\, \dd t \nonumber \\
                           & \phantom{=} - \frac{1}{\rho_{\rm f}}\frac{\partial \lra{\widetilde{P}^{(\chi)}_{\rm f}}}{\partial x_i}\,\dd t + \widetilde{G}^{(\chi)}_{ij} \left[ U^{(\chi)}_{{\rm f},j} - \lra{U^{(\chi)}_{{\rm f},j}} \right]\, \dd t + \sqrt{C_0 \lra{\widetilde{\epsilon}^{(\chi)}_{\rm f}}}\, \dd W_i~. \label{eq: fast variable mean velocity Uf}
\end{align}
\end{subequations}
The rationale behind the decomposition used on the rhs of Eq.~\eqref{eq: fast variable mean velocity Uf} is to introduce a distinction between two potentially different timescales: on the one hand, the timescale governing the rate at which turbulent kinetic energy is dissipated, which is represented by the term $k^{(\chi)}_{\rm f}/\lra{\epsilon^{(\chi)}_{\rm f}}$ on the first line on the rhs of Eq.~\eqref{eq: fast variable mean velocity Uf}; and, on the other hand, the timescale scaling as $k^{(\chi)}_{\rm f}/\lra{\widetilde{\epsilon}^{(\chi)}_{\rm f}}$ with which turbulent kinetic energy is being exchanged, with neither production nor dissipation, between the velocity components, which means that on the second line on the rhs of Eq.~\eqref{eq: fast variable mean velocity Uf} $(\widetilde{\mb{G}}^{(\chi)})^{-1}$ scales as $k^{(\chi)}_{\rm f}/\lra{\widetilde{\epsilon}^{(\chi)}_{\rm f}}$. Since both timescales are expressed as functions of $k_{\rm f}/\lra{\epsilon_{\rm f}}$, this explains indeed that two different turbulent kinetic energy rates are introduced in Eq.~\eqref{eq: fast variable mean velocity Uf}, corresponding to two different scaling with respect to the small parameter $\chi$, as detailed below. Remember that the turbulent kinetic energy dissipation rate $\lra{\epsilon_{\rm f}}$ is external to the description based on $\mb{Z}_{\rm f}=(\mb{X}_{\rm f},\mb{U}_{\rm f})$ and appears therefore as the proper quantity to monitor to introduce different scaling of the timescales as a function of the small parameter $\chi$. 

First, when $\chi=1$ we take the two timescales as equal ($\lra{\epsilon_{\rm f}}=\lra{\epsilon^{(\chi)}_{\rm f}}=\lra{\widetilde{\epsilon}^{(\chi)}_{\rm f}}$) and gather the two pressure gradients terms into one ($\lra{P_{\rm f}}=\lra{P^{(\chi)}_{\rm f}} + \lra{\widetilde{P}^{(\chi)}_{\rm f}}$) to retrieve the classical form of a generalized Langevin model. The Reynolds equation is then
\begin{equation}
\frac{\partial \lra{U_{{\rm f},i}}}{\partial t} + \lra{U_{{\rm f},j}}\frac{\partial \lra{U_{{\rm f},i}}}{\partial x_j} = - \frac{1}{\rho_{\rm f}} \frac{\partial \lra{P_{\rm f}}}{\partial x_i} - \frac{\partial \lra{u_{{\rm f},i}u_{{\rm f},j}}}{\partial x_j}~,
\end{equation}
where the Reynolds stress tensor, $\lra{u_{{\rm f},i}u_{{\rm f},j}}$, is the solution of a transport equation
\begin{multline}
\label{eq: second-order transport for Rij}
\frac{\partial \lra{u_{{\rm f},i}u_{{\rm f},j}}}{\partial t} + \lra{U_{{\rm f},k}}\frac{\partial \lra{u_{{\rm f},i}u_{{\rm f},j}}}{\partial x_k} + \frac{\partial \lra{u_{{\rm f},i}u_{{\rm f},j}u_{{\rm f},k}}}{\partial x_k} = - \lra{u_{{\rm f},i}u_{{\rm f},k}} \frac{\partial \lra{U_{{\rm f},j}}}{\partial x_k} - \lra{u_{{\rm f},j}u_{{\rm f},k}} \frac{\partial \lra{U_{{\rm f},i}}}{\partial x_k} \\
+ G_{ik} \lra{u_{{\rm f},k}u_{{\rm f},j}} + G_{jk} \lra{u_{{\rm f},k}u_{{\rm f},i}} + C_0 \lra{\epsilon_{\rm f}} \delta_{ij}~,
\end{multline}
where the return-to-equilibrium terms of Eq.~\eqref{eq: fast variable mean velocity Uf} have been added, i.e., with $\mb{G}=- 1/2 (\lra{\epsilon_{\rm f}}/k_{\rm f})\bds{\delta} + \widetilde{\mb{G}}$. For our present discussion, the important point is that the Reynolds stress tensor is a non-local quantity and is influenced by previous upstream values. 

We now introduce the specific scaling in terms of a small parameter $\chi$ whose limit to zero is to be taken and make full use of the difference between the two timescales. 

If we write $k_{\rm f}^{(1)}$ and $\lra{\epsilon_{\rm f}^{(1)}}$ to refer to the level of turbulent kinetic energy and dissipation rates, respectively, coming from the `regular' solution, for instance the one obtained with $\chi=1$ and equipped with its corresponding modeled equation for $\lra{\epsilon_{\rm f}^{(1)}}$, we then choose $\lra{\widetilde{\epsilon}^{(\chi)}_{\rm f}}$ to scale as $1/\chi^2$ (or, more precisely, we choose $\lra{\widetilde{\epsilon}^{(\chi)}_{\rm f}}$ so that $\lra{\widetilde{\epsilon}^{(\chi)}_{\rm f}} = 1/\chi^2 \, \lra{\widetilde{\epsilon}_{\rm f}}$ with$\lra{\widetilde{\epsilon}_{\rm f}} \sim \lra{\epsilon_{\rm f}^{(1)}}$). Given that $\widetilde{\mb{G}}^{(\chi)}$ scales as $\lra{\widetilde{\epsilon}^{(\chi)}_{\rm f}}/k^{(\chi)}_{\rm f}$ and that we already know from the results of the fast-variable techniques recalled earlier that we can expect that $k^{(\chi)}_{\rm f} \sim 1/\chi$ (or, more precisely, that $k^{(\chi)}_{\rm f} \simeq 1/\chi \, k_{\rm f}^{(1)}$), then we have $\widetilde{\mb{G}}^{(\chi)}=\lra{\widetilde{\epsilon}^{(\chi)}_{\rm f}}/k^{(\chi)}_{\rm f} \widetilde{\mb{G}} \sim 1/\chi \, \widetilde{\mb{G}}$ where $\widetilde{\mb{G}} \sim \widetilde{\mb{G}}^{(1)}$. As shown earlier, we can also consider that the mean pressure gradient due to $\lra{\widetilde{P}^{(\chi)}_{\rm f}}$ scales as $1/\chi$, as will be confirmed below, and can therefore be written as $\lra{\widetilde{P}^{(\chi)}_{\rm f}} \sim 1/\chi \lra{\widetilde{P}_{\rm f}}$.

On the other hand, for the timescale at which turbulent kinetic energy is dissipated, we choose $\lra{\epsilon^{(\chi)}_{\rm f}}$ to scale on par with $k^{(\chi)}_{\rm f}$, that is as $1/\chi$ (i.e., we take $\lra{\epsilon^{(\chi)}_{\rm f}} \sim 1/\chi \, \lra{\epsilon_{\rm f}^{(1)}}$). This is in line with the notion that $\lra{\epsilon^{(\chi)}_{\rm f}}$ represents the `true' dissipation rate, as can be seen on the rhs of Eq.~\eqref{eq: second-order transport for Rij} which indicates that $\lra{\epsilon^{(\chi)}_{\rm f}}$ varies indeed as the production terms, $\lra{u_{{\rm f},i}u_{{\rm f},k}} \partial \lra{U_{{\rm f},j}}/\partial x_k$, and thus as $1/\chi$ since the mean velocity field and its gradients remain of order 1 (note that, at this level, the correspondence with Eq.~\eqref{eq: second-order transport for Rij} is indicative and not a strict one since, due to the resulting local approximation as a Gaussian variable, the triple correlation vanishes). This implies that the timescale of turbulent kinetic energy dissipation, $k^{(\chi)}_{\rm f}/\lra{\epsilon^{(\chi)}_{\rm f}}$ on the first line on the rhs of Eq.~\eqref{eq: fast variable mean velocity Uf}, is of order 1 with respect to the small parameter $\chi$, i.e., $k^{(\chi)}_{\rm f}/\lra{\epsilon^{(\chi)}_{\rm f}} \sim k^{(1)}_{\rm f}/\lra{\epsilon^{(1)}_{\rm f}}$. Accordingly, we can take that the pressure gradient due to $\lra{P^{(\chi)}_{\rm f}}$ scales as $1$, i.e., $\lra{P^{(\chi)}_{\rm f}}=\lra{P_{\rm f}}$ with $\lra{P_{\rm f}} \sim \lra{P^{(1)}_{\rm f}}$.

Using these scaling relations, we can rewrite Eqs.~\eqref{eq: fast variable mean velocity} as
\begin{subequations}
\label{eq: fast variable mean velocity with chi}
\begin{align}
\dd X^{(\chi)}_{{\rm f},i} & = U^{(\chi)}_{{\rm f},i}\, \dd t~,   \label{eq: fast variable mean velocity with chi Xf}\\
\dd U^{(\chi)}_{{\rm f},i} & = - \frac{1}{\rho_{\rm f}}\frac{\partial \lra{P_{\rm f}}}{\partial x_i}\,\dd t - \frac{1}{2} \frac{\lra{\epsilon^{(1}_{\rm f}}}{k^{(1)}_{\rm f}} \left[ U^{(\chi)}_{{\rm f},i} - \lra{U^{(\chi)}_{{\rm f},i}} \right]\, \dd t \nonumber \\
                           & \phantom{=} - \frac{1}{\chi} \frac{1}{\rho_{\rm f}}\frac{\partial \lra{\widetilde{P}_{\rm f}}}{\partial x_i}\,\dd t + \frac{1}{\chi} \widetilde{G}_{ij} \left[ U^{(\chi)}_{{\rm f},j} - \lra{U^{(\chi)}_{{\rm f},j}} \right]\, \dd t + \frac{1}{\chi} \sqrt{C_0 \lra{\widetilde{\epsilon}_{\rm f}}}\, \dd W_i~, \label{eq: fast variable mean velocity with chi Uf}
\end{align}
\end{subequations}
and apply the fast-variable elimination techniques described above when $\chi \to 0$. The return-to-equilibrium term on the first line on the rhs of Eq.~\eqref{eq: fast variable mean velocity with chi Uf} has been written as $\lra{\epsilon^{(1}_{\rm f}}/k^{(1)}_{\rm f}$ and we have left out a possible constant appearing in front of this expression. This is not a concern since the terms on the first line on the rhs of Eq.~\eqref{eq: fast variable mean velocity with chi Uf} play no role in the elimination process which is entirely governed by the terms scaling as $1/\chi$, that is by the terms on the second line on the rhs of Eq.~\eqref{eq: fast variable mean velocity with chi Uf}. 

The equation satisfied by the mean velocity field $\lra{\mb{U}^{(\chi)}_{\rm f}}$ (the mean Navier-Stokes equation) is
\begin{equation}
\label{eq: mean Navier-Stokes with chi}
\frac{\partial \lra{U^{(\chi)}_{{\rm f},i}}}{\partial t} + \lra{U^{(\chi)}_{{\rm f},j}}\frac{\partial \lra{U^{(\chi)}_{{\rm f},i}}}{\partial x_j} = - \frac{1}{\rho_{\rm f}} \frac{\partial \lra{P_{\rm f}}}{\partial x_i} - \frac{1}{\chi} \left\{ \frac{1}{\rho_{\rm f}}\frac{\partial \lra{\widetilde{P}_{\rm f}}}{\partial x_i} + \frac{\partial \left[ \chi \lra{u^{(\chi)}_{{\rm f},i}u^{(\chi)}_{{\rm f},j}} \right]}{\partial x_j} \right\}~,
\end{equation} 
where the equivalent of the Reynolds stress tensor, $A_{ij}=\chi \lra{u^{(\chi)}_{{\rm f},i}u^{(\chi)}_{{\rm f},j}}$ is now a local term which, in the limit $\chi \to 0$, is the solution of the matrix equation
\begin{equation}
\label{eq: matrix equilibrium with chi}
\widetilde{\mb{G}} \mb{A} + \mb{A} \widetilde{\mb{G}}^{\bot}= - C_0 \lra{\widetilde{\epsilon}_{\rm f}} \bds{\delta}~.
\end{equation}
It is instructive to work out the resulting form of Eq.~\eqref{eq: mean Navier-Stokes with chi} corresponding to two different closures of $\widetilde{\mb{G}}$.

First, if we retain the SLM, then the decomposition used in Eq.~\eqref{eq: fast variable mean velocity Uf} indicates that 
\begin{equation}
\widetilde{G}^{(\chi)}_{ij} = - \frac{3\, C_0}{4} \frac{\lra{\widetilde{\epsilon}^{(\chi)}_{\rm f}}}{k^{(\chi)}_{\rm f}} \delta_{ij} \Longrightarrow \widetilde{G}_{ij} \simeq - \frac{3\, C_0}{4} \frac{\lra{\widetilde{\epsilon}_{\rm f}}}{k^{(1)}_{\rm f}} \delta_{ij} \Longrightarrow A_{ij} = \frac{2}{3} k^{(1)}_{\rm f} \delta_{ij}~.
\end{equation}
Then by taking $\widetilde{P}_{\rm f}$ in Eq.~\eqref{eq: mean Navier-Stokes with chi} so that $\widetilde{P}_{\rm f} + 2/3 k^{(1)}_{\rm f}$ is constant, we get that the equation satisfied by $\lra{\mb{U}_{\rm f}}$ obtained as the limit of the one satisfied by $\lra{U^{(\chi)}_{\rm f}}$ when $\chi \to 0$ is
\begin{equation}
\label{eq: mean Navier-Stokes with chi SLM}
\frac{\partial \lra{U_{{\rm f},i}}}{\partial t} + \lra{U_{{\rm f},j}}\frac{\partial \lra{U_{{\rm f},i}}}{\partial x_j} = - \frac{1}{\rho_{\rm f}} \frac{\partial \lra{P_{\rm f}}}{\partial x_i} ~.
\end{equation}
This is the equivalent of the Euler equations obtained as the first-order approximation in the derivation of the equations for hydrodynamics.

Second, if we consider a more elaborate closure of the matrix $\mb{G}$ in GLMs, such as the IP (Isotropization of Production) model~\cite[chapter 12]{pope2000turbulent}, we have 
\begin{align}
\widetilde{G}^{(\chi)}_{ij} & = \left( - \frac{3\, C_0}{4} - C_2 \frac{\mc{P}}{\lra{\widetilde{\epsilon}^{(\chi)}_{\rm f}}} \right) \frac{\lra{\widetilde{\epsilon}^{(\chi)}_{\rm f}}}{k^{(\chi)}_{\rm f}} \delta_{ij} + C_2 \frac{\partial \lra{U_{{\rm f},i}}}{\partial x_j} \\
& = \frac{\lra{\widetilde{\epsilon}^{(\chi)}_{\rm f}}}{k^{(\chi)}_{\rm f}} \left[ \left( - \frac{3\, C_0}{4} - C_2 \frac{\mc{P}^{(\chi)}}{\lra{\widetilde{\epsilon}^{(\chi)}_{\rm f}}} \right) \delta_{ij} + C_2 \frac{k^{(\chi)}_{\rm f}}{\lra{\widetilde{\epsilon}^{(\chi)}_{\rm f}}}\frac{\partial \lra{U_{{\rm f},i}}}{\partial x_j} \right]~,
\end{align}
where $\mc{P}^{(\chi)}$ is the turbulent kinetic energy production term (varying as $1/\chi$). With the various scaling given above, this gives 
\begin{equation}
\widetilde{G}_{ij} \simeq - \frac{3\, C_0}{4} \frac{\lra{\widetilde{\epsilon}_{\rm f}}}{k^{(1)}_{\rm f}} \left[ \delta_{ij} +  \chi \, \frac{4 C_2}{3 C_0}  \left( \frac{\mc{P}^{(1)}}{\lra{\widetilde{\epsilon}_{\rm f}}} \delta_{ij} - \frac{k^{(1)}_{\rm f}}{\lra{\widetilde{\epsilon}_{\rm f}}} \frac{\partial \lra{U_{{\rm f},i}}}{\partial x_j} \right) \right]~.
\end{equation}
We can then make a first-order approximation of the matrix $\widetilde{G}^{-1}$ in terms of the small parameter $\chi$ to obtain the first-order approximation of $\mb{A}$ from Eq.~\eqref{eq: matrix equilibrium with chi}. Given that $A_{ij}$ is a symmetrical tensor and leaving out the exact calculation of the constants appearing in front of the terms of the series, we obtain  
\begin{equation}
A_{ij} \simeq \frac{2}{3} k^{(1)}_{\rm f} \delta_{ij} + \chi \, C_{\rm A,1} \frac{k^{(1)}_{\rm f}}{\lra{\widetilde{\epsilon}_{\rm f}}} \mc{P}^{(1)} \delta_{ij} - \chi \, C_{\rm A,2} \frac{(k^{(1)}_{\rm f})^2}{\lra{\widetilde{\epsilon}_{\rm f}}} S_{ij}~,
\end{equation} 
where $S_{ij}$ is the strain rate tensor, i.e., $S_{ij}=1/2\left( \partial \lra{U_{{\rm f},i}}/\partial x_j + \partial \lra{U_{{\rm f},j}}/\partial x_i \right)$. By making the same evaluation of $\widetilde{P}_{\rm f}$ as in the zero-order above (the Euler equation) and by using that $\lra{u^{(\chi)}_{{\rm f},i}u^{(\chi)}_{{\rm f},j}} = A_{ij}/\chi $, we obtain from Eq.~\eqref{eq: mean Navier-Stokes with chi} that the equation satisfied by $\lra{\mb{U}_{\rm f}}$ as the limit of the one for $\lra{\mb{U}^{(\chi)}_{\rm f}}$ when $\chi \to 0$ is now
\begin{equation}
\label{eq: mean Navier-Stokes with chi IP}
\frac{\partial \lra{U_{{\rm f},i}}}{\partial t} + \lra{U_{{\rm f},j}}\frac{\partial \lra{U_{{\rm f},i}}}{\partial x_j} = - \frac{1}{\rho_{\rm f}} \frac{\partial \lra{P_{\rm f}}}{\partial x_i} + \nu_{\rm ft} \left( \frac{\partial \lra{U_{{\rm f},i}}}{\partial x_j} + \frac{\partial \lra{U_{{\rm f},j}}}{\partial x_i} \right) ~,
\end{equation}
where the term involving $C_{\rm A,1}$ has been added to the pressure and where $\nu_{\rm ft}$ is referred to as the turbulent viscosity. It is seen that $\nu_{\rm ft}$ is expressed by $\nu_{\rm ft} = C_{\rm ft}\, k_{\rm f}^2/\lra{\epsilon_{\rm f}}$ with $C_{\rm ft}$ a constant, which is a consistent outcome since, with the present scaling of $k_{\rm f}$ (as $1/\chi$) and of $\lra{\epsilon_{\rm f}}$ (as $1/\chi^2$), it is indeed the ratio $k_{\rm f}^2/\lra{\epsilon_{\rm f}}$ that remains of order 1. Therefore, the first-order approximation in terms of the small parameter $\chi$ leads to an equivalent form as the hydrodynamics equation where transport coefficients, such as molecular viscosity, appear in the first-order approximation of the Chapman-Enskog iterative approach. In that sense, the length $k_{\rm f}^{3/2}/\lra{\epsilon_{\rm f}}$ appears as the equivalent of the `mean free path' but this is a mere analogy since the physical picture as free flight and collisions does not apply and is more the characteristic length over which velocities remain significantly correlated. The important point in our context is that we have now a local closure of the Reynolds stress tensor which is consistent with models based on the turbulent-viscosity concept. 

These results are technical in nature and are meant to describe the steps towards the correct diffusive regime as the limit of over-damped Langevin equations for particle velocities. They demonstrate that consistent results are also obtained for velocity statistics (i.e., the mean velocity field) by manipulating carefully the white-noise limit while still handling SDEs. Yet, they should not necessarily be regarded as providing physical arguments for the concept of a turbulent viscosity or diffusivity. Given the length at which we have been obliged to go, it can even be said that they point to their lack of universal validity. There is indeed a notable difference with classical kinetic theory. When going from molecular motions to the Navier-Stokes equations, the ratio of the molecular relaxation timescale to a characteristic fluid one $\mc{L}/\delta U$, where $\delta U$ is a typical velocity difference over $\mc{L}$ the size of the domain considered, is indeed very small and of the order of $Kn\times Ma$, where $Kn$ and $Ma$ are the Knudsen and Mach numbers, respectively. The Newton law is therefore an excellent approximation. Except in some special situations, this is not so in turbulence since the typical timescale ratio, $k_{\rm f}/\lra{\epsilon_{\rm f}}\partial \lra{U_{\rm f}}/\partial x$, is of the order of $1-5$~\cite{pope2000turbulent}. This is a reminder of the importance of carefully sorting out slow and fast variables and of the selection of the particle state vector. In turbulence modeling, it is physically more justified to retain particle velocities.

\subsection{Summary and two-way influences}\label{sec V: summary}
%==========================================

The recognition of the importance of keeping track of the structural variables attached to the solute is characteristic of several complex fluid models, while other models introduce an order parameter to capture long-range order~\cite{ottinger2005beyond}. This is exemplified by the convected Maxwell model for the pressure tensor in polymeric fluids, which is recovered from a kinetic theory for Hookean dumbbells~\cite{venerus2018modern}. In that sense, turbulent dispersed two-phase flow modeling follows in the same footsteps with the reference particle state vector $\mb{Z}_{\rm p}=(\mb{X}_{\rm p},\mb{U}_{\rm p},\mb{U}_{\rm s})$ whose variables are treated as stochastic processes.

We have seen that the introduction of Brownian effects is as straightforward as for complex fluids. Accounting for turbulent fluctuations in non-fully-resolved flows is, however, far more challenging. 

\begin{figure}[ht]
 \centering
 \includegraphics[width=0.9\textwidth]{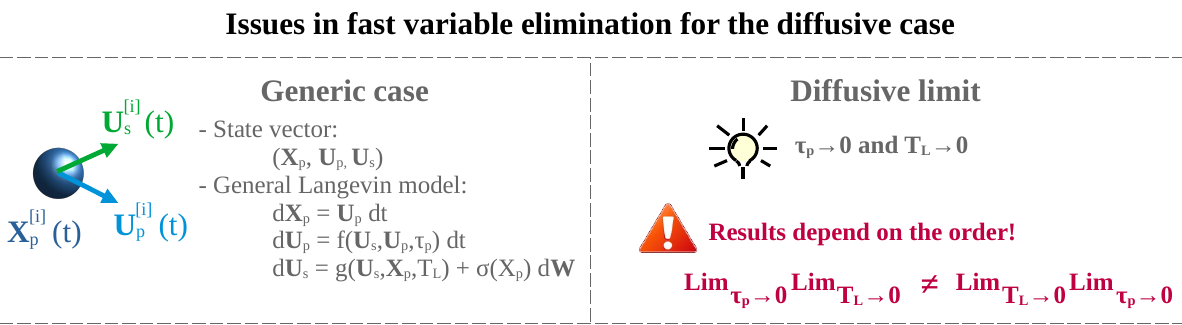}
 \caption{Sketch of the key issue when formulating the diffusive limit of a Langevin model.}
 \label{fig: Overdamped Langevin general}
\end{figure}

Another characteristic of complex fluid models is that they are formulated in the diffusive limit by neglecting particle inertia. In that respect, recent studies of the over-damped Langevin limit have brought more rigorous formulations which shed new light into this question. For example, it is important to notice that the correct diffusive terms appear only when the fast variable to be eliminated depends on the slow one where the white-noise terms emerge (in other words, when $\mb{U}_{\rm f}$ is the first-order derivative of $\mb{X}_{\rm f}$). However, in turbulent two-phase flow modeling $\mb{U}_{\rm s}$ is (loosely speaking) the second-order derivative of $\mb{X}_{\rm p}$. When considering the diffusive limit when both $\tau_{\rm p}$ and $T_{\rm L}^{*}$ become very small, this leads to the surprising result that $\lim_{\tau_{\rm p} \to 0} \lim_{T_{\rm L}^{*} \to 0} \neq \lim_{T_{\rm L}^{*} \to 0} \lim_{\tau_{\rm p} \to 0}$ (as illustrated on Fig.~\ref{fig: Overdamped Langevin general}). 
\begin{subequations}
\label{eq: fast variable 2phi limit}
To manifest this effect, we write the general model for $\mb{Z}_{\rm p}$ as
\begin{align}
\dd X_{{\rm p},i} &= U_{{\rm p},i}\, \dd t~, \label{eq: fast variable 2phi limit Xp} \\
\dd U_{{\rm p},i} &= \frac{1}{\tau_{\rm p}}\left( U_{{\rm s},i} - U_{{\rm p},i} \right) \dd t~, \label{eq: fast variable 2phi limit Up} \\
\dd U_{{\rm s},i} & = G_{ij}^{*}(t,\mb{X}_{\rm p}(t)) \left[ U_{{\rm s},j} - \phi_j^{*}(t,\mb{X}_{\rm p}(t)) \right] \dd t + \sigma_{ij}^{*} (t,\mb{X}_{\rm p}(t))\, \dd W_j~, \label{eq: fast variable 2phi limit s}
\end{align}
\end{subequations}
where the dependencies of $\mb{G}^{*}$, $\bds{\phi}^{*}$ and $\bds{\sigma}^{*}$ in Eq.~\eqref{eq: fast variable 2phi limit s} are indicated to bring out that they depend on $\mb{X}_{\rm p}(t)$ which is typically their twice-integrated variable. In Eq.~\eqref{eq: fast variable 2phi limit s}, the field $\bds{\phi}^{*}$ can be written as
\begin{equation}
\phi_i^{*}(t,\mb{x}) = \lra{U_{{\rm f},i}} (t,\mb{x}) + \left(G_{ij}^{*}\right)^{-1} \left( \frac{1}{\rho_{\rm f}}\frac{\partial \lra{P_{\rm f}}}{\partial x_j} + \Pi_j \right)(t,\mb{x})~,
\end{equation}
where $\bds{\Pi}$ represents the additional mean terms accounting for the crossing-trajectory effect, while it is sufficient to consider that $\mb{G}^{*} \simeq - 1/T_{\rm L}^{*}\, \bds{\delta}$ (without considering explicitly the different Csanady timescales) and $\sigma_{ij}^{*} \sim \sqrt{\lra{\epsilon_{\rm f}}}$ for the sake of simplicity and without any impact on the following outcomes. When we first take the limit $\tau_{\rm p} \to 0$, the two-phase flow model in Eqs.~\eqref{eq: fast variable 2phi limit} reverts by construction to the fluid one, cf. Eqs.~\eqref{eq: general overdamped Langevin}, and the same result obtained in Eq.~\eqref{eq: overdamped Langevin limit turbulence c}, or in Eq.~\eqref{eq: correct limit diffusion with tensor D}, is retrieved when we take the second limit $T_{\rm L} \to 0$ (since $T_{\rm L}^{*}(\tau_{\rm p}=0)=T_{\rm L}$). However, if we first take the limit $T_{\rm L}^{*} \to 0$, the detailed stochastic calculus~\cite[section 9.3.3]{minier2016statistical} shows that the limit system when $\mb{U}_{\rm f}$ is eliminated is
\begin{subequations}
\label{eq: fast variable limit Tl-Tp}
\begin{align}
\dd X_{{\rm p},i}^{(T_{\rm L}^{*})} &= U_{{\rm p},i}^{(T_{\rm L}^{*})}\, \dd t~, \label{eq: fast variable limit Tl-Tp Xp} \\
\dd U_{{\rm p},i}^{(T_{\rm L}^{*})} &= - \frac{1}{\tau_{\rm p}}\, U_{{\rm p},i}^{(T_{\rm L}^{*})} \dd t + \frac{1}{\tau_{\rm p}}\, \phi_i^{*}(t,\mb{X}^{(T_{\rm L}^{*})}_{\rm p}(t)) \dd t + \frac{1}{\tau_{\rm p}}\, \left( (G^{*})^{-1} \sigma^{*} \right)_{ij}(t,\mb{X}^{(T_{\rm L}^{*})}_{\rm p}(t))\, \dd W_j~, \label{eq: fast variable limit Tl-Tp Up} 
\end{align}
\end{subequations}
where the notations $\mb{X}_{\rm p}^{(T_{\rm L}^{*})}$ and $\mb{U}_{\rm p}^{(T_{\rm L}^{*})}$ indicate that the limit $T_{\rm L}^{*} \to 0$ is taken. Note that the rhs of Eq.~\eqref{eq: fast variable limit Tl-Tp Up} corresponds to what the naive formulation (i.e., writing directly $\dd \mb{U}_{\rm s}=0$ in Eq.~\eqref{eq: fast variable 2phi limit s}) would yield. Note also that the field $\bds{\phi}^{*}$ remains regular in that limit case since the mean velocity $\lra{\mb{U}_{\rm f}}$ remains so, as demonstrated above. By taking the second limit $\tau_{\rm p} \to 0$, no extra term appears from this second stochastic calculus and we obtain that the predicted diffusive limit is
\begin{align}
\dd X_{{\rm p},i}^{(\tau_{\rm p},T_{\rm L}^{*})} &= \Phi_i\, \dd t + \left( G^{-1}\sigma \right)_{ij} \dd W_i~, \\
                  &= \lra{U_{{\rm f},i}}\,\dd t + (G_{ij})^{-1} \left( \frac{1}{\rho_{\rm f}}\frac{\partial \lra{P_{\rm f}}}{\partial x_j} \right)\dd t + \left( G^{-1} \sigma^{*} \right)_{ij} \dd W_j
\end{align} 
where the second term on the rhs of Eq.~\eqref{eq: diffusive limit first expression} is missing. In the simple case of isotropic tensors, we have 
\begin{equation}
\label{eq: diffusive limit false form isotropic}
\dd X_{{\rm p},i}^{(\tau_{\rm p},T_{\rm L}^{*})} = \lra{ U_{{\rm f},i} }\, \dd t + T_{\rm L}\frac{\partial}{\partial x_i} \left[ \frac{1}{2} T_{\rm L}\, \sigma^2 \right]\, \dd t + \left( T_{\rm L}\, \sigma \right)\, \dd W_i~,
\end{equation}
which is clearly wrong by comparison with the correct form in Eq.~\eqref{eq: diffusive limit simple form isotropic}. Actually, the non-zero value of the particle relaxation timescale $\tau_{\rm p}$ filters out some of the fluctuations of $\mb{U}_{\rm s}$, when the white-noise limit is considered for the fluid velocity seen, which cannot be retrieved afterward. This leads to a truncated, and ultimately wrong, limit expression. The diffusive limit should therefore be addressed with care when several timescales are involved and it is believed that this is an area where progress in two-phase flow modeling can be of interest for complex fluids.

%===========================================================================================================
\section{Statistical modeling of particle collisions in turbulent flows \label{statistical model collision}}
%===========================================================================================================

Open any textbook on non-equilibrium statistical mechanics and you will find the Boltzmann equation as the archetypal description of particle-particle interactions. The Boltzmann equation is undoubtedly a major achievement and a well-established framework for short-range particle-particle interactions, especially for rarefied gases~\cite{Reif_1985,Liboff_1990,Balescu_1997,keizer2012statistical}. It is therefore justified to refer to it when considering particle collisions in another context. The specific assumptions underlying the classical form of the Boltzmann equation are, however, often taken for granted. This is a regrettable oversight since collisions of particles transported by random media can challenge some aspects of the Boltzmann picture. With a view toward modeling particle collisions in turbulent flows, the purpose of this section is to revisit the key characteristics of the Boltzmann equation to bring out the limitations that need to be addressed. 

\begin{figure}
 \centering
 \includegraphics[width=\textwidth]{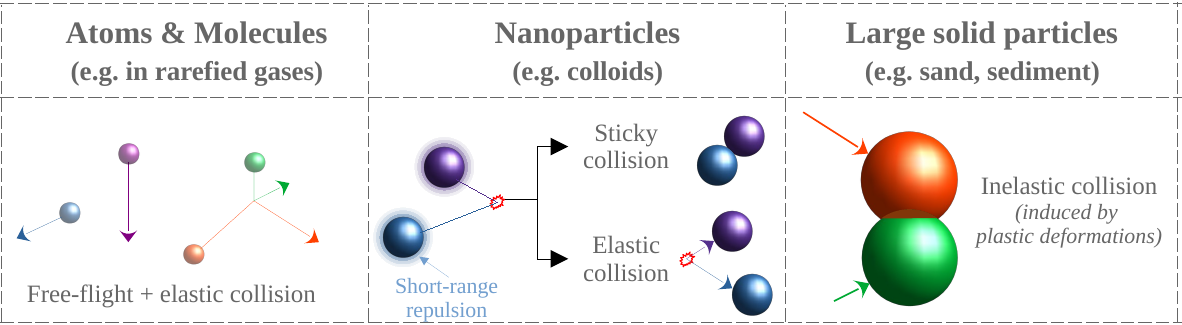}
 \caption{Sketch of the various outcomes of particle collisions depending on the type of forces involved.}
 \label{fig:agglo_correspondence}
\end{figure}

Since there are several detailed accounts of the Boltzmann theory, we limit ourselves to recalling the main points of interest for the present discussion. In its final formulation, the Boltzmann approach is a statistical description developed in terms of the single-particle distribution function $f(t;\mb{y}_{\rm p},\mb{V}_{\rm p},d_{\rm p})$ in which only the kinetic variables $(\mb{X}_{\rm p},\mb{U}_{\rm p})$ are retained as well as the particle diameter $d_{\rm p}$ (or its volume $\mc{V}_{\rm p}$). The physical picture consists in a process where particles undergo free flights and binary collisions. Note that the Boltzmann equation can easily account for the effects of an external and conservative force field (thus, deriving from a potential) so that particles do not necessarily follow straight lines in-between collisions but this does not change our discussion and we can still refer to free flights. In the Boltzmann-Grad limit where particle diameters are very small but with finite cross-section (i.e., in the limit when the number of particles $N_{\rm p} \to +\infty$ and $d_{\rm p} \to 0$, but with $N_{\rm p}\, d_{\rm p}^2$ remaining constant), collisions are regarded as instantaneous and point-wise elastic interactions so that mass, momentum and kinetic energy are conserved during collisions which appear as random redistributive events. When a collision occurs between a particle with a state vector $(\mb{X}_{\rm p},\mb{U}_{\rm p},d_{\rm p})$ and a collisional partner whose variables are noted $(\mb{X}_{\rm c,p},\mb{U}_{\rm c, p},d_{\rm c,p})$, the transformation rules expressing the post-collisional velocities $\mb{U}^{\rm ac}_{\rm p}$ and $\mb{U}^{\rm ac}_{\rm c,p}$ in terms of the pre-collisional ones $\mb{U}_{\rm p}$ and $\mb{U}_{\rm c, p}$ are
\begin{subequations}
\label{eq: collision general transformation rules}
\begin{align}
\mb{U}^{\rm ac}_{\rm p}   & = \mb{U}_{\rm p} - \frac{2 m_{\rm c,p}}{m_{\rm p} + m_{\rm c,p}}\left( \mb{U}_{\rm p} - \mb{U}_{\rm c,p}, \mb{e} \right) \cdot \mb{e}~, \\
\mb{U}^{\rm ac}_{\rm c,p} & = \mb{U}_{\rm c,p} + \frac{2 m_{\rm p}}{m_{\rm p} + m_{\rm c,p}}\left( \mb{U}_{\rm p} - \mb{U}_{\rm c,p}, \mb{e} \right) \cdot \mb{e}~,
\end{align}
\end{subequations}
with $(\mb{a},\mb{b})$ the scalar product between two vectors $\mb{a}$ and $\mb{b}$ in $\mathbb{R}^3$, $\mb{e}$ a unit vector on the unit sphere $\mc{S}^2$ representing the orientation of the collision event, and $m_{\rm p}=\rho_{\rm p} \pi d^3_{\rm p}/6$ (resp. $m_{\rm c,p}=\rho_{\rm p} \pi d^3_{\rm c,p}/6$) the mass of the target particle (resp. its collisional partner). These relations are easily inverted to provide the pre-collisional velocities $\mb{U}^{\rm bc}_{\rm p}$ and $\mb{U}^{\rm bc}_{\rm c,p}$ yielding $\mb{U}_{\rm p}$ and $\mb{U}_{\rm c,p}$ as the outcome of the collisional event, which we write as $\mb{U}^{\rm bc}_{\rm p}=\mb{U}^{\rm bc}_{\rm p}(\mb{U}_{\rm p},\mb{U}_{\rm c,p},\mb{e})$ and $\mb{U}^{\rm bc}_{\rm c,p}=\mb{U}^{\rm bc}_{\rm c,p}(\mb{U}_{\rm p},\mb{U}_{\rm c,p},\mb{e})$. With these notations and leaving out external force fields, the Boltzmann equation is expressed in sample space for the single-particle distribution function $f(t;\mb{y}_{\rm p},\mb{V}_{\rm p},\widehat{d}_{\rm p})$ as
\begin{equation}
\label{eq: general Boltzmann equation}
\frac{\partial f}{\partial t} + V_{{\rm p},k}\frac{\partial f}{\partial y_{{\rm p},k} } = \frac{\partial \phi_{\rm ext}(\mb{y})}{\partial y_k}\frac{\partial f}{\partial V_k} + Q_{\rm coll} (f)~,
\end{equation}
where $\widehat{d}_{\rm p}$ is the sample space variable for $d_{\rm p}$ and $-\partial \phi_{\rm ext}(\mb{x})/\partial x_k$ an external force field (per unit mass) which is left out from now on for the sake of simplicity. In Eq.~\eqref{eq: general Boltzmann equation}, $Q_{\rm coll} (f)$ is a gain and loss term accounting for the collisional events in Eqs.~\eqref{eq: collision general transformation rules} over all possible collisional partners which gives
\begin{equation}
\label{eq: Boltzmann gain-loss term with f2}
Q_{\rm coll} (f) = \int_{\mathbb{R}^3} \dd \mb{V}_{\rm c,p} \int_{\mc{S}^2} B(\mb{V}_{\rm p},\mb{V}_{\rm c,p},\widehat{\mb{e}})\left[ f_2(t;\mb{y}_{\rm p},\mb{V}^{\rm bc}_{\rm p},\widehat{d}_{\rm p},\mb{V}^{\rm bc}_{\rm c,p},\widehat{d}_{\rm c,p}) - f_2(t;\mb{y}_{\rm p},\mb{V}_{\rm p},\widehat{d}_{\rm p},\mb{V}_{\rm c,p},\widehat{d}_{\rm c,p}) \right] \dd \widehat{\mb{e}}
\end{equation}
where $\widehat{\mb{e}}$ is the sample space variable for $\mb{e}$. In Eq.~\eqref{eq: Boltzmann gain-loss term with f2}, $B(\mb{V}_{\rm p},\mb{V}_{\rm c,p},\mb{e})$ is the collision kernel measuring the likelihood of collisions and determined by the interaction potential between particles (thus related to the scattering problem). In the present case, it is sufficient to consider the classical hard-sphere model according to which the collision kernel takes the form 
\begin{equation}
\label{eq: collision kernel kinetic theory}
B(\mb{U}_{\rm p},\mb{U}_{\rm c,p},\mb{e})= \pi \left( \frac{d_{\rm p}}{2} + \frac{d_{\rm c,p}}{2} \right)^2 \vert \left( \mb{U}_{\rm p} - \mb{U}_{\rm c,p}, \mb{e} \right) \vert~. 
\end{equation}
Once a collisional kernel is chosen, it is seen that Eq.~\eqref{eq: Boltzmann gain-loss term with f2} represents the most general expression of the gain and loss term for binary collisions. A key point is that it involves the two-particle distribution function, noted $f_2$, written at the same location (reflecting the hypothesis of point-like collisions for simplicity). As such, the Boltzmann equation in Eqs.~\eqref{eq: general Boltzmann equation} and~\eqref{eq: Boltzmann gain-loss term with f2} is unclosed. It is clear that expressing the evolution equation in sample space for $f_2$ results in an open equation involving the three-particle distribution $f_3$, etc., so that we are faced with a BBGKY-like hierarchy. The closed form of the Boltzmann equation is obtained by resorting to the molecular chaos hypothesis according to which the velocities of colliding partners are uncorrelated, which means that we can write
\begin{equation}
\label{eq: molecular chaos hypothesis}
f_2(t;\mb{y}_{\rm p},\mb{V}_{\rm p},\widehat{d}_{\rm p},\mb{V}_{\rm c,p},\widehat{d}_{\rm c,p}) = f(t;\mb{y}_{\rm p},\mb{V}_{\rm p},\widehat{d}_{\rm p}) \times f(t;\mb{y}_{\rm p},\mb{V}_{\rm c,p},\widehat{d}_{\rm c,p})~.
\end{equation}
This yields the classical expression of the Boltzmann equation, i.e. with a collision term that is a quadratic function of the single-particle distribution function $f$ and, for that reason, is often noted $Q_{\rm coll}(f,f)$ in textbooks. From this brief reminder, it appears that, at least, three characteristics of the Boltzmann description of particle collisions are worth mentioning:
\begin{enumerate}[(a)]
\item It is developed in sample space retaining only the particle kinetic state vector $(\mb{X}_{\rm p},\mb{U}_{\rm p})$; 
\item It usually involves collision kernels based on free flights between particle collisions;
\item It relies on the molecular chaos assumption and assumes Markov behavior.
\end{enumerate}
When considering particle collisions in random media with non-zero space and time correlations, each of these characteristics raises questions that need to be addressed. Furthermore, when dealing with mesoscopic particles such as colloids or discrete inertial particles, we are concerned not only with elastic collisions but also with inelastic ones (for instance, in granular matter) and even with sticky collisions (for example, in particle agglomeration or droplet coalescence). This is illustrated in Fig.~\ref{fig:agglo_correspondence} and it is interesting to discuss the points (a)-(c) above with respect to a wide range of outcomes for particle collisions.

\subsection{Developments in sample and physical spaces}\label{sec: sample or physical spaces}
%=====================================================

The Boltzmann equation was originally developed in sample space and traditional presentations have retained this formulation. It is seen from Eqs.~\eqref{eq: general Boltzmann equation}-\eqref{eq: molecular chaos hypothesis} that the description of the physical process is reduced to a single function which is the solution of a PDE written in a six-dimensional space involving a non-linear collisional term, which makes solving it (in a strong sense) a formidable task. Yet, it has great merit in that it constitutes the main pathway for the derivation of balance equations for continuum fluid mechanics in the hydrodynamical limit through, for instance, the Chapman-Enskog or the Grad approaches. As such, things may look similar for the statistical description of turbulent particle-laden flows with collisions. There are, however, differences that suggest to adopt another point of view. 

First, we have seen that the kinetic description is ill-based for particle transport in non-fully resolved turbulent flows and that it is necessary to introduce a particle state vector that includes, at least, the velocity of the fluid seen $\mb{Z}_{\rm p}=(\mb{X}_{\rm p},\mb{U}_{\rm p},\mb{U}_{\rm s})$. We are then handling a distribution function (or a PDF) that has, at least, ten independent variables if we include the particle diameter $d_{\rm p}$ or its volume $\mc{V}_{\rm p}$. The general evolution equation in sample space for $p(t;\mb{y}_{\rm p},\mb{V}_{\rm p},\mb{V}_{\rm s},\upsilon_{\rm p})$, where $\upsilon_{\rm p}$ is the sample space variable corresponding to $\mc{V}_{\rm p}$, has the form
\begin{equation}
\label{eq: general Boltzmann-FP equation} 
\frac{\partial p}{\partial t} + V_{{\rm p},k}\frac{\partial p}{\partial y_{{\rm p},k}}= -\frac{\partial }{\partial V_{{\rm p},k}}\left[ \left( \frac{V_{{\rm s},k} - V_{{\rm p},k}}{\tau_{\rm p}}\right) p \right] - \frac{\partial \left[A_{{\rm s},k} p \right]}{\partial V_{{\rm s},k}} + \frac{1}{2}\frac{ \partial \left[ \left(B_{\rm s}B_{\rm s}^{\bot}\right)_{kl} p \right] }{\partial V_{{\rm s},k} \partial V_{{\rm s},l}} + Q_{\rm coll}(p)~,
\end{equation}
which makes strong solutions even more difficult to obtain. Second, PDF equations, such as the Boltzmann-Fokker-Planck one in Eq.~\eqref{eq: general Boltzmann-FP equation}, are indeed useful to derive the corresponding transport equations in physical space for statistics of interest. At this stage, there is, however, a crucial difference with the steps leading from the Boltzmann equation to the hydrodynamical ones. In turbulent two-phase flow modeling, we are not interested in reducing the description to a set of PDEs for a small number of one-point moments because these equations are unclosed and, as indicated in Sec.~\ref{sec V: summary}, there is little hope to rely on something like a physically-meaningful turbulent-viscosity concept to help us. In fact, we are led to solving (in a weak sense) the PDF equation by developing stochastic particle systems that act as Monte Carlo solutions. It follows that, just as the Langevin equations are used to model and simulate particle transport, we need to devise stochastic particle systems in physical space that play the same role for the collision term. 

Although more limited than the one on the operator formalism, there is now a large-enough literature dedicated to such stochastic particle systems, often mathematically-oriented~\cite{wagner2004stochastic,eibeck2003stochastic,fournier2005spatially}, as well as more synthetic accounts~\cite{minier2015lagrangian} to which interested readers are referred to. A nice presentation can be found in~\cite{Wagner_2011}.
As indicated in these works, stochastic particle systems are essentially built on jump processes (cf. Sec.~\ref{sec II: Wiener and Poisson}) and can be represented by the generic algorithm for $N_{\rm p}$ particles contained in a small volume $\mc{V}$ and described with a reduced particle state vector $\mb{Z}^{(N)}_{\rm p}=(\mb{U}_{\rm p}^{[1]},\ldots,\mb{U}_{\rm p}^{[N]})$ limited to particle velocities when a locally homogeneous hypothesis is made inside the volume $\mc{V}$. The $N_{\rm p}$ particles interact through the following generalized Poisson process 

\begin{figure}[h]
 \centering
 \includegraphics[width=0.9\textwidth]{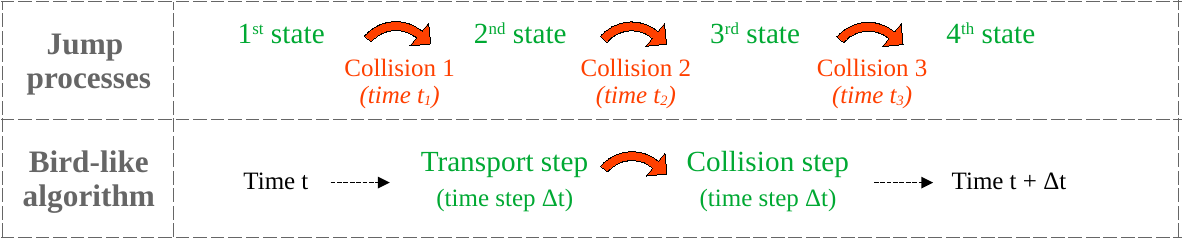}
 \caption{Representation of the two different ways to account for jumps in a generalized Poisson process.}
 \label{fig: jump collision}
\end{figure}

\begin{enumerate}
\item The system waits in the same state during a time which is an exponentially-distributed random variable whose parameter is 
\begin{equation}
\label{SPS collision step 1}
\lambda_c(\mb{Z}^{(N)}_{\rm p})=\frac{1}{2\, N_{\rm p} \, \mc{V}}\sum_{\substack{k,l=1 \\ k \neq l}}^{N_{\rm p}} 
                \int_{\mc{S}^2} B(\mb{U}_{\rm p}^{[k]},\mb{U}_{\rm p}^{[l]},\mb{e})\,\dd \mb{e}~.
\end{equation} 
\item At the time of a jump, two indexes $i$ and $j$ corresponding to two particles in the volume $\mc{V}$ are chosen as well as a direction $\mb{e}$ on the unit sphere $\mc{S}^2$ according to the probability 
\begin{equation}
\label{SPS collision step 2}
p_c(i,j,\mb{e})= \frac{B(\mb{U}_{\rm p}^{[i]},\mb{U}_{\rm p}^{[j]},\mb{e})}{ 2\, N_{\rm p}\, \mc{V}\, \lambda_c(\mb{Z}^{(N)}_{\rm p})}
= \dfrac{B(\mb{U}_{\rm p}^{[i]},\mb{U}_{\rm p}^{[j]},\mb{e})}
{\sum_{\substack{k,l=1 \\ k \neq l}}^{N_{\rm p}} \int_{\mc{S}^2} B(\mb{U}_{\rm p}^{([k]},\mb{U}_{\rm p}^{[l]},\mb{e})\,\dd \mb{e}}.
\end{equation}
\item For the selected particle pair, their velocities are updated with the transformations rules in Eqs.~\eqref{eq: collision general transformation rules}.
\end{enumerate}
This presentation follows one approach to simulate jump processes by generating the random interval times between jumps. As indicated in Sec.~\ref{section: the Wiener and Poisson processes}, another approach consists in using fixed time intervals (or time steps) during which the random numbers of events that take place are sampled in a Poisson distribution whose parameter is determined from the local value of the collision kernel or from the mean free path. Sometimes just the mean numbers of events are considered when they are very large. In the latter formulation, this corresponds to the collision step of the well-known DSMC (Direct Simulation Monte Carlo) method which relies on a time-splitting algorithm where the transport and collision steps are treated sequentially (first, transport without collision; then, collisions without transport) and where the above stochastic particle model is applied in each cell of the calculation~\cite{Bird_1963,Bird_1994} (see also Fig.~\ref{fig: jump collision}).

\begin{figure}
 \centering
 \includegraphics[width=0.9\textwidth]{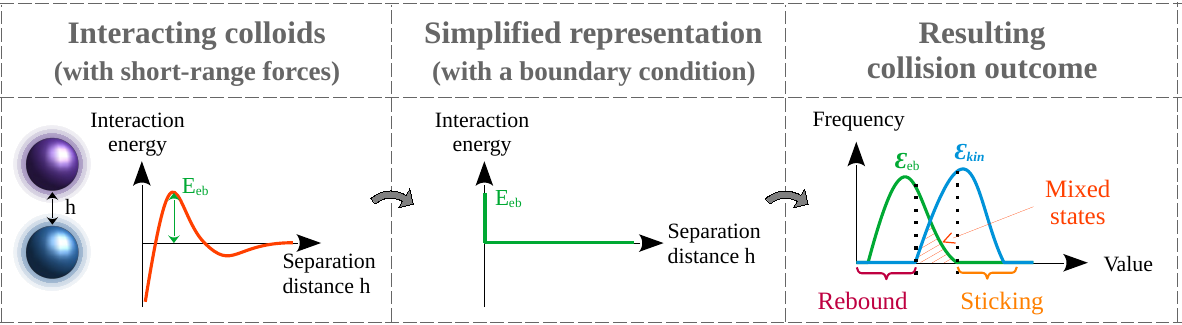}
 \caption{Sketch of the potential interaction between a colloid and a wall, its simplified representation as a boundary condition, and its resulting effect on the collision outcome (deposition or rebound).}
 \label{fig: sketch DLVO}
\end{figure}

Stochastic particle systems are flexible in the sense that they allow more general collision outcomes to be treated. Since we simulate explicitly a number of collisions between selected pairs of particles, it is indeed straightforward to introduce criteria (based on the values of particle-attached variables or properties) to decide whether each collision is elastic, inelastic or sticky. The formulation of these stochastic particle systems is discussed in several works~\cite{minier2015lagrangian} and the corresponding expressions in terms of operators in sample space are well described in~\cite{fournier2005spatially}. To illustrate this point with a specific example, it is interesting to consider the case of colloidal particles interacting with a wall surface where they can deposit or bounce off. Similarly to the colloid-colloid interactions outlined in Sec.~\ref{colloid suspensions and river deltas}, when transported by a fluid flow to the immediate vicinity of a bounding wall surface, colloids are subject to short-ranged potential forces~\cite{israelachvili2011intermolecular} (see also Fig.~\ref{fig: sketch DLVO}): an attractive one (e.g., due to van der Waals forces) and a repulsive one (e.g., due to electrostatic double-layer forces between similar particles). The key observation is that these forces are extremely short-range (typically from a few \si{\angstrom} to a few nanometers). This implies that they act over scales that are a few orders of magnitude below typical hydrodynamical ones. For that reason, the exact potential is replaced by a step function that corresponds to the maximum of the potential, called the energy barrier $E_{\rm eb}$. This step function acts as a boundary condition for colliding particles~\cite{henry2012towards}: deposition occurs when the kinetic energy is higher than the energy barrier; otherwise, an elastic rebound occurs. Since the value of the energy barrier is sensitive to the presence of small protuberances on the particle surface (called roughness), it is actually a random variable whose distribution $p(\mc{E}_{\rm eb})$ needs to be generated. In short, each particle-particle interaction involves a double randomness coming from the particle kinetic energy and the value of the local energy barrier, as sketched in Fig.~\ref{fig: sketch DLVO}. However, as developed in~\cite[section 6.1]{minier2015lagrangian}, the formulation of the same boundary condition in sample space becomes quite intricate with a random combination of absorbing and reflecting conditions which goes far beyond the standard ones usually considered for the PDF equation~\cite{gardiner2009stochastic,Ottinger_1996,risken1996fokker}.

Several remarks can be made:
\begin{enumerate}[(i)]
\item From the descriptions already given in Sec.~\ref{colloid suspensions and river deltas} and the one above, it is clear that the same processes are at play between pairs of colloids in a solution and between colloids and a wall surface. In a way, we are already considering two-particle interactions but with one `particle' being fixed (i.e., the surface wall). This allows to introduce the notions below while still retaining the one-particle PDF framework;
\item It also appears that the modeling approach to colloids deposition follows in the footsteps of DSMC-like methods since transport and particle-wall interactions are treated sequentially, with one step where particles are transported without short-range interactions until they are detected as having hit a surface wall, and a second step where the outcome of the interaction is determined without actual transport;
\item In the frame of one-particle models, the occurrence of a particle-wall impact can be obtained from the resolved particle dynamics through detection algorithms (no collision kernels are needed).
\end{enumerate}

\begin{figure}
 \centering
 \includegraphics[width=0.9\textwidth]{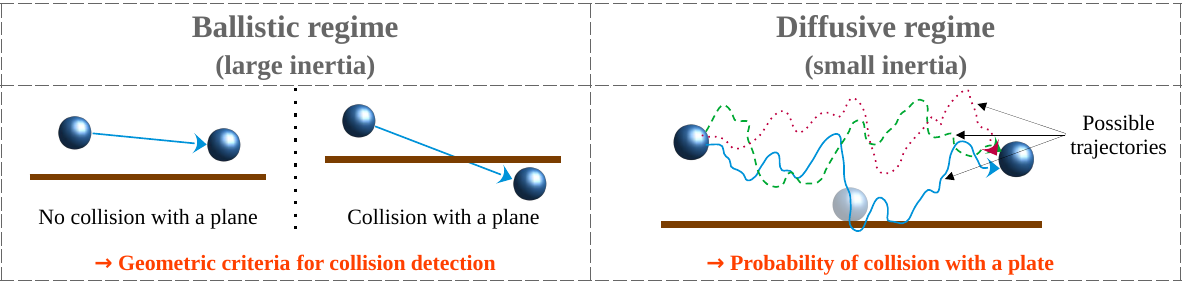}
 \caption{Detection of a particle-wall interaction during a small time interval. In the ballistic regime, the position of the end point of the particle trajectory is enough to detect an interaction. In the diffusive regime, there is a non-zero probability to hit the wall even for start and end points within the domain.}
 \label{fig: Agglo-detection-wall}
\end{figure}
We can elaborate on the last remark, which corresponds to the question: how do we detect a particle-wall interaction? At the moment, there are two situations where solutions can be worked out (cf. Fig.~\ref{fig: Agglo-detection-wall}).
The first one is when particle inertia is high enough so that, within a small time interval (one time step of numerical simulations), the trajectories of these particles are straight lines. Then, the detection criterion consists simply in checking whether the predicted location of a particle at the end of each time interval is still within the domain or lies outside, in which case this implies that an interaction has occurred during the time interval (simple geometrical considerations give the position where the interaction took place and, if necessary, an estimation of the time at which it happened). We can refer to this situation as the `local ballistic regime' (by local, it is meant here local in time since we are considering particle evolution within a small time increment). In the immediate vicinity of a wall, more precisely in the so-called viscous sublayer~\cite{pope2000turbulent}, turbulent fluctuations are strongly dampened by viscous effects so that randomness due to turbulence is lessened. By reducing the observation time intervals (or, conversely, the numerical time steps), it can be believed that the local ballistic regime is always reached. This is not so, however, for particles small enough to be sensitive to Brownian fluctuations. In fact, the second situation where we can come up with a tractable detection criterion for particle-wall interactions is when Brownian effects dominate completely so that we can, locally, neglect fluid velocity and consider that particle motion is governed by Brownian noise and the Stokes-Einstein diffusion coefficient, i.e., $\dd \mb{X}_{\rm p} \simeq \sqrt{2\mc{D}}\, \dd \mb{W}$. In that case, we are in the `local diffusive regime' which is the counterpart of the ballistic one. As illustrated in Fig.~\ref{fig: Agglo-detection-wall}, even if the predicted location of a particle at the end of a time interval is still within the domain (of course, if that position is outside, an interaction has occurred just as in the ballistic regime), there is now a non-zero probability that this particle has touched the wall during the time interval. With the assumption mentioned for the diffusive regime, we can, however, use the distance between a particle and the wall which is a Bessel process. This corresponds to the notion of Brownian (or Bessel) bridges and we can derive an analytical expression for the probability that the distance becomes smaller than the particle radius $R_{\rm p}$
\begin{equation}
P_a^b(R_{\rm p}) = \mathcal{P} \left[ \left. \min_{t<s<t+\Delta t}{(|\mb{X}_{\rm p}(s)|)}\leq R_{\rm p}\, \right| \, |\mb{X}_{\rm p}(t)|=a, |\mb{X}_{\rm p}(t+\Delta t)|=b \, \right]~.
\end{equation}
From the mathematical theory of diffusive bridges~\cite{borodin2015handbook,henry2014astochastic,mohaupt2011new}, we obtain 
\begin{equation}
 P_a^b(R_{\rm p}) = \min\left( 1\, ;\; \frac{\exp\left(\dfrac{R_{\rm p}(a+b-R_{\rm p})}{\mc{D}^2\Delta t}\right)-1}{\exp\left(\dfrac{ab}{\mc{D}^2\Delta t}\right)-1}\right)~.
 \label{eq: Bessel probability formula}
\end{equation}
Apart from these two asymptotic regimes, however, the mathematical derivation appears intractable (to the best of the authors' knowledge, no analytical formula has been obtained), which means that for particles described by a Langevin-type of model, there is no clear detection criterion that we could apply. This is a serious limitation since a wide range of particles belong precisely to that intermediate category. 

\subsection{Limitations of present collision kernels}\label{sec: limited collision kernels}
%====================================================

As mentioned, the processes involved in particle-particle interactions within a flow are the same as the ones between a particle and a wall but for the fact that they involve relative particle motions instead of the dynamics of individual particles. Their treatments depend on the information available. 

\begin{figure}[h]
 \centering
 \includegraphics[width=0.8\textwidth]{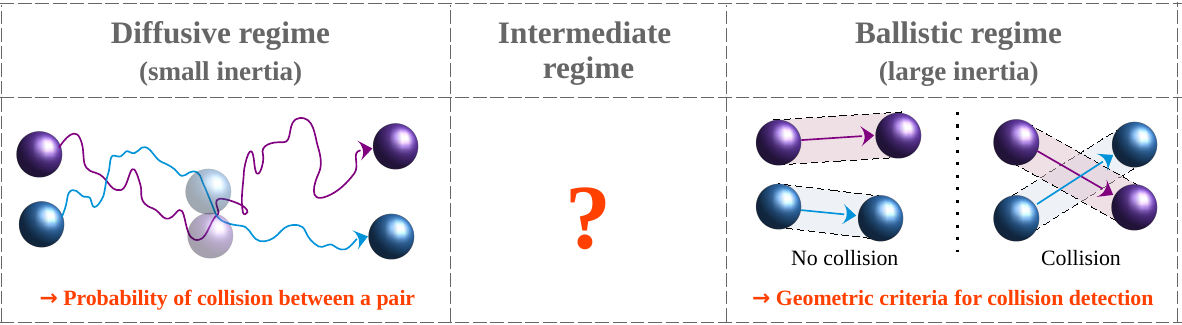}
 \caption{Different particle regimes used to detect a collision between a pair of particles within a small time interval. Current expressions of the detection algorithms are based on cylinder of collisions for the ballistic regime and on diffusive bridges for the diffusive regime but nothing is available in the intermediate regime.}
 \label{fig:agglo detection regime}
\end{figure}

If we have access to the two-particle PDF, we can deduce the relative motion between a pair of particles as well as their relative kinetic energy upon collision to evaluate if the energy barrier between these particles is overcome or not. Note that this criterion is relevant for colloidal particles. For larger or inertial particles, we can use other criteria to determine whether the collision results in a plastic deformation, or involve friction whereby kinetic energy is dissipated into heat inside both particles, or a simple elastic rebound. We are, in fact, in the same situation as the one for a particle interacting with a wall surface and we can use detection algorithms (see Fig.~\ref{fig:agglo detection regime}). For particles in the ballistic regime, this leads to cylinders of collisions or similar deterministic methods based on particle free flights between collisions. In the diffusive regime, similar reasoning based on diffusive bridges can be applied~\cite{henry2014astochastic}. Nevertheless, for particles in the intermediate regime, we are faced with the limitations pointed out at the end of Sec.~\ref{sec: sample or physical spaces}.

If we do not have access to the two-particle PDF, detection algorithms cannot be used and we must rely on pre-determined collision kernels or collision frequencies. Given the amount of work on particle-particle interactions, it might be believed that this is a well-addressed problem. It turns out that this is not the case. The classical collision kernel coming from the usual formulation of kinetic theory is given in Eq.~\eqref{eq: collision kernel kinetic theory}. This is, of course, valid but only insofar as we have particles traveling in straight lines (or deterministic ones), which is the case for atoms or molecules in a void or in an external and conservative force field as in the original Boltzmann picture. For discrete particles in a fluid flow, this remains acceptable but only for `bullet-like' particles in the ballistic regime for which the effects of the underlying fluid can be ignored during the time interval considered, as may the case in granular matter for instance. At first sight, the situation seems also well-established in the diffusive regime, i.e., for very small particles governed by Brownian motion. Indeed, another well-known result is the Brownian collision kernel which, for two classes of particles labeled $i$ and $j$ corresponding to diameters $d_{\rm p}^{[i]}$ and $d_{\rm p}^{[j]}$ respectively, writes
\begin{equation}
\label{eq: Brownian collision kernel}
B^{[i-j]}= \frac{2 k_{\rm B} \Theta_{\rm f}}{3\rho_{\rm f} \nu_{\rm f}} \frac{ \left(d_{\rm p}^{[i]} + d_{\rm p}^{[j]} \right)^2}{d_{\rm p}^{[i]} d_{\rm p}^{[j]}}~.
\end{equation}
The original derivation, which goes back to the original contribution of Smoluchowski, can be found in several textbooks but rests upon the use of a special boundary condition. The above formula is indeed obtained by considering the interaction of Brownian particles with a `target particle' at the surface of which a perfect-sink condition is applied. In other words, every time a particle touches the surface of the target one, it is removed and replaced far from it (to ensure a constant influx of particles toward the target one). In more mathematical terms, this killing condition amounts to enforcing that the first-passage time (the first time when a particle hits the target one) is a stopping-time for the process governing the particle trajectory. It is important to realize that, in the traditional approach where the Brownian collision kernel is deduced from the particle concentration profile around the target particle, the resulting expression depends explicitly on the choice of this boundary condition. If we consider, for instance, that particles are reflected at the surface of the target one, it is known from the mathematical studies of such processes (Brownian or Bessel reflected stochastic diffusion processes) that there is a near-one probability that this particle hits again the boundary condition (i.e., the surface of the target particle) an infinite number of times in any small time interval, in which case we would get that the Brownian collision kernel is infinite! It appears therefore that the expression in Eq.~\eqref{eq: Brownian collision kernel} is still subject to some uncertainty as to its exact validity and that additional studies are needed to reconcile proper mathematical treatment and physical pictures to obtain non-zero but finite predictions of Brownian collision kernels with more support than presently. 

\begin{figure}
 \centering
 \includegraphics[width=0.9\textwidth]{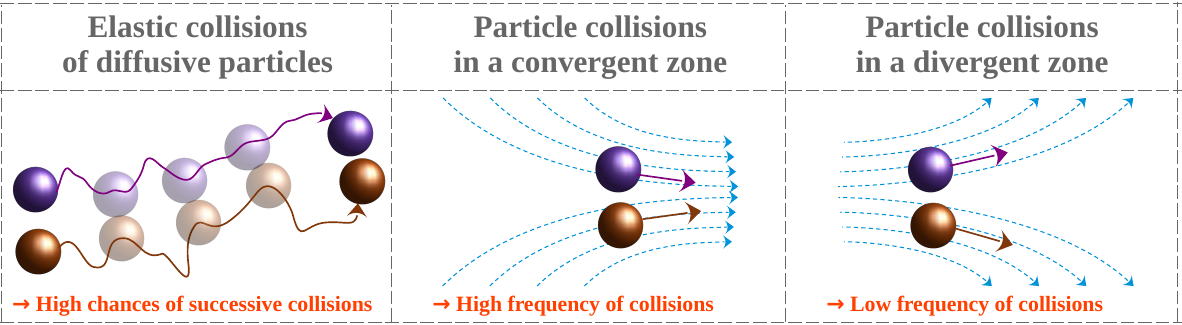}
 \caption{Representation of fluid-related issues for collision kernels. For Brownian particles, elastic interactions without artificially removing one collision partner is predicted to yield an infinite number of successive collisions. For non-Brownian particles, the influence of the underlying fluid flow can prevent or induce collisions at rates different than what criteria based only on particle velocities predict.}
 \label{fig: Agglo-collision-kernel}
\end{figure}

For the wide range of particles whose behavior is intermediate between the diffusive and ballistic regimes, the situation is even more puzzling but, quite surprisingly, the formulation of collision kernels accounting for the influence of fluid flows has received scarce attention. Yet, we cannot do without the presence of an underlying fluid flow since particles in that range are correlated with fluid patterns. For example, two particles which seem on a collision path can avoid one another due to lubrication forces between them as they come closer. The presence of some typical fluid flow structures can have also a marked influence on the resulting collision kernels. For instance, particles caught within a convergent zone interact more frequently than they would have done without the fluid bringing them together. In a mirror situation, particles in a divergent zone, such as a vortex, can be dragged away and prevented from interacting even though they seem on a course to do so had we just considered their ballistic regime. These different qualitative situations are pictured in Fig.~\ref{fig: Agglo-collision-kernel}. 

Note that we are not referring to particle preferential concentration, which is related to local particle accumulation or depletion in some typical fluid structures (cf. Sec.~\ref{sec: particle preferential concentration effect}), but to the fluid-induced changes of particle encounter rates. Though preferential concentration modifies the number of interactions in some zones due to an increase or decrease of the local particle concentration, we are here concerned with the probability of such interactions. On the other hand, it must be noted that the collision probability itself can vary significantly depending on the instantaneous flow structures in which particles are trapped and on the intermittent nature of turbulent flows~\cite{Bec_2014,Bec_2016}, so that even in a statistically homogeneous flow collisions can take place at different rates in different zones of the flow.

\subsection{The molecular chaos assumption}
%==========================================

For discrete particles in the intermediate regime between the ballistic and diffusive ones, the presence of the fluid flow has far-reaching consequences that go beyond questioning present closures of collision kernels: it challenges the molecular chaos hypothesis. Indeed, the velocities of particles with inertia neither very large nor negligible are correlated with the fluid velocity seen. Since they collide, or interact at distances small compared to hydrodynamical scales, they are at the same location and therefore see the same fluid velocity. Given the turbulent flow non-zero time and length correlations and that these fluid velocities are the driving forces for particle motion, it follows that the velocities of two colliding particles must be correlated. There is no avoiding the fact that the hypothesis of `molecular chaos', through which colliding particle velocities are assumed independent (cf. Eq.~\eqref{eq: molecular chaos hypothesis}), is not satisfied. 

To the authors' knowledge, this issue does not seem to be well-recognized, apart from some interesting considerations in~\cite{ramkrishna2000population} for agglomeration problems in the context of population balance methods which rely on the same assumption. Yet, the failure of the molecular chaos hypothesis implies that modifications must be brought to current formulations in physical as well as in sample spaces. In physical space, for example, we cannot apply anymore the Bird-like algorithm as described in Eqs.~\eqref{SPS collision step 2}-\eqref{SPS collision step 2} with Eqs.~\eqref{eq: collision general transformation rules}. Indeed, even if we select a given collision kernel or collision frequency, the molecular chaos assumption is present in the treatment of the $N_{\rm p}$-PDF as the product of the $N_{\rm p}$ single-particle ones 
\begin{equation}
p^{(N)}(t;\mb{V}_{\rm p}^{[1]}, \ldots, \mb{V}_{\rm p}^{[N_{\rm p}]})= \prod_{i=1}^{N_{\rm p}} p^{(1)}(t;\mb{V}_{\rm p}^{[i]})~,
\end{equation}
as manifested by the number of particle pairs taken to be $N_{\rm p} \times N_{\rm p}$ (note that particles are sampled independently to make up a pair). We need to generate `true pairs' in the sense that their velocities must be correctly correlated. In sample space, this is clearly translated by the fact that in the Boltzmann equation, cf. Eq.~\eqref{eq: general Boltzmann equation}, we must retain the gain and loss term as in Eq.~\eqref{eq: Boltzmann gain-loss term with f2} (without resorting to Eq.~\eqref{eq: molecular chaos hypothesis}).

What is needed to close the system is to build a model for the two-particle PDF. Recognizing the role played by the velocity of the fluid seen, it is proposed to do so by devising a model for the joint state vectors $\mb{Z}_{\rm p}=(\mb{X}_{\rm p},\mb{U}_{\rm p},\mb{U}_{\rm s})$ of two discrete particles labeled $i$ and $j$, that is for the joint PDF
\begin{equation}
p=p(t;\mb{y}^{[i]}_{\rm p},\mb{V}^{[i]}_{\rm p},\mb{V}^{[i]}_{\rm s},\upsilon^{[i]}_{\rm p};\mb{y}^{[j]}_{\rm p},\mb{V}^{[j]}_{\rm p},\mb{V}^{[j]}_{\rm s},\upsilon^{[j]}_{\rm p})~.
\end{equation}
Since the two velocities of the fluid seen, $\mb{U}_{\rm s}^{[i]}$ and $\mb{U}_{\rm s}^{[j]}$, are driving the two-particle motion, this indicates that the actual challenge is to develop a two-fluid-particle-PDF model that would therefore contain time and length information. At the moment, the development of such models for general non-homogeneous turbulent flows remains an unexplored domain. 

\subsection{Jump-diffusion processes as modeling tools} \label{Jump-diffusion processes for collisions}
%======================================================

In classical kinetic theory, the driving picture is one in which particles (i.e., atoms and molecules) undergo free flights and collisions. In contrast, the overriding image in dispersed turbulent two-phase flow is one where particles (i.e., small discrete elements) evolve through random motions and collisions. This is well captured by diffusion-jump stochastic processes and most of the physical aspects discussed previously can be cast into this framework provided that we retain the extended particle state vector $\mb{Z}_{\rm p}=(\mb{X}_{\rm p}, \mb{U}_{\rm p},\mb{U}_{\rm s})$. 

\begin{figure}
 \centering
 \includegraphics[width=0.9\textwidth]{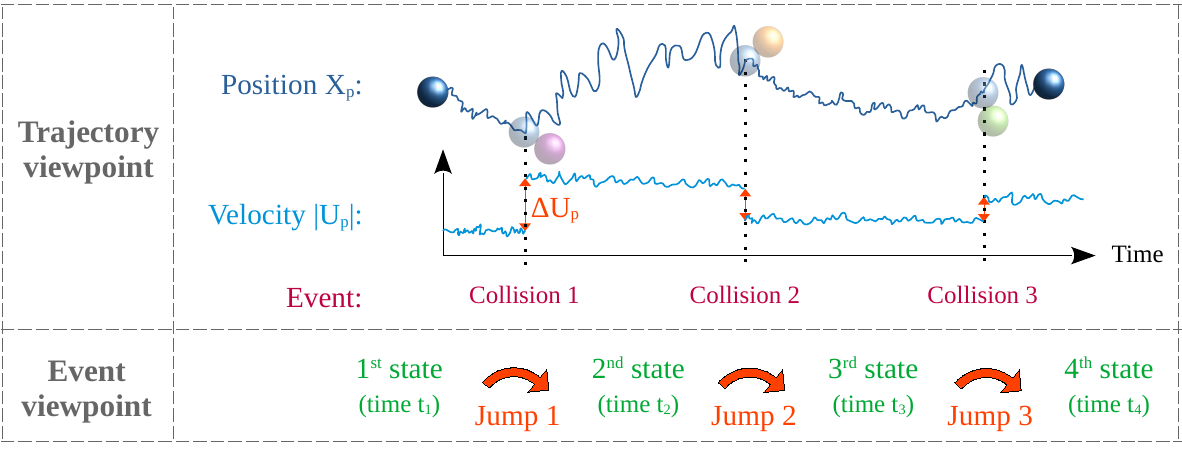}
 \caption{Sketch of a diffusion-jump process on a particle position and velocity. The velocity jumps at random times due to collisions with other particles and follows a diffusion process in between. The position trajectory remains continuous but exhibits different fluctuations due to velocity jumps.}
 \label{fig: Agglo-jump-collision}
\end{figure}

With respect to the issue of particle collisions, the formulation depends on the level of description at which PDF models are developed. If we can write a two-particle PDF model in turbulent flows, we can build (in principle!) a detection algorithm for binary collisions to generate the probability of a collision event and its corresponding characteristics (angle of collision, outcome of the collision, etc.). On the other hand, if we stay at the level of one-particle PDF models, we cannot do without assuming a collision kernel, or a collision frequency, and also conditional statistical information (conditional angle of collision and velocity of the collision partner). Provided that this information is available, the particle equations have the form
\begin{subequations}
\label{eq: general diffusion-jump process}
\begin{align}
\dd \mb{X}_{\rm p} &= \mb{U}_{\rm p} \dd t~, \label{eq: general diffusion-jump process a} \\
\dd \mb{U}_{\rm p} &= \frac{\mb{U}_{\rm s} - \mb{U}_{\rm p}}{\tau_{\rm p}}\, \dd t + \sqrt{\frac{2 k_{\rm B} \Theta_{\rm f}}{m_{\rm p} \tau_{\rm p}}}\,\dd \mb{W}_{\rm p} + \mb{C}\, \dd N_{\rm t}~, \label{eq: general diffusion-jump process b}\\
\dd \mb{U}_{\rm s} &= \mb{A}_{\rm s}\dd t + \mb{B}_{\rm s}\cdot d\mb{W}_{\rm s}~, \label{eq: general diffusion-jump process c}
\end{align}
\end{subequations}
where we have retained only the drag force in the particle momentum equation, Eq.~\eqref{eq: general diffusion-jump process b}, but have accounted for Brownian motion as in Eqs.~\eqref{eq: Brownian motion simple X-U}-\eqref{eq: F6D theorem basic Brownian}, with $\mb{W}_{\rm p}$ and $\mb{W}_{\rm s}$ two independent vectorial Wiener processes. The last term on the rhs of Eq.~\eqref{eq: general diffusion-jump process b} involves a non-homogeneous Poisson process $N_{\rm t}$, which is a scalar process applied to the velocity components of that particle. This Poisson process accounts for the number of collisions along a particle trajectory, since $\dd N_{\rm t}$ is non-zero only at the times when a collision takes place, and is defined by its parameter which is equal to the local value of the collision frequency $\lambda_{\rm c}(\mb{X}_{\rm p})$. The process $\mb{C}$ governs the amplitude of the different jumps of the particle velocity due to collision events and is a vectorial process since it affects the particle velocity components differently. It is seen that we are actually following the main lines of the Bird-like algorithm given in Eqs.~\eqref{SPS collision step 1}-\eqref{SPS collision step 2} and the transformation rules in Eqs.~\eqref{eq: collision general transformation rules}. Indeed, the first step is simulated by the increments of the non-homogeneous Poisson process $N_{\rm t}$ (hence the choice of its frequency $\lambda_{\rm c}(\mb{X}_{\rm p})$, which plays the same role as in Eq.~\eqref{SPS collision step 1}). The vectorial process $\mb{C}$ represents the second step where we select a collision partner, with the difference that the velocity of this partner particle must be sampled in the conditional velocity PDF (conditioned on the pre-collision velocity $\mb{U}_{\rm p}$), before applying the same transformation rules in Eqs.~\eqref{eq: collision general transformation rules}. This is illustrated in Fig.~\ref{fig: Agglo-jump-collision}.

To connect with previous developments, it is illustrative to retain the same collision kernel, i.e. $B(\mb{U}_{\rm p},\mb{U}_{\rm c,p},\mb{e})$, and to work out the resulting conditional probability, noted $W(t;\mb{U}^{\rm ac}_{\rm p} \, \vert \, \mb{U}_{\rm p})$, for a particle velocity to jump from $\mb{U}_{\rm p}$ to $\mb{U}^{\rm ac}_{\rm p}$ at time $t$ due to a collision with a collisional partner. In sample space, this gives the following expression 
\begin{equation}
\label{eq: expression of jumps W first}
W(t;\mb{V}^{\rm ac}_{\rm p} \, \vert \, \mb{V}_{\rm p})\, \dd \mb{V}^{\rm ac}_{\rm p} = \left( \int_{\mc{S}^2} \lambda_{\rm c}(\mb{X}_{\rm p})\, \left( \frac{ B(\mb{V}_{\rm p},\mb{V}_{\rm c,p},\widehat{\mb{e}}) }{ \lambda_{\rm c}(\mb{X}_{\rm p}) } \right) \, \widetilde{p} (t; \mb{V}_{\rm c,p} \,\vert\, \mb{V}_{\rm p}) \, \dd \widehat{\mb{e}} \right)\, \dd \mb{V}_{\rm c,p}~,
\end{equation}
where the term in the integral corresponds to the probability that a jump takes place multiplied by the probability that a specific event governed by the choice of the collision angle $\mb{e}$ occurs. Indeed, for two chosen values of the particle velocity before and after the jump, $\mb{U}_{\rm p}$ and $\mb{U}^{\rm ac}_{\rm p}$ respectively, and a chosen value of the angle $\mb{e}$, the transformation rules in Eqs.~\eqref{eq: collision general transformation rules} show that the velocity of the collision partner $\mb{U}_{\rm c,p}$ is fully determined $\mb{U}_{\rm c,p}=\mb{U}_{\rm c,p}(\mb{U}^{\rm ac}_{\rm p},\mb{U}_{\rm p},\mb{e})$. Note that the conditional velocity PDF is noted $\widetilde{p}$ to indicate that this is an assumed PDF since we are in the frame of single-particle PDF where this information is not accessible. With this understanding of the (pre)-collisional velocity of the collision partner, the integrand in Eq.~\eqref{eq: expression of jumps W first} is a function of $\mb{e}$ and the integral is thus over all possible values of $\mb{e}$ on the unit sphere $\mc{S}^2$. When we integrate $W(t;\mb{V}^{\rm ac}_{\rm p} \, \vert \, \mb{V}_{\rm p})$ over all possible velocity jumps, we obtain
\begin{equation}
\label{eq: expression of jumps W second}
\int_{\mathbb{R}^3} W( \mb{V}^{\rm ac}_{\rm p} \, \vert \, \mb{V}_{\rm p})\, \dd \mb{V}^{\rm ac}_{\rm p} = \int_{\mathbb{R}^3} \int_{\mc{S}^2} B(\mb{V}_{\rm p},\mb{V}_{\rm c,p},\widehat{\mb{e}})\, \widetilde{p} (t; \mb{V}_{\rm c,p} \,\vert\, \mb{V}_{\rm p}) \, \dd \widehat{\mb{e}} \, \dd \mb{V}_{\rm c,p}~.
\end{equation}
If we consider, for the sake of simplicity, the loss term in the Chapman-Kolmogorov equation due to all possible jumps we have simply to multiply by the probability of the particle velocity $p(t;\mb{V}_{\rm p})$ 
\begin{equation}
I_{\rm loss}(t;\mb{V}_{\rm p}) = - \int_{\mathbb{R}^3} W( \mb{V}^{\rm ac}_{\rm p} \, \vert \, \mb{V}_{\rm p})\, p(t;\mb{V}_{\rm p}) \, \dd \mb{V}^{\rm ac}_{\rm p}~,
\end{equation}
which gives
\begin{equation}
\label{eq: diffusion-jump form of Iloss}
I_{\rm loss}(t;\mb{V}_{\rm p}) = \int_{\mathbb{R}^3} \int_{\mc{S}^2} B(\mb{V}_{\rm p},\mb{V}_{\rm c,p},\widehat{\mb{e}})\, \widetilde{p} (t; \mb{V}_{\rm c,p} \,\vert\, \mb{V}_{\rm p}) \, p(t;\mb{V}_{\rm p}) \, \dd \widehat{\mb{e}} \, \dd \mb{V}_{\rm c,p}~. 
\end{equation}
It is straightforward to show that similar results are obtained with the gain term. Should the assumed conditional PDF $\widetilde{p}$ be equal to the true conditional one, we would retrieve $p(t;\mb{V}_{\rm p},\mb{V}_{\rm c,p})$ in the previous expressions, which would indicate that we have the Fokker-Planck-Boltzmann equation, cf. Eq.~\eqref{eq: general Boltzmann-FP equation}, with the correct form of the collision terms, as in Eq.~\eqref{eq: Boltzmann gain-loss term with f2}. Actually, such a statement would be overreaching but the expressions of the collision terms as in Eq.~\eqref{eq: diffusion-jump form of Iloss} show that we are nevertheless consistent with a `linearized version' of the Boltzmann collision integral. In the frame of single-particle PDF models, it seems difficult to achieve better results but these formulations confirm that diffusion-jump stochastic processes are useful tools to capture the physics of turbulent dispersed two-phase flows with collisions.

%========================================================
\section{Conclusions and perspectives \label{conclusion}}
%========================================================

The purpose of this review was to bring out the issues in statistical physics that appear when modeling discrete particle dynamics in non-fully resolved turbulent flows since, as analyzed for the fluid-particle and particle-particle interactions, the non-zero time and space correlations of turbulent flows challenges the framework of kinetic theory. To respect the Markovian approach, models based on sets of particle-attached variables that include fluid flow characteristics sampled along particle trajectories, such as the velocity of the fluid seen, need to be developed. As modeling tools, we have relied on the modern formulation of stochastic processes. In particular, we have mostly followed the trajectory point of view to revisit classical ways to account for noise but also to present recent developments on fast-variable elimination and the derivation of the diffusive limit. There is a special benefit in doing so. Contrary to the classical vision where PDFs are used to derive a limited number of mean fields (the hydrodynamical level) which are the quantities to be resolved, there is no such interest in turbulent dispersed two-phase flows since the set of mean fields corresponding to the first and second-order moments extracted from the PDF models presented in this review involves a large number of coupled partial differential equations that become quickly intractable~\cite{minier2001pdf,minier2014guidelines}. This means that we are interested in modeling as well as in simulating the PDFs. To that extent, the formulation in terms of trajectories is attractive since it leads directly to dynamical Monte Carlo methods~\cite{pope1985pdf,pope1994lagrangian}.

In the course of the discussion, the analyses concerned essentially the statistical content of various models more than their predictive capacities. The latter is, of course, an important aspect but distinct from the actual objective of the present work which put the emphasis on the modeling approach itself rather than on this or that proposition. In that sense, we have tried to detach ourselves from what we believe to be an excessive urge in turbulence modeling to `see some computational results' or `let's run a code', sometimes skipping the necessary assessment of the models on which they are based. Along the same line, it is believed that playing with statistical formulations `\`{a} la Kolmogorov' is still useful and some of the new propositions put forward in this review are meant to rekindle the interest for such modeling attempts. 

What are the roads ahead? While the development of particle transport models is perhaps more advanced, setting up a new (or an extended) framework for particle collisions which takes into account fluid-induced correlations between particle velocities appears as an important issue to address. As mentioned on a few occasions, some of the current difficulties would be overcome by two-particle PDF models, or alternative formulations containing spatial information around discrete particle positions. This is definitely a direction worth considering and where new ideas would have a strong impact on the ability of PDF models to capture key properties of turbulent dispersed two-phase flows. A related theme of research is the construction of unified PDF models that include both the fluid and the discrete-particle phases into a single particle stochastic system (containing therefore two sets of stochastic particles, one for the fluid-phase description and one for the dispersed-phase one). This was first developed in~\cite{minier2001pdf} with specific models to account for discrete-particle back-effects on the fluid (a subject not addressed in the present review) but these propositions need to be taken up and put into practice to test their limits and improve their formulations. At the moment, PDF models exist separately for the fluid phase and, as shown here, for the particle phase but rarely for the complete fluid-and-particle system. Finally, even though we are operating `above the hydrodynamical level of description', the expression of such unified two-phase PDF models in terms of stochastic particles opens the door to connections with the particle-based stochastic methods used `below the hydrodynamical level of description', that is the various methods such as Molecular Dynamics, Dissipative Particle Dynamics, Smoothed Dissipative Particle Dynamics~\cite{berendsen2007simulating}. Devising coarse-graining methods to drive us from one scale, or one level of description, to a higher one (or a reduced level of description) is indeed challenging even if the overriding image is one when we somehow `lump' particles together to form a new particle for the next level. There is, however, an additional difficulty in that we need to go from a N-particle PDF model in which particle-particle interactions are directly accounted for (in MD, DPD, SDPD) to a one-particle PDF model where particle-particle interactions are handled through mean-field approaches. Nevertheless, the formulation of all these methods in terms of particle-based stochastic systems allows such possibilities to be considered and it is hoped that turbulent dispersed two-phase flow modeling will find its place among all these well-recognized domains of physics. 

%\newpage
\vspace{30pt}

\bibliography{biblio}

%apsrmp4-2.bst 2018-12-27 (MD) hand-edited version of apsrmp4-1.bst
%Control: key (0)
%Control: author (3) reversed first dotless
%Control: editor formatted (0) differently from author
%Control: production of article title (0) allowed
%Control: page (1) range
%Control: year (0) verbatim
%Control: production of eprint (0) enabled
\begin{thebibliography}{116}%
\makeatletter
\providecommand \@ifxundefined [1]{%
 \@ifx{#1\undefined}
}%
\providecommand \@ifnum [1]{%
 \ifnum #1\expandafter \@firstoftwo
 \else \expandafter \@secondoftwo
 \fi
}%
\providecommand \@ifx [1]{%
 \ifx #1\expandafter \@firstoftwo
 \else \expandafter \@secondoftwo
 \fi
}%
\providecommand \natexlab [1]{#1}%
\providecommand \enquote  [1]{``#1''}%
\providecommand \bibnamefont  [1]{#1}%
\providecommand \bibfnamefont [1]{#1}%
\providecommand \citenamefont [1]{#1}%
\providecommand \href@noop [0]{\@secondoftwo}%
\providecommand \href [0]{\begingroup \@sanitize@url \@href}%
\providecommand \@href[1]{\@@startlink{#1}\@@href}%
\providecommand \@@href[1]{\endgroup#1\@@endlink}%
\providecommand \@sanitize@url [0]{\catcode `\\12\catcode `\$12\catcode
  `\&12\catcode `\#12\catcode `\^12\catcode `\_12\catcode `\%12\relax}%
\providecommand \@@startlink[1]{}%
\providecommand \@@endlink[0]{}%
\providecommand \url  [0]{\begingroup\@sanitize@url \@url }%
\providecommand \@url [1]{\endgroup\@href {#1}{\urlprefix }}%
\providecommand \urlprefix  [0]{URL }%
\providecommand \Eprint [0]{\href }%
\providecommand \doibase [0]{https://doi.org/}%
\providecommand \selectlanguage [0]{\@gobble}%
\providecommand \bibinfo  [0]{\@secondoftwo}%
\providecommand \bibfield  [0]{\@secondoftwo}%
\providecommand \translation [1]{[#1]}%
\providecommand \BibitemOpen [0]{}%
\providecommand \bibitemStop [0]{}%
\providecommand \bibitemNoStop [0]{.\EOS\space}%
\providecommand \EOS [0]{\spacefactor3000\relax}%
\providecommand \BibitemShut  [1]{\csname bibitem#1\endcsname}%
\let\auto@bib@innerbib\@empty
%</preamble>
\bibitem [{\citenamefont {Allen}(2017)}]{allen2017sediment}%
  \BibitemOpen
  \bibfield  {author} {\bibinfo {author} {\bibnamefont {Allen}, \bibfnamefont
  {PA}}} (\bibinfo {year} {2017}),\ \href@noop {} {\emph {\bibinfo {title}
  {Sediment Routing Systems: The Fate of Sediment from Source to Sink}}}\
  (\bibinfo  {publisher} {Cambridge University Press})\BibitemShut {NoStop}%
\bibitem [{\citenamefont {Arnold}(1974)}]{arnold1974stochastic}%
  \BibitemOpen
  \bibfield  {author} {\bibinfo {author} {\bibnamefont {Arnold}, \bibfnamefont
  {L}}} (\bibinfo {year} {1974}),\ \href@noop {} {\emph {\bibinfo {title}
  {Stochastic differential equations: Theory and applications(Book)}}}\
  (\bibinfo  {publisher} {New York, Wiley-Interscience})\BibitemShut {NoStop}%
\bibitem [{\citenamefont {Bagnold}(2005)}]{bagnold2005physics}%
  \BibitemOpen
  \bibfield  {author} {\bibinfo {author} {\bibnamefont {Bagnold}, \bibfnamefont
  {R~A}}} (\bibinfo {year} {2005}),\ \href@noop {} {\emph {\bibinfo {title}
  {The Physics of Blown Sand and Desert Dunes}}},\ Dover Earth Science\
  (\bibinfo  {publisher} {Dover Publications})\BibitemShut {NoStop}%
\bibitem [{\citenamefont {Balachandar}\ and\ \citenamefont
  {Eaton}(2010)}]{balachandar2010turbulent}%
  \BibitemOpen
  \bibfield  {author} {\bibinfo {author} {\bibnamefont {Balachandar},
  \bibfnamefont {S}}, and\ \bibinfo {author} {\bibfnamefont {John~K}\
  \bibnamefont {Eaton}}} (\bibinfo {year} {2010}),\ \bibfield  {title}
  {\enquote {\bibinfo {title} {Turbulent dispersed multiphase flow},}\
  }\href@noop {} {\bibfield  {journal} {\bibinfo  {journal} {Annual review of
  fluid mechanics}\ }\textbf {\bibinfo {volume} {42}},\ \bibinfo {pages}
  {111--133}}\BibitemShut {NoStop}%
\bibitem [{\citenamefont {Balescu}(1997)}]{Balescu_1997}%
  \BibitemOpen
  \bibfield  {author} {\bibinfo {author} {\bibnamefont {Balescu}, \bibfnamefont
  {R}}} (\bibinfo {year} {1997}),\ \href@noop {} {\emph {\bibinfo {title}
  {Statistical Dynamics: Matter out of equilibrium}}}\ (\bibinfo  {publisher}
  {Imperial College Press, London})\BibitemShut {NoStop}%
\bibitem [{\citenamefont {Bec}\ \emph {et~al.}(2014)\citenamefont {Bec},
  \citenamefont {Hommann},\ and\ \citenamefont {Ray}}]{Bec_2014}%
  \BibitemOpen
  \bibfield  {author} {\bibinfo {author} {\bibnamefont {Bec}, \bibfnamefont
  {J}}, \bibinfo {author} {\bibfnamefont {H.}~\bibnamefont {Hommann}}, and\
  \bibinfo {author} {\bibfnamefont {S.~S.}\ \bibnamefont {Ray}}} (\bibinfo
  {year} {2014}),\ \bibfield  {title} {\enquote {\bibinfo {title}
  {Gravity-driven enhancement of heavy particle clustering in turbulent
  flow},}\ }\href@noop {} {\bibfield  {journal} {\bibinfo  {journal} {Phys.
  Rev. Lett.}\ }\textbf {\bibinfo {volume} {112}}~(\bibinfo {number}
  {184501})}\BibitemShut {NoStop}%
\bibitem [{\citenamefont {Bec}\ \emph {et~al.}(2016)\citenamefont {Bec},
  \citenamefont {Ray}, \citenamefont {Saw},\ and\ \citenamefont
  {Hommann}}]{Bec_2016}%
  \BibitemOpen
  \bibfield  {author} {\bibinfo {author} {\bibnamefont {Bec}, \bibfnamefont
  {J}}, \bibinfo {author} {\bibfnamefont {S.~S.}\ \bibnamefont {Ray}}, \bibinfo
  {author} {\bibfnamefont {E.~W.}\ \bibnamefont {Saw}}, and\ \bibinfo {author}
  {\bibfnamefont {H.}~\bibnamefont {Hommann}}} (\bibinfo {year} {2016}),\
  \bibfield  {title} {\enquote {\bibinfo {title} {Abrupt growth of large
  aggregates by correlated coalescences in turbulent flows},}\ }\href@noop {}
  {\bibfield  {journal} {\bibinfo  {journal} {Physical Review E}\ }\textbf
  {\bibinfo {volume} {93}}~(\bibinfo {number} {031102(R)})}\BibitemShut
  {NoStop}%
\bibitem [{\citenamefont {Bentkamp}\ \emph {et~al.}(2019)\citenamefont
  {Bentkamp}, \citenamefont {Lalescu},\ and\ \citenamefont
  {Wilczek}}]{bentkamp2019persistent}%
  \BibitemOpen
  \bibfield  {author} {\bibinfo {author} {\bibnamefont {Bentkamp},
  \bibfnamefont {Lukas}}, \bibinfo {author} {\bibfnamefont {Cristian~C}\
  \bibnamefont {Lalescu}}, and\ \bibinfo {author} {\bibfnamefont {Michael}\
  \bibnamefont {Wilczek}}} (\bibinfo {year} {2019}),\ \bibfield  {title}
  {\enquote {\bibinfo {title} {Persistent accelerations disentangle lagrangian
  turbulence},}\ }\href {https://doi.org/10.1038/s41467-019-11060-9} {\bibfield
   {journal} {\bibinfo  {journal} {Nature Communications}\ }\textbf {\bibinfo
  {volume} {10}}~(\bibinfo {number} {1}),\ \bibinfo {pages} {3550}}\BibitemShut
  {NoStop}%
\bibitem [{\citenamefont {Berendsen}(2007)}]{berendsen2007simulating}%
  \BibitemOpen
  \bibfield  {author} {\bibinfo {author} {\bibnamefont {Berendsen},
  \bibfnamefont {H~J~C}}} (\bibinfo {year} {2007}),\ \href@noop {} {\emph
  {\bibinfo {title} {Simulating the physical world: hierarchical modeling from
  quantum mechanics to fluid dynamics.}}}\ (\bibinfo  {publisher} {Cambridge
  University Press})\BibitemShut {NoStop}%
\bibitem [{\citenamefont {Bird}(1963)}]{Bird_1963}%
  \BibitemOpen
  \bibfield  {author} {\bibinfo {author} {\bibnamefont {Bird}, \bibfnamefont
  {G~A}}} (\bibinfo {year} {1963}),\ \bibfield  {title} {\enquote {\bibinfo
  {title} {Approach to translational equilibrium in a rigid sphere gas},}\
  }\href@noop {} {\bibfield  {journal} {\bibinfo  {journal} {Phys. Fluids}\
  }\textbf {\bibinfo {volume} {6}},\ \bibinfo {pages} {1518}}\BibitemShut
  {NoStop}%
\bibitem [{\citenamefont {Bird}(1994)}]{Bird_1994}%
  \BibitemOpen
  \bibfield  {author} {\bibinfo {author} {\bibnamefont {Bird}, \bibfnamefont
  {G~A}}} (\bibinfo {year} {1994}),\ \href@noop {} {\emph {\bibinfo {title}
  {Molecular gas dynamics and the direct simulation of gas flows}}}\ (\bibinfo
  {publisher} {Clarendon Oxford})\BibitemShut {NoStop}%
\bibitem [{\citenamefont {Bird}\ \emph {et~al.}(2002)\citenamefont {Bird},
  \citenamefont {Steward},\ and\ \citenamefont
  {Lightfoot}}]{bird2002transport}%
  \BibitemOpen
  \bibfield  {author} {\bibinfo {author} {\bibnamefont {Bird}, \bibfnamefont
  {R~Byron}}, \bibinfo {author} {\bibfnamefont {Warren~E.}\ \bibnamefont
  {Steward}}, and\ \bibinfo {author} {\bibfnamefont {Edwin~N.}\ \bibnamefont
  {Lightfoot}}} (\bibinfo {year} {2002}),\ \href@noop {} {\emph {\bibinfo
  {title} {Dynamics of polymeric liquids}}},\ \bibinfo {edition} {2nd}\ ed.\
  (\bibinfo  {publisher} {John Wiley and Sons, Singapore})\BibitemShut
  {NoStop}%
\bibitem [{\citenamefont {Borodin}\ and\ \citenamefont
  {Salminen}(2015)}]{borodin2015handbook}%
  \BibitemOpen
  \bibfield  {author} {\bibinfo {author} {\bibnamefont {Borodin}, \bibfnamefont
  {Andrei~N}}, and\ \bibinfo {author} {\bibfnamefont {Paavo}\ \bibnamefont
  {Salminen}}} (\bibinfo {year} {2015}),\ \href@noop {} {\emph {\bibinfo
  {title} {Handbook of Brownian motion-facts and formulae}}}\ (\bibinfo
  {publisher} {Springer Science \& Business Media})\BibitemShut {NoStop}%
\bibitem [{\citenamefont {Bragg}\ \emph {et~al.}(2012)\citenamefont {Bragg},
  \citenamefont {Swailes},\ and\ \citenamefont {Skartlien}}]{bragg2012drift}%
  \BibitemOpen
  \bibfield  {author} {\bibinfo {author} {\bibnamefont {Bragg}, \bibfnamefont
  {A}}, \bibinfo {author} {\bibfnamefont {DC}~\bibnamefont {Swailes}}, and\
  \bibinfo {author} {\bibfnamefont {R}~\bibnamefont {Skartlien}}} (\bibinfo
  {year} {2012}),\ \bibfield  {title} {\enquote {\bibinfo {title} {Drift-free
  kinetic equations for turbulent dispersion},}\ }\href@noop {} {\bibfield
  {journal} {\bibinfo  {journal} {Physical Review E}\ }\textbf {\bibinfo
  {volume} {86}}~(\bibinfo {number} {5}),\ \bibinfo {pages}
  {056306}}\BibitemShut {NoStop}%
\bibitem [{\citenamefont {Brandt}\ and\ \citenamefont
  {Coletti}(2022)}]{brandt2022particle}%
  \BibitemOpen
  \bibfield  {author} {\bibinfo {author} {\bibnamefont {Brandt}, \bibfnamefont
  {Luca}}, and\ \bibinfo {author} {\bibfnamefont {Filippo}\ \bibnamefont
  {Coletti}}} (\bibinfo {year} {2022}),\ \bibfield  {title} {\enquote {\bibinfo
  {title} {Particle-laden turbulence: progress and perspectives},}\ }\href
  {https://doi.org/10.1146/annurev-fluid-030121-021103} {\bibfield  {journal}
  {\bibinfo  {journal} {Annual Review of Fluid Mechanics}\ }\textbf {\bibinfo
  {volume} {54}},\ \bibinfo {pages} {159--189}}\BibitemShut {NoStop}%
\bibitem [{\citenamefont {Brennen}(2005)}]{Brennen_2005}%
  \BibitemOpen
  \bibfield  {author} {\bibinfo {author} {\bibnamefont {Brennen}, \bibfnamefont
  {C~E}}} (\bibinfo {year} {2005}),\ \href@noop {} {\emph {\bibinfo {title}
  {Fundamentals of Multiphase Flows}}}\ (\bibinfo  {publisher} {Cambridge
  University Press})\BibitemShut {NoStop}%
\bibitem [{\citenamefont {Chen}\ and\ \citenamefont
  {Kim}(2004)}]{chen2004brownian}%
  \BibitemOpen
  \bibfield  {author} {\bibinfo {author} {\bibnamefont {Chen}, \bibfnamefont
  {Jim~C}}, and\ \bibinfo {author} {\bibfnamefont {Albert~S}\ \bibnamefont
  {Kim}}} (\bibinfo {year} {2004}),\ \bibfield  {title} {\enquote {\bibinfo
  {title} {Brownian dynamics, molecular dynamics, and monte carlo modeling of
  colloidal systems},}\ }\href {https://doi.org/10.1016/j.cis.2004.10.001}
  {\bibfield  {journal} {\bibinfo  {journal} {Advances in colloid and interface
  science}\ }\textbf {\bibinfo {volume} {112}}~(\bibinfo {number} {1-3}),\
  \bibinfo {pages} {159--173}}\BibitemShut {NoStop}%
\bibitem [{\citenamefont {Chibbaro}\ and\ \citenamefont
  {Minier}(2011)}]{chibbaro2011note}%
  \BibitemOpen
  \bibfield  {author} {\bibinfo {author} {\bibnamefont {Chibbaro},
  \bibfnamefont {Sergio}}, and\ \bibinfo {author} {\bibfnamefont {Jean-Pierre}\
  \bibnamefont {Minier}}} (\bibinfo {year} {2011}),\ \bibfield  {title}
  {\enquote {\bibinfo {title} {A note on the consistency of hybrid
  eulerian/lagrangian approach to multiphase flows},}\ }\href@noop {}
  {\bibfield  {journal} {\bibinfo  {journal} {International Journal of
  Multiphase Flow}\ }\textbf {\bibinfo {volume} {37}}~(\bibinfo {number} {3}),\
  \bibinfo {pages} {293--297}}\BibitemShut {NoStop}%
\bibitem [{\citenamefont {Chibbaro}\ and\ \citenamefont
  {Minier}(2014)}]{chibbaro2014stochastic}%
  \BibitemOpen
  \bibfield  {author} {\bibinfo {author} {\bibnamefont {Chibbaro},
  \bibfnamefont {Sergio}}, and\ \bibinfo {author} {\bibfnamefont {Jean~Pierre}\
  \bibnamefont {Minier}}} (\bibinfo {year} {2014}),\ \href
  {https://doi.org/10.1007/978-3-7091-1622-7} {\emph {\bibinfo {title}
  {Stochastic methods in fluid mechanics}}},\ CISM International Centre for
  Mechanical Sciences\ (\bibinfo  {publisher} {Springer Vienna})\BibitemShut
  {NoStop}%
\bibitem [{\citenamefont {Chorin}(2013)}]{chorin2013vorticity}%
  \BibitemOpen
  \bibfield  {author} {\bibinfo {author} {\bibnamefont {Chorin}, \bibfnamefont
  {Alexandre~J}}} (\bibinfo {year} {2013}),\ \href@noop {} {\emph {\bibinfo
  {title} {Vorticity and turbulence}}},\ Vol.\ \bibinfo {volume} {103}\
  (\bibinfo  {publisher} {Springer Science \& Business Media})\BibitemShut
  {NoStop}%
\bibitem [{\citenamefont {Clift}\ \emph {et~al.}(1978)\citenamefont {Clift},
  \citenamefont {Grace},\ and\ \citenamefont {Weber}}]{clift1978bubbles}%
  \BibitemOpen
  \bibfield  {author} {\bibinfo {author} {\bibnamefont {Clift}, \bibfnamefont
  {R}}, \bibinfo {author} {\bibfnamefont {J.~R.}\ \bibnamefont {Grace}}, and\
  \bibinfo {author} {\bibfnamefont {M.E.}\ \bibnamefont {Weber}}} (\bibinfo
  {year} {1978}),\ \href@noop {} {\emph {\bibinfo {title} {Bubbles, Drops and
  Particles}}}\ (\bibinfo  {publisher} {Academic Press})\BibitemShut {NoStop}%
\bibitem [{\citenamefont {Doi}\ and\ \citenamefont
  {Edwards}(1988)}]{doi1988theory}%
  \BibitemOpen
  \bibfield  {author} {\bibinfo {author} {\bibnamefont {Doi}, \bibfnamefont
  {Masao}}, and\ \bibinfo {author} {\bibfnamefont {Samuel~Frederick}\
  \bibnamefont {Edwards}}} (\bibinfo {year} {1988}),\ \href@noop {} {\emph
  {\bibinfo {title} {The theory of polymer dynamics}}},\ Vol.~\bibinfo {volume}
  {73}\ (\bibinfo  {publisher} {oxford university press})\BibitemShut {NoStop}%
\bibitem [{\citenamefont {Donsker}(1964)}]{donsker1964function}%
  \BibitemOpen
  \bibfield  {author} {\bibinfo {author} {\bibnamefont {Donsker}, \bibfnamefont
  {M~D}}} (\bibinfo {year} {1964}),\ \bibfield  {title} {\enquote {\bibinfo
  {title} {On function space integrals},}\ }in\ \href@noop {} {\emph {\bibinfo
  {booktitle} {Analysis in Function Space}}},\ \bibinfo {editor} {edited by\
  \bibinfo {editor} {\bibfnamefont {W.~T.}\ \bibnamefont {Martin}}\ and\
  \bibinfo {editor} {\bibfnamefont {I.}~\bibnamefont {Segal}}}\ (\bibinfo
  {publisher} {the MIT Press})\ pp.\ \bibinfo {pages} {17--30}\BibitemShut
  {NoStop}%
\bibitem [{\citenamefont {Dreeben}\ and\ \citenamefont
  {Pope}(1997)}]{dreeben1997probability}%
  \BibitemOpen
  \bibfield  {author} {\bibinfo {author} {\bibnamefont {Dreeben}, \bibfnamefont
  {Thomas~D}}, and\ \bibinfo {author} {\bibfnamefont {Stephen~B}\ \bibnamefont
  {Pope}}} (\bibinfo {year} {1997}),\ \bibfield  {title} {\enquote {\bibinfo
  {title} {Probability density function and reynolds-stress modeling of
  near-wall turbulent flows},}\ }\href@noop {} {\bibfield  {journal} {\bibinfo
  {journal} {Physics of Fluids}\ }\textbf {\bibinfo {volume} {9}}~(\bibinfo
  {number} {1}),\ \bibinfo {pages} {154--163}}\BibitemShut {NoStop}%
\bibitem [{\citenamefont {Dreeben}\ and\ \citenamefont
  {Pope}(1998)}]{dreeben1998probability}%
  \BibitemOpen
  \bibfield  {author} {\bibinfo {author} {\bibnamefont {Dreeben}, \bibfnamefont
  {Thomas~D}}, and\ \bibinfo {author} {\bibfnamefont {Stephen~B}\ \bibnamefont
  {Pope}}} (\bibinfo {year} {1998}),\ \bibfield  {title} {\enquote {\bibinfo
  {title} {Probability density function/{M}onte {C}arlo simulation of near-wall
  turbulent flows},}\ }\href@noop {} {\bibfield  {journal} {\bibinfo  {journal}
  {Journal of Fluid Mechanics}\ }\textbf {\bibinfo {volume} {357}},\ \bibinfo
  {pages} {141--166}}\BibitemShut {NoStop}%
\bibitem [{\citenamefont {Eaton}\ and\ \citenamefont
  {Fessler}(1994)}]{eaton1994preferential}%
  \BibitemOpen
  \bibfield  {author} {\bibinfo {author} {\bibnamefont {Eaton}, \bibfnamefont
  {JK}}, and\ \bibinfo {author} {\bibfnamefont {J.R.}\ \bibnamefont {Fessler}}}
  (\bibinfo {year} {1994}),\ \bibfield  {title} {\enquote {\bibinfo {title}
  {Preferential concentration of particles by turbulence},}\ }\href
  {https://doi.org/10.1016/0301-9322(94)90072-8} {\bibfield  {journal}
  {\bibinfo  {journal} {International Journal of Multiphase Flow}\ }\textbf
  {\bibinfo {volume} {20}},\ \bibinfo {pages} {169--209}}\BibitemShut {NoStop}%
\bibitem [{\citenamefont {Eibeck}\ and\ \citenamefont
  {Wagner}(2003)}]{eibeck2003stochastic}%
  \BibitemOpen
  \bibfield  {author} {\bibinfo {author} {\bibnamefont {Eibeck}, \bibfnamefont
  {Andreas}}, and\ \bibinfo {author} {\bibfnamefont {Wolfgang}\ \bibnamefont
  {Wagner}}} (\bibinfo {year} {2003}),\ \bibfield  {title} {\enquote {\bibinfo
  {title} {Stochastic interacting particle systems and nonlinear kinetic
  equations},}\ }\href@noop {} {\bibfield  {journal} {\bibinfo  {journal} {The
  Annals of Applied Probability}\ }\textbf {\bibinfo {volume} {13}}~(\bibinfo
  {number} {3}),\ \bibinfo {pages} {845--889}}\BibitemShut {NoStop}%
\bibitem [{\citenamefont {Einstein}\ \emph {et~al.}(1905)\citenamefont
  {Einstein} \emph {et~al.}}]{einstein1905motion}%
  \BibitemOpen
  \bibfield  {author} {\bibinfo {author} {\bibnamefont {Einstein},
  \bibfnamefont {Albert}},  \emph {et~al.}} (\bibinfo {year} {1905}),\
  \bibfield  {title} {\enquote {\bibinfo {title} {On the motion of small
  particles suspended in liquids at rest required by the molecular-kinetic
  theory of heat},}\ }\href@noop {} {\bibfield  {journal} {\bibinfo  {journal}
  {Annalen der physik}\ }\textbf {\bibinfo {volume} {17}}~(\bibinfo {number}
  {549-560}),\ \bibinfo {pages} {208}}\BibitemShut {NoStop}%
\bibitem [{\citenamefont {Elimelech}\ \emph {et~al.}(2013)\citenamefont
  {Elimelech}, \citenamefont {Gregory},\ and\ \citenamefont
  {Jia}}]{elimelech2013particle}%
  \BibitemOpen
  \bibfield  {author} {\bibinfo {author} {\bibnamefont {Elimelech},
  \bibfnamefont {Menachem}}, \bibinfo {author} {\bibfnamefont {John}\
  \bibnamefont {Gregory}}, and\ \bibinfo {author} {\bibfnamefont {Xiadong}\
  \bibnamefont {Jia}}} (\bibinfo {year} {2013}),\ \href
  {https://doi.org/10.1016/C2013-0-04548-3} {\emph {\bibinfo {title} {Particle
  deposition and aggregation: measurement, modelling and simulation}}}\
  (\bibinfo  {publisher} {Butterworth-Heinemann})\BibitemShut {NoStop}%
\bibitem [{\citenamefont {Elperin}\ \emph {et~al.}(2007)\citenamefont
  {Elperin}, \citenamefont {Kleeorin}, \citenamefont {Liberman}, \citenamefont
  {L’vov},\ and\ \citenamefont {Rogachevskii}}]{elperin2007clustering}%
  \BibitemOpen
  \bibfield  {author} {\bibinfo {author} {\bibnamefont {Elperin}, \bibfnamefont
  {Tov}}, \bibinfo {author} {\bibfnamefont {Nathan}\ \bibnamefont {Kleeorin}},
  \bibinfo {author} {\bibfnamefont {Michael~A}\ \bibnamefont {Liberman}},
  \bibinfo {author} {\bibfnamefont {Victor~S}\ \bibnamefont {L’vov}}, and\
  \bibinfo {author} {\bibfnamefont {Igor}\ \bibnamefont {Rogachevskii}}}
  (\bibinfo {year} {2007}),\ \bibfield  {title} {\enquote {\bibinfo {title}
  {Clustering of aerosols in atmospheric turbulent flow},}\ }\href@noop {}
  {\bibfield  {journal} {\bibinfo  {journal} {Environmental Fluid Mechanics}\
  }\textbf {\bibinfo {volume} {7}},\ \bibinfo {pages} {173--193}}\BibitemShut
  {NoStop}%
\bibitem [{\citenamefont {Elperin}\ \emph {et~al.}(2002)\citenamefont
  {Elperin}, \citenamefont {Kleeorin}, \citenamefont {L’vov}, \citenamefont
  {Rogachevskii},\ and\ \citenamefont {Sokoloff}}]{elperin2002clustering}%
  \BibitemOpen
  \bibfield  {author} {\bibinfo {author} {\bibnamefont {Elperin}, \bibfnamefont
  {Tov}}, \bibinfo {author} {\bibfnamefont {Nathan}\ \bibnamefont {Kleeorin}},
  \bibinfo {author} {\bibfnamefont {Victor~S}\ \bibnamefont {L’vov}},
  \bibinfo {author} {\bibfnamefont {Igor}\ \bibnamefont {Rogachevskii}}, and\
  \bibinfo {author} {\bibfnamefont {Dmitry}\ \bibnamefont {Sokoloff}}}
  (\bibinfo {year} {2002}),\ \bibfield  {title} {\enquote {\bibinfo {title}
  {Clustering instability of the spatial distribution of inertial particles in
  turbulent flows},}\ }\href@noop {} {\bibfield  {journal} {\bibinfo  {journal}
  {Physical Review E}\ }\textbf {\bibinfo {volume} {66}}~(\bibinfo {number}
  {3}),\ \bibinfo {pages} {036302}}\BibitemShut {NoStop}%
\bibitem [{\citenamefont {Fournier}\ and\ \citenamefont
  {Mischler}(2005)}]{fournier2005spatially}%
  \BibitemOpen
  \bibfield  {author} {\bibinfo {author} {\bibnamefont {Fournier},
  \bibfnamefont {Nicolas}}, and\ \bibinfo {author} {\bibfnamefont
  {St{\'e}phane}\ \bibnamefont {Mischler}}} (\bibinfo {year} {2005}),\
  \bibfield  {title} {\enquote {\bibinfo {title} {A spatially homogeneous
  boltzmann equation for elastic, inelastic and coalescing collisions},}\
  }\href@noop {} {\bibfield  {journal} {\bibinfo  {journal} {Journal de
  math{\'e}matiques pures et appliqu{\'e}es}\ }\textbf {\bibinfo {volume}
  {84}}~(\bibinfo {number} {9}),\ \bibinfo {pages} {1173--1234}}\BibitemShut
  {NoStop}%
\bibitem [{\citenamefont {Frisch}\ and\ \citenamefont
  {Kolmogorov}(1995)}]{frisch1995turbulence}%
  \BibitemOpen
  \bibfield  {author} {\bibinfo {author} {\bibnamefont {Frisch}, \bibfnamefont
  {U}}, and\ \bibinfo {author} {\bibfnamefont {A~N}\ \bibnamefont
  {Kolmogorov}}} (\bibinfo {year} {1995}),\ \href
  {https://books.google.fr/books?id=-JcjT4wYgfgC} {\emph {\bibinfo {title}
  {Turbulence: {T}he {L}egacy of {A. N. K}olmogorov}}}\ (\bibinfo  {publisher}
  {Cambridge University Press})\BibitemShut {NoStop}%
\bibitem [{\citenamefont {Furutsu}(1963)}]{furutsu1963statistical}%
  \BibitemOpen
  \bibfield  {author} {\bibinfo {author} {\bibnamefont {Furutsu}, \bibfnamefont
  {Koichi}}} (\bibinfo {year} {1963}),\ \bibfield  {title} {\enquote {\bibinfo
  {title} {On the statistical theory of electromagnetic waves in a fluctuating
  medium},}\ }\href@noop {} {\bibfield  {journal} {\bibinfo  {journal} {J. Res.
  Nat. Bur. Standards D}\ }\textbf {\bibinfo {volume} {67}},\ \bibinfo {pages}
  {303--323}}\BibitemShut {NoStop}%
\bibitem [{\citenamefont {Gardiner}(2009)}]{gardiner2009stochastic}%
  \BibitemOpen
  \bibfield  {author} {\bibinfo {author} {\bibnamefont {Gardiner},
  \bibfnamefont {Crispin}}} (\bibinfo {year} {2009}),\ \href@noop {} {\emph
  {\bibinfo {title} {Stochastic methods}}},\ \bibinfo {edition} {{F}ourth}\
  ed.\ (\bibinfo  {publisher} {Springer Berlin})\BibitemShut {NoStop}%
\bibitem [{\citenamefont {Gatignol}(1983)}]{gatignol1983faxen}%
  \BibitemOpen
  \bibfield  {author} {\bibinfo {author} {\bibnamefont {Gatignol},
  \bibfnamefont {Ren{\'e}e}}} (\bibinfo {year} {1983}),\ \bibfield  {title}
  {\enquote {\bibinfo {title} {The {F}ax{\'e}n formulae for a rigid particle in
  an unsteady non-uniform {S}tokes flow},}\ }\href@noop {} {\bibfield
  {journal} {\bibinfo  {journal} {Journal de M\'{e}canique Th\'{e}orique et
  Appliqu\'{e}e}\ }\textbf {\bibinfo {volume} {2}}~(\bibinfo {number} {2}),\
  \bibinfo {pages} {143--160}}\BibitemShut {NoStop}%
\bibitem [{\citenamefont {Gicquel}\ \emph {et~al.}(2002)\citenamefont
  {Gicquel}, \citenamefont {Givi}, \citenamefont {Jaberi},\ and\ \citenamefont
  {Pope}}]{Gicquel_2002}%
  \BibitemOpen
  \bibfield  {author} {\bibinfo {author} {\bibnamefont {Gicquel}, \bibfnamefont
  {L~Y~M}}, \bibinfo {author} {\bibfnamefont {P.}~\bibnamefont {Givi}},
  \bibinfo {author} {\bibfnamefont {F.~A.}\ \bibnamefont {Jaberi}}, and\
  \bibinfo {author} {\bibfnamefont {S.~B.}\ \bibnamefont {Pope}}} (\bibinfo
  {year} {2002}),\ \bibfield  {title} {\enquote {\bibinfo {title} {Velocity
  filtered density function for large eddy simulation of turbulent flows},}\
  }\href@noop {} {\bibfield  {journal} {\bibinfo  {journal} {Phys. Fluids}\
  }\textbf {\bibinfo {volume} {14}}~(\bibinfo {number} {3}),\ \bibinfo {pages}
  {1196--1213}}\BibitemShut {NoStop}%
\bibitem [{\citenamefont {Guingo}\ and\ \citenamefont
  {Minier}(2008)}]{guingo2008stochastic}%
  \BibitemOpen
  \bibfield  {author} {\bibinfo {author} {\bibnamefont {Guingo}, \bibfnamefont
  {Mathieu}}, and\ \bibinfo {author} {\bibfnamefont {Jean-Pierre}\ \bibnamefont
  {Minier}}} (\bibinfo {year} {2008}),\ \bibfield  {title} {\enquote {\bibinfo
  {title} {A stochastic model of coherent structures for particle deposition in
  turbulent flows},}\ }\href@noop {} {\bibfield  {journal} {\bibinfo  {journal}
  {Physics of Fluids}\ }\textbf {\bibinfo {volume} {20}}~(\bibinfo {number}
  {5})}\BibitemShut {NoStop}%
\bibitem [{\citenamefont {H{\"a}unggi}\ and\ \citenamefont
  {Jung}(1994)}]{haunggi1994colored}%
  \BibitemOpen
  \bibfield  {author} {\bibinfo {author} {\bibnamefont {H{\"a}unggi},
  \bibfnamefont {Peter}}, and\ \bibinfo {author} {\bibfnamefont {Peter}\
  \bibnamefont {Jung}}} (\bibinfo {year} {1994}),\ \bibfield  {title} {\enquote
  {\bibinfo {title} {Colored noise in dynamical systems},}\ }\href@noop {}
  {\bibfield  {journal} {\bibinfo  {journal} {Advances in chemical physics}\
  }\textbf {\bibinfo {volume} {89}},\ \bibinfo {pages} {239--326}}\BibitemShut
  {NoStop}%
\bibitem [{\citenamefont {Haworth}\ and\ \citenamefont
  {Pope}(1986)}]{haworth1986generalized}%
  \BibitemOpen
  \bibfield  {author} {\bibinfo {author} {\bibnamefont {Haworth}, \bibfnamefont
  {Daniel~C}}, and\ \bibinfo {author} {\bibfnamefont {Stephen~B}\ \bibnamefont
  {Pope}}} (\bibinfo {year} {1986}),\ \bibfield  {title} {\enquote {\bibinfo
  {title} {A generalized langevin model for turbulent flows},}\ }\href@noop {}
  {\bibfield  {journal} {\bibinfo  {journal} {The Physics of fluids}\ }\textbf
  {\bibinfo {volume} {29}}~(\bibinfo {number} {2}),\ \bibinfo {pages}
  {387--405}}\BibitemShut {NoStop}%
\bibitem [{\citenamefont {Haworth}(2010)}]{haworth2010progress}%
  \BibitemOpen
  \bibfield  {author} {\bibinfo {author} {\bibnamefont {Haworth}, \bibfnamefont
  {Daniel~Connell}}} (\bibinfo {year} {2010}),\ \bibfield  {title} {\enquote
  {\bibinfo {title} {Progress in probability density function methods for
  turbulent reacting flows},}\ }\href@noop {} {\bibfield  {journal} {\bibinfo
  {journal} {Progress in Energy and combustion Science}\ }\textbf {\bibinfo
  {volume} {36}}~(\bibinfo {number} {2}),\ \bibinfo {pages}
  {168--259}}\BibitemShut {NoStop}%
\bibitem [{\citenamefont {Henry}\ \emph {et~al.}(2018)\citenamefont {Henry},
  \citenamefont {Krstulovic},\ and\ \citenamefont {Bec}}]{henry2018tumbling}%
  \BibitemOpen
  \bibfield  {author} {\bibinfo {author} {\bibnamefont {Henry}, \bibfnamefont
  {Christophe}}, \bibinfo {author} {\bibfnamefont {Giorgio}\ \bibnamefont
  {Krstulovic}}, and\ \bibinfo {author} {\bibfnamefont {J{\'e}r{\'e}mie}\
  \bibnamefont {Bec}}} (\bibinfo {year} {2018}),\ \bibfield  {title} {\enquote
  {\bibinfo {title} {Tumbling dynamics of inertial inextensible chains in
  extensional flow},}\ }\href
  {https://doi.org/https://doi.org/10.1103/PhysRevE.98.023107} {\bibfield
  {journal} {\bibinfo  {journal} {Physical Review E}\ }\textbf {\bibinfo
  {volume} {98}}~(\bibinfo {number} {2}),\ \bibinfo {pages}
  {023107}}\BibitemShut {NoStop}%
\bibitem [{\citenamefont {Henry}\ and\ \citenamefont
  {Minier}(2014)}]{henry2014progress}%
  \BibitemOpen
  \bibfield  {author} {\bibinfo {author} {\bibnamefont {Henry}, \bibfnamefont
  {Christophe}}, and\ \bibinfo {author} {\bibfnamefont {Jean-Pierre}\
  \bibnamefont {Minier}}} (\bibinfo {year} {2014}),\ \bibfield  {title}
  {\enquote {\bibinfo {title} {Progress in particle resuspension from rough
  surfaces by turbulent flows},}\ }\href
  {https://doi.org/10.1016/j.pecs.2014.06.001} {\bibfield  {journal} {\bibinfo
  {journal} {Progress in Energy and Combustion Science}\ }\textbf {\bibinfo
  {volume} {45}},\ \bibinfo {pages} {1--53}}\BibitemShut {NoStop}%
\bibitem [{\citenamefont {Henry}\ \emph {et~al.}(2023)\citenamefont {Henry},
  \citenamefont {Minier},\ and\ \citenamefont {Brambilla}}]{henry2023particle}%
  \BibitemOpen
  \bibfield  {author} {\bibinfo {author} {\bibnamefont {Henry}, \bibfnamefont
  {Christophe}}, \bibinfo {author} {\bibfnamefont {Jean-Pierre}\ \bibnamefont
  {Minier}}, and\ \bibinfo {author} {\bibfnamefont {Sara}\ \bibnamefont
  {Brambilla}}} (\bibinfo {year} {2023}),\ \bibfield  {title} {\enquote
  {\bibinfo {title} {Particle resuspension: challenges and perspectives for
  future models},}\ }\href@noop {} {\bibfield  {journal} {\bibinfo  {journal}
  {Physics Reports}\ }\textbf {\bibinfo {volume} {1007}},\ \bibinfo {pages}
  {1--98}}\BibitemShut {NoStop}%
\bibitem [{\citenamefont {Henry}\ \emph {et~al.}(2012)\citenamefont {Henry},
  \citenamefont {Minier},\ and\ \citenamefont
  {Lef{\`e}vre}}]{henry2012towards}%
  \BibitemOpen
  \bibfield  {author} {\bibinfo {author} {\bibnamefont {Henry}, \bibfnamefont
  {Christophe}}, \bibinfo {author} {\bibfnamefont {Jean-Pierre}\ \bibnamefont
  {Minier}}, and\ \bibinfo {author} {\bibfnamefont {Gr{\'e}gory}\ \bibnamefont
  {Lef{\`e}vre}}} (\bibinfo {year} {2012}),\ \bibfield  {title} {\enquote
  {\bibinfo {title} {Towards a description of particulate fouling: From single
  particle deposition to clogging},}\ }\href
  {https://doi.org/10.1016/j.cis.2012.10.001} {\bibfield  {journal} {\bibinfo
  {journal} {Advances in Colloid and Interface Science}\ }\textbf {\bibinfo
  {volume} {185}},\ \bibinfo {pages} {34--76}}\BibitemShut {NoStop}%
\bibitem [{\citenamefont {Henry}\ \emph {et~al.}(2014)\citenamefont {Henry},
  \citenamefont {Minier}, \citenamefont {Mohaupt}, \citenamefont {Profeta},
  \citenamefont {Pozorski},\ and\ \citenamefont
  {Tani{\`e}re}}]{henry2014astochastic}%
  \BibitemOpen
  \bibfield  {author} {\bibinfo {author} {\bibnamefont {Henry}, \bibfnamefont
  {Christophe}}, \bibinfo {author} {\bibfnamefont {Jean-Pierre}\ \bibnamefont
  {Minier}}, \bibinfo {author} {\bibfnamefont {Mika{\"e}l}\ \bibnamefont
  {Mohaupt}}, \bibinfo {author} {\bibfnamefont {Christophe}\ \bibnamefont
  {Profeta}}, \bibinfo {author} {\bibfnamefont {Jacek}\ \bibnamefont
  {Pozorski}}, and\ \bibinfo {author} {\bibfnamefont {Anne}\ \bibnamefont
  {Tani{\`e}re}}} (\bibinfo {year} {2014}),\ \bibfield  {title} {\enquote
  {\bibinfo {title} {A stochastic approach for the simulation of collisions
  between colloidal particles at large time steps},}\ }\href
  {https://doi.org/10.1016/j.ijmultiphaseflow.2014.01.007} {\bibfield
  {journal} {\bibinfo  {journal} {International Journal of Multiphase Flow}\
  }\textbf {\bibinfo {volume} {61}},\ \bibinfo {pages} {94--107}}\BibitemShut
  {NoStop}%
\bibitem [{\citenamefont {Hunt}\ \emph {et~al.}(1991)\citenamefont {Hunt},
  \citenamefont {Phillips},\ and\ \citenamefont
  {Williams}}]{hunt1991turbulence}%
  \BibitemOpen
  \bibfield  {author} {\bibinfo {author} {\bibnamefont {Hunt}, \bibfnamefont
  {Julian~CR}}, \bibinfo {author} {\bibfnamefont {Owen~M}\ \bibnamefont
  {Phillips}}, and\ \bibinfo {author} {\bibfnamefont {David}\ \bibnamefont
  {Williams}}} (\bibinfo {year} {1991}),\ \href@noop {} {\emph {\bibinfo
  {title} {Turbulence and stochastic processes: Kolmogorov's ideas 50 years
  on}}}\ (\bibinfo  {publisher} {Royal Society London})\BibitemShut {NoStop}%
\bibitem [{\citenamefont {Hunter}(2001)}]{hunter2001foundations}%
  \BibitemOpen
  \bibfield  {author} {\bibinfo {author} {\bibnamefont {Hunter}, \bibfnamefont
  {RJ}}} (\bibinfo {year} {2001}),\ \href@noop {} {\emph {\bibinfo {title}
  {Foundations of colloid science}}}\ (\bibinfo  {publisher}
  {Oxford})\BibitemShut {NoStop}%
\bibitem [{\citenamefont
  {Israelachvili}(2011)}]{israelachvili2011intermolecular}%
  \BibitemOpen
  \bibfield  {author} {\bibinfo {author} {\bibnamefont {Israelachvili},
  \bibfnamefont {J~N}}} (\bibinfo {year} {2011}),\ \href
  {https://doi.org/10.1016/C2011-0-05119-0} {\emph {\bibinfo {title}
  {Intermolecular \& Surface forces}}},\ \bibinfo {edition} {{T}hird}\ ed.\
  (\bibinfo  {publisher} {Academic press})\BibitemShut {NoStop}%
\bibitem [{\citenamefont {Jimenez}\ and\ \citenamefont
  {Wray}(1998)}]{jimenez1998characteristics}%
  \BibitemOpen
  \bibfield  {author} {\bibinfo {author} {\bibnamefont {Jimenez}, \bibfnamefont
  {Javier}}, and\ \bibinfo {author} {\bibfnamefont {Alan~A}\ \bibnamefont
  {Wray}}} (\bibinfo {year} {1998}),\ \bibfield  {title} {\enquote {\bibinfo
  {title} {On the characteristics of vortex filaments in isotropic
  turbulence},}\ }\href@noop {} {\bibfield  {journal} {\bibinfo  {journal}
  {Journal of Fluid Mechanics}\ }\textbf {\bibinfo {volume} {373}},\ \bibinfo
  {pages} {255--285}}\BibitemShut {NoStop}%
\bibitem [{\citenamefont {Jim{\'e}nez}\ \emph {et~al.}(1993)\citenamefont
  {Jim{\'e}nez}, \citenamefont {Wray}, \citenamefont {Saffman},\ and\
  \citenamefont {Rogallo}}]{jimenez1993structure}%
  \BibitemOpen
  \bibfield  {author} {\bibinfo {author} {\bibnamefont {Jim{\'e}nez},
  \bibfnamefont {Javier}}, \bibinfo {author} {\bibfnamefont {Alan~A}\
  \bibnamefont {Wray}}, \bibinfo {author} {\bibfnamefont {Philip~G}\
  \bibnamefont {Saffman}}, and\ \bibinfo {author} {\bibfnamefont {Robert~S}\
  \bibnamefont {Rogallo}}} (\bibinfo {year} {1993}),\ \bibfield  {title}
  {\enquote {\bibinfo {title} {The structure of intense vorticity in isotropic
  turbulence},}\ }\href@noop {} {\bibfield  {journal} {\bibinfo  {journal}
  {Journal of Fluid Mechanics}\ }\textbf {\bibinfo {volume} {255}},\ \bibinfo
  {pages} {65--90}}\BibitemShut {NoStop}%
\bibitem [{\citenamefont {Johnson}\ and\ \citenamefont
  {Meneveau}(2018)}]{johnson2018predicting}%
  \BibitemOpen
  \bibfield  {author} {\bibinfo {author} {\bibnamefont {Johnson}, \bibfnamefont
  {Perry~L}}, and\ \bibinfo {author} {\bibfnamefont {Charles}\ \bibnamefont
  {Meneveau}}} (\bibinfo {year} {2018}),\ \bibfield  {title} {\enquote
  {\bibinfo {title} {Predicting viscous-range velocity gradient dynamics in
  large-eddy simulations of turbulence},}\ }\href@noop {} {\bibfield  {journal}
  {\bibinfo  {journal} {Journal of Fluid Mechanics}\ }\textbf {\bibinfo
  {volume} {837}},\ \bibinfo {pages} {80--114}}\BibitemShut {NoStop}%
\bibitem [{\citenamefont {Jones}(2002)}]{jones2002soft}%
  \BibitemOpen
  \bibfield  {author} {\bibinfo {author} {\bibnamefont {Jones}, \bibfnamefont
  {Richard~AL}}} (\bibinfo {year} {2002}),\ \href@noop {} {\emph {\bibinfo
  {title} {Soft condensed matter}}},\ Vol.~\bibinfo {volume} {6}\ (\bibinfo
  {publisher} {Oxford University Press})\BibitemShut {NoStop}%
\bibitem [{\citenamefont {van Kampen}(1998)}]{van1998remarks}%
  \BibitemOpen
  \bibfield  {author} {\bibinfo {author} {\bibnamefont {van Kampen},
  \bibfnamefont {Nicolaas~G}}} (\bibinfo {year} {1998}),\ \bibfield  {title}
  {\enquote {\bibinfo {title} {Remarks on non-markov processes},}\ }\href@noop
  {} {\bibfield  {journal} {\bibinfo  {journal} {Brazilian Journal of Physics}\
  }\textbf {\bibinfo {volume} {28}},\ \bibinfo {pages} {90--96}}\BibitemShut
  {NoStop}%
\bibitem [{\citenamefont {Karatzas}\ and\ \citenamefont
  {Shreve}(1991)}]{karatzas1991brownian}%
  \BibitemOpen
  \bibfield  {author} {\bibinfo {author} {\bibnamefont {Karatzas},
  \bibfnamefont {Ioannis}}, and\ \bibinfo {author} {\bibfnamefont {Steven}\
  \bibnamefont {Shreve}}} (\bibinfo {year} {1991}),\ \href@noop {} {\emph
  {\bibinfo {title} {Brownian motion and stochastic calculus}}},\ Vol.\
  \bibinfo {volume} {113}\ (\bibinfo  {publisher} {Springer Science \& Business
  Media})\BibitemShut {NoStop}%
\bibitem [{\citenamefont {Keizer}(2012)}]{keizer2012statistical}%
  \BibitemOpen
  \bibfield  {author} {\bibinfo {author} {\bibnamefont {Keizer}, \bibfnamefont
  {Joel}}} (\bibinfo {year} {2012}),\ \href@noop {} {\emph {\bibinfo {title}
  {Statistical thermodynamics of nonequilibrium processes}}}\ (\bibinfo
  {publisher} {Springer Science \& Business Media})\BibitemShut {NoStop}%
\bibitem [{\citenamefont {Kerstein}(1999)}]{kerstein1999one}%
  \BibitemOpen
  \bibfield  {author} {\bibinfo {author} {\bibnamefont {Kerstein},
  \bibfnamefont {Alan~R}}} (\bibinfo {year} {1999}),\ \bibfield  {title}
  {\enquote {\bibinfo {title} {One-dimensional turbulence: model formulation
  and application to homogeneous turbulence, shear flows, and buoyant
  stratified flows},}\ }\href@noop {} {\bibfield  {journal} {\bibinfo
  {journal} {Journal of Fluid Mechanics}\ }\textbf {\bibinfo {volume} {392}},\
  \bibinfo {pages} {277--334}}\BibitemShut {NoStop}%
\bibitem [{\citenamefont {Kjelstrup}\ and\ \citenamefont
  {Bedeaux}(2008)}]{kjelstrup2008non}%
  \BibitemOpen
  \bibfield  {author} {\bibinfo {author} {\bibnamefont {Kjelstrup},
  \bibfnamefont {Signe}}, and\ \bibinfo {author} {\bibfnamefont {Dick}\
  \bibnamefont {Bedeaux}}} (\bibinfo {year} {2008}),\ \href@noop {} {\emph
  {\bibinfo {title} {Non-equilibrium thermodynamics of heterogeneous
  systems}}}\ (\bibinfo  {publisher} {World Scientific})\BibitemShut {NoStop}%
\bibitem [{\citenamefont {Kjelstrup}\ \emph {et~al.}(2010)\citenamefont
  {Kjelstrup}, \citenamefont {Bedeaux}, \citenamefont {Johannessen},\ and\
  \citenamefont {Gross}}]{kjelstrup2010non}%
  \BibitemOpen
  \bibfield  {author} {\bibinfo {author} {\bibnamefont {Kjelstrup},
  \bibfnamefont {Signe}}, \bibinfo {author} {\bibfnamefont {Dick}\ \bibnamefont
  {Bedeaux}}, \bibinfo {author} {\bibfnamefont {Eivind}\ \bibnamefont
  {Johannessen}}, and\ \bibinfo {author} {\bibfnamefont {Joachim}\ \bibnamefont
  {Gross}}} (\bibinfo {year} {2010}),\ \href@noop {} {\emph {\bibinfo {title}
  {Non-equilibrium thermodynamics for engineers}}}\ (\bibinfo  {publisher}
  {World Scientific})\BibitemShut {NoStop}%
\bibitem [{\citenamefont {Klebaner}(2012)}]{klebaner2012introduction}%
  \BibitemOpen
  \bibfield  {author} {\bibinfo {author} {\bibnamefont {Klebaner},
  \bibfnamefont {Fima~C}}} (\bibinfo {year} {2012}),\ \href@noop {} {\emph
  {\bibinfo {title} {Introduction to stochastic calculus with applications}}}\
  (\bibinfo  {publisher} {World Scientific Publishing Company})\BibitemShut
  {NoStop}%
\bibitem [{\citenamefont {Kok}\ \emph {et~al.}(2012)\citenamefont {Kok},
  \citenamefont {Parteli}, \citenamefont {Michaels},\ and\ \citenamefont
  {Karam}}]{kok2012physics}%
  \BibitemOpen
  \bibfield  {author} {\bibinfo {author} {\bibnamefont {Kok}, \bibfnamefont
  {Jasper~F}}, \bibinfo {author} {\bibfnamefont {Eric~JR}\ \bibnamefont
  {Parteli}}, \bibinfo {author} {\bibfnamefont {Timothy~I}\ \bibnamefont
  {Michaels}}, and\ \bibinfo {author} {\bibfnamefont {Diana~Bou}\ \bibnamefont
  {Karam}}} (\bibinfo {year} {2012}),\ \bibfield  {title} {\enquote {\bibinfo
  {title} {The physics of wind-blown sand and dust},}\ }\href
  {https://doi.org/10.1088/0034-4885/75/10/106901} {\bibfield  {journal}
  {\bibinfo  {journal} {Reports on Progress in Physics}\ }\textbf {\bibinfo
  {volume} {75}}~(\bibinfo {number} {10}),\ \bibinfo {pages}
  {106901}}\BibitemShut {NoStop}%
\bibitem [{\citenamefont {Kuerten}(2016)}]{kuerten2016point}%
  \BibitemOpen
  \bibfield  {author} {\bibinfo {author} {\bibnamefont {Kuerten}, \bibfnamefont
  {J~G~M}}} (\bibinfo {year} {2016}),\ \bibfield  {title} {\enquote {\bibinfo
  {title} {Point-particle {DNS} and {LES} of particle-laden turbulent flow-a
  state-of-the-art review},}\ }\href
  {https://doi.org/10.1007/s10494-016-9765-y} {\bibfield  {journal} {\bibinfo
  {journal} {Flow, Turbulence and Combustion}\ }\textbf {\bibinfo {volume}
  {97}}~(\bibinfo {number} {3}),\ \bibinfo {pages} {689--713}}\BibitemShut
  {NoStop}%
\bibitem [{\citenamefont {Langevin}(1908)}]{langevin1908theory}%
  \BibitemOpen
  \bibfield  {author} {\bibinfo {author} {\bibnamefont {Langevin},
  \bibfnamefont {Paul}}} (\bibinfo {year} {1908}),\ \bibfield  {title}
  {\enquote {\bibinfo {title} {On the theory of brownian motion.}}\ }\href@noop
  {} {\bibfield  {journal} {\bibinfo  {journal} {CR Acad Sci (Paris)}\ }\textbf
  {\bibinfo {volume} {146}},\ \bibinfo {pages} {530}}\BibitemShut {NoStop}%
\bibitem [{\citenamefont {Lanotte}\ \emph {et~al.}(2013)\citenamefont
  {Lanotte}, \citenamefont {Biferale}, \citenamefont {Boffetta},\ and\
  \citenamefont {Toschi}}]{lanotte2013new}%
  \BibitemOpen
  \bibfield  {author} {\bibinfo {author} {\bibnamefont {Lanotte}, \bibfnamefont
  {Alessandra~S}}, \bibinfo {author} {\bibfnamefont {Luca}\ \bibnamefont
  {Biferale}}, \bibinfo {author} {\bibfnamefont {Guido}\ \bibnamefont
  {Boffetta}}, and\ \bibinfo {author} {\bibfnamefont {Federico}\ \bibnamefont
  {Toschi}}} (\bibinfo {year} {2013}),\ \bibfield  {title} {\enquote {\bibinfo
  {title} {A new assessment of the second-order moment of lagrangian velocity
  increments in turbulence},}\ }\href@noop {} {\bibfield  {journal} {\bibinfo
  {journal} {Journal of Turbulence}\ }\textbf {\bibinfo {volume}
  {14}}~(\bibinfo {number} {7}),\ \bibinfo {pages} {34--48}}\BibitemShut
  {NoStop}%
\bibitem [{\citenamefont {Liboff}(1998)}]{Liboff_1990}%
  \BibitemOpen
  \bibfield  {author} {\bibinfo {author} {\bibnamefont {Liboff}, \bibfnamefont
  {R~L}}} (\bibinfo {year} {1998}),\ \href@noop {} {\emph {\bibinfo {title}
  {Kinetic Theory. Classical, Quantum, and Relativistic descriptions}}},\
  \bibinfo {edition} {{S}econd}\ ed.\ (\bibinfo  {publisher} {Prentice-Hall
  International})\BibitemShut {NoStop}%
\bibitem [{\citenamefont {Liger-Belair}(2014)}]{liger2014many}%
  \BibitemOpen
  \bibfield  {author} {\bibinfo {author} {\bibnamefont {Liger-Belair},
  \bibfnamefont {G{\'e}rard}}} (\bibinfo {year} {2014}),\ \bibfield  {title}
  {\enquote {\bibinfo {title} {How many bubbles in your glass of bubbly?}}\
  }\href {https://doi.org/10.1021/jp500295e} {\bibfield  {journal} {\bibinfo
  {journal} {The Journal of Physical Chemistry B}\ }\textbf {\bibinfo {volume}
  {118}}~(\bibinfo {number} {11}),\ \bibinfo {pages} {3156--3163}}\BibitemShut
  {NoStop}%
\bibitem [{\citenamefont {Lindner}\ and\ \citenamefont
  {Shelley}(2015)}]{lindner2015elastic}%
  \BibitemOpen
  \bibfield  {author} {\bibinfo {author} {\bibnamefont {Lindner}, \bibfnamefont
  {Anke}}, and\ \bibinfo {author} {\bibfnamefont {Michael}\ \bibnamefont
  {Shelley}}} (\bibinfo {year} {2015}),\ \bibfield  {title} {\enquote {\bibinfo
  {title} {Elastic fibers in flows},}\ }in\ \href
  {http://pubs.rsc.org/en/content/chapter/bk9781849738132-00168/978-1-78262-849-1}
  {\emph {\bibinfo {booktitle} {Fluid-Structure Interactions in
  Low-Reynolds-Number Flows}}}\ (\bibinfo  {publisher} {Royal Society of
  Chemistry})\ pp.\ \bibinfo {pages} {168--192}\BibitemShut {NoStop}%
\bibitem [{\citenamefont {Marchioli}(2017)}]{marchioli2017large}%
  \BibitemOpen
  \bibfield  {author} {\bibinfo {author} {\bibnamefont {Marchioli},
  \bibfnamefont {Cristian}}} (\bibinfo {year} {2017}),\ \bibfield  {title}
  {\enquote {\bibinfo {title} {Large-eddy simulation of turbulent dispersed
  flows: a review of modelling approaches},}\ }\href@noop {} {\bibfield
  {journal} {\bibinfo  {journal} {Acta Mechanica}\ }\textbf {\bibinfo {volume}
  {228}}~(\bibinfo {number} {3}),\ \bibinfo {pages} {741--771}}\BibitemShut
  {NoStop}%
\bibitem [{\citenamefont {Martin}\ \emph {et~al.}(2008)\citenamefont {Martin},
  \citenamefont {Baross}, \citenamefont {Kelley},\ and\ \citenamefont
  {Russell}}]{martin2008hydrothermal}%
  \BibitemOpen
  \bibfield  {author} {\bibinfo {author} {\bibnamefont {Martin}, \bibfnamefont
  {William}}, \bibinfo {author} {\bibfnamefont {John}\ \bibnamefont {Baross}},
  \bibinfo {author} {\bibfnamefont {Deborah}\ \bibnamefont {Kelley}}, and\
  \bibinfo {author} {\bibfnamefont {Michael~J}\ \bibnamefont {Russell}}}
  (\bibinfo {year} {2008}),\ \bibfield  {title} {\enquote {\bibinfo {title}
  {Hydrothermal vents and the origin of life},}\ }\href
  {https://doi.org/10.1038/nrmicro1991} {\bibfield  {journal} {\bibinfo
  {journal} {Nature Reviews Microbiology}\ }\textbf {\bibinfo {volume}
  {6}}~(\bibinfo {number} {11}),\ \bibinfo {pages} {805--814}}\BibitemShut
  {NoStop}%
\bibitem [{\citenamefont {Maxey}\ and\ \citenamefont
  {Riley}(1983)}]{maxey1983equation}%
  \BibitemOpen
  \bibfield  {author} {\bibinfo {author} {\bibnamefont {Maxey}, \bibfnamefont
  {Martin~R}}, and\ \bibinfo {author} {\bibfnamefont {James~J}\ \bibnamefont
  {Riley}}} (\bibinfo {year} {1983}),\ \bibfield  {title} {\enquote {\bibinfo
  {title} {Equation of motion for a small rigid sphere in a nonuniform flow},}\
  }\href {https://doi.org/10.1063/1.864230} {\bibfield  {journal} {\bibinfo
  {journal} {The Physics of Fluids}\ }\textbf {\bibinfo {volume}
  {26}}~(\bibinfo {number} {4}),\ \bibinfo {pages} {883--889}}\BibitemShut
  {NoStop}%
\bibitem [{\citenamefont {McLaughlin}(1991)}]{mcLaughlin1991inertial}%
  \BibitemOpen
  \bibfield  {author} {\bibinfo {author} {\bibnamefont {McLaughlin},
  \bibfnamefont {J~P}}} (\bibinfo {year} {1991}),\ \bibfield  {title} {\enquote
  {\bibinfo {title} {Inertial migration of a small sphere in linear shear
  flows},}\ }\href@noop {} {\bibfield  {journal} {\bibinfo  {journal} {J. Fluid
  Mech.}\ }\textbf {\bibinfo {volume} {224}},\ \bibinfo {pages}
  {261--274}}\BibitemShut {NoStop}%
\bibitem [{\citenamefont {Minier}(2015)}]{minier2015lagrangian}%
  \BibitemOpen
  \bibfield  {author} {\bibinfo {author} {\bibnamefont {Minier}, \bibfnamefont
  {Jean-Pierre}}} (\bibinfo {year} {2015}),\ \bibfield  {title} {\enquote
  {\bibinfo {title} {On {L}agrangian stochastic methods for turbulent
  polydisperse two-phase reactive flows},}\ }\href
  {https://doi.org/10.1016/j.pecs.2015.02.003} {\bibfield  {journal} {\bibinfo
  {journal} {Progress in Energy and Combustion Science}\ }\textbf {\bibinfo
  {volume} {50}},\ \bibinfo {pages} {1--62}}\BibitemShut {NoStop}%
\bibitem [{\citenamefont {Minier}(2016)}]{minier2016statistical}%
  \BibitemOpen
  \bibfield  {author} {\bibinfo {author} {\bibnamefont {Minier}, \bibfnamefont
  {Jean-Pierre}}} (\bibinfo {year} {2016}),\ \bibfield  {title} {\enquote
  {\bibinfo {title} {Statistical descriptions of polydisperse turbulent
  two-phase flows},}\ }\href {https://doi.org/10.1016/j.physrep.2016.10.007}
  {\bibfield  {journal} {\bibinfo  {journal} {Physics reports}\ }\textbf
  {\bibinfo {volume} {665}},\ \bibinfo {pages} {1--122}}\BibitemShut {NoStop}%
\bibitem [{\citenamefont {Minier}(2021)}]{minier2021methodology}%
  \BibitemOpen
  \bibfield  {author} {\bibinfo {author} {\bibnamefont {Minier}, \bibfnamefont
  {Jean-Pierre}}} (\bibinfo {year} {2021}),\ \bibfield  {title} {\enquote
  {\bibinfo {title} {A methodology to devise consistent probability density
  function models for particle dynamics in turbulent dispersed two-phase
  flows},}\ }\href {https://doi.org/10.1063/5.0039249} {\bibfield  {journal}
  {\bibinfo  {journal} {Physics of Fluids}\ }\textbf {\bibinfo {volume}
  {33}}~(\bibinfo {number} {2}),\ \bibinfo {pages} {023312}}\BibitemShut
  {NoStop}%
\bibitem [{\citenamefont {Minier}\ \emph {et~al.}(2014)\citenamefont {Minier},
  \citenamefont {Chibbaro},\ and\ \citenamefont {Pope}}]{minier2014guidelines}%
  \BibitemOpen
  \bibfield  {author} {\bibinfo {author} {\bibnamefont {Minier}, \bibfnamefont
  {Jean-Pierre}}, \bibinfo {author} {\bibfnamefont {Sergio}\ \bibnamefont
  {Chibbaro}}, and\ \bibinfo {author} {\bibfnamefont {Stephen~B}\ \bibnamefont
  {Pope}}} (\bibinfo {year} {2014}),\ \bibfield  {title} {\enquote {\bibinfo
  {title} {Guidelines for the formulation of {L}agrangian stochastic models for
  particle simulations of single-phase and dispersed two-phase turbulent
  flows},}\ }\href {https://doi.org/10.1063/1.4901315} {\bibfield  {journal}
  {\bibinfo  {journal} {Physics of Fluids}\ }\textbf {\bibinfo {volume}
  {26}}~(\bibinfo {number} {11}),\ \bibinfo {pages} {113303}}\BibitemShut
  {NoStop}%
\bibitem [{\citenamefont {Minier}\ and\ \citenamefont
  {Peirano}(2001)}]{minier2001pdf}%
  \BibitemOpen
  \bibfield  {author} {\bibinfo {author} {\bibnamefont {Minier}, \bibfnamefont
  {Jean-Pierre}}, and\ \bibinfo {author} {\bibfnamefont {Eric}\ \bibnamefont
  {Peirano}}} (\bibinfo {year} {2001}),\ \bibfield  {title} {\enquote {\bibinfo
  {title} {The pdf approach to turbulent polydispersed two-phase flows},}\
  }\href {https://doi.org/10.1016/S0370-1573(01)00011-4} {\bibfield  {journal}
  {\bibinfo  {journal} {Physics reports}\ }\textbf {\bibinfo {volume}
  {352}}~(\bibinfo {number} {1-3}),\ \bibinfo {pages} {1--214}}\BibitemShut
  {NoStop}%
\bibitem [{\citenamefont {Minier}\ \emph {et~al.}(2004)\citenamefont {Minier},
  \citenamefont {Peirano},\ and\ \citenamefont {Chibbaro}}]{minier2004pdf}%
  \BibitemOpen
  \bibfield  {author} {\bibinfo {author} {\bibnamefont {Minier}, \bibfnamefont
  {Jean-Pierre}}, \bibinfo {author} {\bibfnamefont {Eric}\ \bibnamefont
  {Peirano}}, and\ \bibinfo {author} {\bibfnamefont {Sergio}\ \bibnamefont
  {Chibbaro}}} (\bibinfo {year} {2004}),\ \bibfield  {title} {\enquote
  {\bibinfo {title} {Pdf model based on langevin equation for polydispersed
  two-phase flows applied to a bluff-body gas-solid flow},}\ }\href@noop {}
  {\bibfield  {journal} {\bibinfo  {journal} {Physics of fluids}\ }\textbf
  {\bibinfo {volume} {16}}~(\bibinfo {number} {7}),\ \bibinfo {pages}
  {2419--2431}}\BibitemShut {NoStop}%
\bibitem [{\citenamefont {Minier}\ and\ \citenamefont
  {Profeta}(2015)}]{minier2015kinetic}%
  \BibitemOpen
  \bibfield  {author} {\bibinfo {author} {\bibnamefont {Minier}, \bibfnamefont
  {Jean-Pierre}}, and\ \bibinfo {author} {\bibfnamefont {Christophe}\
  \bibnamefont {Profeta}}} (\bibinfo {year} {2015}),\ \bibfield  {title}
  {\enquote {\bibinfo {title} {Kinetic and dynamic probability-density-function
  descriptions of disperse turbulent two-phase flows},}\ }\href
  {https://doi.org/10.1103/PhysRevE.92.053020} {\bibfield  {journal} {\bibinfo
  {journal} {Physical Review E}\ }\textbf {\bibinfo {volume} {92}}~(\bibinfo
  {number} {5}),\ \bibinfo {pages} {053020}}\BibitemShut {NoStop}%
\bibitem [{\citenamefont {Mohaupt}\ \emph {et~al.}(2011)\citenamefont
  {Mohaupt}, \citenamefont {Minier},\ and\ \citenamefont
  {Tani{\`e}re}}]{mohaupt2011new}%
  \BibitemOpen
  \bibfield  {author} {\bibinfo {author} {\bibnamefont {Mohaupt}, \bibfnamefont
  {M}}, \bibinfo {author} {\bibfnamefont {J-P}\ \bibnamefont {Minier}}, and\
  \bibinfo {author} {\bibfnamefont {A}~\bibnamefont {Tani{\`e}re}}} (\bibinfo
  {year} {2011}),\ \bibfield  {title} {\enquote {\bibinfo {title} {A new
  approach for the detection of particle interactions for large-inertia and
  colloidal particles in a turbulent flow},}\ }\href@noop {} {\bibfield
  {journal} {\bibinfo  {journal} {International journal of multiphase flow}\
  }\textbf {\bibinfo {volume} {37}}~(\bibinfo {number} {7}),\ \bibinfo {pages}
  {746--755}}\BibitemShut {NoStop}%
\bibitem [{\citenamefont {Monchaux}\ \emph {et~al.}(2012)\citenamefont
  {Monchaux}, \citenamefont {Bourgoin},\ and\ \citenamefont
  {Cartellier}}]{monchaux2012analyzing}%
  \BibitemOpen
  \bibfield  {author} {\bibinfo {author} {\bibnamefont {Monchaux},
  \bibfnamefont {Romain}}, \bibinfo {author} {\bibfnamefont {Mickael}\
  \bibnamefont {Bourgoin}}, and\ \bibinfo {author} {\bibfnamefont {Alain}\
  \bibnamefont {Cartellier}}} (\bibinfo {year} {2012}),\ \bibfield  {title}
  {\enquote {\bibinfo {title} {Analyzing preferential concentration and
  clustering of inertial particles in turbulence},}\ }\href@noop {} {\bibfield
  {journal} {\bibinfo  {journal} {International Journal of Multiphase Flow}\
  }\textbf {\bibinfo {volume} {40}},\ \bibinfo {pages} {1--18}}\BibitemShut
  {NoStop}%
\bibitem [{\citenamefont {Monin}\ and\ \citenamefont
  {Yaglom}(2013)}]{monin2013statistical}%
  \BibitemOpen
  \bibfield  {author} {\bibinfo {author} {\bibnamefont {Monin}, \bibfnamefont
  {AS}}, and\ \bibinfo {author} {\bibfnamefont {A.M.}\ \bibnamefont {Yaglom}}}
  (\bibinfo {year} {2013}),\ \href
  {https://books.google.fr/books?id=6xPEAgAAQBAJ} {\emph {\bibinfo {title}
  {Statistical Fluid Mechanics, Volume II: Mechanics of Turbulence}}},\
  \bibinfo {series} {Dover Books on Physics}, Vol.~\bibinfo {volume} {2}\
  (\bibinfo  {publisher} {Dover Publications})\BibitemShut {NoStop}%
\bibitem [{\citenamefont {Novikov}(1965)}]{novikov1965functionals}%
  \BibitemOpen
  \bibfield  {author} {\bibinfo {author} {\bibnamefont {Novikov}, \bibfnamefont
  {Evgenii~A}}} (\bibinfo {year} {1965}),\ \bibfield  {title} {\enquote
  {\bibinfo {title} {Functionals and the random-force method in turbulence
  theory},}\ }\href@noop {} {\bibfield  {journal} {\bibinfo  {journal} {Sov.
  Phys. JETP}\ }\textbf {\bibinfo {volume} {20}}~(\bibinfo {number} {5}),\
  \bibinfo {pages} {1290--1294}}\BibitemShut {NoStop}%
\bibitem [{\citenamefont {{\O}ksendal}(2003)}]{oksendal2003stochastic}%
  \BibitemOpen
  \bibfield  {author} {\bibinfo {author} {\bibnamefont {{\O}ksendal},
  \bibfnamefont {Bernt}}} (\bibinfo {year} {2003}),\ \href@noop {} {\emph
  {\bibinfo {title} {Stochastic differential equations}}}\ (\bibinfo
  {publisher} {Springer})\BibitemShut {NoStop}%
\bibitem [{\citenamefont {\"{O}ttinger}(1996)}]{Ottinger_1996}%
  \BibitemOpen
  \bibfield  {author} {\bibinfo {author} {\bibnamefont {\"{O}ttinger},
  \bibfnamefont {H~C}}} (\bibinfo {year} {1996}),\ \href@noop {} {\emph
  {\bibinfo {title} {Stochastic {P}rocesses in {P}olymeric {F}luids. Tools and
  Examples for Developing Simulation Algorithms}}}\ (\bibinfo  {publisher}
  {Springer, Berlin})\BibitemShut {NoStop}%
\bibitem [{\citenamefont {{\"O}ttinger}(2012)}]{ottinger2012stochastic}%
  \BibitemOpen
  \bibfield  {author} {\bibinfo {author} {\bibnamefont {{\"O}ttinger},
  \bibfnamefont {Hans~C}}} (\bibinfo {year} {2012}),\ \href@noop {} {\emph
  {\bibinfo {title} {Stochastic processes in polymeric fluids: tools and
  examples for developing simulation algorithms}}}\ (\bibinfo  {publisher}
  {Springer Science \& Business Media})\BibitemShut {NoStop}%
\bibitem [{\citenamefont {{\"O}ttinger}(2005)}]{ottinger2005beyond}%
  \BibitemOpen
  \bibfield  {author} {\bibinfo {author} {\bibnamefont {{\"O}ttinger},
  \bibfnamefont {Hans~Christian}}} (\bibinfo {year} {2005}),\ \href@noop {}
  {\emph {\bibinfo {title} {Beyond equilibrium thermodynamics}}}\ (\bibinfo
  {publisher} {John Wiley \& Sons})\BibitemShut {NoStop}%
\bibitem [{\citenamefont {Peirano}\ and\ \citenamefont
  {Minier}(2002)}]{peirano2002probabilistic}%
  \BibitemOpen
  \bibfield  {author} {\bibinfo {author} {\bibnamefont {Peirano}, \bibfnamefont
  {Eric}}, and\ \bibinfo {author} {\bibfnamefont {Jean-Pierre}\ \bibnamefont
  {Minier}}} (\bibinfo {year} {2002}),\ \bibfield  {title} {\enquote {\bibinfo
  {title} {Probabilistic formalism and hierarchy of models for polydispersed
  turbulent two-phase flows},}\ }\href@noop {} {\bibfield  {journal} {\bibinfo
  {journal} {Physical Review E}\ }\textbf {\bibinfo {volume} {65}}~(\bibinfo
  {number} {4}),\ \bibinfo {pages} {046301}}\BibitemShut {NoStop}%
\bibitem [{\citenamefont {Pope}(1991)}]{pope1991application}%
  \BibitemOpen
  \bibfield  {author} {\bibinfo {author} {\bibnamefont {Pope}, \bibfnamefont
  {SB}}} (\bibinfo {year} {1991}),\ \bibfield  {title} {\enquote {\bibinfo
  {title} {Application of the velocity-dissipation probability density function
  model to inhomogeneous turbulent flows},}\ }\href@noop {} {\bibfield
  {journal} {\bibinfo  {journal} {Physics of Fluids A: Fluid Dynamics}\
  }\textbf {\bibinfo {volume} {3}}~(\bibinfo {number} {8}),\ \bibinfo {pages}
  {1947--1957}}\BibitemShut {NoStop}%
\bibitem [{\citenamefont {Pope}(2000)}]{pope2000turbulent}%
  \BibitemOpen
  \bibfield  {author} {\bibinfo {author} {\bibnamefont {Pope}, \bibfnamefont
  {SB}}} (\bibinfo {year} {2000}),\ \href
  {https://books.google.fr/books?id=HZsTw9SMx-0C} {\emph {\bibinfo {title}
  {Turbulent {F}lows}}}\ (\bibinfo  {publisher} {Cambridge University
  Press})\BibitemShut {NoStop}%
\bibitem [{\citenamefont {Pope}(1985)}]{pope1985pdf}%
  \BibitemOpen
  \bibfield  {author} {\bibinfo {author} {\bibnamefont {Pope}, \bibfnamefont
  {Stephen~B}}} (\bibinfo {year} {1985}),\ \bibfield  {title} {\enquote
  {\bibinfo {title} {Pdf methods for turbulent reactive flows},}\ }\href@noop
  {} {\bibfield  {journal} {\bibinfo  {journal} {Progress in energy and
  combustion science}\ }\textbf {\bibinfo {volume} {11}}~(\bibinfo {number}
  {2}),\ \bibinfo {pages} {119--192}}\BibitemShut {NoStop}%
\bibitem [{\citenamefont {Pope}(1994{\natexlab{a}})}]{pope1994lagrangian}%
  \BibitemOpen
  \bibfield  {author} {\bibinfo {author} {\bibnamefont {Pope}, \bibfnamefont
  {Stephen~B}}} (\bibinfo {year} {1994}{\natexlab{a}}),\ \bibfield  {title}
  {\enquote {\bibinfo {title} {Lagrangian {PDF} methods for turbulent flows},}\
  }\href@noop {} {\bibfield  {journal} {\bibinfo  {journal} {Annual review of
  fluid mechanics}\ }\textbf {\bibinfo {volume} {26}}~(\bibinfo {number} {1}),\
  \bibinfo {pages} {23--63}}\BibitemShut {NoStop}%
\bibitem [{\citenamefont {Pope}(1994{\natexlab{b}})}]{pope1994relationship}%
  \BibitemOpen
  \bibfield  {author} {\bibinfo {author} {\bibnamefont {Pope}, \bibfnamefont
  {Stephen~B}}} (\bibinfo {year} {1994}{\natexlab{b}}),\ \bibfield  {title}
  {\enquote {\bibinfo {title} {On the relationship between stochastic
  lagrangian models of turbulence and second-moment closures},}\ }\href@noop {}
  {\bibfield  {journal} {\bibinfo  {journal} {Physics of Fluids}\ }\textbf
  {\bibinfo {volume} {6}}~(\bibinfo {number} {2}),\ \bibinfo {pages}
  {973--985}}\BibitemShut {NoStop}%
\bibitem [{\citenamefont {Pozorski}\ and\ \citenamefont
  {Apte}(2009)}]{Pozorski_2009}%
  \BibitemOpen
  \bibfield  {author} {\bibinfo {author} {\bibnamefont {Pozorski},
  \bibfnamefont {J}}, and\ \bibinfo {author} {\bibfnamefont {S.V.}\
  \bibnamefont {Apte}}} (\bibinfo {year} {2009}),\ \bibfield  {title} {\enquote
  {\bibinfo {title} {Filtered particle tracking in isotropic turbulence and
  stochastic modeling of subgrid-scale dispersion},}\ }\href@noop {} {\bibfield
   {journal} {\bibinfo  {journal} {Int. J. Multiphase Flow}\ }\textbf {\bibinfo
  {volume} {35}},\ \bibinfo {pages} {118--128}}\BibitemShut {NoStop}%
\bibitem [{\citenamefont {Pozorski}\ and\ \citenamefont
  {Minier}(1999)}]{pozorski1999probability}%
  \BibitemOpen
  \bibfield  {author} {\bibinfo {author} {\bibnamefont {Pozorski},
  \bibfnamefont {Jacek}}, and\ \bibinfo {author} {\bibfnamefont {Jean-Pierre}\
  \bibnamefont {Minier}}} (\bibinfo {year} {1999}),\ \bibfield  {title}
  {\enquote {\bibinfo {title} {Probability density function modeling of
  dispersed two-phase turbulent flows},}\ }\href@noop {} {\bibfield  {journal}
  {\bibinfo  {journal} {Physical Review E}\ }\textbf {\bibinfo {volume}
  {59}}~(\bibinfo {number} {1}),\ \bibinfo {pages} {855}}\BibitemShut {NoStop}%
\bibitem [{\citenamefont {Ramkrishna}(2000)}]{ramkrishna2000population}%
  \BibitemOpen
  \bibfield  {author} {\bibinfo {author} {\bibnamefont {Ramkrishna},
  \bibfnamefont {Doraiswami}}} (\bibinfo {year} {2000}),\ \href@noop {} {\emph
  {\bibinfo {title} {Population balances: Theory and applications to
  particulate systems in engineering}}}\ (\bibinfo  {publisher} {Academic
  Press})\BibitemShut {NoStop}%
\bibitem [{\citenamefont {Reif}(1985)}]{Reif_1985}%
  \BibitemOpen
  \bibfield  {author} {\bibinfo {author} {\bibnamefont {Reif}, \bibfnamefont
  {F}}} (\bibinfo {year} {1985}),\ \href@noop {} {\emph {\bibinfo {title}
  {Fundamentals of Statistical and Thermal Physics}}}\ (\bibinfo  {publisher}
  {McGraw Hill International Editions, Singapore})\BibitemShut {NoStop}%
\bibitem [{\citenamefont {Risken}(1996)}]{risken1996fokker}%
  \BibitemOpen
  \bibfield  {author} {\bibinfo {author} {\bibnamefont {Risken}, \bibfnamefont
  {Hannes}}} (\bibinfo {year} {1996}),\ \href@noop {} {\emph {\bibinfo {title}
  {Fokker-planck equation}}}\ (\bibinfo  {publisher} {Springer})\BibitemShut
  {NoStop}%
\bibitem [{\citenamefont {Rouse~Jr}(1953)}]{rouse1953theory}%
  \BibitemOpen
  \bibfield  {author} {\bibinfo {author} {\bibnamefont {Rouse~Jr},
  \bibfnamefont {Prince~E}}} (\bibinfo {year} {1953}),\ \bibfield  {title}
  {\enquote {\bibinfo {title} {A theory of the linear viscoelastic properties
  of dilute solutions of coiling polymers},}\ }\href
  {https://doi.org/10.1063/1.1699180} {\bibfield  {journal} {\bibinfo
  {journal} {The Journal of Chemical Physics}\ }\textbf {\bibinfo {volume}
  {21}}~(\bibinfo {number} {7}),\ \bibinfo {pages} {1272--1280}}\BibitemShut
  {NoStop}%
\bibitem [{\citenamefont {Saffman}(1995)}]{saffman1995vortex}%
  \BibitemOpen
  \bibfield  {author} {\bibinfo {author} {\bibnamefont {Saffman}, \bibfnamefont
  {Philip~G}}} (\bibinfo {year} {1995}),\ \href@noop {} {\emph {\bibinfo
  {title} {Vortex dynamics}}}\ (\bibinfo  {publisher} {Cambridge university
  press})\BibitemShut {NoStop}%
\bibitem [{\citenamefont {Sa{\"\i}d}\ \emph {et~al.}(2005)\citenamefont
  {Sa{\"\i}d}, \citenamefont {Mhiri}, \citenamefont {Le~Palec},\ and\
  \citenamefont {Bournot}}]{said2005experimental}%
  \BibitemOpen
  \bibfield  {author} {\bibinfo {author} {\bibnamefont {Sa{\"\i}d},
  \bibfnamefont {Nejla~Mahjoub}}, \bibinfo {author} {\bibfnamefont {Hatem}\
  \bibnamefont {Mhiri}}, \bibinfo {author} {\bibfnamefont {Georges}\
  \bibnamefont {Le~Palec}}, and\ \bibinfo {author} {\bibfnamefont {Philippe}\
  \bibnamefont {Bournot}}} (\bibinfo {year} {2005}),\ \bibfield  {title}
  {\enquote {\bibinfo {title} {Experimental and numerical analysis of pollutant
  dispersion from a chimney},}\ }\href
  {https://doi.org/10.1016/j.atmosenv.2004.11.040} {\bibfield  {journal}
  {\bibinfo  {journal} {Atmospheric Environment}\ }\textbf {\bibinfo {volume}
  {39}}~(\bibinfo {number} {9}),\ \bibinfo {pages} {1727--1738}}\BibitemShut
  {NoStop}%
\bibitem [{\citenamefont {Sheikhi}\ \emph {et~al.}(2003)\citenamefont
  {Sheikhi}, \citenamefont {Drozda}, \citenamefont {Givi},\ and\ \citenamefont
  {Pope}}]{Sheikhi_2003}%
  \BibitemOpen
  \bibfield  {author} {\bibinfo {author} {\bibnamefont {Sheikhi}, \bibfnamefont
  {M~R~H}}, \bibinfo {author} {\bibfnamefont {T.~G.}\ \bibnamefont {Drozda}},
  \bibinfo {author} {\bibfnamefont {P.}~\bibnamefont {Givi}}, and\ \bibinfo
  {author} {\bibfnamefont {S.~B.}\ \bibnamefont {Pope}}} (\bibinfo {year}
  {2003}),\ \bibfield  {title} {\enquote {\bibinfo {title} {Velocity-scalar
  filtered density function for large eddy simulation of turbulent flows},}\
  }\href@noop {} {\bibfield  {journal} {\bibinfo  {journal} {Phys. Fluids}\
  }\textbf {\bibinfo {volume} {15}}~(\bibinfo {number} {8}),\ \bibinfo {pages}
  {2321--2337}}\BibitemShut {NoStop}%
\bibitem [{\citenamefont {Sheikhi}\ \emph {et~al.}(2007)\citenamefont
  {Sheikhi}, \citenamefont {Givi},\ and\ \citenamefont {Pope}}]{Sheikhi_2007}%
  \BibitemOpen
  \bibfield  {author} {\bibinfo {author} {\bibnamefont {Sheikhi}, \bibfnamefont
  {M~R~H}}, \bibinfo {author} {\bibfnamefont {P.}~\bibnamefont {Givi}}, and\
  \bibinfo {author} {\bibfnamefont {S.~B.}\ \bibnamefont {Pope}}} (\bibinfo
  {year} {2007}),\ \bibfield  {title} {\enquote {\bibinfo {title}
  {Velocity-scalar filtered mass density function for large eddy simulation of
  turbulent reactive flows},}\ }\href@noop {} {\bibfield  {journal} {\bibinfo
  {journal} {Phys. Fluids}\ }\textbf {\bibinfo {volume} {19}},\ \bibinfo
  {pages} {095106}}\BibitemShut {NoStop}%
\bibitem [{\citenamefont {Sheikhi}\ \emph {et~al.}(2009)\citenamefont
  {Sheikhi}, \citenamefont {Givi},\ and\ \citenamefont {Pope}}]{Sheikhi_2009}%
  \BibitemOpen
  \bibfield  {author} {\bibinfo {author} {\bibnamefont {Sheikhi}, \bibfnamefont
  {M~R~H}}, \bibinfo {author} {\bibfnamefont {P.}~\bibnamefont {Givi}}, and\
  \bibinfo {author} {\bibfnamefont {S.~B.}\ \bibnamefont {Pope}}} (\bibinfo
  {year} {2009}),\ \bibfield  {title} {\enquote {\bibinfo {title}
  {Frequency-velocity-scalar filtered mass density function for large eddy
  simulation of turbulent flows},}\ }\href@noop {} {\bibfield  {journal}
  {\bibinfo  {journal} {Phys. Fluids}\ }\textbf {\bibinfo {volume} {21}},\
  \bibinfo {pages} {075102}}\BibitemShut {NoStop}%
\bibitem [{\citenamefont {Somasi}\ \emph {et~al.}(2002)\citenamefont {Somasi},
  \citenamefont {Khomami}, \citenamefont {Woo}, \citenamefont {Hur},\ and\
  \citenamefont {Shaqfeh}}]{somasi2002brownian}%
  \BibitemOpen
  \bibfield  {author} {\bibinfo {author} {\bibnamefont {Somasi}, \bibfnamefont
  {Madan}}, \bibinfo {author} {\bibfnamefont {Bamin}\ \bibnamefont {Khomami}},
  \bibinfo {author} {\bibfnamefont {Nathanael~J}\ \bibnamefont {Woo}}, \bibinfo
  {author} {\bibfnamefont {Joe~S}\ \bibnamefont {Hur}}, and\ \bibinfo {author}
  {\bibfnamefont {Eric~SG}\ \bibnamefont {Shaqfeh}}} (\bibinfo {year} {2002}),\
  \bibfield  {title} {\enquote {\bibinfo {title} {Brownian dynamics simulations
  of bead-rod and bead-spring chains: numerical algorithms and coarse-graining
  issues},}\ }\href {https://doi.org/10.1016/S0377-0257(02)00132-5} {\bibfield
  {journal} {\bibinfo  {journal} {Journal of Non-Newtonian Fluid Mechanics}\
  }\textbf {\bibinfo {volume} {108}}~(\bibinfo {number} {1-3}),\ \bibinfo
  {pages} {227--255}}\BibitemShut {NoStop}%
\bibitem [{\citenamefont {Stephen}\ and\ \citenamefont
  {Straley}(1974)}]{stephen1974physics}%
  \BibitemOpen
  \bibfield  {author} {\bibinfo {author} {\bibnamefont {Stephen}, \bibfnamefont
  {Michael~J}}, and\ \bibinfo {author} {\bibfnamefont {Joseph~P}\ \bibnamefont
  {Straley}}} (\bibinfo {year} {1974}),\ \bibfield  {title} {\enquote {\bibinfo
  {title} {Physics of liquid crystals},}\ }\href
  {https://doi.org/10.1103/RevModPhys.46.617} {\bibfield  {journal} {\bibinfo
  {journal} {Reviews of Modern Physics}\ }\textbf {\bibinfo {volume}
  {46}}~(\bibinfo {number} {4}),\ \bibinfo {pages} {617}}\BibitemShut {NoStop}%
\bibitem [{\citenamefont {Van~Kampen}(2007)}]{van2007stochastic}%
  \BibitemOpen
  \bibfield  {author} {\bibinfo {author} {\bibnamefont {Van~Kampen},
  \bibfnamefont {N~G}}} (\bibinfo {year} {2007}),\ \href@noop {} {\emph
  {\bibinfo {title} {Stochastic Proceses in Physics and Chemistry}}},\ \bibinfo
  {edition} {{T}hird}\ ed.\ (\bibinfo  {publisher} {North Holland})\BibitemShut
  {NoStop}%
\bibitem [{\citenamefont {Van~Kampen}(1989)}]{van1989langevin}%
  \BibitemOpen
  \bibfield  {author} {\bibinfo {author} {\bibnamefont {Van~Kampen},
  \bibfnamefont {Nico~G}}} (\bibinfo {year} {1989}),\ \bibfield  {title}
  {\enquote {\bibinfo {title} {Langevin-like equation with colored noise},}\
  }\href@noop {} {\bibfield  {journal} {\bibinfo  {journal} {Journal of
  statistical physics}\ }\textbf {\bibinfo {volume} {54}},\ \bibinfo {pages}
  {1289--1308}}\BibitemShut {NoStop}%
\bibitem [{\citenamefont {Venerus}\ and\ \citenamefont
  {{\"O}ttinger}(2018)}]{venerus2018modern}%
  \BibitemOpen
  \bibfield  {author} {\bibinfo {author} {\bibnamefont {Venerus}, \bibfnamefont
  {David~C}}, and\ \bibinfo {author} {\bibfnamefont {Hans~Christian}\
  \bibnamefont {{\"O}ttinger}}} (\bibinfo {year} {2018}),\ \href@noop {} {\emph
  {\bibinfo {title} {A modern course in transport phenomena}}}\ (\bibinfo
  {publisher} {Cambridge University Press})\BibitemShut {NoStop}%
\bibitem [{\citenamefont {Vincent}\ and\ \citenamefont
  {Meneguzzi}(1991)}]{vincent1991spatial}%
  \BibitemOpen
  \bibfield  {author} {\bibinfo {author} {\bibnamefont {Vincent}, \bibfnamefont
  {Albert}}, and\ \bibinfo {author} {\bibfnamefont {Maria}\ \bibnamefont
  {Meneguzzi}}} (\bibinfo {year} {1991}),\ \bibfield  {title} {\enquote
  {\bibinfo {title} {The spatial structure and statistical properties of
  homogeneous turbulence},}\ }\href@noop {} {\bibfield  {journal} {\bibinfo
  {journal} {Journal of Fluid Mechanics}\ }\textbf {\bibinfo {volume} {225}},\
  \bibinfo {pages} {1--20}}\BibitemShut {NoStop}%
\bibitem [{\citenamefont {Wac{\l}awczyk}\ \emph {et~al.}(2004)\citenamefont
  {Wac{\l}awczyk}, \citenamefont {Pozorski},\ and\ \citenamefont
  {Minier}}]{waclawczyk2004probability}%
  \BibitemOpen
  \bibfield  {author} {\bibinfo {author} {\bibnamefont {Wac{\l}awczyk},
  \bibfnamefont {Marta}}, \bibinfo {author} {\bibfnamefont {Jacek}\
  \bibnamefont {Pozorski}}, and\ \bibinfo {author} {\bibfnamefont
  {Jean-Pierre}\ \bibnamefont {Minier}}} (\bibinfo {year} {2004}),\ \bibfield
  {title} {\enquote {\bibinfo {title} {Probability density function computation
  of turbulent flows with a new near-wall model},}\ }\href@noop {} {\bibfield
  {journal} {\bibinfo  {journal} {Physics of Fluids}\ }\textbf {\bibinfo
  {volume} {16}}~(\bibinfo {number} {5}),\ \bibinfo {pages}
  {1410--1422}}\BibitemShut {NoStop}%
\bibitem [{\citenamefont {Wagner}(2011)}]{Wagner_2011}%
  \BibitemOpen
  \bibfield  {author} {\bibinfo {author} {\bibnamefont {Wagner}, \bibfnamefont
  {W}}} (\bibinfo {year} {2011}),\ \bibfield  {title} {\enquote {\bibinfo
  {title} {Stochastic models in kinetic theory},}\ }\href@noop {} {\bibfield
  {journal} {\bibinfo  {journal} {Phys. Fluids}\ }\textbf {\bibinfo {volume}
  {23}},\ \bibinfo {pages} {030602}}\BibitemShut {NoStop}%
\bibitem [{\citenamefont {Wagner}(2004)}]{wagner2004stochastic}%
  \BibitemOpen
  \bibfield  {author} {\bibinfo {author} {\bibnamefont {Wagner}, \bibfnamefont
  {Wolfgang}}} (\bibinfo {year} {2004}),\ \bibfield  {title} {\enquote
  {\bibinfo {title} {Stochastic models and monte carlo algorithms for boltzmann
  type equations},}\ }in\ \href@noop {} {\emph {\bibinfo {booktitle} {Monte
  Carlo and Quasi-Monte Carlo Methods 2002: Proceedings of a Conference held at
  the National University of Singapore, Republic of Singapore, November 25--28,
  2002}}}\ (\bibinfo {organization} {Springer})\ pp.\ \bibinfo {pages}
  {129--153}\BibitemShut {NoStop}%
\bibitem [{\citenamefont {Wang}\ \emph {et~al.}(1997)\citenamefont {Wang},
  \citenamefont {Squires}, \citenamefont {Chen},\ and\ \citenamefont
  {McLaughlin}}]{wang1997role}%
  \BibitemOpen
  \bibfield  {author} {\bibinfo {author} {\bibnamefont {Wang}, \bibfnamefont
  {Q}}, \bibinfo {author} {\bibfnamefont {KD}~\bibnamefont {Squires}}, \bibinfo
  {author} {\bibfnamefont {M}~\bibnamefont {Chen}}, and\ \bibinfo {author}
  {\bibfnamefont {JB}~\bibnamefont {McLaughlin}}} (\bibinfo {year} {1997}),\
  \bibfield  {title} {\enquote {\bibinfo {title} {On the role of the lift force
  in turbulence simulations of particle deposition},}\ }\href@noop {}
  {\bibfield  {journal} {\bibinfo  {journal} {International Journal of
  Multiphase Flow}\ }\textbf {\bibinfo {volume} {23}}~(\bibinfo {number} {4}),\
  \bibinfo {pages} {749--763}}\BibitemShut {NoStop}%
\bibitem [{\citenamefont {Wiegel}(1986)}]{wiegel1986introduction}%
  \BibitemOpen
  \bibfield  {author} {\bibinfo {author} {\bibnamefont {Wiegel}, \bibfnamefont
  {Frederik~W}}} (\bibinfo {year} {1986}),\ \href@noop {} {\emph {\bibinfo
  {title} {Introduction to path-integral methods in physics and polymer
  science}}}\ (\bibinfo  {publisher} {World Scientific})\BibitemShut {NoStop}%
\bibitem [{\citenamefont {Yeung}(2002)}]{yeung2002lagrangian}%
  \BibitemOpen
  \bibfield  {author} {\bibinfo {author} {\bibnamefont {Yeung}, \bibfnamefont
  {PK}}} (\bibinfo {year} {2002}),\ \bibfield  {title} {\enquote {\bibinfo
  {title} {Lagrangian investigations of turbulence},}\ }\href@noop {}
  {\bibfield  {journal} {\bibinfo  {journal} {Annual review of fluid
  mechanics}\ }\textbf {\bibinfo {volume} {34}}~(\bibinfo {number} {1}),\
  \bibinfo {pages} {115--142}}\BibitemShut {NoStop}%
\bibitem [{\citenamefont {Yeung}\ and\ \citenamefont
  {Pope}(1989)}]{yeung1989lagrangian}%
  \BibitemOpen
  \bibfield  {author} {\bibinfo {author} {\bibnamefont {Yeung}, \bibfnamefont
  {Pui-Kuen}}, and\ \bibinfo {author} {\bibfnamefont {Stephen~B}\ \bibnamefont
  {Pope}}} (\bibinfo {year} {1989}),\ \bibfield  {title} {\enquote {\bibinfo
  {title} {Lagrangian statistics from direct numerical simulations of isotropic
  turbulence},}\ }\href@noop {} {\bibfield  {journal} {\bibinfo  {journal}
  {Journal of Fluid Mechanics}\ }\textbf {\bibinfo {volume} {207}},\ \bibinfo
  {pages} {531--586}}\BibitemShut {NoStop}%
\end{thebibliography}%

\end{document}